\documentclass[12pt,dvips]{report}

\usepackage{epsfig}
\usepackage{amsmath,amssymb}
\usepackage{float}
\usepackage{verbatim}
\usepackage{graphicx}
\usepackage{bm}

\usepackage[overload]{textcase}
\usepackage{graphics}
\usepackage{multirow}

\usepackage{color}

\newcommand{\rr}{{\mathbb{R}}}
\newcommand{\R}{{\mathbb R}}
\newcommand{\calA}{{\cal A}}

\newcommand{\ot}{\otimes}
\newcommand{\op}{\oplus}

\newcommand {\sr}{\stackrel}

\newcommand {\eqv}{\equiv}

\newcommand {\ora}{\overrightarrow}
\newcommand {\ola}{\overleftarrow}

\newcommand {\lang}{\langle}
\newcommand {\rang}{\rangle}
\newcommand {\td}{\tilde}
\newcommand {\dg}{\dagger}
\newcommand {\ra}{\rightarrow}
\newcommand {\Ra}{\Rightarrow}

\newcommand {\La}{\Leftarrow}
\newcommand {\lra}{\leftrightarrow}
\newcommand {\Lra}{\Leftrightarrow}
\newcommand {\ral}{\longrightarrow}

\newcommand{\tht}{\theta}

\newcommand {\vep}{\varepsilon}
\newcommand {\vphi}{\varphi}
\newcommand {\del}{\partial}
\newcommand {\Ld}{\Lambda}
\newcommand {\ld}{\lambda}

\newcommand {\om}{\omega}
\newcommand {\al}{\alpha}
\newcommand {\bi}{\beta}

\newcommand{\wt}{\widetilde}

\newcommand {\vs} {\vspace{1cm}}
\newcommand {\bfr}{\begin{flushright}}
\newcommand {\efr}{\end{flushright}}
\newcommand {\bfl}{\begin{flushleft}}
\newcommand {\efl}{\end{flushleft}}

\newcommand{\tc}{\textcolor}
\newcommand {\nn} {\nonumber}

\newcommand {\txt}{\textrm}
\newcommand {\bd}{\begin{document}}
\newcommand {\ed}{\end{document}}

\newcommand {\be}{\begin{equation}}
\newcommand {\ee}{\end{equation}}
\newcommand {\bea}{\begin{eqnarray}}
\newcommand {\eea}{\end{eqnarray}}
\newcommand {\ba}{\begin{array}}
\newcommand {\ea}{\end{array}}

\newcommand{\bbib}{\begin{thebibliography}{99}}
\newcommand{\ebib}{\end{thebibliography}}
\newcommand {\bab}{\begin{abstract}}
\newcommand {\eab}{\end{abstract}}

\newcommand {\bc}{\begin{center}}
\newcommand {\ec}{\end{center}}
\newcommand {\bit}{\begin{itemize}}
\newcommand {\eit}{\end{itemize}}
\newcommand {\ul}{\underline}
\newcommand {\txtc}{\textcolor}

\newcommand {\ad}{\textrm{ad}}
\newcommand {\Ad}{\textrm{Ad}}
\def\A{{\cal A}}\def\B{{\cal B}}\def\C{{\cal C}}\def\D{{\cal D}}\def\E{{\cal E}}\def\F{{\cal F}}\def\G{{\cal G}}\def\H{{\cal H}}\def\I{{\cal I}}
\def\J{{\cal J}}\def\K{{\cal K}}\def\L{{\cal L}}\def\M{{\cal M}}\def\N{{\cal N}}\def\O{{\cal O}}\def\P{{\cal P}}\def\Q{{\cal Q}}\def\R{{\cal R}}
\def\S{{\cal S}}\def\T{{\cal T}}\def\U{{\cal U}}\def\V{{\cal V}}\def\W{{\cal W}}\def\X{{\cal X}}\def\Y{{\cal Y}}\def\Z{{\cal Z}}

\def\xbar{\bar{x}}\def\ybar{\bar{y}}\def\zbar{\bar{z}}\def\kbar{\bar{k}}\def\pbar{\bar{p}}

\def\mplane{Moyal plane ~$\M_\theta=\{x;~x^\dg=x,~x^\mu x^\nu-x^\nu x^\mu=i\theta^{\mu\nu}=-i\theta^{\nu\mu}\}$~}
\def\moyalplane{Moyal plane ~$\M_\theta=\{x;~x^\dg=x,~x^\mu x^\nu-x^\nu x^\mu=i\theta^{\mu\nu}=-i\theta^{\nu\mu}\}$~}
\def\gmplane{Moyal plane ~$\M_\theta=\{x;~x^\dg=x,~x^\mu x^\nu-x^\nu x^\mu=i\theta^{\mu\nu}=-i\theta^{\nu\mu}\}$~}
\def\superspace{superspace ~$\S=\{x_I=(x_\mu,\theta_a);~x_I x_J=(-1)^{|x_I||x_J|}~x_Jx_I,~~~~ |x_\mu|=0,~~|\theta_a|=1,\}$~}
\def\LT{Lorentz transformation ~$x\ra \Ld x$~} 
\def\PT{Poincar$\acute{\txt{e}}$ transformation ~$x\ra \Ld x+b$~} 
\def\RLT{representations equation~$U(\Ld)U(\Ld')=f(\Ld,\Ld')~U(\Ld\Ld')$~}
\def\RPT{representations equation~$U(\Ld,b)U(\Ld',b')=f(\Ld,\Ld',b,b')~U(\Ld\Ld',b+\Ld b')$~}
\def\xxmstar{$e^{{i\over 2}\ola{\del}\wedge\ora{\del}}$}
\def\ppmstar{$e^{-{i\over 2}p\wedge p'}$}
\def\pkmstar{$e^{-{i\over 2}p\wedge k}$}
\def\kpmstar{$e^{-{i\over 2}k\wedge p}$}
\def\kkmstar{$e^{-{i\over 2}k\wedge k'}$}
\def\PPmstar{$e^{-{i\over 2}P\wedge P'}$}
\def\xPmstar{$e^{{1\over 2}\ola{\del}\wedge P}$}
\def\fxxmstar{$e^{{i\over 2}\ola{\del}_\mu\theta^{\mu\nu}\ora{\del}_\nu}$}
\def\cqc{$~[q_i,p_j]=i\hbar\delta_{ij},~[p_i,p_j]=[q_i,q_j]=0~$}
\def\Tr{\txt{Tr}}
\def\tr{\txt{tr}}

\renewcommand{\theequation}{\arabic{equation}}
\numberwithin{equation}{section}

\renewcommand{\baselinestretch}{1.5}

\begin{document}
%
\pagestyle{plain}

\begin{center}
\textbf{\large Quantum Theory, Noncommutativity and Heuristics}
\vspace{0.2cm}\\
{Earnest Akofor}\\
\emph{\small{eakofor@physics.syr.edu}}\\
\emph{\small Department of Physics, Syracuse University, Syracuse, NY
13244-1130, USA}\\
(http://users.aims.ac.za/$\sim$akofor/academics/academics.html)
\end{center}
\vspace{0.2cm}
\begin{center}
{\large Abstract}
\end{center}
 Noncommutative field theories are a class of theories beyond the standard model of elementary particle physics. Their importance may be summarized in two facts. Firstly as field theories on noncommutative spacetimes they come with natural regularization parameters. Secondly they are related in a natural way to theories of quantum gravity which typically give rise to noncommutative spacetimes. Therefore noncommutative field theories can shed light on the problem of quantizing gravity. An attractive aspect of noncommutative field theories is that they can be formulated so as to preserve spacetime symmetries and to avoid the introduction of irrelevant degrees freedom and so they provide models of consistent fundamental theories.

 In these notes we review the formulation of symmetry aspects of noncommutative field theories on the simplest type of noncommutative spacetime, the Moyal plane. We discuss violations of Lorentz, P, CP, PT and CPT symmetries as well as causality. Some experimentally detectable signatures of these violations involving Planck scale physics of the early universe and linear response finite temperature field theory are also presented.
\newpage
\tableofcontents
\listoffigures

\chapter{\bf {Introduction}} \label{ch:one}
Quantum theory and the theory of general relativity do not appear to be compatible at very short distance scales due to the following argument. One generally expects that at very short length scales the general relativistic theory of gravity needs to become a quantum field theory due to the high energies that are required to probe such short distances. However, standard quantization methods do not suffice because the quantization of classical gravity theories results in quantum theories lacking in renormalizability which is one of the requirements for a consistent fundamental quantum field theory.

In quantum field theory (QFT) renormalization is an attempt to understand the physical reasons for the UV or short distance divergences that occur in the naturally expected contributions of energetically unrestricted intermediate processes to the potential or probability amplitude of a given energetically restricted physical process in spacetime. Renormalization procedures naturally start with some kind of regulator, a set of regularizing parameters, followed by the isolation of regulator dependent contributions into finite and purely divergent pieces. A theory is said to be renormalizable if the divergences can be understood with a finite regulator; one containing a manageable number of regularizing parameters, without the need of introducing or allowing an arbitrarily large number of extra fundamental degrees of freedom, otherwise the theory is inconsistent and is said to be nonrenormalizable.

A physical process, an isolation of part of the course of dynamics of a physical system, is one whose potential survives any induced and intrinsic isomorphic transformations, ie. symmetries, of both spacetime and the spaces of configurations or auxiliary variables of the system in spacetime. The potential depends on the configuration variables and on the way these variables couple in the classical action that describes the dynamics of the system through a least action principle. The way the variables couple is in turn determined by symmetries. Our configuration variables shall be fields which include matter or half integer spin fields and gauge or integer spin fields which are thought to mediate fundamental interactions between the matter fields.
Symmetries may be separated into nonlocal and local symmetries. Nonlocal symmetries are homogeneous in that the value of the transformation parameter is the same at each point or infinitesimal region of spacetime and/or spaces of configurations. The definition of a gauge field allows it to have some physically irrelevant components. Gauge symmetries are special local symmetries often used as a tool or standard for tracking the number of irrelevant components in a gauge field in addition to their normal use as symmetries; to determine how gauge fields couple among themselves, and to other fields, in the classical action. The use of gauge symmetries to determine the coupling of gauge fields is due to the assumption that they should correspond to some global symmetry when their transformation parameters are made homogeneous and vice versa.

 A finite regulator may or may not survive all of the symmetries of a quantum theory. Anomalies are unexpected (nonsymmetric) contributions, from the intermediate processes, which are found to be due to the nonexistence of a finite regulator that can survive all symmetries of the action for the underlying theory. The anomalies can be presented as the failure of a conserved (Noether) current of a symmetry of the classical action to remain conserved after quantization.
 The underlying theory in this case is said to be anomalous. Renormalization by definition must also account for the anomalies as well.
 Following symmetries anomalies may be global or local.
 Global anomalies do not introduce any extra degrees of freedom and so do not spoil renormalizability. However the theory will be nonrenormalizable if the gauge anomalies from all possible intermediate processes do not sum to zero.
 This is because the unphysical degrees of freedom that the gauge symmetry represents will contribute to a supposedly physical intermediate process implying an inconsistency.
 The nonrenormalizability of the quantized version of any classically successful theory such as the theory of gravity indicates that such a theory is only an effective theory that can be obtained in the classical limit of a more fundamental quantum theory.
 Theories on noncommutative spacetime come with a natural symmetry surviving regulator and can therefore serve as bases for testing consistent quantum theories of gravity.

We will review quantum theory and quantization of spacetime in this chapter. In chapter \ref{ch:two}, mostly \cite{qft-us} with minor changes, we will review quantum field theory on the Moyal plane and some of its physical implications including results of investigations on discrete spacetime symmetries and locality. Chapter \ref{cmb1}, mostly \cite{cmbpaper}, involves a theoretical model for a possible effect of noncommutativity on the CMB power spectrum meanwhile  chapter \ref{cmb2}, mostly \cite{numerical}, presents results on the analysis of possible effects of noncommutativity from anisotropy in the CMB radiation. Investigations on causality violating effects in finite temperature field theory appear in chapter \ref{fdt}, mostly \cite{fdt..}. Chapter \ref{ch:conclus} is the concluding chapter and the appendices contain indispensable information that is mostly in heuristic form.

\section{Quantum Theory}
A classical theory, in the description of a physical system, assumes that any underlying characteristic of the physical system can undergo only (deterministic) continuous changes. Noncontinuous changes (which can be nondeterministic) are attributed to statistically averaged characteristics, of a given physical system placed, in an ensemble (ie. a large collection) of physical systems.

Quantum theory involves extensions, of the classical theoretical description of a physical system, in which some of the underlying characteristics of the physical system instead undergo noncontinuous changes (which may be deterministic, nondeterministic or partially deterministic). The classical description can be obtained from the quantum description in the limit where the noncontinuous changes are small enough to be approximately considered as continuous changes.

 The effects of noncontinuous changes are expected to be observed when the system is involved in high energy interactions, where dissociations are most likely to occur.
\subsection{Quantum mechanics}
Mechanics describes the characteristic changes of a given mechanical system (any physical system involved in mostly nondestructive interactions). Quantum mechanics focuses on an extension of the classical mechanical description to include also those underlying characteristics (electrical charge, radiative energy, angular momentum, etc) of the mechanical system that undergo noncontinuous changes.

Quantum mechanics resulted from efforts that either predicted or explained observed phenomena such as the energy distribution in a black body's spectrum, the photoelectric effect, the Compton effect, electron diffraction, atomic spectra, etc. Early quantization ideas were presented by Planck, Einstein, Bohr, De Broglie, Hiesenberg and Schrodinger.

Planck had to assume that the \emph{blackbody} consisted of oscillators that could emit or absorb energy only in fixed amounts $\vep$ that needed to depend linearly on the frequency only. That is $\vep=hf$, where $h$ is a constant. Similarly Einstein in order to explain the \emph{photoelectric effect} (the ejection of electrons from the light-illuminated surface of a metal, with the kinetic energy of the electrons depending linearly on frequency but not on the intensity of the light) assumed that the energy of light was quantized (distributed in space as localized lumps each of which can be produced, transported or absorbed only as a whole) so that the energy of a particle of light may be written as $E=hf$ and hence deduced a corresponding momentum with $|\vec{p}|={h\over c}f={h\over\ld}$. Thus the wave phase of light could then be rewritten as $e^{2\pi i(f t-\vec{k}\cdot \vec{x})}=e^{i{2\pi\over h}(Et-\vec{p}\cdot \vec{x})}$, in terms of its particles' states $(\vec{x},\vec{p}),\\
~~\vec{p}=h\vec{k}={h\over\ld}\hat{v}={h\over\ld}{\dot{\vec{x}}\over |\dot{\vec{x}}|}$,~~ or~~ $(x^\mu,p^\mu)\eqv (ct,\vec{x},{E\over c},\vec{p})$. Since the energy and momentum of a massive free particle are related by $E^2=\vec{p}^2c^2+m^2c^4$ the particle of light is therefore a massless free particle.
De Broglie postulated that the wave phase relation be applied also to massive free particles $E^2=\vec{p}^2c^2+m^2c^4$ in which case these particles should also display wave-like properties with \\
~~$f={E\over h},~~\ld={h\over |\vec{p}|},~~\vec{p}={E\over c^2}\vec{v}$. This was confirmed in \emph{electron diffraction} experiments. It was then straightforward to write down ``wave'' or ``field'' equations (eg. the nonrelativistic Schrodinger equation $(i\del_t+{1\over 2m}\vec{\del}^2-V(\vec{x},t))\psi(\vec{x},t)=0$) for massive particles in an external potential $V(\vec{x},t)$, where a ``field'' $\psi(\vec{x},t)$ is a superposition or linear sum of ``waves''. Light quantization also explains the \emph{Compton effect}: the observed shift in wavelength of light when it scatters off free electrons.

On the other hand, it was realized by Bohr and others that it is not possible to map out a clear path or orbit for the electron in an atom.   In the continuum theory, the Fourier transform of the electron's electric dipole moment $eq$ predicted a continuous frequency spectrum for radiation with the Fourier coefficients of $eq$ giving the intensities associated with each radiated frequency. However, the observed frequencies were discrete implying that the Fourier representation was not an appropriate way to represent $eq$. A matrix representation was finally chosen by Heisenberg and others as an appropriate representation for $eq$, where the components of the matrix may be interpreted as ``transition probabilities'' among the discrete frequencies in analogy to the classical Fourier coefficients which were normally interpreted as radiation intensities associated with the continuous frequencies.

Empirical results in \emph{atomic spectroscopy}, eg. Rydberg's wavelength formula ~${1\over\ld_{ij}} ={R\over n_i}-{R\over n_j}$ where $n_i,n_j$ are integers and $R$ a constant, indicate that the energy levels of an electron in a physical atom may be represented by the eigenvalues of a matrix called Hamiltonian $H$. The Hamiltonian $H$ is a matrix-valued ``function'' of equally matrix-valued \footnote{instead of a Fourier sequence} observable \footnote{In ``observable'' or ``measurable'', measurement of a quantity $U$ refers to an assignment of a number to the quantity $U$. An observable will randomly take on one value of its spectrum each time it is measurement.}  quantities $q,p$ that represent the canonical position and  momentum from classical Hamiltonian mechanics.
The relations may be expressed as follows


\bea
\label{fn:hamiltonian1}&&H=H({p},q,t)=(H_{mn}),\\
&&\label{heisenbergeq}{dF\over dt}={i\over \hbar}(HF-FH)+{\del F\over\del t},~~~~F=F({p},q,t)=(F_{mn}),\\
&&{q}p-p{q}=i\hbar,~~~~q=q(t)=(q_{mn}),~~{p}={p}(t)=({p}_{mn}),\\
&& H\psi_\nu =\hbar\nu ~\psi_\nu~~~(\txt{Eigenvalue problem for the matrix $H$}),\\
\label{fn:hamiltonian2}&&H=S\Ld S^{-1},~~\Ld_{mn}=\hbar\omega_m\delta_{mn},~~\omega_m=2\pi\nu_m,
\eea
where the commutator $[H,F]=HF-FH$ may be interpreted as a quantum mechanical analogue of the classical Poisson bracket $\{h,f\}=\del_qh\del_pf-\del_ph\del_qf$.

The equations above come from an empirically  deduced form for the coordinate $q$ given by
\bea
&&q_{mn}(t)=q^0_{mn}~e^{i\omega_{mn}t},~~\omega_{mn}=2\pi(\nu_m-\nu_n),\nn\\
&&{d q_{mn}(t)\over dt}={i\over \hbar}(\Ld q-q\Ld)_{mn}=i\omega_{mn}~q_{mn}(t).
\eea
where $\nu_{mn}=\nu_m-\nu_n$ is the frequency of a photon emitted by an electron that ``drops'' from a higher energy level $m$ to a lower energy level $n$ (the energy of the photon is $h\nu$). The canonical quantization conditions \\ ~~$[{q}_i,p_j]=i\hbar\delta_{ij},~[{p}_i,{p}_j]=[q_i,q_j]=0$~ are an extension (see eqn (\ref{BScondition2}) of appendix \ref{cano-quant}) of the Bohr-Sommerfeld quantization\footnote{This model considers (planar) elliptical rather than (planar) circular orbits of Bohr's model for the electronic orbits of Hydrogen. This quantization condition is merely an additional constraint (to the usual classical equations of motion) imposed in order to obtain a discrete rather than a continuous set of orbits, energies, angular momenta and related quantities. It may also be written as ~$\oint_C(p_jdq^j-q_jdp^j)=2nh$~ or as ~$\oint_C\bar{z}_jdz^j=2nhi,~~z_j=q_j+ip_j$.} condition
\bea
\label{BScondition1}\oint_C {p}_idq^i=nh.
\eea

The time evolution equation (\ref{heisenbergeq}) generates a one parameter time translation group $\{e^{iHt}\}$ with the Hamiltonian $H$ as the sole generator. The spectrum (from the eigenvalue problem $H\psi_\nu=h\nu\psi_\nu$ for $H$) of $H$ is preserved by this time translation symmetry and consequently each atom has a unique emission or absorption spectrum that characterizes (or serves as a thumbprint for) the type of chemical element the atoms of that type produce. The eigenvalue problem for $H=H(\vec{q},\vec{p})$ may be seen as the problem of finding the irreducible representations of the one parameter time translation group and so each frequency represents an irreducible or elementary attributes (a single excitation, or energy, level of an electron of the atom) of a non-rotating atomic electron system. Naturally, the electron system can be free to rotate around or relative to the nucleus in which case we have invariance under the time translation plus rotation group whose irreducible representations would give the elementary attributes of the system.
The canonical quantization condition for a system with several canonical degrees of freedom is  $[{q}_i,p_j]=i\delta_{ij}\hbar,~[q_i,q_j]=[{p}_i,{p}_j]=0$. For a system with Hamiltonian $H=H(\vec{{p}}^2,\vec {{p}}\cdot\vec{q},\vec{q}^2)$ and angular momentum $L_{ij}={1\over 2}(q_i{p}_j-q_j{p}_i)$, $H$ commutes with $L_{ij}$ and $\{H,L^2=L_{ij}L_{ij}\}$ generate the center of the algebra of the symmetry group.
\bea
[L_{ij},L_{{k}{l}}]=-{1\over 2}(\delta_{i{k}}L_{j{l}}+\delta_{j{l}}L_{i{k}} )+{1\over 2}(\delta_{i{l}}L_{j{k}}+\delta_{j{k}}L_{i{l}} ).
\eea
 All parts of the atomic system can also be displaced by the same amount in ``free'' space without disturbing the spectrum of the atomic system. Thus one needs to consider a Hamiltonian of the form
$H=\sum_{ab}h(\vec{p}_a\cdot\vec{p}_b,\vec{p}_a(\vec{q}_a-\vec{q}_b),(\vec{q}_a-\vec{q}_b)^2)$ where $a,b$ label the various pieces or particles of the system. Then $H$ also commutes with the total momentum operator ~$\vec{P}=\sum_a\vec{p}_a$ ~ which is the generator of spatial translations. The canonical commutation relations are
\bea
&&[{q}^i_a,p^j_b]=i\delta^{ij}\delta_{ab}\hbar,~[q^i_a,q^j_b]=[p^i_a,p^j_b]=0\nn\\
\eea
and the angular momentum operator will be the sum of the individual ones:
\bea
&&L^{ij}=\sum_aL^{ij}_a=\sum_a{1\over{2}}(q^i_aP^j-q^j_aP^i)={1\over{2}}(Q^iP^j-Q^jP^i),\nn\\
&&\vec{Q}=\sum_a\vec{q}_a.
\eea
The center of the algebra of the symmetry group of the atomic system is now generated by $(H,L^2,\vec{P})$.
At this point one realizes that the problem of quantizing the atomic system includes the problem of finding the irreducible representations of its symmetry group (or equivalently of the algebra of the symmetry group) generated by $H,P^i,L^{ij}$. To include relativistic effects, one needs to replace the (spatial rotation plus spatial translation plus time translation) group with the Poincare group (spacetime rotation plus translation group). Then relativistic quantum mechanics involves the problem of finding the spectrum of the center of the group generated by the operators $\P^\mu,J^{\mu\nu}$ which have the canonical representation
\bea
&&\P^\mu=(\P^0(H),\vec{\P}(\vec{P})),~\Q^\mu=(\Q^0(Q^0),\vec{\Q}(\vec{Q})),\nn\\
&&\P^0(H)=H=H(\gamma^0,\vec{\gamma},t,\vec{Q},\vec{P}),~~\vec{\P}(\vec{P})=\vec{P},\nn\\
&&\Q^0(Q^0)=Q^0,~~\vec{\Q}(\vec{Q})=\vec{Q},\nn\\
&&J^{\mu\nu}={1\over{2}}(\Q^\mu \P^\nu-\Q^\nu \P^\mu)+{i\over 4}(\gamma^\mu\gamma^\nu-\gamma^\nu\gamma^\mu)+...\eqv L^{\mu\nu}\otimes 1_{S}+1_L\otimes S^{\mu\nu}+...,\nn\\
&&~~~~=J^{\mu\nu}_L+J^{\mu\nu}_S,\nn\\
&&[J_{{\mu}{\nu}},J_{\al\beta}]=-{1\over 2}(\eta_{{\mu}\al}J_{{\nu}\beta}+\eta_{{\nu}\beta}J_{{\mu}\al} )+{1\over 2}(\eta_{{\mu}\beta}J_{{\nu}\al}+\eta_{{\nu}\al}J_{{\mu}\beta} ),\nn\\
&&[J_L^{{\mu}{\nu}},J_S^{\al\beta}]=0,\nn\\
&&[P^\mu,Q^\nu]=i\eta^{\mu\nu},~~Q^\mu=(Q^0,\vec{Q}),\nn\\
&&\{\gamma^\mu,\gamma^\nu\}=2\eta^{\mu\nu}.
\eea
In the Schrodinger representation~~ $Q^\mu\ra \mu_{x^\mu},~P_\mu\ra i{\del\over\del x^\mu}$ (here $\mu_{x^\mu}$ denotes ordinary multiplication by the spacetime coordinates $x^\mu$), one then has the consistency condition
\bea
\label{schrodinger-eqn} i{\del\over\del t}=H(\gamma^0,\vec{\gamma},t,\vec{x},i{\del\over\del\vec{ x}})
\eea
 on the space of sections $E/\mathbb{\mathbb{R}}^{d+1}=\{\psi:\mathbb{R}^{d+1}\ra E\simeq O(\mathbb{C}^M\ot\mathbb{C}^N)\times(\mathbb{C}^M\ot\mathbb{C}^N)\}$ of a vector bundle $E$ over $\mathbb{R}^{d+1}$ where $\psi=\psi_L\otimes \psi_S$ is the product of the orbital and spin angular momentum wavefunctions and $O(\mathbb{C}^M\ot\mathbb{C}^N)$ is the space of linear operators on $\mathbb{C}^M\ot\mathbb{C}^N$.

Even though it is not possible to say precisely where the atomic electron's orbit is, it is however possible to say that it is mostly around the nucleus of the atom; that is, the electron's orbit is localized in the region around the nucleus.
A basic quantity introduced for the study of localization\footnote{A system is localized in a certain region $\D$ at a particular time if the total probability of finding it in $\D$ at that time is $1$. Alternatively, the region $\D$ is dense in the support of the probability density function of the system.} was Schrodinger's wavefunction in wave mechanics which is any function satisfying the consistency condition (\ref{schrodinger-eqn}). Schrodinger's wave mechanics is equivalent to Heisenberg's matrix mechanics which was discussed earlier. In general the wave function is a complex-valued function(al) of the quantized configuration variables such as canonical coordinate in quantum mechanics or fields in quantum field theory, whose absolute value can be interpreted as a joint probability density function for the quantized canonical variables on which it depends.

When the quantum, ie. quantized classical, configuration variables are represented as elements of an algebra $\O(\H)$ of operators on a Hilbert space\footnote{A Hilbert space is a vector space completed into a metric space by a norm that is induced by an inner product measure defined on the vector space.} $\H$ then the wavefunction would be the value of a chosen linear functional\footnote{Wavefunctions of physical systems and probability amplitudes for various physical processes are examples of (values of) linear functionals on $\O(\H)$. The wavefunction for a physical system is a time-dependent linear functional whose value on a given quantum configuration is the probability amplitude for finding the system in that quantum configuration and it satisfies Schrodinger's equation.} on the quantum configuration variable in question.  Thus the time evolution equation may also be written either in terms of the wavefunction or in terms of a corresponding vector in the Hilbert space $\H$. The time evolution equation in terms of the wavefunction is known as Schrodinger's equation. More specifically the sole irreducible representation, up to unitary equivalence, of the relations (\ref{fn:hamiltonian1}) through (\ref{fn:hamiltonian2}) on a Hilbert space is known as Schrodinger's representation.

\subsection{Quantum field theory}
Quantum field theory is a relativistic quantum theory of systems with arbitrary numbers and types of degrees of freedom. Quantum mechanics treats a system of $N$ (interacting) particles using a fixed number and type of $N$ (coupled) equations.  However not all interacting systems have a fixed number and species of particles. Particle transformations and relativistic quantum effects such as particle creation and annihilation may occur. Particles of a kind are now regarded as localizable disturbances (ie. perturbations or fluctuations) in a field of that kind. In particular the field description treats elementary particles as (Fourier) modes of the oscillatory part of an associated field in direct analogy to the electromagnetic field, the modes of whose oscillatory part correspond to the various frequencies of the electromagnetic spectrum. One has an analog of the canonical quantization condition;
\bea
&&\vec{q}_n(t)\ra q_p^\al(t)=\sum_{\vec{x}}\psi^\al(\vec{x},t)u_p(\vec{x},t),\nn\\
&&\vec{p}_n(t)\ra \pi_p^\al(t)=\sum_{\vec{x}}\Pi^\al(\vec{x},t)u_p(\vec{x},t),\nn\\
&&q_p^\al(t)\pi_{p'}^\beta(t)-(-1)^{2s}\pi_{p'}^\beta(t)q_p^\al(t)=i\hbar\delta^{\al\beta}\delta_{pp'},\nn\\
&&\sum_{p}u^\ast_p(\vec{x},t)u_p(\vec{y},t)=\delta^3(\vec{x}-\vec{y}),\nn\\
&&\Pi^\al(\vec{x},t)\psi^\beta(\vec{y},t)-(-1)^{2s}\psi^\beta(\vec{y},t)\Pi^\al(\vec{x},t)=i\hbar\delta^{\al\beta}\delta(\vec{x}-\vec{y}),\nn\\
&&\Pi(\vec{x},t)={\del\L\over\del\del_t\psi }(\vec{x},t),\nn\\
&&S[\psi]=\int\L(x,dx,\psi,d\psi),
\eea
where $n$ is a discrete label for a collection of particles and the value of $x$ needs to be chosen in such a way as to obtain a consistent theory for the field $\psi$. For example Pauli's exclusion principle\footnote{The exclusion principle associates the shell structure of atomic electron systems, space occupying/shape forming properties of matter, stability of astronomical objects such as neutron stars, etc to the difficulty for two elementary matter systems to have exactly the same set of fundamental quantum labels. Electromagnetic fields for example and other force fields do not appear to exhibit these properties. The exclusion principle is connected to the idea of spin angular momentum by the requirement that the probability amplitude of a composite physical process must be a rotationally invariant/covariant functional of the probability amplitudes for the individual elementary processes of which it is composed.} requires that s be a half integer for matter fields and and integer for interaction mediation fields. $p$ is a characteristic or typical value (an eigenvalue for a corresponding momentum operator as a Noether charge associated with translational invariance) for the momentum of an individual mode. This is because $\vec{q}_1$ and $\vec{q}_2$ (corresponding to $q^\al_{p_1},~q^\al_{p_2}$) denote different positions in space. $\al$ is a ``spin'' index which is an extension of the spatial vector index. The differential action or Lagrangian $\L(x,dx,\psi,d\psi)$ is a differential form on spacetime.

Thus an individual mode is described by the triple $(q_p^\al(t),\pi_p^\al(t),u_p(\vec{x},t))$, where $|u_p(\vec{x},t)|^2d^3x$ is the probability of finding the mode in an infinitesimal neighborhood of $\vec{x}$ of volume $d^3x$ at any given time $t$. This means that the role of the point $\vec{x}$ is now being played by the linear functional \\
~~$u_p:(q^\al_p(t),\psi_\al(\vec{x},t))\mapsto u_p(\vec{x},t)=\langle q^\al_p(t)\psi_\al(\vec{x},t)\rangle $. The field $\psi^\al$ can also be directly interpreted as the particle coordinate, where the particle is constrained to move along a time-parametrized path $\vec{q}:[0,1]\ra \C$ of the configuration space \\
~~$\C=\bigcup_n\C_n\eqv\bigcup_n\{\vec{q}_n\}$  in many particle quantum mechanics meanwhile the particle is constrained to move along a spacetime-parametrized hypersurface\\
~~ $\psi^\al:\M\simeq([0,1]^4,g)\ra \U=\bigcup_{x\in\M}\{\psi^\al(x)\}\subset\C$ of the configuration space \\
~~$\C=\bigcup_{p}\{q^\al_p\} $ in quantum field theory and similarly the particle is constrained to move along a $(\sigma,\tau)$-parametrized two dimensional surface \\
~~$X^\al:([0,1]^2,h)\ra \C=\mathbb{R}^{d+1}$ in string theory.

In quantum field theory the role originally played by the Hamiltonian $H$ alone in quantum mechanics is now played by the 4-momentum operator \\
~~$P_\mu=(H,\vec{P})$~~(A component $T^{\mu 0}$ of the Energy-momentum tensor \\
~~$T^{\mu\nu}=\int d^3x~\T^{\mu\nu},~~~~\del_\mu \T^{\mu\nu}=0 $, a Noether charge corresponding to spacetime translation symmetry). The eigen-value problem for $H$, and any other quantities that commute with $H$, is replaced by the problem of finding the solutions $U$ of the equation \\
~~$U(\Ld_1,b_1)U(\Ld_2,b_2)=U(\Ld_1\Ld_2,b_1+\Ld_1b_2)$~ which is Wigner's method of classifying elementary particle states. That is, finding the irreducible representations  of the Lorentz-Poincare transformation \\
~~$LP:{{\mathbb{R}^{3+1}}}\ra{{\mathbb{R}^{3+1}}},~ x\mapsto\Ld x+b,~\Ld^T=\Ld^{-1} ,~~x=(x_\mu)=(x_0,\vec{x})$, the automorphism or symmetry group of the spacetime ${{\mathbb{R}^{3+1}}}$. The irreducible representations correspond to free elementary point particles that can be localized in ${{\mathbb{R}^{3+1}}}$. In addition to reparametrization symmetry the Lorentz-Poincare transformation is a symmetry and thus a canonical transformation\footnote{Section \ref{cano-quant}} of the relativistic point particle action
\bea
S[x,\Gamma]=m\int_\Gamma\sqrt{\eta_{\mu\nu}dx^\mu dx^\nu}
\eea
since this Lagrangian involves only the metric $ds^2=\eta_{\mu\nu}dx^\mu dx^\nu$ which is the defining structure of the Minkowski spacetime.

Thus given any space $\S$, one can also consider the problem of finding the irreducible representations of the
automorphism group $G(\S):\S\ra\S$ of $\S$ so as to be able to characterize/classify all the possible elementary physical systems that can be localized in $\S$. Examples of spaces include topological metric spaces, manifolds (which also include Lie groups), fiber bundles, etc and other spaces derived from these using various mathematical constructs. Here it is also important to note that the symmetry group of a topologically nontrivial\footnote{A space is topologically nontrivial if any two of its subspaces cannot always be continuously deformed into each other. Topology is the study of invariance under continuous shape change or deformation (ie. geometry) transformations. Physically interesting geometries would be the fixed points of these geometry transformations.} space (as compared to the flat spacetime ${{\mathbb{R}^{3+1}}}$) is ``enlarged'' mainly due to additional discrete transformation channels leading to various periodicity types and therefore one expects additional distinct physical properties induced on the elementary systems in $\S$ by its nontrivial topology. Conversely, if the elementary systems in $\S$ are observed to display unexpected additional properties, say through experimentation, that do not seem to depend on the geometry, ie. shape/size structures, on $\S$ then they may be investigated by introducing nontrivial topology. Ways of introducing nontrivial topology include employing nondynamical constraints (like quotienting of a [topologically trivial] space by the actions of [discrete] transformation groups ) as well as dynamical constraints such as postulating the presence of unknown forms of ``elementary'' systems that can couple to the known elementary systems in a way that can explain the additional properties and also gives possible explanations as to whether the unknown forms of elementary systems could be experimentally detectable or not.

For example, consider the variational problem for an electron (considered as the less physically realistic case, a point particle, so that it can only trace 1-dimensional paths) with action $S[q]=\int_0^1 L(t,dt,q,dq),~~q:[0,1]\ra {{\mathbb{R}^{3+1}}}$. If there is a very strong magnetic field confined in a thin infinitely long tube through the space ${{R^{3+1}}}$, then since any electron (and hence its path) with insufficient energy cannot penetrate this tube, it means that for such an electron the variational problem will have more than one solution as a path on one side of the magnetic tube cannot be continuously varied to a path on the opposite side of the tube. For the same reason a path that wraps around the tube n times cannot be varied to a path that wraps around it any $m$ times in the opposite sense or $m\neq n$ times in the same sense. Hence to every path is associated an integer parameter labeling the number of times and sense in which its path winds around the tube. Therefore if identical electrons of insufficient energy are produced at some point and later interact then one expects to observe the effect of the difference in the topological charges (winding numbers) they gained during their individual journeys. This effect may be included in the action by adding a non trivial but smooth path deformation independent term
\bea
&&\nu[q]=\int_0^1B_idq^i,~~\nn\\
&&\delta\nu[q]=\delta (\int_0^1B_idq^i)= \int_{0}^{1}(\delta B_i~dq^i+B_i\delta dq^i)= \int_0^1(\delta q^i\del_i B_j~dq^j+B_i{\del_j(\delta q^i)dq^j})\nn\\
&&~~~~=\int_0^1\delta q^i(\del_iB_j-\del_jB_i)dq^j+\int_0^1\del_i(B_j\delta q^j)dq^i\nn\\
&&~~~~=\int_0^1\delta q^i(\del_iB_j-\del_jB_i)dq^j+[B_j\delta q^j]|_0^1=0,
 \eea
 where the path $q(t)$ can be smoothly deformed to the path $q(t)+\delta q(t)$.
 That is, smooth path deformation independence requires $dB=0$~ in the region between between any two paths, with common end points, that can be continuously deformed into each other, ~$\oint_\Gamma B=n(\Gamma)\in \mathbb{Z},~~B=B_idq^i$. Alternatively, let $\Gamma_+$ ($\Gamma_-$) be the path oriented from $t=0$ to $t=1$ ($t=1$ to $t=0$),~ $(\Gamma+\delta\Gamma)_+$ be the varied (with end fixed $\delta q(0)=0=\delta q(1)$) path oriented from $t=1$ to $t=0$ and $\Gamma$ be the closed path $(\Gamma+\delta\Gamma)_++\Gamma_-$. Then Stokes' theorem implies that
\bea
&&\delta\nu[q]=\nu_{(\Gamma+\delta\Gamma)_+}[q]-\nu_{\Gamma_+}[q]=\nu_{(\Gamma+\delta\Gamma)_+}[q]+\nu_{\Gamma_-}[q]=\nu_{(\Gamma+\delta\Gamma)_++\Gamma_-}[q]\nn\\
&&~~~~=\nu_{\Gamma}[q]=\oint_\Gamma B=\int_{\txt{int}(\Gamma)} dB.
\eea
Therefore, $\delta\nu[q]=0$ unless the variation takes the path across the tube since $dB|_{{{\mathbb{R}^{3+1}}}\backslash\txt{tube}}=0$. $B$ may be normalized so that any non-zero contribution from $\nu[q]$ is an integer.

 One notes obviously that $B$ (as well as the tube) can also be a dynamical field. The configuration space of the electron is $\S\simeq{{\mathbb{R}^{3+1}}}\backslash\txt{tube}$ instead of $\mathbb{R}^{3+1}$. This same analysis can be carried out for the physically more realistic systems such as strings, $p$-branes, and fields in general as well; which can be sensitive to several other kinds of topologies. The additional terms $\nu(q)$ are known as Wess-Zumino terms and their gauge non-invariance can be adapted to cancel gauge anomalies and so they may be used to define gauge invariant functional integrals in quantum field theory\footnote{See for example \cite{nair} for a review of quantum field theory.}.

\section{Quantization of spacetime}
 It is estimated \cite{doplicher} that in order to satisfy the uncertainty principle in quantum theory and also prevent the undesirable phenomenon of blackhole formation in the general relativistic theory of gravity during a high energy experiment, the length scales being probed by the experiment must not be much smaller than the Planck length $l_p=10^{-35}m$. Information will be lost if blackholes are allowed to form during the experiment. It follows that one cannot, by such careful experiments, distinguish between two local events in spacetime whose separation is much smaller than $l_p$.

 Thus the physical spacetime is expected to be quantized with cells of size of the order of $l_p$. Nonrenormalizable field theories, including the theory of gravity, are expected to be regularized in the physical spacetime. Here the minimum length scale naturally provides the UV cutoff needed to regulate otherwise divergent integrals encountered in the computation of probability amplitudes of certain scattering processes.

 Various methods of quantizing spacetime include the following.
 \begin{enumerate}
\item\emph{Lattice regularization} methods. Space time is given the structure of a lattice with a lattice constant of the order $l_p$. These methods preserve gauge symmetries but break spacetime symmetries.

\item \emph{Canonical quantization} methods. Here the local coordinates of spacetime become noncommutative in such a way that their spectra\footnote{See section \ref{spectral-theory}. If $a$ describes a physically measurable quantity then its spectrum would contain the possible values that one can obtain when the quantity is measured.} are still invariant under the original classical symmetry group of the spacetime.

 \item \emph{Deformation quantization} methods.  The local coordinates of spacetime become noncommutative and the symmetry group of spacetime is also modified so that it can preserve the spectra of the coordinates. Because of the observation that the elementary physical systems that live in a space $\S$ are given by the irreducible representations of the symmetry group of $\S$, one could rather directly quantize/deform the group algebras of the symmetry group of $\S$ and study their consequences. Mathematically these deformed groups or quantum groups may be considered as belonging to a certain class of Hopf algebras\footnote{See \cite{mack1,drinfeld,majid}}.

\item \emph{Noncommutative geometry}\footnote{\cite{masson} for example gives an informal introduction to noncommutative geometry.}. The noncommutative algebra of spacetime coordinates is first introduced. One then constructs any possible $C^\ast$-algebras\footnote{See equation (\ref{c-star-algebra})} from the universal algebra generated by the algebra of coordinates.

 Noncommutative geometry involves extensions of classical geometric structures to an arbitrary $\ast$-algebra $\A$ and its dual space of linear functionals $\A^\ast=\{\phi:\A\ra \mathbb{C}\}$. The extensions are based on duality between a classical space \footnote{A classical space is a set of points (ie. imaginary objects) with one or more classical structures, such as special mapping or transformation structures, group structures, topological or continuity structures, differential structures and so on,  defined on it.} $X$ and the point-wise product algebra $\A=(\C(X),+,\txt{pt-wise})$ of complex classical functions $\C(X)=\{f:X\ra \mathbb{C}\}$ on $X$. The algebra $\A$ plays the role of a (non)commutative space of functions on the dual space $\A^\ast$.  The symmetry groups of noncommutative spaces are known as quantum groups. Noncommutative geometry provides a unifying framework for various methods of quantizing spacetime.
\end{enumerate}

\section{Motivation for noncommutative field theory}
Apart from the fact that spacetime quantization arose historically due to the need for regularization in quantum field theory, noncommutative spaces also arise naturally in various physical and mathematical theories. This fact lends support to the construction of general spacetime quantization schemes and noncommutative geometry. We will now list some of the noncommutative spaces that are often encountered in physics.

\subsection{Phase space in quantum mechanics}
In quantum mechanics the canonical quantization conditions \\
~~\cqc imply that the quantum phase space is a naturally noncommutative space.
\subsection{Superspace in supersymmetric field theory}
Geometrically, supersymmetric theories are theories on a noncommutative space (known as superspace) with graded coordinates $x_I=(x_\mu,\theta_a)$
\bea
&&x_I x_J=(-1)^{|x_I||x_J|}~x_Jx_I,~~~~ |x_\mu|=0,~~|\theta_a|=1,\nn\\
&&|x_Ix_J|=|x_I|+|x_J|.
\eea
\subsection{The center of motion of an electron in a magnetic field}
When an electron moves in a constant magnetic field the coordinates of the center of its circular motion (ie. guiding center)  become noncommutative when the system is quantized canonically.
The solution to
\bea
&&m{d \vec{v}\over
dt}=e\vec{v}\times \vec{B},~~\vec{v}={d\vec{x}\over dt}\nn\\
\eea
(see appendix \ref{electron-motion}) shows that the center of circular motion is
\bea
 &&\vec{x}_c(t)=\vec{x}_0+{m\over
eB}~{\hat{\vec{B}}\times\vec{v}_0}+\hat{\vec{B}}~(\hat{\vec{B}}\cdot{\vec{v}_0})(t-t_0),\nn\\
\eea
which upon quantization (see appendix \ref{electron-motion}) satisfies the relation
\bea
 [x^i_c(t),x^j_c(t)]=i\theta^{ij}=i{\hbar\over eB}\vep^{ikj}\hat{B}^k~~\forall t.
\eea

\subsection{Phase space of a Landau problem with a strong magnetic field}
Consider the problem where an electron moves in a plane $\vec{x}=(x,y,0)$ subject to a constant magnetic field $\vec{B}=(0,0,B)$ perpendicular to the plane, then the Lagrangian is
\bea L={1\over 2}m\vec{v}^2-e\vec{v}\cdot\vec{A},~~\vec{v}=(\dot{x},\dot{y},0),~~\vec{A}=-{1\over 2}\vec{x}\times\vec{B}.
\eea
When the magnetic field is strong, ie. ${eB}\gg mc$, we have ~$\L\simeq-e\vec{v}\cdot\vec{A}$~ giving ~$p_x={eB\over 2}y$~ and ~$p_y=-{eB\over 2} x$~ so that canonical quantization yields
\bea
[x,y]\label{nc-landau}\simeq i{\hbar\over eB}.
\eea

\subsection{Fundamental strings and D-branes}
Consider the open string sigma model given by
\bea
S={1\over2\pi\ld}\{\int_{\D} ~{1\over2}(g_{\mu\nu}dx^\mu \vee dx^\nu+i\ld B_{\mu\nu}dx^\mu\wedge dx^\nu)+i\int_{\del\D} dx^\mu A_\mu\},
\eea
where $\D$ is the string's worldsheet and $g_{\mu\nu},~B_{\mu\nu}$ are constant.
Then the second term is a surface term and so the noncommutativity that arises will be on a  D-brane at the end of the string. In string theory\footnote{String theory is a quantum field theory in which elementary particles states arise as dynamical fluctuations of the trajectory of a one dimensional object, known as the fundamental string, in superspacetime. The fundamental strings as well as the elementary particle states can interact and/or condense to produce charged $p$ dimensional classical or bound states, known as $p$-branes, the existence of some of which is a prediction of the theory. The charged $p$-branes interact with one another as well as with the elementary particle states of the fundamental string.} a $D-$brane is the spacetime hypersurface on which the end of an open string can move freely (ie. the end of the string is confined to this hypersurface) as allowed by a nontrivial choice of boundary conditions in the variational principle that determines the dynamics of the string. More precisely, a $Dp$-brane (or Dirichlet $p-$brane) is the hypersurface of dimensions $p$ (or $p+1$ when the time direction is included) defined by Dirichlet boundary conditions \\ $\delta X^{r_{\sigma_0}}(\tau,\sigma_0)=0,~~\sigma_0\in \{0,\pi\},~~r_{\sigma_0}:\{0,1,2,...,D\}\ra \{1,2,...,D-p\}$\\
$~~\txt{or}~\{1,2,...,D-p-1\}$.

One may write the Fourier expansion
\bea
&&x^\mu(\sigma,\tau)=\sum_kx_k^\mu(\tau)e^{-ik\sigma},
\eea
where the modes $x_k^\mu(\tau)$ may be regarded as individual particles. Then one has a Landau problem for each mode and noncommutativity of the coordinates $x_k^\mu(\tau)$ of the type $(\ref{nc-landau})$ arises when $B_{\mu\nu}$ is large.

\subsection{Myers Effect}
The action principle for a collection of $N~~D0$-branes in the presence of background fields leads to Lie algebra-type noncommutativity
\bea
&&[x^i,x^j]=f^{ij}{}_kx^k
\eea
for the coordinates of the system of $D0$-branes in space-time. This corresponds to static configurations $(\dot{x}^i=0)$ of the $D0$-brane system that extremize the action as required by the least action principle.

The coordinates $x^i$ are $N\times N$ matrices in the adjoint representation of $U(N)$.
In the absence of the background field $f$~ one has $[x^i,x^j]=0$ and these matrices can be simultaneously diagonalized and the $N$ eigen-values represent the positions of the $N$ $D0$-branes~\cite{myers,BBS}.


\begin{center}
Summary of Chapter \ref{ch:two}
\end{center}

\begin{enumerate}
\item Noncommutativity of spacetime $\mathbb{R}^{d+1}$ coordinates is implied by a noncommutative spacetime function algebra $\A_\theta(\mathbb{R}^{d+1})=(\F(\mathbb{R}^{d+1}),\ast)$; with multiplication or self action $\mu:\A\ot\A\ra\A,~~(f,h)\mapsto f\ast g$ which induces left,right multiplicative self representations $\mu^L,\mu^R:\A\ra O(V(\A))$ on $\A$ regarded as a vector space $V(\A)=\A$. Group action needs to preserve the noncommutative product structure $\mu$; ie.
\bea
g\circ \mu=\mu\circ \Delta(g),~~g\in G
\eea
(where $\Delta(g)=g\ot g$, or $\Delta(e^{\al T})=e^{\al\Delta(T)},~\Delta(T)=T\ot 1+1\ot T$, in the undeformed case)
which requires a \textbf{\emph{deformed group action}} and hence a \textbf{\emph{deformed group algebra}}.

\item The group action equally needs to preserve the spectrum
\bea
&&\A^\ast_\tau=\{\tau^\ast:\A_\tau\ra \mathbb{C},~\tau^\ast(ab)=\tau^\ast(a)\tau^\ast(b),~~\tau^\ast(a+b)=\tau^\ast(a)+\tau^\ast(b) \}\nn\\
  \eea
  of the algebra $\A_\tau$ of the statistics or particle interchange operators $\tau_i$. It suffices for the action of $g$ to commute with each $\tau\in \A_\tau$
    \bea
    \Delta(g)\circ\tau=\tau \circ \Delta(g),~~\forall\tau\in \A_\tau\nn\\
    \eea
    which implies a \textbf{\emph{deformed statistics operator}}. Here
    \bea
    &&\A_\tau=\A\{\tau_i,~i\in \mathbb{N};~~\tau_i\tau_{i+1}\tau_i=\tau_{i+1}\tau_i\tau_{i+1},~\tau_i\tau_i=1\}
     \eea
     is the group algebra of the permutation group (a subalgebra of the automorphism algebra of any tensor product algebra just as the permutation group is a subgroup of the automorphism group of any homogeneous \emph{tensor product} algebra, so named after a homogeneous \emph{polynomial} algebra).
    \bea
    &&\tau_i\tau_{i+1}\tau_i=\tau_{i+1}\tau_i\tau_{i+1}~~\Ra~~\tau^\ast(\tau_i)=\tau^\ast(\tau_{i+1}),\nn\\ &&\tau_i^2=e~~\Ra~~\tau_{\pm}^\ast(\tau_i)=\pm \tau_{\pm}^\ast(e)\nn\\
    \eea  and
    \bea
    &&e^2=e~~\Ra~~\tau_{\pm}^\ast(e)=1~~\Ra~~\tau_{\pm}^\ast(\tau_i)=\pm 1~~\forall i.
    \eea

\item  Let $\A_\tau\ra O(\H)$ be a representation of $\A_\tau$ as an algebra of operators $O(\H)$ on a Hilbert space $\H=\T^\ast(\Phi)$ (the dual space of the tensor algebra $\T(\Phi)$ of quantum fields $\Phi=\{\phi\}$). Since the associated spectrum of eigenfunctions \\ ( fermion/boson or pure identical many-particle wavefunctions \\
    $\{\psi:\T(\Phi)\ra \mathbb{C}^N\}$ ) must be preserved in the same manner, a \textbf{\emph{deformed algebra of quantum operators}} in the quantum fields is required in addition to the star-product deformation of the localization functions of the fields.

\item The above process is reversible in that a deformed algebra of quantum fields necessarily leads to a noncummutative algebra of functions.

\item Consequences of deformed statistics of quantum fields include

1) Modification of the statistical interparticle force and hence degeneracy pressure that determines the fate of galactic nuclei after fuel burning seizes.

2) Pauli forbidden transitions may be observable,

3) Lorentz, P, PT, CP, CPT and causality violations can occur.

\item In scattering theory the $S$-operator contains time ordering $T$ and only interaction terms. Therefore the twist factor $e^{{1\over 2}\ola{\del}\wedge P}$ does not always drop out directly as surface terms in the action $S_I$. However it can be checked that the twist factor drops out from all terms in the expansion of the $S$-operator in abelian gauge theories with or without matter fields as well as in pure nonabelian gauge theories.
\item In the (Schrodinger) representation of the noncommutative Moyal algebra $\A_\theta(\mathbb{R}^D)=(\F(\mathbb{R}^D),\ast)$ as an algebra of multiplication operators
    \bea
    &&m(\H_\theta)=\{\mu_f: \H_\theta\ra \H_\theta,~f\in\A_\theta(\mathbb{R}^D),~\xi\ra \mu_f\xi\},\nn
     \eea
     on the Schrodinger Hilbert space $\H_\theta=(\A_\theta(\mathbb{R}^D),\lang\rang)$, with the coordinates $\hat{x}^\mu$ acting as multiplication operators and the momenta $\hat{p}$ acting as derivations,  one encounters two possible independent multiplication representations ~$\mu^L,~\mu^L_f\mu^L_g=\mu^L_{f\ast g}$~ and~ $\mu^R,~\mu^R_f\mu^R_g=\mu^R_{g\ast f},~[\mu^L_f,\mu^R_g]=0$,~
     corresponding to left and right multiplication
     \bea
     &&\mu_f^L\xi=f\ast \xi,~~~~\mu_f^R\xi=\xi\ast f,
     \eea
     and this is the case for any noncommutative algebra. The Moyal case is special in that a commutative representation $\mu_{\hat{x}^\mu}^c={1\over 2}(\mu_{\hat{x}^\mu}^L+\mu_{\hat{x}^\mu}^R)$ can be found for the algebra of the coordinates as one can check that
      \bea
      &&[\mu^L_f\pm \mu^R_f,\mu^L_g\pm \mu^R_g ]=\mu^L_{f\ast g-g\ast f}\pm \mu^R_{g\ast f-f\ast g}~~~~\forall f,g\in \A_\theta(\mathbb{R}^D).\nn
      \eea
      Thus some of the fields in physics can be associated with the commutative sector $\A_0(\mathbb{R}^D)$ generated by the commutative coordinates $\{\hat{x}^\mu_c\}$ which are defined by \\
      $\mu_{\hat{x}^\mu}^c=\mu_{\hat{x}_c^\mu}$.
    Owing to the commutativity of momenta \\
    ~~(~$\mu_{\hat{p}_\mu}={1\over 2} \theta^{-1}_{\mu\nu}\ad_{\hat{x}^\nu}\eqv {1\over 2}\theta^{-1}_{\mu\nu}(\mu_{\hat{x}^\nu}^L-\mu_{\hat{x}^\nu}^R)$~) and the principle of minimal coupling, gauge fields (including Yang-Mills and Gravity fields) may be associated with the commutative sector $\A_0(\mathbb{R}^D)$.

If gauge fields are commutative while matter fields are noncommutative then the matter-gauge interaction terms will inherit (via the choice of covariant derivative $D_\mu$) a twist factor from the matter sector meanwhile the pure gauge interaction terms will lack this factor leading to \\
~~$S=T(e^{-iS_I}e^{-\frac{i}{2}\overleftarrow{\txt{ad}}_{P_0} \theta^{0i} P^{\textrm{in}}_{i}})\neq S_0,$~ where
$P^{\txt{in}}_i$ represents the anticipated total incident momentum when the matrix elements $\lang f|S|i\rang$ of $S$ are finally taken. This will lead to $P,CPT$ noninvariance of the $S$-operator.

\item The direct Poincare transformation of products of deformed or twisted quantum fields $\phi=\phi_0 e^{{1\over 2}\ola{\del}\wedge P}$ takes into account the use of the coproduct to transform local products of fields. The $S$ operator $S=T e^{-i\int\H_I}$ is invariant under this transformation even though the causality or locality condition, which is required for Lorentz invariance in commutative theories, does not hold:
    \bea
    &&[\H_I(x),\H_I(y)]\neq 0~~~~ \txt{for}~~~~ (x_0-y_0)^2<(\vec{x}-\vec{y})^2.
    \eea

\end{enumerate}
\chapter{Introduction to Noncommutative geometry}\label{ch:two}
We give an introductory review of quantum physics on the noncommutative spacetime called the Groenewold-Moyal plane. Basic ideas like star products, twisted statistics, second quantized fields and discrete symmetries are discussed. We also outline some of the recent developments in these fields and mention where one can search for experimental signals.

\section{Introduction}
Quantum electrodynamics is not free from divergences. The calculation of Feynman diagrams involves a cut-off $\Lambda$ on the momentum variables in the integrands. In this case, the theory will not see length scales smaller than $\Lambda^{-1}$. The theory fails to explain physics in the regions of spacetime volume less than $\Lambda^{-4}$.

Heisenberg proposed in the 1930's that an effective cut-off can be introduced in quantum field theories by introducing an effective lattice structure for the underlying spacetime. A lattice structure of spacetime takes care of the divergences in quantum field theories, but a lattice breaks Lorentz invariance.

Heisenberg's proposal to obtain an effective lattice structure was to make the spacetime noncommutative. The noncommutative spacetime structure is point-less on small length scales. Noncommuting spacetime coordinates introduce a fundamental length scale. This fundamental length  can be taken to be of the order of the Planck length. The notion of point below this length scale has no operational meaning.

We can explain Heisenberg's ideas by recalling the quantization of a classical system. The point of departure from classical to quantum physics is the algebra of functions on the phase space. The classical phase space, a symplectic manifold $M$, consists of ``points" forming the pure states of the system. Every observable physical quantity on this manifold $M$ is specified by a function $f$. The Hamiltonian $H$ is a function on $M$, which measures energy. The evolution of $f$ on the manifold is specified by $H$ by the equation
\bea
\dot{f}=\{f, H\}
\eea
where $\dot{f} = df/dt$ and $\{ \; ,\;  \}$ is the Poisson bracket.

The quantum phase space is a ``noncommutative space" where the algebra of functions is replaced by the algebra of linear operators. The algebra $ {\mathcal{F}}(T^*Q) $ of functions on the classical phase space $T^*Q$, associated with a given spacetime $Q$, is a commutative algebra. According to Dirac, quantization can be achieved by replacing a function $f$ in this algebra by an operator $ \hat f $ and equating $i\hbar$ times the Poisson bracket between functions to the commutator between the corresponding operators. In classical physics, the functions $f$ commute, so $ {\mathcal{F}}(T^*Q)$ is a commutative algebra. But the corresponding quantum algebra $\hat{\mathcal{F}}$ is not commutative. Dynamics is on $\hat{\mathcal{F}}$. So quantum physics is $\textit{noncommutative dynamics}$.

A particular aspect of this dynamics is $\textit{fuzzy}$ $\textit{phase space}$ where we cannot localize points, and which has an attendent effective ultraviolet cutoff. A fuzzy phase space can still admit the action of a continuous symmetry group such as the spatial rotation group as the automorphism group \cite{madore}. For example, one can quantize functions on a sphere $S^{2}$ to obtain a fuzzy sphere \cite{fuzzybook}. It admits $SO(3)$ as an automorphism group. The fuzzy sphere can be identified with the algebra $M_{n}$ of $n \times n$ complex matrices. The volume of phase space in this case becomes finite. Semiclassically there are a finite number of cells on the fuzzy sphere, each with a finite area \cite{madore}.

Thus in quantum physics, the commutative algebra of functions on phase space is deformed to a noncommutative algebra, leading to a ``noncommutative phase space''. Such deformations, characteristic of quantization, are now appearing in different approaches to fundamental physics. Examples are the following:

1.) Noncommutative geometry has made its appearance as a method for regularizing quantum field theories (qft's) and in studies of deformation quantization.

2.) It has turned up in string physics as quantized $D$-branes.

3.) Certain approaches to canonical gravity \cite{Aschieri} have used noncommutative geometry with great effectiveness.

4.) There are also plausible arguments based on the uncertainty principle \cite{doplicher} that indicate a noncommutative spacetime in the presence of gravity.

5.) It has been conjuctered by `t Hooft \cite{thooft} that the horizon of a black hole should have a fuzzy 2-sphere structure to give a finite entropy.

6.) A noncommutative structure emerges naturally in quantum Hall effect \cite{qhe}.

\section{Noncommutative Spacetime}
\subsection{A Little Bit of History}
The idea that spacetime geometry may be noncommutative is old and goes back as far as the 30's. In 1947 Snyder used the noncommutative structure of spacetime to introduce a small length scale cut-off in field theory without breaking Lorentz invariance \cite{Snyder}. In the same year, Yang \cite{yang} also published a paper on quantized spacetime, extending Snyder's work. The term `noncommutative geometry' was introduced by von Neumann \cite{madore}. He used it to describe in general a geometry in which the algebra of noncommuting linear operators replaces the algebra of functions.

Snyder's idea was forgotten with the successful development of the renormalization program. Later, in the 1980's Connes \cite{connesA} and Woronowicz \cite{woronowicz} revived noncommutative geometry by introducing a differential structure in the noncommutative framework.

\subsection{Spacetime Uncertaintities}
It is generally believed that the picture of spacetime as a manifold of points breaks down at distance scales of the order of the Planck length: Spacetime events cannot be localized with an accuracy given by Planck length.

The following argument can be found in Doplicher {\it et al.} \cite{doplicher}. In order to probe physics at a fundamental length scale $L$ close to the Planck scale, the Compton wavelength $\frac{\hbar}{Mc}$ of the probe must fulfill
\begin{equation}
\frac{\hbar}{Mc}\,\leq\, L\ \ \textrm{or}\ \
M\,\geq\,\frac{\hbar}{Lc}\,\simeq\,\textrm{Planck mass}.
\end{equation}
Such high mass in the small volume $L^3$ will strongly affect gravity and can cause black holes and their horizons to form. This suggests a fundamental length limiting spatial localization. That is, there is a space-space uncertainty,
\bea
\Delta x_{1}\Delta x_{2} + \Delta x_{2}\Delta x_{3} + \Delta x_{3}\Delta x_{1} \ \gtrsim L^{2}
\eea

Similar arguments can be made about time localization. Observation of very short time scales requires very high energies. They can produce black holes and black hole horizons will then limit spatial resolution suggesting
\bea
\Delta x_{0}(\Delta x_{1} + \Delta x_{2} + \Delta x_{3}) \geq L^{2}.
\eea

The above uncertainty relations suggest that spacetime ought to be described as a noncommutative manifold just as classical phase space is replaced by noncommutative phase space in quantum physics which leads to Heisenberg's uncertainty relations. The points on the classical commutative manifold should then be replaced by states on a noncommutative algebra.
\subsection{The Groenewold-Moyal Plane}
The noncommutative Groenewold-Moyal (GM) spacetime is a deformation of ordinary spacetime in which the spacetime coordinate functions $\widehat{x}_\mu$ do not commute \cite{ConnesC, Madore, Landi, Bondia}:
\bea
\label{non-commu-coordinates}
[\widehat{x}^{\mu}, \widehat{x}^{\nu}]=i \theta^{\mu \nu},~~~\theta^{\mu \nu}=-\theta^{\nu \mu}=\textrm{constants},
\eea
where the coordinate functions $\widehat{x}_\mu$ give Cartesian coordinates $x_{\mu}$ of (flat) spacetime:
\bea
\widehat{x}_{\mu}(x)=x_{\mu}.
\eea
The deformation matrix $\theta$ is taken to be a real and antisymmetric constant matrix \cite{RGCai}. Its elements have the dimension of (length)$^{2}$, thus a scale for the smallest patch of area in the ${\mu}$ - ${\nu}$ plane. They also give a measure of the strength of noncommutativity. One cannot probe spacetime with a resolution below this scale. That is, spacetime is ``fuzzy'' \cite{Ydri} below this scale. In the limit $\theta_{\mu \nu} \to 0$, one recovers ordinary spacetime.
\section{The Star Products}
In this part we will go into more details of the GM plane. The GM plane incorporates spacetime uncertainties. Such an introduction of spacetime noncommutativity replaces point-by-point multiplication of two fields by a type of ``smeared'' product. This type of product is called a star product.
\subsection{Deforming an Algebra}
There is a general way of deforming the algebra of functions on a manifold $M$ \cite{queiroz}. The GM plane, ${\cal A}_{\theta}({\mathbb R}^{d+1})$, associated with spacetime ${\mathbb R}^{d+1}$ is an example of such a deformed algebra.

Consider a Riemannian manifold $(M,g)$ with metric $g$. If the group \\
$\R^N\;(N\geq 2)$ acts as a group of isometries on $M$, then it acts on the Hilbert space $L^2(M,d\mu_g)$ of square integrable functions on $M$. The volume form $d\mu_g$ for the scalar product on $L^2(M,d\mu_g)$ is induced from $g$.

If $\Big\{\lambda = (\lambda_1, \ldots, \lambda_N)\Big\}$ denote the unitary irreducible representations (UIR's) of $\R^N$, then we can write
\begin{equation}\label{eq:1}
L^2(M, d\mu_g) = \bigoplus_\lambda {\cal H}^{(\lambda)} \;,
\end{equation}
where $\R^N$ acts by the UIR $\lambda$ on ${\cal H}^{(\lambda)}$.

We choose $\lambda$ such that
\bea
\lambda : a \longrightarrow e^{i \lambda a}
\eea
where $a=(a_{1}, a_{2}, \cdots, a_{N}) \in {\mathbb R}^{N}$.

Choose two smooth functions $f_{\lambda}$ and $f_{\lambda '}$ in ${\cal H}^{(\lambda)}$ and ${\cal H}^{(\lambda ')}$. Then under the pointwise multiplication
\bea
f_{\lambda} \otimes f_{\lambda '} \rightarrow f_{\lambda} f_{\lambda '}
\eea
where, if $p$ is a point on $M$,
\bea
(f_{\lambda} f_{\lambda '})(p) = f_{\lambda}(p) f_{\lambda '}(p).
\eea
Also
\bea
\label{eq:2}
f_{\lambda} f_{\lambda '} \in {\cal H}^{(\lambda + \lambda ')}
\eea
where we have taken the group law as addition.

Let $\theta^{\mu \nu}$ be an antisymmetric constant matrix in the space of UIR's of ${\mathbb R}^N$. The above algebra with pointwise multiplication can be deformed into a new deformed algebra. The pointwise product becomes a $\theta$ dependent ``smeared" product $*_{\theta}$ in the deformed algebra,
\bea
\label{eq:3}
f_\lambda *_\theta f_{\lambda'} =  f_\lambda \;  f_{\lambda'} \; e^{-\frac{i}{2} \lambda_\mu \theta^{\mu \nu} \lambda'_\nu}\;.
\eea
This deformed algebra is also associative because of eqn. (\ref{eq:2}). The GM plane, ${\cal A}_\theta({\mathbb R}^{d+1})$, is a special case of this algebra.

In the case of the GM plane, the group ${\mathbb R}^{d+1}$ acts on \\
${\cal A}_{\theta}({\mathbb R}^{d+1})$ $\{={\cal C}^\infty({\mathbb R}^{d+1}) \; \textrm{as a set} \}$ by translations leaving the flat Euclidean metric invariant. The IRR's are labelled by the ``momenta" $\lambda = p = (p^{0}, p^{1}, \ldots, p^{d})$. A basis for the Hilbert space ${\cal H}^{(p)}$ is formed by plane waves $e_{p}$ with $e_{p}(x) = e^{-ip_{\mu}  x^{\mu}}$, $x = (x^0,x^1, \ldots, x^d)$ being a point of ${\mathbb R}^{d+1}$. The $*$-product for the GM plane follows from eqn. (\ref{eq:3}),
\bea
\label{eq:4}
e_{p} \ast_\theta e_{q} = e_{p} \; e_{q} \; e^{-\frac{i}{2}p_\mu \theta^{\mu \nu}q_\nu}\;.
\eea
This $*$-product defines the Moyal plane ${\cal A}_\theta({\mathbb R}^{d+1})$.

In the limit $\theta_{\mu \nu} \rightarrow 0$, the operators $e_{p}$  and $e_{q}$ become commutative functions on ${\mathbb R}^{N}$.

\subsection{The Voros and Moyal Star Products}
This section is based on the book \cite{fuzzybook}.

The algebra ${\cal A}_{0}$ of smooth functions on a manifold $M$ under point-wise multiplication is a  commutative algebra. In the previous section we saw that ${\cal A}_{0}$ can be deformed into a new algebra ${\cal A}_{\theta}$ in which the point-wise product is deformed to a noncommutative (but still associative) product called the $*$-product.

Such deformations were studied by Weyl, Wigner, Groenewold and Moyal \cite{lipkin, weyl, grosse-pres}. The $*$-product has a central role in many discussions of noncommutative geometry. It appears in other branches of physics like quantum optics.

The $*$-product can be obtained from the algebra of creation and annihilation operators. It is explained below.
\subsubsection{Coherent States}
The dynamics of a quantum harmonic oscillator most closely resembles that of a classical harmonic oscillator when the oscillator quantum state is a coherent state. Consider a quantum oscillator with annihilation and creation operators $a$, $a^\dagger,~aa^\dagger=a^\dagger a+1$. The coherent states $|z\rangle$ defined by
\bea
a |z \rangle = z |z \rangle
\eea
are given by
\bea
|z \rangle = e^{z a^\dagger - {\bar z} a} |0 \rangle = e^{-\frac{1}{2}|z|^2} e^{z a^\dagger} |0 \rangle \,, \quad z \in {\mathbb C}.\nn
\eea
They also have the property
\bea
\quad \quad \langle z^\prime | z \rangle = e^{\frac{1}{2} |z - z^\prime|^2} \,.
\label{eq:csp1}
\eea

The coherent states are overcomplete, with the resolution of identity
\bea
{\bf 1} = \int \frac{d^2 z}{\pi} |z \rangle \langle z|  \,, \quad d^2 z
= dx_1 dx_2 \,,
\label{eq:overcomp}
\eea
where
\bea
z=\frac{x_1 + i x_2}{\sqrt2} \,.\nn
\eea

Consider an operator ${\hat A}$. The ``symbol" (or ``representation'') of ${\hat A}$ is a function $A$ on ${\mathbb C}$ with values $A ( z \,, {\bar z}) = \langle z| {\hat A} | z \rangle$. A central property of coherent states is that an operator ${\hat A}$ is determined just by its diagonal matrix elements, that is, by the symbol $A$ of ${\hat A}$.
\subsubsection{The Coherent State or Voros $*$-product on the GM Plane}
As indicated above, we can map an operator $\hat A$ to a function $A$ using coherent states as follows:
\bea
{\hat A} \longrightarrow A \,, \quad A(z \,, {\bar z}) = \langle z | {\hat
  A} |z \rangle.
\eea
This is a bijective linear map and induces a product $*_C$ on functions ($C$ indicating ``coherent state''). With this product, we get an algebra $(C^\infty({\mathbb C}) \,, *_C)$ of functions. Since the map ${\hat A} \rightarrow A$ has the property $({\hat A})^* \rightarrow A^* \equiv {\bar A}$, this map is a $*$-morphism from operators to $(C^\infty({\mathbb C}) \,, *_C)$ where $*$ on functions is complex conjugation.

Let us get familiar with this new function algebra.

The image of $a$ is the function $\alpha$ where $\alpha(z\,,{\bar z}) =z$. The image of $a^n$ has the value $z^n$ at $(z \,, {\bar z})$, so by definition,
\bea
(\alpha *_C \alpha \ldots *_C \alpha) (z \,, {\bar z}) = z^n \,.
\eea

The image of $a^* \equiv a^\dagger$ is ${\bar \alpha}$ where ${\bar \alpha}(z, {\bar z}) = {\bar z}$ and that of $(a^*)^n$ is ${\bar \alpha} *_C {\bar \alpha} \cdots *_C {\bar \alpha}$ where
\bea
{\bar \alpha} *_C {\bar \alpha} \cdots *_C {\bar \alpha}(z\,, {\bar
  z}) = {\bar z}^n \,.
\eea

Since $\langle z | a^* a | z \rangle = {\bar z} z$ and $\langle z | a a^* | z \rangle = {\bar z} z + 1$, we get
\bea
{\bar \alpha} *_C \alpha = {\bar \alpha} \alpha \,, \quad \quad
\alpha *_C {\bar \alpha} = \alpha {\bar \alpha} + {\bf 1} \,,
\eea
where $ {\bar \alpha} \alpha =  \alpha {\bar \alpha}$ is the pointwise product of $\alpha$ and ${\bar \alpha}$, and ${\bf 1}$ is the constant function with value $1$ for all $z$.

For general operators ${\hat f}$, the construction proceeds as follows. Consider
\bea
: e^{\xi a^\dagger - {\bar \xi} a}:
\eea
where the normal ordering symbol $: \cdots :$ means as usual that $a^\dagger$'s are to be put to the left of $a$'s. Thus
\bea
: a a^\dagger a^\dagger a : &=& a^\dagger a^\dagger a a \,, \nonumber \\
: e^{\xi a^\dagger - {\bar \xi} a}: &=& e^{\xi a^\dagger} e^{-{\bar
      \xi} a} \,. \nn
\eea

Hence
\bea
\langle z | :e^{\xi a^\dagger - {\bar \xi} a}: |z \rangle = e^{\xi
  {\bar z} - {\bar \xi} z} \,.
\eea

Writing ${\hat f}$ as a Fourier transform,
\bea
{\hat f} = \int \frac{d^2 \xi}{\pi} : e^{\xi a^\dagger - {\bar \xi}
  a}: {\tilde f}(\xi \,, {\bar \xi}) \,, \quad \quad {\tilde f}
(\xi \,, {\bar \xi}) \in {\mathbb C} \,,
\eea
its symbol is seen to be
\bea
f = \int \frac{d^2 \xi}{\pi} e^{\xi {\bar z} - {\bar \xi} z} {\tilde
  f}(\xi \,, {\bar \xi}) \,.
\eea

This map is invertible since $f$ determines ${\tilde f}$. Consider also the second operator
\bea
{\hat g} = \int \frac{d^2 \eta}{\pi} : e^{\eta a^\dagger - {\bar \eta}
  a}: {\tilde g}(\eta \,, \bar {\eta}) \,,
\eea
and its symbol
\bea
g = \int \frac{d^2 \eta}{\pi} e^{\eta {\bar z} - {\bar \eta} z}
{\tilde g}(\eta \,, \bar {\eta}) \,.
\eea

The task is to find the symbol $f *_C g$ of ${\hat f}{\hat g}$. Let us first find
\bea
e^{\xi {\bar z} - {\bar \xi} z} *_C  e^{\eta {\bar z} - {\bar \eta} z} \,.
\eea

We have
\bea
:e^{\xi a^\dagger - {\bar \xi} a}: \,  : e^{\eta a^\dagger - {\bar
      \eta} a}: = : e^{\xi a^\dagger - {\bar \xi} a} \,
e^{\eta a^\dagger - {\bar \eta} a}: e^{-{\bar \xi}{\eta}}
\eea
and hence
\bea
\label{eq:exp1}
e^{\xi {\bar z} - {\bar \xi} z} *_C e^{\eta {\bar z} - {\bar \eta} z}
&=& e^{-{\bar \xi} \eta} e^{\xi {\bar z} - {\bar \xi} z} \,
e^{\eta {\bar z} - {\bar \eta} z} \nonumber \\
&=& e^{\xi {\bar z} - {\bar \xi} z} e^{{\overleftarrow \partial}_z \,
  {\overrightarrow \partial}_{\bar z}}
e^{\eta {\bar z} - {\bar \eta} z} \,.
\eea

The bidifferential operators $\big ({\overleftarrow \partial}_z \, {\overrightarrow \partial}_{\bar z} \big )^k \,, (k= 1,2,...)$ have the definition
\bea
\alpha \big ({\overleftarrow \partial}_z \, {\overrightarrow
  \partial}_{\bar z} \big )^k \beta \, (z \,, {\bar z}) =
\frac{\partial^k \alpha (z \,, {\bar z})}{\partial z^k}
\frac{\partial^k \beta (z \,, {\bar z})}{\partial {\bar z}^k} \,.
\eea

The exponential in (\ref{eq:exp1}) involving them can be defined using the power series.

The coherent state $*$-product $f *_C g$ follows from (\ref{eq:exp1}):
\bea
f *_C g \,(z \,, {\bar z}) = \big ( f  e^{{\overleftarrow \partial}_z
  \, {\overrightarrow \partial}_{\bar z}} g \big ) (z \,, {\bar z})  \,.
\label{eq:csstar1}
\eea

We can explicitly introduce a deformation parameter $\theta > 0 $ in the discussion by changing (\ref{eq:csstar1}) to
\bea
f *_C g \, (z \,, {\bar z}) = \big ( f  e^{ \theta \, {\overleftarrow
    \partial}_z \, {\overrightarrow \partial}_{\bar z}} g \big )
(z \,, {\bar z}) \,.
\label{eq:csstar2}
\eea

After rescaling $z^\prime = \frac{z}{\sqrt{\theta}}$, (\ref{eq:csstar2}) gives (\ref{eq:csstar1}). As $z^\prime$ and ${\bar z}^\prime$ after quantization become $a \,, a^\dagger$, $z$ and ${\bar z}$ become the scaled oscillators $a_\theta \,, a_\theta^\dagger$
\bea
\lbrack a_\theta \,, a_\theta \rbrack = \lbrack a_\theta^\dagger  \,,
a_\theta^\dagger \rbrack = 0 \,, \quad  \lbrack a_\theta \,,
a_\theta^\dagger \rbrack = \theta \,.
\label{eq:tetacom}
\eea

Equation (\ref{eq:tetacom}) is associated with the Moyal plane with Cartesian coordinate functions $x_1 \,, x_2$. If $a_\theta = \frac{x_1 + i x_2}{\sqrt2} \,, a_\theta^\dagger = \frac{x_1 - i x_2}{\sqrt2}$,
\bea
\lbrack x_i \,, x_j \rbrack = i \theta \varepsilon_{ij} \,, \quad
\varepsilon_{ij} = - \varepsilon_{ji} \,, \quad \varepsilon_{12} = 1 \,.
\label{eq:deform1}
\eea

The Moyal plane is the plane ${\mathbb R}^2$, but with its function algebra deformed in accordance with eqn. (\ref{eq:deform1}). The deformed algebra has the product eqn. (\ref{eq:csstar2}) or equivalently the Moyal product derived below.
\subsubsection{The Moyal Product on the GM Plane}
We get this by changing the map ${\hat f} \rightarrow f$ from operators to functions. For a given function $f$, the operator ${\hat f}$ is thus different for the coherent state and Moyal $*$'s. The $*$-product on two functions is accordingly also different.

Let us introduce the Weyl map and the Weyl symbol. The Weyl map of the operator
\bea
{\hat f} = \int \frac{d^2 \xi}{\pi} {\tilde f}(\xi \,, {\bar \xi})
e^{\xi a^\dagger - {\bar \xi} a}
\label{eq:Weyl1}
\eea
to the function $f$ is defined by
\bea
f(z\,,{\bar z}) = \int \frac{d^2 \xi}{\pi} {\tilde f}(\xi \,,
{\bar \xi}) e^{\xi {\bar z} - {\bar \xi} z} \,.
\label{eq:Weyl2}
\eea

Equation (\ref{eq:Weyl2}) makes sense since ${\tilde f}$ is fully determined by ${\hat f}$ as follows:
\bea
\langle z| {\hat f} | z \rangle = \int \frac{d^2 \xi}{\pi} {\tilde
  f}(\xi \,, \bar {\xi}) e^{-\frac{1}{2} \xi {\bar \xi} }
e^{\xi {\bar z} - {\bar \xi} z} \,.    \nn
\eea
${\tilde f}$ can be calculated from here by Fourier transformation.

The map is invertible since ${\tilde f}$ follows from $f$ by the Fourier transform of eqn. (\ref{eq:Weyl2}) and ${\tilde f}$ fixes ${\hat f}$ by eqn. (\ref{eq:Weyl1}). $f$ is called the {\it Weyl symbol} of ${\hat f}$.

As the Weyl map is bijective, we can find a new $*$ product, call it $*_W$, between functions by setting $f*_W g = \, \mbox{Weyl symbol of} \, {\hat f}{\hat g}$.

For
\bea
{\hat f}(\xi, \bar{\xi}) =  e^{\xi a^\dagger - {\bar \xi} a} \,, \quad {\hat g}(\eta, \bar{\eta}) =
e^{\eta a^\dagger - {\bar \eta} a} \,,\nn
\eea
to find $f*_W g$, we first rewrite ${\hat f}{\hat g}$ according to
\bea
{\hat f}{\hat g} = e^{\frac{1}{2}(\xi {\bar \eta} - {\bar \xi} \eta)}  e^{(\xi +
  \eta) a^\dagger - ({\bar \xi} +{\bar \eta}) a} \,.\nn
\eea
Hence
\bea
f*_W g \,(z\,, {\bar z}) &=& e^{\xi {\bar z}-{\bar \xi} z}
e^{\frac{1}{2}(\xi {\bar \eta} - {\bar \xi} \eta)}
e^{\eta {\bar z}-{\bar \eta} z}
\nonumber \\
&=& f e^{\frac{1}{2} \big ( {\overleftarrow \partial}_z \,
{\overrightarrow \partial}_{\bar z} - {\overleftarrow \partial}_{\bar
  z} \, {\overrightarrow \partial}_z
\big )} g \, (z \,,{\bar z}) \,.
\label{eq:Weyl3}
\eea

Multiplying by ${\tilde f}$, ${\tilde g}$ and integrating, we get eqn. (\ref{eq:Weyl3}) for arbitrary functions:
\bea
f*_W g \, (z\,, {\bar z}) = \Big ( f e^{\frac{1}{2} \big (
  {\overleftarrow \partial}_z \, {\overrightarrow \partial}_{\bar z} -
{\overleftarrow \partial}_{\bar z} \, {\overrightarrow \partial}_z
\big )} g \Big ) (z \,,{\bar z}) \,.
\eea

Note that
\bea
{\overleftarrow \partial}_z \, {\overrightarrow \partial}_{\bar z}
-{\overleftarrow \partial}_{\bar z} \, {\overrightarrow \partial}_z
= i ( {\overleftarrow \partial}_1 \, {\overrightarrow \partial}_2
-{\overleftarrow \partial}_2 \, {\overrightarrow \partial}_1 )
= i \varepsilon_{ij}  {\overleftarrow \partial}_i \, {\overrightarrow
  \partial}_j  \,.\nn
\eea

Introducing also $\theta$, we can write the $*_W$-product as
\bea
f *_W g = f e^{i \frac{\theta}{2} \varepsilon_{ij}  {\overleftarrow \partial}_i
  \, {\overrightarrow \partial}_j} g \,.
\label{eq:Weyl4}
\eea

By eqn. (\ref{eq:deform1}), $\theta \varepsilon_{ij} = \omega_{ij}$ fixes the Poisson brackets, or the Poisson structure on the Moyal plane. Eqn. (\ref{eq:Weyl4}) is customarily written as
\bea
f *_W g = f  e^{\frac{i}{2} \omega_{ij}  {\overleftarrow \partial}_i \,
  {\overrightarrow \partial}_j} g \nn
\eea
using the Poisson structure. (But we have not cared to position the indices so as to indicate their tensor nature and to write $\omega^{ij}$.)
\subsection{Properties of the $*$-Products}
A $*$-product without a subscript indicates that it can be either a $*_C$ or a $*_W$.
\subsubsection{Cyclic Invariance}
The trace of operators, $\textrm{Tr}:\hat{A}\mapsto\int{d^2z\over\pi}\langle z|\hat{A}|z\rangle$, has the fundamental property $Tr {\hat A} {\hat B} = Tr {\hat B} {\hat A}$, which leads to the general cyclic identities
\bea
Tr \, {\hat A}_1 \ldots {\hat A}_n = Tr \, {\hat A}_n {\hat A}_1
\ldots {\hat A}_{n-1} \,.
\label{eq:ctr1}
\eea

We now show that
\bea
Tr \, {\hat A} {\hat B} = \int \frac{d^2 z}{\pi} \,  A * B \, (z \,,
{\bar z}) \,, \quad \quad * = *_C \quad \mbox{or} \quad *_W \,.
\label{eq:ctr2}
\eea

(The functions on the right hand side are different for $*_C$ and $*_W$ if ${\hat A} \,, {\hat B}$ are fixed). From this follows the analogue of (\ref{eq:ctr1}):
\bea
\int \frac{d^2 z}{\pi} \, \big ( A_1 * A_2 * \cdots * A_n) \, (z \,, {\bar
  z} \big ) = \int \frac{d^2 z}{\pi} \big ( A_n * A_1 * \cdots *
A_{n-1}) \, (z \,, {\bar z} \big ) \,.
\label{eq:ctr3}
\eea

For $*_C$, eqn. (\ref{eq:ctr2}) follows from eqn. (\ref{eq:overcomp}). The coherent state image of $ e^{\xi a^\dagger - {\bar \xi} a} $ is the function with value
\bea
e^{\xi {\bar z} - {\bar \xi} z} e^{-\frac{1}{2}{\bar \xi}{\xi}}
\label{eq:csf1}
\eea
at $z$, with a similar correspondence if $\xi \rightarrow \eta$. So
\bea
Tr  \, e^{\xi a^\dagger - {\bar \xi} a} \, e^{\eta a^\dagger - {\bar
    \eta} a} = \int {\frac{d^2 z}{\pi}} \, \Big (
e^{\xi {\bar z} - {\bar \xi} z} e^{-\frac{1}{2}{\bar \xi}{\xi}} \Big)
    \Big ( e^{\eta {\bar z} - {\bar \eta} z}
e^{-\frac{1}{2}{\bar \eta}{\eta}} \Big ) e^{-{\bar \xi}{\eta}}\nn
\eea

The integral produces the $\delta$-function
\bea
\prod_i 2 \delta (\xi_i + \eta_i) \,, \quad \quad  \xi_i = \frac{\xi_1 +
  \xi_2}{\sqrt{2}} \,, \quad \eta_i = \frac{\eta_1 + \eta_2} {\sqrt{2}} \,.\nn
\eea

We can hence substitute $e^{- \big ( \frac{1}{2}{\bar \xi}{\xi} +  \frac{1}{2}{\bar \eta} {\eta} + {\bar \xi}{\eta} \big)}$ by $e^{\frac{1}{2} (\xi {\bar \eta} - {\bar \xi} \eta)}$ and get eqn. (\ref{eq:ctr2}) for Weyl $*$ for these exponentials and so for general functions by using  eqn. (\ref{eq:Weyl1}).
\subsubsection{A Special Identity for the Weyl Star}
The above calculation also gives the identity
\bea
\int \frac{d^2 z}{\pi} A *_W B \, (z \,, {\bar z}) = \int \frac{d^2
  z}{\pi} A (z \,, {\bar z}) \, B \, (z \,, {\bar z}) \,.\nn
\eea

That is because
\bea
\prod_i \delta(\xi_i + \eta_i) \, e^{\frac{1}{2} (\xi {\bar \eta} -
{\bar \xi} \eta)} = \prod_i \, \delta(\xi_i + \eta_i) \,.\nn
\eea

In eqn. (\ref{eq:ctr3}), $A$ and $B$ in turn can be Weyl $*$-products of other functions. Thus in integrals of Weyl $*$-products of functions, one $*_W$ can be replaced by the pointwise (commutative) product:
\bea
&&\int \frac{d^2 z}{\pi} \big ( A_1 *_W A_2 *_W \cdots A_K \big ) *_W
  ( B_1 *_W B_2 *_W \cdots B_L \big ) \, (z \,, {\bar z})
\nonumber \\
&& \quad \quad \quad \quad = \int \frac{d^2 z}{\pi} \big ( A_1 *_W A_2
*_W \cdots A_K \big ) \,( B_1 *_W B_2 *_W  \cdots B_L \big ) \, (z \,,
{\bar z}) \,.\nn
\eea

This identity is frequently useful.
\subsubsection{Equivalence of $*_C$ and $*_W$}
For the operator
\bea
{\hat A} = e^{\xi a^\dagger -{\bar \xi} a} \,,
\eea
the coherent state function $A_C$ has the value (\ref{eq:csf1}) at $z$, and the Weyl symbol $A_W$ has the value
\bea
A_W(z \,, {\bar z}) = e^{\xi {\bar z} - {\bar \xi} z} \,.\nn
\eea

As both $\big ( C^\infty({\mathbb R}^2) \,, *_C \big )$ and $\big (C^\infty({\mathbb R}^2) \,, *_W \big )$ are isomorphic to the operator algebra, they too are isomorphic. The isomorphism is established by the maps
\bea
A_C \longleftrightarrow A_W \nn
\eea
and their extension via Fourier transform to all operators and functions ${\hat A} \,, A_{C \,, W}$.

Clearly
\bea
A_W = e^{-\frac{1}{2} \partial_z \partial_{\bar z}} A_C \,, \quad
A_C = e^{\frac{1}{2} \partial_z \partial_{\bar z}} A_W  \,, \nonumber \\
~~A_C *_C B_C \longleftrightarrow A_W *_W B_W \,.\nn
\eea

The mutual isomorphism of these three algebras is a $*$-isomorphism since
$({\hat A} {\hat B})^\dagger \longrightarrow {\bar B}_{C \,, W} *_{C \,, W} {\bar A}_{C \,, W}$.
\subsubsection{Integration and Tracial States}
This is a good point to introduce the ideas of a state and a tracial state on a $*$-algebra ${\cal A}$ with unity ${\bf 1}$.

A state $\omega$ is a linear map from ${\cal A}$ to ${\mathbb C}$, $\omega (a) \in {\mathbb C}$ for all $a \in {\cal A}$ with the following properties:
\bea
\omega(a^*) &=& \overline{\omega(a)} \,, \nonumber \\
\omega (a^*a) & \geq & 0 \,, \nonumber \\
\omega({\bf 1}) & = & 1 \,.\nn
\label{eq:ts1}
\eea

If ${\cal A}$ consists of operators on a Hilbert space and $\rho$ is a density matrix, it defines a state $\omega_\rho$ via
\bea
\omega_\rho (a) =  Tr (\rho a) \,.
\label{eq:ts2}
\eea

If $\rho = e^{- \beta H}/ Tr (e^{-\beta H})$ for a Hamiltonian $H$, it gives a Gibbs state via eqn. (\ref{eq:ts2}).

Thus the concept of a state on an algebra ${\cal A}$ generalizes the notion of a density matrix. There is a remarkable construction, the Gel'fand- Naimark-Segal (GNS) construction, which shows how to associate any state with a rank-$1$ density matrix \cite{Haag}.

A state is {\it tracial} if it has cyclic invariance:
\bea
\omega (ab) = \omega (ba) \,.
\label{eq:tracial1}
\eea

The Gibbs state is not tracial, but fulfills an identity generalizing eqn. (\ref{eq:tracial1}). It is a Kubo-Martin-Schwinger (KMS) state \cite{Haag}.

A positive map $\omega^\prime$ is in general an unnormalized state: It must fulfill all the conditions that a state fulfills, but is not obliged to fulfill the condition $\omega^\prime({\bf 1}) = 1$.

Let us define a positive map $\omega^\prime$ on $(C^\infty({\mathbb R}^{2}) \,, *)$ ($*
= *_C \, \mbox{or} \, *_W$) using integration:
\bea
\omega^\prime(A) = \int \frac{d^2 z}{\pi} \, {\hat A}
(z \,, {\bar z}) \,.\nn
\eea

It is easy to verfy that $\omega^\prime$ fulfills the properties of a positive map. A {\it tracial} positive map $\omega^\prime$ also has the cyclic invariance, eqn. (\ref{eq:tracial1}).

The cyclic invariance (\ref{eq:tracial1}) of $\omega^\prime (A * B)$ means that
it is a tracial positive map.

\subsubsection{The $\theta$-Expansion}

On introducing $\theta$, we have (\ref{eq:csstar2}) and
\bea
f *_W g (z \,, {\bar z}) =  f e^{\frac{\theta}{2} \big (
  {\overleftarrow \partial}_z \, {\overrightarrow \partial}_{\bar z} -
{\overleftarrow \partial}_{\bar z} \, {\overrightarrow \partial}_z
  \big )} g \, (z \,,{\bar z}) \,.\nn
\eea

The series expansion in $\theta$ is thus
\bea
f *_C g \, (z \,, {\bar z}) = f g \, (z \,, {\bar z}) + \theta \,
\frac{\partial f}{\partial z} (z \,, {\bar z}) \frac{\partial g}
{\partial {\bar z}} (z \,, {\bar z}) + {\cal O} (\theta^2) \,,\nn
\eea
\bea
f *_W g \, (z \,, {\bar z}) =  f g (z \,, {\bar z}) + \frac{\theta}{2}
\Big ( \frac{\partial f}{\partial z} \frac{\partial g}
{\partial {\bar z}} - \frac{\partial f}{\partial {\bar z}}
\frac{\partial g} {\partial z} \Big ) \, (z \,, {\bar z}) +  {\cal O} (\theta^2) \,.\nn
\eea

Introducing the notation
\bea
\lbrack f \,, g \rbrack_* = f * g - g * f \,, \quad *=*C \quad
\mbox{or} \quad *_W \,,
\label{eq:ps1}
\eea
we see that
\bea
\lbrack f \,, g \rbrack_{*_C} &=& \theta \Big ( \frac{\partial
  f}{\partial z} \frac{\partial g} {\partial {\bar z}} -
\frac{\partial f}{\partial {\bar z}} \frac{\partial g} {\partial z}
\Big ) (z \,, {\bar z}) +  {\cal O} (\theta^2) \,, \nonumber \\
\lbrack f \,, g \rbrack_{*_W} &=& \theta \Big ( \frac{\partial
  f}{\partial z} \frac{\partial g} {\partial {\bar z}} -
\frac{\partial f}{\partial {\bar z}} \frac{\partial g} {\partial z}
\Big ) (z \,, {\bar z}) +  {\cal O} (\theta^2) \,.\nn
\eea

We thus see that
\bea
\lbrack f \,, g \rbrack_* = i \theta \{f \,, g \}_{P.B.} + {\cal O} (\theta^2) \,,
\label{eq:ps2}
\eea
where $\{f \,, g \}$ is the Poisson bracket of $f$ and $g$ and the ${\cal O}(\theta^2)$ term depends on $*_{C \,, W}$. Thus the $*$-product is an associative product that to leading order in the deformation parameter (``Planck's constant") $\theta$ is compatible with the rules of quantization of Dirac. We can say that with the $*$-product, we have deformation quantization of the classical commutative algebra of functions.

But it should be emphasized that even to leading order in $\theta$, $f*_C g$ and $f *_W g$ do not agree. Still the algebras $\big (C^\infty({\mathbb R}^2 \,, *_C) \big )$ and $\big ( C^\infty({\mathbb R}^2 \,, *_W) \big )$ are $*$-isomorphic.

If a Poisson structure on a manifold $M$ with Poisson bracket $\{. \,, .\}$ is given, then one can have a $*$-product $f * g$ as a formal power series in $\theta$ such that eqn. (\ref{eq:ps2}) holds \cite{Kontsevich}.
\section{Spacetime Symmetries on Noncommutative Plane}
In this section we address how to implement spacetime symmetries on the noncommutative spacetime algebra ${\cal A}_{\theta}({\mathbb R}^{N})$, where functions are multiplied by a $*$-product. In section 2, we modelled the spacetime noncommutativity using the commutation relations given by eqn. (\ref{non-commu-coordinates}). Those relations are clearly not invariant under naive Lorentz transformations. That is, the noncommutative structure we have modelled breaks Lorentz symmetry. Fortunately, there is a way to overcome this difficulty: one can interpret these relations in a Lorentz-invariant way by implementing a deformed Lorentz group action \cite{chaichian}.
\subsection{The Deformed Poincar\'e Group Action}
The single particle states in quantum mechanics can be identified with the carrier (or representation) space of the one-particle unitary irreducible representations (UIRR's) of the identity component of the Poincar\'e group, $P_{+}^{\uparrow}$ or rather its two-fold cover $\bar{P}^{\uparrow}_{+}$. Let $U(g)$, $g \in \bar{P}^{\uparrow}_{+}$, be the UIRR for a spinless particle of mass $m$ on a Hilbert space ${\cal H}$. Then ${\cal H}$ has the basis $\{ |k \rangle\}$ of momentum eigenstates, where $k=(k_{0}, {\bf k})$, $k_{0}=|\sqrt{{\bf k}^{2}+m^{2}}|$. $U(g)$ transforms $|k\rangle$ according to
\bea
U(g) |k\rangle = |g k\rangle.
\eea
Then conventionally $\bar{P}^{\uparrow}_{+}$ acts on the two-particle Hilbert space ${\cal H} \otimes {\cal H}$ in the following way:
\bea
U(g) \otimes U(g)~~|k\rangle \otimes |q\rangle = |g k\rangle \times |g q\rangle.
\eea
There are similar equations for multiparticle states.

Note that we can write $U(g) \otimes U(g) = [U \otimes U](g \times g)$.

Thus while defining the group action on multi-particle states, we see that we have made use of the isomorphism $G \rightarrow G \times G$ defined by $g \rightarrow g \times g$. This map is essential for the group action on multi-particle states. It is said to be a coproduct on $G$. We denote it by $\Delta$:
\bea
\Delta : G \rightarrow G \times G,
\eea
\bea
\Delta(g) = g \times g.
\eea

The coproduct exists in the algebra level also. Tensor products of representations of an algebra are in fact determined by $\Delta$ \cite{mack1, mack2}. It is a homomorphism from the group algebra (generalization of the Fourier transform, the group algebra of the group ${\mathbb{R}}^{n}$) $G^*$ to $G^* \otimes G^*$. A coproduct map need not be unique: Not all choices of $\Delta$ are equivalent. In particular the Clebsch-Gordan coefficients, that occur in the reduction of group representations, can depend upon $\Delta$. Examples of this sort occur for $\bar{P}^{\uparrow}_{+}$. In any case, it must fulfill
\bea
\label{eq:g1g2Action}
\Delta(g_{1})\Delta(g_{2}) &=& \Delta (g_{1}g_{2}), \; \; g_{1}, g_{2} \in G
\eea

Note that eqn. (\ref{eq:g1g2Action}) implies the coproduct on the group algebra $G^{*}$ by linearity. If $\alpha, \beta : G \rightarrow {\mathbb C}$ are smooth compactly supported functions on $G$, then the group algebra $G^{*}$ contains the generating elements
\bea
\int d\mu(g) \alpha(g) g,~~~~~~\int d\mu(g') \alpha(g') g',
\eea
where $d\mu$ is the measure in $G$. The coproduct action on $G^{*}$ is then
\bea
\Delta : G^{*} &\rightarrow& G^{*} \otimes G^{*} \nn \\
\int d\mu(g) \alpha(g) g &\rightarrow&  \int d\mu(g) \alpha(g) \Delta(g).
\eea

The representations $U_{k}$ of $G^{*}$ on ${\cal H}_{k} (k = i, j)$,
\bea
U_{k} : \int d\mu(g) \alpha(g) g &\rightarrow&  \int d\mu(g) \alpha(g) U_{k}(g)
\eea
induced by those of $G$ also extend to the representation $U_{i} \otimes U_{j}$ on ${\cal H}_{i} \otimes {\cal H}_{j}$:
\bea
U_{i} \otimes U_{j} : \int d\mu(g) \alpha(g) g &\rightarrow&  \int d\mu(g) \alpha(g) (U_{i} \otimes U_{j})\Delta(g).
\eea

Thus the action of a symmetry group on the tensor product of representation spaces carrying any two representations $\rho_1$ and $\rho_2$ is determined by $\Delta$:
\bea
g \triangleright (\alpha \otimes \beta) = (\rho_1 \otimes \rho_2)\Delta(g)(\alpha \otimes \beta).
\eea

If the representation space is itself an algebra ${\cal A}$, we have a rule for taking products of elements of ${\cal A}$ that involves the multiplication map $m$:
\bea
&& m : {\cal A} \otimes {\cal A} \rightarrow {\cal A},\\
&& \alpha \otimes \beta \rightarrow m(\alpha \otimes \beta)=\alpha \beta,
\eea
where $\alpha, \beta \in {\cal A}$.

It is now essential that $\Delta$ be compatible with $m$. That is
\bea
\label{eq:compatibility}
m\Big[(\rho \otimes \rho)\Delta(g)(\alpha \otimes \beta)\Big]=\rho(g)m(\alpha \otimes \beta),
\eea
where $\rho$ is a representation of the group acting on the algebra.

The compatibility condition (\ref{eq:compatibility}) is encoded in the commutative diagram:
\begin{equation}
\begin{array}{ccc}
\alpha \otimes \beta & \stackrel{g\triangleright}{\longrightarrow} & ( \rho \otimes \rho ) \Delta
(g) \alpha \otimes \beta \\
& &  \\ m \,\, \downarrow  & & \downarrow \,\, m \\ &  & \\ m(\alpha
\otimes
\beta) &
\stackrel{g\triangleright}{\longrightarrow} & \rho(g) m (\alpha \otimes \beta)
\end{array}
\end{equation}
If such a $\Delta$ can be found, $G$ is an automorphism of ${\cal A}$. In the absence of such a $\Delta$, $G$ does not act on ${\cal A}$.

Let us consider the action of $P_{+}^{\uparrow}$ on the nocommutative spacetime algebra (GM plane)  ${\cal A}_{\theta}({\mathbb R}^{d+1})$. The algebra ${\cal A}_{\theta}({\mathbb R}^{d+1})$ consists of smooth functions on ${\mathbb R}^{d+1}$ with the multiplication map
\bea
m_{\theta}: {\cal A}_{\theta}({\mathbb R}^{d+1}) \otimes {\cal A}_{\theta}({\mathbb R}^{d+1}) \rightarrow {\cal A}_{\theta}({\mathbb R}^{d+1}).
\eea
For two functions $\alpha$ and $\beta$ in the algebra ${\cal A}_{\theta}$, the multiplication map is not a point-wise multiplication, it is the $*$-multiplication:
\bea
m_{\theta} (\alpha \otimes \beta) (x) = (\alpha * \beta)(x).
\eea
Explicitly the $*$-product between two functions $\alpha$ and $\beta$ is written as
\bea
(\alpha * \beta)(x) = \textrm{exp}\Big(\frac{i}{2}\theta^{\mu \nu}\frac{\partial}{\partial x^{\mu}}\frac{\partial}{\partial y^{\nu}}\Big)\alpha(x)\beta(y)\Big|_{x=y}.
\eea

Before implementing the Poincar\'e group action on ${\cal A}_{\theta}$, we write down a useful expression for $m_{\theta}$ in terms of the commutative multiplication map $m_{0}$,
\bea
m_{\theta} = m_{0} {\cal F}_{\theta},
\eea
where
\bea
{\cal F}_{\theta} = \textrm{exp}(-\frac{i}{2}\theta^{\alpha \beta}P_{\alpha} \otimes P_{\beta}), \; \; \; P_{\alpha} = -i \partial_{\alpha}
\eea
is called the ``Drinfel'd twist" or simply the ``twist". The indices here are raised or lowered with the Minkowski metric with signature ($+, -, -, -$).

It is easy to show from this equation that the Poincar\'e group action through the coproduct $\Delta(g)$ on the noncommutative algebra of functions is not compatible with the $*$-product. That is, $P_{+}^{\uparrow}$ does not act on ${\cal A}_{\theta}({\mathbb R}^{d+1})$ in the usual way. There is a way to implement Poincar\'e symmetry on noncommuative algebra. Using the twist element, the coproduct of the universal enveloping algebra ${\cal U}({\cal P})$ of the Poincar\'e algebra can be deformed in such a way that it is compatible with the above $*$-multiplication. The deformed coproduct, denoted by $\Delta_{\theta}$ is:
\bea
\Delta_{\theta} = {\cal F}^{-1}_{\theta} \Delta {\cal F}_{\theta}
\eea

We can check compatibility of the twisted coproduct $\Delta_{\theta}$ with the twisted multiplication $m_{\theta}$ as follows
\begin{eqnarray}
m_{\theta} \left( (\rho \otimes \rho) \Delta_{\theta}(g) ( \alpha \otimes
\beta ) \right) &=&
m_{0} \left( {\cal F}_{\theta} ({\cal F}_{\theta}^{-1} \rho(g) \otimes
\rho(g) {\cal F}_{\theta}) \alpha \otimes \beta \right) \nonumber \\
&=& \rho(g) \left( \alpha * \beta \right), \quad
 \alpha,\beta \in {\cal A}_\theta(\rr^{d+1})
\label{proofcomp}
\end{eqnarray}
as required.
This compatibility is encoded in the commutative diagram
\begin{equation}
\begin{array}{ccc}
\alpha \otimes \beta & \stackrel{g \triangleright}{\longrightarrow} & ( \rho \otimes \rho ) \Delta_{\theta}
(g) \alpha \otimes \beta \\
& &  \\ m_{\theta} \,\, \downarrow  & & \downarrow \,\, m_{\theta} \\ &  & \\ \alpha * \beta &
\stackrel{g \triangleright}{\longrightarrow} & \rho(g) (\alpha * \beta)
\end{array}
\end{equation}
Thus $G$ is an automorphism of ${\cal A}_{\theta}$ if the coproduct is $\Delta_{\theta}$.

It is easy to see that the coproduct for the generators $P_{\alpha}$ of the Lie algebra of the translation group are not deformed,
\bea
\Delta_{\theta}(P_{\alpha}) = \Delta(P_{\alpha})
\eea
while the coproduct for the generators of the Lie algebra of the Lorentz group are deformed:
\bea
\Delta_{\theta}(M_{\mu \nu}) &=& 1 \otimes M_{\mu \nu} + M_{\mu \nu} \otimes 1 - \frac{1}{2}\Big[(P\cdot \theta)_{\mu}\otimes P_{\nu}-P_{\nu}\otimes (P\cdot \theta)_{\mu} - (\mu \leftrightarrow \nu) \Big], \nn \\
(P \cdot \theta)_{\lambda} &=& P_{\rho}\theta^{\rho}_{\lambda}.
\eea

The idea of twisting the coproduct in noncommutative spacetime algebra is due to \cite{chaichian, drinfeld, majid, fiore2, fiore1, fioresolo1, fioresolo2, watts1, Oeckl:2000eg, watts2, gms, Dimitrijevic:2004rf, matlock, Aschieri}. But its origins can be traced back to Drinfel'd \cite{drinfeld} in mathematics. This Drinfel'd twist leads naturally to deformed $R$-matrices and statistics for quantum groups, as discussed by Majid \cite{majid}. Subsequently, Fiore and Schupp \cite{fiore1} and Watts \cite{watts1,watts2} explored the significance of the Drinfel'd twist and $R$-matrices while Fiore \cite{fioresolo1, fioresolo2} and Fiore and Schupp \cite{fiore2}, Oeckl \cite{Oeckl:2000eg} and Grosse {\it et al.} \cite{gms} studied the importance of $R$-matrices for statistics. Oeckl \cite{Oeckl:2000eg} and Grosse {\it et al.} \cite{gms} also developed quantum field theories using different and apparently inequivalent approaches, the first on the Moyal plane and the second on the $q$-deformed fuzzy sphere. In \cite{Aschieri, Dimitrijevic:2004rf} the authors focused on the diffiomorphism group $\mathcal{D}$ and developed Riemannian geometry and gravity theories based on $\Delta_\theta$, while \cite{chaichian} focused on the Poincar\'{e} subgroup $\mathcal{P}$ of $\mathcal{D}$ and explored the consequences of $\Delta_\theta$ for quantum field theories. Twisted conformal symmetry was discussed by \cite{matlock}. Recent work, including ours \cite{bal-unitary, bal, uv-ir, bal-sasha-babar, bal-stat, cpt-paper, twistd}, has significant overlap with the earlier literature.
\subsection{The Twisted Statistics}
In the previous section, we discussed how to implement the Poincar\'e group action in the noncommutative framework. We changed the ordinary coproduct to a twisted coproduct $\Delta_{\theta}$ to make it compatible with the multiplication map $m_{\theta}$. This very process of twisting the coproduct has an impact on statistics. In this section we discuss how the deformed Poincar\'e symmetry leads to a new kind of statistics for the particles.

Consider a two-particle system in quantum mechanics for the case $\theta^{\mu \nu}=0$. A two-particle wave function is a function of two sets of variables, and lives in ${\cal A}_{0} \otimes {\cal A}_{0}$. It transforms according to the usual coproduct $\Delta$. Similarly in the noncommutative case, the two-particle wave function lives in ${\cal A}_{\theta} \otimes {\cal A}_{\theta}$ and transforms according to the twisted coproduct $\Delta_{\theta}$.

In the commutative case, we require that the physical wave functions describing identical particles are either symmetric (bosons) or antisymmetric (fermions), that is, we work with either the symmetrized or antisymmetrized tensor product,
\bea
\phi \otimes_{S} \chi &\equiv& \frac{1}{2}\left(\phi \otimes \chi
+ \chi
\otimes \phi \right),\\
\phi \otimes_{A} \chi &\equiv& \frac{1}{2}\left(\phi \otimes \chi
- \chi
\otimes \phi \right).
\eea
which satisfies
\bea
\phi \otimes_{S} \chi &=& + \chi
\otimes_{S} \phi,\\
\phi \otimes_{A} \chi &=& - \chi
\otimes_{A} \phi.
\eea
These relations have to hold in all frames of reference in a Lorentz-invariant theory. That is, symmetrization and antisymmetrization must commute with the Lorentz group action.

Since $\Delta(g)=g \times g$, we have
\bea
\tau_0 (\rho \otimes \rho) \Delta(g)=(\rho \times \rho)\Delta(g) \tau_0, \; \; g \in P_{+}^{\uparrow}
\eea
where $\tau_{0}$ is the flip operator:
\bea
\tau_{0} (\phi \otimes \chi) = \chi \otimes \phi.
\eea

Since
\bea
\phi \otimes_{S, A} \chi =\frac{1 \pm \tau_{0}}{2}~\phi \otimes \chi,
\eea
we see that Lorentz transformations preserve symmetrization and anti-symmetrization.

The twisted coproduct action of the Lorentz group is not compatible with the usual symmetrization and anti-symmetrization. The origin of this fact can be traced to the fact that the coproduct is not cocommutative except when $\theta^{\mu \nu}={0}$. That is,
\bea
\tau_{0}{\cal F}_{\theta} &=& {\cal F}_{\theta}^{-1}\tau_{0}, \\
\tau_{0} (\rho \otimes \rho) \Delta_{\theta}(g) &=& (\rho \otimes \rho)\Delta_{-\theta}(g)\tau_{0}
\eea

One can easily construct an appropriate deformation $\tau_{\theta}$ of the operator $\tau_{0}$ using the twist operator ${\cal F}_{\theta}$ and the definition of the twisted coproduct, such that it commutes with $\Delta_{\theta}$. Since $\Delta_{\theta}(g)={\cal F}_{\theta}^{-1}\Delta(g){\cal F}_{\theta}$, it is
\bea
\tau_{\theta} &=& {\cal F}_{\theta}^{-1} \tau_{0} {\cal F}_{\theta}.
\eea
It has the property,
\bea
(\tau_\theta)^2 &=& {\bf 1}\otimes {\bf 1}.
\eea

The states constructed according to
\bea
\phi \otimes_{S_\theta} \chi \equiv
\left(\frac{1\,+ \tau_\theta}{2}\right)\,
(\phi\,\otimes\,\chi),
\eea
\bea
\phi \otimes_{A_\theta} \chi \equiv
\left(\frac{1\,-\tau_\theta}{2}\right)\,(\phi\,\otimes\,\chi)
\eea
form the physical two-particle Hilbert spaces of (generalized) bosons and fermions obeying twisted statistics.
\subsection{Statistics of Quantum Fields}
The very act of implementing Poincar\'e symmetry on a noncommutative spacetime algebra leads to twisted fermions and bosons. In this section we look at the second quantized version of the theory and we encounter another surprise on the way.

We can connect an operator in Hilbert space and a quantum field in the following way. A quantum field on evaluation at a spacetime point gives an operator-valued distribution acting on a Hilbert space. A quantum field at a spacetime point $x_{1}$ acting on the vacuum gives a one-particle state centered at $x_{1}$. Similarly we can construct a two-particle state in the Hilbert space. The product of two quantum fields at spacetime points $x_{1}$ and $x_{2}$ when acting on the vacuum generates a two-particle state where one particle is centered at $x_{1}$ and the other at $x_{2}$.

In the commutative case, a free spin-zero quantum scalar field $\varphi_{0}(x)$ of mass $m$ has the mode expansion
\bea
\varphi_{0}(x) =\int d \mu(p) \; (c_{\bf p}\; \textrm{e}_{p}(x) + d_{\bf p}^{\dagger} \; \textrm{e}_{-p}(x))
\eea
where
\bea
\textrm{e}_{p}(x) = \textrm{e}^{-i\; p\cdot x}, \; \; p \cdot x = p_{0}x_{0} - {\bf p}\cdot {\bf x}, \; \; d \mu(p) = \frac{1}{(2\pi)^{3}}\frac{d^{3}p}{2p_{0}}, \; \; \; p_{0} = \sqrt{{\bf p}^{2} + m^{2}} > 0. \nn
\eea

The annihilation-creation operators $c_{{\bf p}}$, $c^{\dagger}_{{\bf p}}$, $d_{{\bf p}}$,
$d_{{\bf p}}^{\dagger}$ satisfy the standard commutation relations,
\bea
\label{eq:standard}
c_{{\bf p}}c_{{\bf q}}^{\dagger} \pm c_{{\bf q}}^{\dagger}c_{{\bf p}} &=& 2p_{0} \;\delta^{3}({\bf p}-{\bf q})\\
d_{{\bf p}}d_{{\bf q}}^{\dagger} \pm d_{{\bf q}}^{\dagger}d_{{\bf p}} &=& 2p_{0} \;\delta^{3}({\bf p}-{\bf q}).
\eea
The remaining commutators involving these operators vanish.

If $c_{\bf p}$ is the annihilation operator of the second-quantized field $\varphi_{0}(x)$, an elementary calculation tells us that
\bea
\label{eq:qft-qm}
\langle 0 |\varphi_{0}(x) c^\dagger_{\bf p} |0\rangle &=& e_p(x)=
e^{-i p \cdot x}.\nonumber
\eea
\bea
\frac{1}{2}\langle 0 |\varphi_{0}(x_1) \varphi_{0}(x_2)
c^\dagger_{\bf q}
c^\dagger_{\bf p} |0\rangle &=&  \nonumber
\left(\frac{{\bf 1} \pm
\tau_{0}}{2}\right)(e_p \otimes e_q)(x_1,x_2) \\
&\equiv& (e_p \otimes_{S_0,A_0} e_q)(x_1,x_2)\nn\\
&\equiv& \langle x_{1}, x_{2}|p, q\rangle_{S_0,A_0}.
\eea
where we have used the commutation relation
\bea
c_{\bf p}^\dagger~c_{\bf q}^\dagger\,= \pm~c_{\bf q}^\dagger~c_{\bf p}^\dagger~.
\eea

From the previous section we have learned that the two-particle states in noncommutative spacetime should be constructed in such a way that they obey twisted symmetry. That is,
\bea
|p, q\rangle_{S_0,A_0}~\rightarrow~|p, q\rangle_{S_\theta,A_\theta}.
\eea
This can happen only if we modify the quantum field $\varphi_{0}(x)$ in such a way that the analogue of eqn. (\ref{eq:qft-qm}) in the noncommutative framework gives us $|p, q\rangle_{S_\theta,A_\theta}$. Let us denote the modified quantum field by $\varphi_{\theta}$. It has a mode expansion
\bea
\varphi_{\theta}(x) =\int d \mu(p) \; (a_{\bf p}\; \textrm{e}_{p}(x) + b_{\bf p}^{\dagger} \; \textrm{e}_{-p}(x))
\eea

Noncommutativity of spacetime does not change the dispersion relation for the quantum field in our framework. It will definitely change the operator coefficients of the plane wave basis. Here we denote the new $\theta$-deformed annihilation-creation operators by $a_{{\bf p}}$, $a^{\dagger}_{{\bf p}}$, $b_{{\bf p}}$, $b_{{\bf p}}^{\dagger}$. Let us try to connect the quantum field in noncommutative spacetime with its counterpart in commutative spacetime, keeping in mind that they should coincide in the limit $\theta^{\mu \nu} \rightarrow 0$.

The two-particle state $|p, q\rangle_{S_{\theta}, A_{\theta}}$ for bosons and fermions obeying deformed statistics is constructed as follows:
\bea
|p, q\rangle_{S_{\theta}, A_{\theta}} &\equiv& |p\rangle \otimes_{_{S_{\theta}, A_{\theta}}} |q\rangle =\Big(\frac{1 \pm \tau_{\theta}}{2}\Big) (|p\rangle \otimes |q\rangle)\nn \\
&=& \frac{1}{2}\Big(|p\rangle \otimes |q\rangle \pm \textrm{e}^{-i q_{\mu}\theta^{\mu \nu}p_{\nu}}|q\rangle \otimes |p\rangle\Big).
\eea

Exchanging $p$ and $q$ in the above, one finds
\bea
\label{eq:pq-qp}
|p, q\rangle_{S_{\theta}, A_{\theta}} = \pm \; \textrm{e}^{i p_{\mu}\theta^{\mu \nu}q_{\nu}}|q, p\rangle_{S_{\theta}, A_{\theta}}.
\eea

In Fock space the above two-particle state is constructed from the modified second-quantized field $\varphi_{\theta}$ according to
\bea
\frac{1}{2}\langle0|\varphi_{\theta}(x_{1})\varphi_{\theta}(x_{2}) a_{\bf q}^{\dagger}a_{\bf p}^{\dagger}|0\rangle &=& \Big(\frac{1\pm \tau_{\theta}}{2}\Big) (e_{p} \otimes e_{q})(x_{1}, x_{2})\nn \\
&=& (e_{p} \otimes_{S_{\theta}, A_{\theta}} e_{q})(x_{1}, x_{2})\nn \\
&=& \langle x_1, x_2|p, q\rangle_{S_{\theta}, A_{\theta}}.
\eea

On using eqn. (\ref{eq:pq-qp}), this leads to the relation
\bea
\label{non-commu-1}
a_{\bf p}^{\dagger}a_{\bf q}^{\dagger} =  \pm \; \textrm{e}^{i p_{\mu}\theta^{\mu \nu}q_{\nu}}\; a_{\bf q}^{\dagger}a_{\bf p}^{\dagger}.
\eea

It implies
\bea
\label{non-commu-2}
a_{\bf p}a_{\bf q} &=&  \pm \; \textrm{e}^{i p_{\mu}\theta^{\mu \nu}q_{\nu}}\; a_{\bf q}a_{\bf p.}\eea

Thus we have a new type of bilinear relations reflecting the deformed quantum symmetry.

This result shows that while constructing a quantum field theory on noncommutative spacetime, we should twist the creation and annihilation operators in addition to the $*$-multiplication between the fields.

In the limit $\theta^{\mu \nu}=0$, the twisted creation and annihilation operators should match with their counterparts in the commutative case. There is a way to connect these operators in the two cases. The transformation connecting the twisted operators,  $a_{{\bf p}}$, $b_{{\bf p}}$, and the untwisted operators, $c_{{\bf p}}$, $d_{{\bf p}}$, is called the ``dressing transformation" \cite{Grosse, Faddeev-Zamolodchikov}. It is defined as follows:
\bea
\label{dressingT}
a_{{\bf p}} = c_{{\bf p}} \; e^{-\frac{i}{2}p_{\mu} \theta^{\mu\nu}P_{\nu}},\; \; \;
b_{{\bf p}} = d_{{\bf p}}\; e^{-\frac{i}{2}p_{\mu}\theta^{\mu\nu}P_{\nu}},
\eea
where $P_{\mu}$ is the four-momentum operator,
\bea
P_{\mu} = \int \frac{d^{3}p}{2p_{0}}\; (c_{{\bf p}}^\dagger c_{{\bf p}} +
d^{\dagger}_{{\bf p}}
d_{{\bf p}})\; p_{\mu}.
\eea

The Grosse-Faddeev-Zamolodchikov algebra is the above twisted or dressed algebra \cite{Grosse, Faddeev-Zamolodchikov}. (See also \cite{queiroz1} in this connection.)

Note that the four-momentum operator $P_{\mu}$ can also be written in terms of the twisted operators:
\bea
\label{eq:pmu1}
P_{\mu} =\int \frac{d^{3}p}{2p_{0}}\; (a_{{\bf p}}^{\dagger} a_{{\bf p}} +
b^{\dagger}_{{\bf p}}
b_{{\bf p}}) \; p_{\mu}.
\eea
That is because $p_{\mu} \theta^{\mu\nu}P_{\nu}$ commutes with any of the operators for momentum $p$. For example
\bea
[P_{\mu},a_{{\bf p}}]=-p_{\mu} a_{{\bf
p}},
\eea
so that
\bea
[p_{\nu} \theta^{\nu\mu}P_{\mu}, a_{{\bf p}}]= p_{\nu}\theta^{\nu\mu}p_{\mu}=0,
\eea
$\theta$ being antisymmetric.

The antisymmetry of $\theta^{\mu \nu}$ allows us to write
\bea
c_{\bf p}e^{-\frac{i}{2}p_{\mu} \theta^{\mu \nu} P_{\nu}} = e^{-\frac{i}{2}p_{\mu}
\theta^{\mu \nu} P_{\nu}} c_{\bf p},
\eea
\bea
c^{\dagger}_{\bf p}e^{\frac{i}{2}p_{\mu} \theta^{\mu \nu} P_{\nu}} = e^{\frac{i}{2}p_{\mu} \theta^{\mu \nu} P_{\nu}}c^{\dagger}_{\bf p}.
\eea
Hence the ordering of factors here is immeterial.

It should also be noted that the map from the $c$- to the $a$-operators is invertible,
\bea
c_{\bf p} = a_{\bf p} \; e^{\frac{i}{2}p_{\mu} \theta^{\mu\nu}P_{\nu}},\; \; \;
d_{\bf p} =
b_{\bf p}\; e^{\frac{i}{2}p_{\mu}\theta^{\mu\nu}P_{\nu}},\nn
\eea
where $P_{\mu}$ is written as in eqn.~(\ref{eq:pmu1}).

The $\star$-product between the modified (twisted) quantum fields is
\bea
(\varphi_{\theta} \star \varphi_{\theta})(x) = \varphi_{\theta}(x) e^{\frac{i}{2} \overleftarrow{\partial}
\wedge \overrightarrow{\partial}} \varphi_{\theta}(y) |_{x=y},
\eea
\bea
\overleftarrow{\partial} \wedge \overrightarrow{\partial} :=
\overleftarrow{\partial}_{\mu} \theta^{\mu \nu} \overrightarrow{\partial}_{\nu}.\nn
\eea

The twisted quantum field $\varphi_{\theta}$ differs from the untwisted quantum field
$\varphi_{0}$ in two ways:
\begin{center}
$i.)$ $e_{p} \in {\cal A}_\theta(\rr^{d+1})~~~~~~~~~~~~~~~~~~~~~~~~~~~~~~~~~~~~~~~~~~~~~~~~$
\end{center}
and
\begin{center}
$ii.)$ $a_{{\bf p}}$ is twisted by statistics. $~~~~~~~~~~~~~~~~~~~~~~~~~~~~~~~~$
\end{center}
The twisted statistics can be accounted by writing \cite{bal-sasha-babar}
\bea
\varphi_{\theta} = \varphi_{0} \; e^{\frac{1}{2} \overleftarrow{\partial} \wedge P},
\label{eq:gaugefield}
\eea
where $P_{\mu}$ is the total momentum operator. From this follows that the
$\star$-product of an
arbitrary number of fields $\varphi_{\theta}^{(i)}$ ($i$ = 1, 2, 3, $\cdots$) is
\bea
\varphi_{\theta}^{(1)} \star \varphi_{\theta}^{(2)} \star {\cdots} =(\varphi^{(1)}_{0}\varphi^{(2)}_{0} {\cdots}) \; e^{\frac{1}{2} \overleftarrow{\partial} \wedge P}.
\label{eq:productfields}
\eea

Similar deformations occur for all tensorial and spinorial quantum fields.

In \cite{cmb}, a noncommutative cosmic microwave background (CMB) power spectrum is calculated by promoting the quantum fluctuations $\varphi_{0}$ of the scalar field driving inflation (the inflaton) to a twisted quantum field $\varphi_{\theta}$. The power spectrum becomes direction-dependent, breaking the statistical anisotropy of the CMB. Also, $n$-point correlation functions become non-Gaussian when the fields are noncommutative, assuming that they are Gaussian in their commutative limits. These effects can be tested experimentally.

In this chapter we discuss field theory with spacetime noncommutativity. It should also be noted that there is another approach in which noncommutativity is encoded in the degrees of freedom of the fields while keeping spacetime commutative \cite{Carmona, Carmona1}. Such noncommutativity can also be interpreted in terms of twisted statistics. In \cite{queiroz1} a noncommutative black body spectrum is calculated using this approach (which is based on \cite{Carmona, Carmona1}). Also, a noncommutative-gas driven inflation is considered in \cite{queiroz1} along this formulation.

\subsection{From Twisted Statistics to Noncommutative Spacetime}
Noncommutative spacetime leads to twisted statistics. It is also possible to start from a twisted statistics and end up with a noncommutative spacetime \cite{queiroz, Bal-Queiroz}. Consider the commutative version $\varphi_{0}$ of the above quantum field $\varphi_{\theta}$. The creation and annihilation operators of this field fulfill the standard commutation relations as given in eqn. (\ref{eq:standard}).

Let us twist statistics by deforming the creation-annihilation operators $c_{\bf p}$ and $c_{\bf p}^\dagger$ to
\bea
a_{\bf p} =c_{\bf p}\;e^{-\frac{i}{2} \; p_\mu \;\theta^{\mu \nu}\;P_\nu}\;, \hspace{10mm}
a^\dagger_{\bf p} = c^\dagger_{\bf p}\;e^{\frac{i}{2} \; p_\mu \;\theta^{\mu \nu}\;P_\nu}
\eea

Now statistics is twisted since $a$'s and $a^\dagger$'s no longer fulfill standard relations. They obey the relations given in eqn. (\ref{non-commu-1}) and eqn. (\ref{non-commu-2}) This twist affects the usual symmetry of particle interchange. The $n$-particle wave function $\psi_{k_1 \cdots k_n}$,
\bea
\psi_{k_1, \cdots, k_n}(x_1, \ldots, x_n) = \langle 0|\varphi(x_1)\varphi(x_2)\ldots \varphi(x_n)~a^\dagger_{{\bf k}_n}a^\dagger_{{\bf k}_{n-1}}\ldots
a^\dagger_{{\bf k}_1}\;|0\rangle
\eea
is no longer symmetric under the interchange of $k_i$. It fullfils a twisted symmetry given by
\bea
\label{eq:42.0}
\psi_{k_1 \cdots k_i \; k_{i+1} \cdots k_n} = \textrm{exp}\Big(-i k_{i}^{\mu} \; \theta_{\mu \nu} \; k_{i+1}^{\nu}\Big) \; \psi_{k_1 \cdots k_{i+1} \; k_i \cdots k_n }
\eea
showing that statistics is twisted. We can show that this in fact leads to a noncommutative spacetime if we require Poincar\'e invariance. It is explained below.

In the commutative case, the elements $g$ of $P_{+}^{\uparrow}$ acts on $\psi_{k_1 \cdots k_n}$ by the representative $U(g)\otimes U(g) \otimes \cdots \otimes U(g)$ ($n$ factors) compatibly with the symmetry of $\psi_{k_1 \cdots k_n}$. This action is based on the coproduct
\bea
\Delta(g) = g \times g\;.
\eea

But for $\theta^{\mu \nu} \neq 0$, and for $g\neq\textrm{identity}$, already for the case $n=2$,
\bea
\Delta(g) \psi_{p, q} &=& \psi_{gp, gq}\nn \\
&=& e^{-i p_{\mu} \theta^{\mu \nu} q_{\nu}} \Delta(g) \psi_{q, p}\nn \\
&=& e^{-i p_{\mu} \theta^{\mu \nu} q_{\nu}} \psi_{gq, gp}\nn \\
&\neq& e^{-i (gp)_{\mu}\theta^{\mu \nu}(gq)_{\nu}} \psi_{gq, gp}.
\eea

Thus the usual coproduct $\Delta_0$ is not compatible with the statistics (\ref{eq:42.0}). It has to be twisted to
\bea
\label{eq:43}
\Delta_\theta(g) = {\cal F}^{-1}_\theta \Delta (g) {\cal F}_\theta,~~\Delta(g)=(g\times g)
\eea
to be compatible with the new statistics. At this point $\Delta_\theta(g)$ is not compatible with $m_{0}$, the commutative (point-wise) multiplication map. So we are forced to change the multiplication map to $m_{\theta}$,
\bea
m_{\theta} = m_{0} \; {\cal F}_\theta
\eea
for this compatibility. Since
\bea
m_\theta(\alpha \otimes \beta) = \alpha * \beta,
\eea
we end up with noncommutative spacetime. Thus twisted statistics can lead to spacetime noncommutativity.
\subsection{Violation of the Pauli Principle}
In section 4.3, we wrote down the twisted commutation relations. In the fermionic sector, these relations read
\bea
\label{phaseFermion1}
a_{\bf p}^{\dagger}a_{\bf q}^{\dagger} + \; \textrm{e}^{i p_{\mu}\theta^{\mu \nu}q_{\nu}}\; a_{\bf q}^{\dagger}a_{\bf p}^{\dagger}&=&0 \\
\label{phaseFermion2}
a_{\bf p}a_{\bf q}^{\dagger} + \; \textrm{e}^{-i p_{\mu}\theta^{\mu \nu}q_{\nu}}\; a_{\bf q}^{\dagger}a_{\bf p} &=& 2q_{0}\delta^{3}({\bf p} - {\bf q}).
\eea

In the commutative case, above relations read
\bea
c_{\bf p}^{\dagger}c_{\bf q}^{\dagger} + c_{\bf q}^{\dagger}c_{\bf p}^{\dagger}&=&0 \\
c_{\bf p}c_{\bf q}^{\dagger} + c_{\bf q}^{\dagger}c_{\bf p} &=& 2q_{0}\delta^{3}({\bf p} - {\bf q}).
\eea
The phase factor appearing in eqn (\ref{phaseFermion1}) and eqn. (\ref{phaseFermion2}) while exchanging the operators has a nontrivial physical consequence which forces us to reconsider the Pauli exclusion principle. A modification of Pauli principle compatible with the twisted statistics can lead to Pauli forbidden processess and they can be subjected to stringent experimental tests.

For example, there are results from SuperKamiokande \cite{sk} and Borexino \cite{borexino} putting limits on the violation of Pauli exclusion principle in nucleon systems. These results are based on non-observed transition from Pauli-allowed states to Pauli-forbidden states with $\beta^{\pm}$ decays or $\gamma$, $p$, $n$ emission. A bound for $\theta$ as strong as $10^{11}$ Gev is obtained from these results \cite{Gianpiero}.
\subsection{Statisitcal Potential}
Twisting the statistics can modify the spatial correlation functions of fermions and bosons and thus affect the statistical potential existing between any two particles.

Consider a canonical ensemble, a system of $N$ indistinguishable, non-interacting particles confined to a three-dimensional cubical box of volume $V$, characterized by the inverse temperature $\beta$. In the coordinate representation, we write down the density matrix of the system \cite{pathria}
\bea
\langle {\bf r}_{1}, \cdots {\bf r}_{N}|\hat{\rho}|{\bf r}'_{1}, \cdots {\bf r}'_{N}\rangle = \frac{1}{Q_{N}(\beta)}\langle {\bf r}_{1}, \cdots {\bf r}_{N}|\textrm{e}^{-\beta \hat{H}}|{\bf r}'_{1}, \cdots {\bf r}'_{N}\rangle,
\eea
where $Q_{N}(\beta)$ is the partition function of the system given by
\bea
Q_{N}(\beta) = \textrm{Tr}(\textrm{e}^{-\beta \hat{H}}) = \int d^{3N} r \langle {\bf r}_{1}, \cdots {\bf r}_{N}|\textrm{e}^{-\beta \hat{H}}|{\bf r}'_{1}, \cdots {\bf r}'_{N}\rangle .
\eea

Since the particles are non-interacting, we may write down the eigenfunctions and eigenvalues of the system in terms of the single-particle wave functions and single-particle energies.

For free non-relativistic particles, we have the energy eigenvalues
\bea
E = \frac{\hbar^{2}}{2m}\sum_{i=1}^{N} k_{i}^{2}
\eea
where $k_{i}$ is the magnitude of the wave vector of the $i$-th particle. Imposing periodic boundary conditions, we write down the normalized single-particle wave function
\bea
u_{{\bf k}}({\bf r}) = V^{-1/2} \textrm{e}^{i {\bf k} \cdot {\bf r}}
\eea
with ${\bf k} = 2 \pi V^{-1/3} {\bf n}$ and ${\bf n}$ is a three-dimensional vector whose components take values $0, \pm 1, \pm 2, \cdots$.

Following the steps given in \cite{pathria}, we write down the diagonal elements of the density matrix for the simplest relevant case with $N=2$,
\bea
\label{approxTwoPart}
\langle {\bf r}_{1}, {\bf r}_{2}|\hat{\rho}|{\bf r}_{1}, {\bf r}_{2}\rangle \approx \frac{1}{V^{2}} \big(1 \pm \textrm{exp}(-2 \pi r_{12}^{2}/\lambda^{2})\Big)
\eea
where the plus and the minus signs indicate bosons and fermions respectively, $r_{12}=|{\bf r}_{1} - {\bf r}_{2}|$ and $\lambda$ is the mean thermal wavelength,
\bea
\lambda = \hbar \sqrt{\frac{2 \pi \beta}{m}},~~~~~~\beta = \frac{1}{k_{B}T}.
\eea

Note that eqn. (\ref{approxTwoPart}) is obtained under the assumption that the mean interparticle distance $(V/N)^{1/3}$ in the system is much larger than the mean thermal wavelength $\lambda$. Eqn. (\ref{approxTwoPart}) indicates that spatial correlations are non-zero even when the particles are non-interacting. These correlations are purely due to statistics: They emerge from the symmetrization or anti-symmetrization of the wave functions describing the particles. Particles obeying Bose statistics give a positive spatial correlation and particles obeying Fermi statistics give a negative spatial correlation.

We can express spatial correlations between particles by introducing a statistical potential $v_{s}(r)$ and thus treat the particles classically \cite{Uhlenbeck}. The statistical potential corresponding to the spatial correlation given in eqn. (\ref{approxTwoPart}) is
\bea
v_{s}(r) = -k_{B}T \; \textrm{ln} \Big(1 \pm \textrm{exp}(-2\pi r_{12}^{2}/\lambda^{2})\Big)
\eea

From this equation, it follows that two bosons always experience a ``statistical attraction" while two fermions always experience a ``statistical repulsion". In both cases, the potential decays rapidly when $r > \lambda$.

So far our discussion focussed on particles in commutative spacetime. We can derive an expression for the statistical potential between two particles living in a noncommutative spacetime. The results \cite{correlation} are interesting. In a noncommutative spacetime with 2+1 dimensions and for the case $\theta^{0i} = 0$, we write down the answer for the spatial correlation between two non-interacting particles from \cite{correlation}
\bea
\label{two-P-theta}
\langle {\bf r}_{1}, {\bf r}_{2}|\hat{\rho}|{\bf r}_{1}, {\bf r}_{2}\rangle_{\theta} \approx \frac{1}{A^2}\left(1\pm\frac{1}{1+\frac{\theta^2}{\lambda^4}}
e^{- 2 \pi\, r_{12}^2/(\lambda^{2}(1+\frac{\theta^2}{\lambda^4}))}\right)
\eea
Here $A$ is the area of the system. This result can be generalized to higher dimensions by replacing $\theta^2$ by an appropriate sum of $(\theta^{ij})^2$ \cite{correlation}. It reduces to the standard (untwisted) result given in eqn. (\ref{approxTwoPart}) in the limit $\theta\rightarrow0$.
\begin{figure}
\centerline{\epsfig{figure=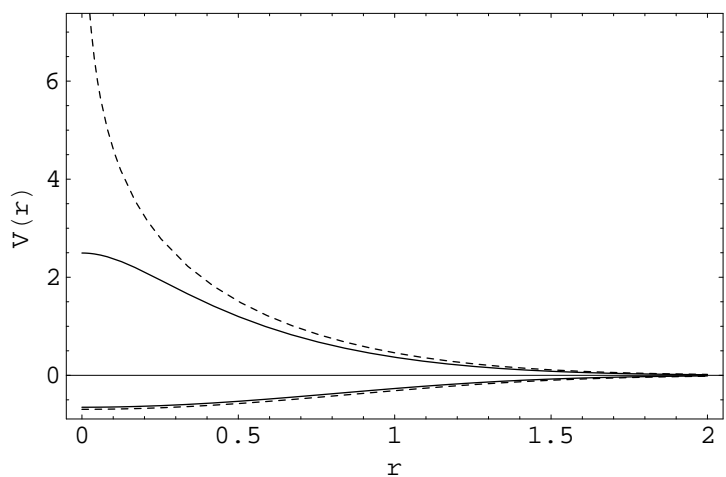}}
\caption {Statistical potential $v(r)$ measured in units of $k_BT$.  An irrelevant additive constant has been set zero.  The upper two curves represent the fermionic cases and the lower curves the bosonic cases.  The solid line shows the noncommutative result and the dashed line the commutative case. The curves are drawn for the value $\frac{\theta}{\lambda^2}=0.3$. The separation $r$ is measured in units of the thermal length $\lambda$. \cite{correlation}}
\label{fig:potential}
\end{figure}
Notice that the spatial correlation function for fermions does not vanish in the limit $r\rightarrow 0$ (See Fig. \ref{fig:potential}).  That means that there is a finite probability that fermions may come very close to each other. This probability is determined by the noncommutativity parameter $\theta$. Also notice that the assumptions made in \cite{correlation} are valid for low temperature and low density limits. At high temperature and high density limits a much more careful analysis is required to investigate the noncommutative effects.
\section{Matter Fields, Gauge Fields and Interactions}
In section 4, we discussed the statistics of quantum fields by taking a simple example of a massive, spin-zero quantum field. In this section, we discuss how matter and gauge fields are constructed in the noncommutative formulation and their interactions. We also explain some interesting results which can be verified experimentally.
\subsection{Pure Matter Fields}
Consider a second quantized real Hermitian field of mass $m$,
\bea
\Phi = \Phi^{-} + \Phi^{+}
\eea
where the creation and annihilation fields are constructed from the creation and annihilation operators:
\bea
\Phi^{-}(x) &=& \int d\mu(p) \; e^{ipx} \; a_{\bf p}^{\dagger}\\
\Phi^{+}(x) &=& \int d\mu(p) \; e^{-ipx} \; a_{\bf p}
\eea
The deformed quantum field $\Phi$ can be written in terms of the un-deformed quantum field $\Phi_{0}$,
\bea
\label{Theta-to-zero}
\Phi(x) = \Phi_{0}(x) e^{\frac{1}{2}\overleftarrow{\partial}^{\mu}\theta_{\mu \nu}P^{\nu}}
\eea
where the creation and annihilation fields of the un-deformed quantum field is constructed from the usual creation and annihilation operators
\bea
\Phi_{0}^{-}(x) &=& \int d\mu(p) \; e^{ipx} \; c_{\bf p}^{\dagger}, \\
\Phi_{0}^{+}(x) &=& \int d\mu(p) \; e^{-ipx} \; c_{\bf p}
\eea

When evaluating the product of $\Phi$'s at the same point, we must take $*$-product of the $e_{p}$'s since $e_{p} \in {\cal A}_{\theta}({\mathbb R}^{N})$. We can make use of eqn. (\ref{Theta-to-zero}) to simplify the $*$-product of $\Phi$'s at the same point to a commutative (point-wise) product of $\Phi_{0}$'s. For the $*$-product of $n$ $\Phi$'s,
\bea
\label{Theta-to-zero2}
\Phi(x) * \Phi(x) * \cdots * \Phi(x) = \Big(\Phi_{0}(x)\Big)^{n} e^{\frac{1}{2}\overleftarrow{\partial}^{\mu}\theta_{\mu \nu}P^{\nu}}
\eea
This is a very important result. Using this result, we can prove that there is no UV-IR mixing in a noncommutative field theory with matter fields and no gauge interactions \cite{Oeckl:2000eg, uv-ir}.

The interaction Hamiltonian density is built out of quantum fields. It transforms like a single scalar field in the noncommutative theory also. (This is the case only when we choose a $*$-product between the fields to write down the Hamiltonian density.) Thus a generic interaction Hamiltonian density ${\cal H}_{I}$ involving only $\Phi$'s (for simplicity) is given by
\bea
\label{Hi}
{\cal H}_{I}(x) = \Phi(x) * \Phi(x) * \cdots * \Phi(x)
\eea
This form of the Hamiltonian and the twisted statistics of the fields is all that is required to show that there is no UV-IR mixing in this theory. This happens because the $S$-matrix becomes independent of $\theta^{\mu\nu}$.

We illustrate this result for the first nontrivial term $S^{(1)}$ in the expansion of the $S$-matrix. It is
\bea
S^{(1)} = \int d^{4}x \; {\cal H}_{I} (x).
\eea

Using eqn. (\ref{Theta-to-zero}) we write down the interaction Hamiltonian density given in eqn. (\ref{Hi}) as
\bea
{\cal H}_{I}(x) = \Big(\Phi_{0}(x)\Big)^{n} e^{\frac{1}{2}\overleftarrow{\partial}^{\mu}\theta_{\mu \nu}P^{\nu}}
\eea
Assuming that the fields behave ``nicely" at infinity, the integration over ~$x$~ gives
\bea
\int d^{4}x \Big(\Phi_{*}(x)\Big)^{n} = \int d^{4}x \Big(\Phi_{0}(x)\Big)^{n} e^{\frac{1}{2}\overleftarrow{\partial}^{\mu}\theta_{\mu \nu}P^{\nu}} = \int d^{4}x \Big(\Phi_{0}(x)\Big)^{n}.
\eea

Thus $~S^{(1)}~$ is independent of $~\theta^{\mu\nu}$.
By similar calculations we can show that the $S$-operator is independent of $~\theta^{\mu \nu}~$ to all orders \cite{bal, uv-ir, bal-sasha-babar, bal-stat}.

\subsection{Covariant Derivatives of Quantum Fields}

In this section we briefly discuss how to choose appropriate covariant derivatives $D_{\mu}$ of a quantum field associated with~ ${\cal A}_{\theta}({\mathbb R}^{3+1})$.

To define the desirable properties of covariant derivatives $D_{\mu}$, let us first look at ways of multiplying the field ~$\Phi_{\theta}$~ by a function ~$\alpha_{0} \in {\cal A}_{0}({\mathbb R}^{3+1})$. There are two possibilities \cite{bal-sasha-babar}:
\bea
\Phi &\rightarrow& (\Phi_{0} \alpha_{0}) e^{\frac{1}{2}\overleftarrow{\partial}\wedge P} \equiv T_{0} (\alpha_{0})\Phi,\\
\Phi &\rightarrow& (\Phi_{0} *_{\theta} \alpha_{0}) e^{\frac{1}{2}\overleftarrow{\partial}\wedge P} \equiv T_{\theta} (\alpha_{0})\Phi
\eea
where ~$T_{0}$~ gives a representation of the commutative algebra of functions and ~$T_{\theta}$~ gives that of a $*$-algebra.

A ~$D_{\mu}$~ that can qualify as the covariant derivative of a quantum field associated with ~${\cal A}_{0}({\mathbb R}^{3+1})$~ should preserve statistics, Poincar\'e and gauge invariance and must obey the Leibnitz rule
\bea
\label{Leibnitz}
D_{\mu}(T_{0}(\alpha_{0})\Phi) = T_{0}(\alpha_{0})(D_{\mu}\Phi) + T_{0} (\partial_{\mu} \alpha_{0})\Phi
\eea
The requirement given in eqn. (\ref{Leibnitz}) reflects the fact that $D_{\mu}$ is associated with the commutative algebra ${\cal A}_{0}({\mathbb R}^{3+1})$.

There are two immediate choices for $D_{\mu}\Phi$:
\bea
\label{firstChoice}
&&1. ~~ D_{\mu}\Phi = ((D_{\mu})_{0}\Phi_{0})e^{\frac{1}{2}\overleftarrow{\partial}\wedge P},\\
&&2. ~~ D_{\mu}\Phi = ((D_{\mu})_{0}e^{\frac{1}{2}\overleftarrow{\partial}\wedge P})(\Phi_{0})e^{\frac{1}{2}\overleftarrow{\partial}\wedge P}
\eea
where $(D_{\mu})_{0} = \partial_{\mu} + (A_{\mu})_{0}$ and $(A_{\mu})_{0}$ is the commutative gauge field, a function only of the commutative coordinates $x_{c}$.

Both the choices preserve statistics, Poincar\'e and gauge invariance, but the second choice does not satisfy eqn. (\ref{Leibnitz}). Thus we identify the correct covariant derivative in our formalism as the one given in the first choice, eqn. (\ref{firstChoice}).

\subsection{Matter fields with gauge interactions}
We assume that gauge (and gravity) fields are commutative fields, which means that they are functions only of $x^{\mu}_{c}$. For Aschieri et al. \cite{Aschieri}, instead, they are associated with ${\cal A}_\theta(\rr^{3+1})$. Matter fields on ${\cal A}_{\theta}({\mathbb R}^{3+1})$ must be transported by the connection compatibly with  eqn.~(\ref{Theta-to-zero}), so from the previous section, we see that the natural choice for  covariant derivative is
\bea
D_{\mu} \Phi = (D_{\mu}^{c} \Phi_{0}) \; e^{\frac{i}{2}
\overleftarrow{\partial} \wedge P},
\label{eq:covariant}
\eea
where
\bea
D_{\mu}^{c} \Phi_{0} = \partial_{\mu}\Phi_{0} + A_{\mu}\Phi_{0}\; ,
\eea
$P_{\mu}$ is the total momentum operator for all the fields and the fields $A_{\mu}$ and $\Phi_{0}$ are multiplied point-wise,
\bea
A_{\mu}\Phi_{0}(x)=A_{\mu}(x)\Phi_{0}(x).
\eea

Having identified the correct covariant derivative, it is simple to write down the Hamiltonian for gauge theories. The commutator of two covariant derivatives gives us the curvature. On using eqn. (\ref{eq:covariant}),
\begin{eqnarray}
[D_{\mu}, D_{\nu}] \Phi &=& \Big([D^{c}_{\mu}, D^{c}_{\nu}]\Phi_{0}\Big)e^{\frac{i}{2}\overleftarrow{\partial} \wedge P} \\
&=&\Big(F_{\mu \nu}^{c}\Phi_{0}\Big)e^{\frac{i}{2}\overleftarrow{\partial} \wedge
P}.
\end{eqnarray}
As $F_{\mu \nu}^{c}$ is the standard $\theta^{\mu \nu}=0$ curvature, our gauge field
is associated with ${\cal A}_{0}({\mathbb R}^{3+1})$. Thus pure gauge theories on the GM plane are identical to their counterparts on commutative spacetime. (For Aschieri et al. \cite{Aschieri} the
curvature would be the $\star$-commutator of $D_{\mu}$'s.)

The gauge theory formulation we adopt here is fully explained in \cite{bal-sasha-babar}. It differs from the formulation of Aschieri et al. \cite{Aschieri} (where covariant derivative is defined using star product) and has the advantage of being able to accommodate any gauge group and not just $U(N)$ gauge groups and their direct products. The gauge theory formulation we adopt here thus avoids multiplicity of fields that the expression for covariant derivatives with $\star$ product entails.

In the single-particle sector (obtained by taking the matrix element of eqn.~(\ref{eq:covariant}) between vacuum and one-particle states), the $P$ term can be dropped and we get for a single particle wave function $f$ of a particle associated with $\Phi$,
\bea
D_{\mu}f(x) = \partial_{\mu}f(x)+A_{\mu}(x)f(x).
\eea
Note that we can also write $D_{\mu}\Phi$ using $\star$-product:
\bea
D_{\mu}\Phi = \Big(D_{\mu}^{c} e^{\frac{i}{2} \overleftarrow{\partial} \wedge
P}\Big)\star \Big(\Phi_{0}e^{\frac{i}{2} \overleftarrow{\partial} \wedge
P}\Big).
\eea
Our choice of covariant derivative allows us to write the interaction Hamiltonian density
for pure gauge fields as follows:
\bea
{\cal H}_{I \theta}^{^G} = {\cal H}_{I 0}^{^G}.
\eea

For a theory with matter and gauge fields, the interaction Hamiltonian density splits into two parts,
\bea
{\cal H}_{I \theta} = {\cal H}^{^{M, G}}_{I \theta}+{\cal H}^{^G}_{I \theta},
\eea
where
\bea
{\cal H}^{^{M, G}}_{I \theta}&=&{\cal H}^{^{M, G}}_{I 0} \; e^{\frac{i}{2}
\overleftarrow{\partial} \wedge P},\nn \\
{\cal H}^{^G}_{I \theta}&=&{\cal H}^{^G}_{I 0}.
\eea
The matter-gauge field couplings are also included in ${\cal H}^{^{M, G}}_{I \theta}$.

In quantum electrodynamics ($QED$), ${\cal H}^{^G}_{I \theta}=0$. Thus the $S$-operator for the twisted $QED$ is the same for the untwisted $QED$:
\bea
S^{^{QED}}_{\theta}=S^{^{QED}}_{0}.
\eea

\begin{figure}
\centerline{\epsfig{figure=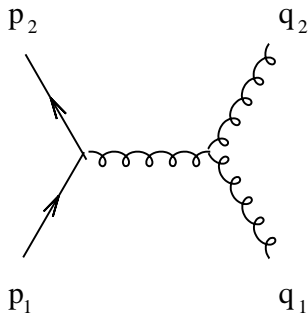,clip=4cm,width=4cm}}
\caption{A Feynman diagram in QCD with non-trivial $\theta$-dependence. The twist of ${\cal H}^{^{M, G}}_{I 0}$ changes the gluon propagator. The propagator is different from the usual one by its dependence on terms of the form $\vec{\theta}^{0} \cdot {\bf P}_{in}$, where $(\vec{\theta}^{0})_{i} = \theta^{0i}$ and ${\bf P}_{in}$ is the total momentum of the incoming particles. Such a frame-dependent modification violates Lorentz invariance.}
\label{fig:qcd}
\end{figure}

In a non-abelian gauge theory, ${\cal H}^{^G}_{\theta}={\cal H}^{^G}_{0} \neq 0$, so that in the presence of nonsinglet matter fields \cite{bal-sasha-babar},
\bea
S^{^{M, G}}_{\theta} \neq S^{^{M, G}}_{0},
\eea
because of the cross-terms between ${\cal H}^{^{M, G}}_{I \theta}$ and ${\cal H}^{^{G}}_{I \theta}$. In particular, this inequality happens in QCD. One such example is the quark-gluon scattering through a gluon exchange. The Feynman diagram for this process is given in Fig. \ref{fig:qcd}.

\subsection{Causality and Lorentz Invariance}
The very process of replacing the point-wise multiplication of functions at the same point by a $*$-multiplication makes the theory non-local. The $*$-product contains an infinite number of space-time derivatives and this in turn affects the fundamental causal structure on which all local, point-like quantum field theories are built upon.

Let ${\cal H}_{I}$ be the interaction Hamiltonian density in the interaction representation. The interaction representation $S$-matrix is
\bea
S = \textrm{T exp} \Big(-i \int d^{4}x \; {\cal H}_{I}(x)\Big).
\eea

In a commutative theory, the interaction Hamiltonian density ${\cal H}_{I}$ satisfies the Bogoliubov - Shirkov \cite{Bogoliubov} causality
\bea
\label{causality1}
[{\cal H}_{I}(x), {\cal H}_{I}(y)] = 0, \; \; \; x \sim y
\eea
where $x \sim y$ means $x$ and $y$ are space-like separated.

This causality relation plays a crucial role in maintaining the Lorentz invariance in all the local, point-like quantum field theories. Weinberg \cite{Weinberg1, Weinberg2} has discussed the fundamental significance of this equation in connection with the relativistic invariance of the $S$-matrix. If eqn. (\ref{causality1}) fails, $S$ cannot be relativistically invariant.

To see why this is the case, we consider the lowest term ~$S^{(2)}$~ of the ~$S$-matrix containing non-trivial time ordering. It is ~$S^{(2)}=-\frac{1}{2}\int d^4xd^4y~T(~{\cal H}_{I}(x){\cal H}_{I}(y)~),$ where
\begin{eqnarray}
&&T(~{\cal H}_{I}(x){\cal H}_{I}(y)~):=\theta(x^0-y^0){\cal H}_{I}(x){\cal H}_{I}(y) +
\theta(y^0-x^0){\cal H}_{I}(y){\cal H}_{I}(x)\nonumber\\
&&~~~~= {\cal H}_{I}(x){\cal H}_{I}(y)+ ( \theta(x^0-y^0)-1){\cal H}_{I}(x){\cal H}_{I}(y)+ \theta(y^0-x^0){\cal H}_{I}(y){\cal H}_{I}(x)\nonumber\\
&&~~~~= {\cal H}_{I}(x){\cal H}_{I}(y)-\theta(y^0-x^0)[{\cal H}_{I}(x),{\cal H}_{I}(y)].
\end{eqnarray}

If $U(\Lambda)$ is the unitary operator  on the quantum Hilbert space for implementing the Lorentz transformation ~$\Lambda$~ connected to the identity, that is, ~$\Lambda\in {P}^\uparrow_+$, then

$$U(\Lambda)T( {\cal H}_I(x){\cal H}_I(y) )U(\Lambda)^{-1}={\cal H}_I(\Lambda x){\cal H}_I(\Lambda y)-\theta(y^0-x^0)[{\cal H}_I(\Lambda x),{\cal H}_I(\Lambda y)].$$
If this is equal to~ $T( {\cal H}_I(\Lambda x){\cal H}_I(\Lambda y) )$, that is, if

$$  \theta( y^0- x^0 )[{\cal H}_I(\Lambda x),{\cal H}_I(\Lambda y)]= \theta((\Lambda y)^0-(\Lambda x)^0 )[( {\cal H}_I(\Lambda x),{\cal H}_I(\Lambda y) ],$$
then ~$S^{(2)}$~ is invariant under ~$\Lambda\in {P}^\uparrow_+$. It is clearly invariant under translations. Hence the invariance of $S^{(2)}$ under $ {P}^\uparrow_+$ requires that either  ~$\theta(y^0-x^0)$~ is invariant or that ~$[{\cal H}_{I}(x),{\cal H}_{I}(y)]=0$.

When  ~$x\nsim y$, the time step function ~$\theta(y^0-x^0)$~ is invariant under  ~$ P_{+}^{\uparrow}$~ since ~$\Lambda\in  {P}^\uparrow_+$~ cannot reverse the direction of time.

However, when ~$x\sim y$, ~$\Lambda\in  {P}^\uparrow_+$~ can reverse the direction of time and so ~$\theta(y^0-x^0)$~ is not invariant. One therefore requires that ~$[{\cal H}_{I}(x),{\cal H}_{I}(y)]=0$ if $x\sim y$. Therefore a commonly imposed condition for the invariance of ~$S^{(2)}$~ under ~${P}^\uparrow_+$~ is

\bea \label{causality2}[{\cal H}_{I}(x),{\cal H}_{I}(y)]=0 ~~~~\text{whenever}~~~~ x\sim y.\eea

One can show by similar arguments that it is natural to impose the causality condition $(\ref{causality2})$ to maintain the ~${P}^\uparrow_+$~ invariance of of the general term

$$S^{(n)}=\frac{(-i)^n}{n!}\int d^4x_1d^4x_2...d^4x_n~T(~{\cal H}_{I}(x_1){\cal H}_{I}(x_2)...{\cal H}_{I}(x_n)~),$$
in $S$.  Here

{\small\begin{eqnarray}
 &&T(~{\cal H}_{I}(x_1)...{\cal H}_{I}(x_n)~)\nonumber\\
 &&~~=\sum_{p\in S_n}~\theta(x^0_{p(1)}-x^0_{p(2)})\theta(x^0_{p(2)}-x^0_{p(3)})  ...\theta(x^0_{p(n-1)}-x^0_{p(n)}) ~{\cal H}_{I}(x_{p(1)})...{\cal H}_{I}(x_{p(n)}).\nonumber
\end{eqnarray}}

In a noncommutative theory, due to twisted statistics, the interaction Hamiltonian density might not satisfy (\ref{causality2}) but $S$ can still be Lorentz-invariant. For example, consider the interaction Hamiltonian density for the electron-photon system
\bea
{\cal H}_{I}(x) = i e \; (\bar{\psi} \ast \gamma^{\rho}A_{\rho}\psi)(x).
\eea

For simplicity, we consider the case where $\theta^{0i}=0$ and $\theta^{ij} \neq 0$. We write down the $S$-matrix
\bea
S &=& \textrm{T exp} \Big(-i \int d^{3}x {\cal H}_{I}(x)\Big)
\eea
where ${\cal H}_{I}(x) = i e \; (\bar{\psi}\gamma^{\rho}A_{\rho}\psi)(x).$ Here we have used the property of the Moyal product to remove the $*$ in ${\cal H}_{I}$ while integrating over the spatial variables. The fields $\psi$ and $\bar{\psi}$ are still noncommutative as their oscillator modes contain $\theta^{\mu \nu}$.

We can write down ${\cal H}_{I}(x)$ in the form
\bea
{\cal H}_{I}(x) ={\cal H}^{(0)}_I(x)e^{\frac{1}{2}\overleftarrow{\partial}\wedge\overrightarrow{P} }
\eea
where ${\cal H}^{(0)}_I$ gives the interaction Hamiltonian for $\theta^{\mu\nu}=0$ and satisfies the causality condition $(\ref{causality2})$. It follows that ${\cal H}_I$ does not fulfill the causality condition $(\ref{causality2})$. Still, as shown in $\cite{bal-sasha-babar}$, $S$ is Lorentz invariant. (For further discussion, see $\cite{bal-sasha-babar}$.)
\section{Discrete Symmetries - ${\bf C}$, ${\bf P}$, ${\bf T}$ and ${\bf CPT}$}
So far our discussion was centered around the identity component $P_{+}^{\uparrow}$ of the Lorentz group $P$. In this section we investigate the symmetries of our noncommutative theory under the action of discrete symmetries - parity ${\bf P}$, time reversal ${\bf T}$, charge conjugation ${\bf C}$ and their combined operation ${\bf CPT}$. The ${\bf CPT}$ theorem \cite{pct, cpt} is very fundamental in nature and all local relativistic quantum field theories are ${\bf CPT}$ invariant. Quantum field theories on the GM plane are non-local and so it is important to investigate the validity of the ${\bf CPT}$
theorem in these theories.

\subsection{Transformation of Quantum Fields Under ${\bf C}$, ${\bf P}$ and ${\bf T}$ }
Under \textbf{C}, the Poincar\'e  group ~$P_+^{\uparrow}$, the creation and annihilation operators ~$c_{\bf k}$, $c^\dagger_{\bf k}$, $d_{\bf k}$, $d^{\dagger}_{\bf k}$ of a second quantized field transform in the same way as their counterparts in an untwisted theory \cite{bal-sasha-babar}. Using the dressing transformation $\cite{Grosse, Faddeev-Zamolodchikov}$, we can then deduce the transformation laws for ~$a_\textbf{k}, ~a^\dagger_\textbf{k},~ b_\textbf{k},~ b^\dagger_\textbf{k}$, and the quantum fields. They automatically imply the appropriate twisted coproduct in the matter sector (and of course the untwisted coproduct for gauge fields.) It then implies the transformation laws for the fields under the full group generated by ${\bf C}$ and ${\cal P}$ by the group properties of that group: they are all induced from those of $c_{\bf k}$, $c^\dagger_{\bf k}$, $d_{\bf k}$, $d^{\dagger}_{\bf k}$ in the above fashion. (We always try to preserve such group properties.) We make use of this observation when we discuss the transformation properties of quantum fields under discrete symmetries.

So far we have not mentioned the transformaton property of the noncommutativity parameter $\theta^{\mu \nu}$. The matrix $\theta^{\mu \nu}$ is a constant antisymmetric matrix. In the approach using the twisted coproduct for the Poincar\'e group, $\theta^{\mu \nu}$ is {\it not} transformed by Poincar\'e transformations or in fact by any other symmetry: they are truly constants. Nevertheless Poincar\'e invariance and other symmetries can be certainly recovered for interactions invariant under the twisted symmetry actions at the level of classical theory and also for Wightman functions \cite{drinfeld, bal-stat, Aschieri, dimitrijevic}.

We discuss the transformation of quantum fields under the action of discrete symmetries below.

\subsubsection{Charge conjugation ${\bf C}$}
The charge conjugation operator is not a part of the Lorentz group and commutes with $P_{\mu}$ (and in fact with the full Poincar\'e group). This implies that the coproduct \cite{chaichian, Aschieri} for the charge conjugation operator ${\bf C}$ in the twisted case is the same as the coproduct for ${\bf C}$ in the untwisted case. So, we write
\bea
\label{coproduct-for-C}\Delta_{\theta}({\bf C}) = \Delta_{0}({\bf C}) = {\bf C} \otimes {\bf C},
\eea
with the understanding that \textbf{C} is an element of the group algebra $ G^\ast$, where $G=\{\textrm{\textbf{C}}\}\times P^\uparrow_+$. (This is why we use $\otimes$ and not $\times$  in (\ref{coproduct-for-C}).)

Under charge conjugation,
\bea
c_{{\bf k}} \stackrel{{\bf C}}\longrightarrow d_{{\bf k}}, \; \; \; a_{{\bf k}} \stackrel{{\bf C}}\longrightarrow b_{{\bf k}}
\eea
where the twisted operators are related to the untwisted ones by the dressing transformation \cite{Grosse, Faddeev-Zamolodchikov}: ~$a_{{\bf k}}=c_{{\bf k}} \; e^{-\frac{i}{2} k \wedge P}$ and  ~$b_{{\bf k}}=d_{{\bf k}} \; e^{-\frac{i}{2} k \wedge P}$.

It follows that
\bea
\varphi_{\theta} \stackrel{{\bf C}}\longrightarrow \varphi^{_{\bf C}}_{0}\; e^{\frac{1}{2} \overleftarrow{\partial} \wedge P}, \; \; \varphi_{0}^{_{\bf C}} = {\bf C} \varphi_{0} {\bf C}^{-1}.
\eea
while the $\ast$-product of two such fields $\varphi_\theta$ and $\chi_\theta$ transforms according to
\bea
\varphi_{\theta} \star \chi_{\theta} &=&(\varphi_{0} \chi_{0})\; e^{\frac{1}{2} \overleftarrow{\partial} \wedge P} \nn \\ &\stackrel{{\bf C}}\longrightarrow& ({\bf C} \varphi_{0}\chi_{0}{\bf C}^{-1})\; e^{\frac{1}{2} \overleftarrow{\partial} \wedge P} \nn \\
&=&(\varphi_{0}^{{\bf C}} \chi_{0}^{{\bf C}})\; e^{\frac{1}{2}\overleftarrow{\partial} \wedge P}.
\eea

\subsubsection{Parity ${\bf P}$}
Parity is a unitary operator on ${\cal A}_{0}({\mathbb R}^{3+1})$. But parity transformations do not induce automorphisms of ${\cal A}_{\theta}({\mathbb R}^{3+1})$ \cite{bal-unitary} if its coproduct is
\bea
\Delta_{0}({\bf P})={\bf P} \otimes {\bf P}.
\eea
That is, this coproduct is not compatible with the $\star$-product. Hence the coproduct for parity is not the same as that for the $\theta^{\mu \nu}=0$ case.

But the twisted coproduct $\Delta_{\theta}$, where
\bea
\Delta_\theta({\bf P}) = {\cal F}_{\theta}^{-1} \; \Delta_{0} ({\bf P}) \; {\cal
F}_{\theta},
\eea
{\it is} compatible with the $\star$-product. So, for ${\bf P}$ as well, compatibility with the $\star$-product fixes the coproduct \cite{bal}.

Under parity,
\bea
c_{{\bf k}} \stackrel{{\bf P}}\longrightarrow c_{-{\bf k}}, \; \; \; \; d_{{\bf k}}
\stackrel{{\bf P}}\longrightarrow d_{-{\bf k}}
\label{eq:cdP}
\eea
and hence
\bea
a_{{\bf k}} \stackrel{{\bf P}}\longrightarrow a_{-{\bf k}} \;
e^{i(k_{0}\theta^{0i}P_{i}-k_{i}\theta^{i0}P_{0})}, \; \; \; \; b_{{\bf k}}
\stackrel{{\bf P}}\longrightarrow b_{-{\bf k}} \;
e^{i(k_{0}\theta^{0i}P_{i}-k_{i}\theta^{i0}P_{0})}.
\label{eq:abP}
\eea
By an earlier remark \cite{bal-sasha-babar}, eqns. (\ref{eq:cdP}) and (\ref{eq:abP}) imply the transformation law for twisted scalar fields. A twisted complex scalar field $\varphi_{\theta}$ transforms under parity as follows,
\bea
\varphi_{\theta}&=& \varphi_{0} \; e^{\frac{1}{2}\overleftarrow{\partial} \wedge P}~\stackrel{{\bf P}}\longrightarrow~ {\bf P}\Big(\varphi_{0} \; e^{\frac{1}{2}\overleftarrow{\partial} \wedge P}\Big){\bf P}^{-1}
~=~\varphi_{0}^{_{\bf P}}\; e^{\frac{1}{2} \overleftarrow{\partial} \wedge (P_{0}, -\overrightarrow{P})},
\eea
where ~$\varphi_{0}^{_{\bf P}} = {\bf P} \varphi_{0} {\bf P}^{-1}$~ and ~$\overleftarrow{\partial} \wedge (P_{0}, -\overrightarrow{P}) := -\overleftarrow{\partial}_{0}
\theta^{0i} P_{i} -\overleftarrow{\partial}_{i} \theta^{ij}P_{j}+ \overleftarrow{\partial}_{i} \theta^{i0}P_{0}$.

The product of two such fields $\varphi_{\theta}$ and $\chi_{\theta}$ transforms according to
\bea
\varphi_{\theta} \star \chi_{\theta} &=&(\varphi_{0} \chi_{0})\; e^{\frac{1}{2} \overleftarrow{\partial} \wedge P}\stackrel{{\bf P}}\longrightarrow (\varphi^{_{\bf P}}_{0} \chi^{_{\bf P}}_{0})\; e^{\frac{1}{2} \overleftarrow{\partial} \wedge (P_{0}, -\overrightarrow{P})}
\eea

Thus fields transform under ${\bf P}$ with an extra factor $e^{-(\overleftarrow{\partial}_{0}\theta^{0i}P_{i} + \partial_{i}\theta^{ij}P_{j})} = e^{-\overleftarrow{\partial}_{\mu}\theta^{\mu j}P_{j}}$ when $\theta^{\mu \nu} \neq 0$.

\subsubsection{Time reversal ${\bf T}$}
Time reversal ~${\bf T}$~ is an anti-linear operator. Due to antilinearity, ~${\bf T}$~ induces automorphisms on ~${\cal A}_{\theta}({\mathbb R}^{3+1})$~ for the coproduct

$$\Delta_0(T)=T\otimes T~~~~\textrm{if}~~\theta^{ij}=0,$$
but not otherwise.

Under time reversal,
\bea
c_{{\bf k}} \stackrel{{\bf T}}\longrightarrow c_{-{\bf k}}, \; \; \; \; d_{{\bf k}}
\stackrel{{\bf T}}\longrightarrow d_{-{\bf k}}
\eea
\bea
a_{{\bf k}} \stackrel{{\bf T}}\longrightarrow a_{-{\bf k}} \;
e^{-i(k_{i}\theta^{ij}P_{j})}, \; \; \; \; b_{{\bf k}} \stackrel{{\bf T}}\longrightarrow
b_{-{\bf k}} \; e^{-i(k_{i}\theta^{ij}P_{j})}.
\eea

When ~$\theta^{\mu\nu}\neq 0$,~ compatibility with the $\star$-product fixes the coproduct for ~${\bf T}$~ to be
\bea
\Delta_\theta({\bf T}) = {\cal F}_{\theta}^{-1} \; \Delta_{0} ({\bf T}) \; {\cal F}_{\theta}.
\eea

This coproduct is also required in order to maintain the group properties of ${\cal P}$,
the full Poincar\'e group.

A twisted complex scalar field $\varphi_{\theta}$ hence transforms under time reversal as
follows,
\bea
\varphi_{\theta}&=& \varphi_{0} \; e^{\frac{1}{2}\overleftarrow{\partial} \wedge P}~~\stackrel{{\bf T}}\longrightarrow ~~\varphi_{0}^{_{\bf T}}\; e^{\frac{1}{2} \overleftarrow{\partial} \wedge (P_{0}, -\overrightarrow{P})},
\eea
where $\varphi_0^T=T\varphi_0T^{-1}$,
while the product of two such fields $\varphi_{\theta}$ and $\chi_{\theta}$ transforms according to
\bea
\varphi_{\theta} \star \chi_{\theta} &=&(\varphi_{0} \chi_{0})\; e^{\frac{1}{2} \overleftarrow{\partial} \wedge P}~~\stackrel{{\bf T}}\longrightarrow~~ (\varphi^{_{\bf T}}_{0} \chi^{_{\bf T}}_{0})\; e^{\frac{1}{2} \overleftarrow{\partial} \wedge (P_{0}, -\overrightarrow{P})}
\eea

Thus the time reversal operation as well induces an extra factor $e^{-\overleftarrow{\partial}_{i}\theta^{ij}P_{j}}$ in the transformation
property of fields when $\theta^{\mu \nu} \neq 0$.

\subsubsection{${\bf CPT}$}
When ${\bf CPT}$ is applied,
\bea
c_{{\bf k}} \stackrel{{\bf CPT}}\longrightarrow d_{{\bf k}}, \; \; \; \; d_{{\bf k}}
\stackrel{{\bf CPT}}\longrightarrow c_{{\bf k}},
\eea
\bea
a_{{\bf k}} \stackrel{{\bf CPT}}\longrightarrow b_{{\bf k}}e^{i(k \wedge P)}, \; \; \; \;
b_{{\bf k}} \stackrel{{\bf CPT}}\longrightarrow a_{{\bf k}}e^{i(k \wedge P)}.
\eea

The coproduct for ${\bf CPT}$ is of course
\bea
\Delta_\theta({\bf CPT}) = {\cal F}_{\theta}^{-1} \; \Delta_{0} ({\bf CPT}) \; {\cal F}_{\theta}.
\eea

A twisted complex scalar field $\varphi_{\theta}$ transforms under ${\bf CPT}$ as follows,
\bea
\varphi_{\theta}&=& \varphi_{0} \; e^{\frac{1}{2}\overleftarrow{\partial} \wedge P}\nn \\
&\stackrel{{\bf CPT}}\longrightarrow& {\bf CPT}\Big(\varphi_{0} \; e^{\frac{1}{2}\overleftarrow{\partial} \wedge P}\Big) ({\bf CPT})^{-1}\nn \\
&=& \varphi_{0}^{_{{\bf CPT}}}\; e^{\frac{1}{2} \overleftarrow{\partial} \wedge P},
\eea
while the product of two such fields $\varphi_{\theta}$ and $\chi_{\theta}$ transforms according to
\bea
\varphi_{\theta} \star \chi_{\theta} &=&(\varphi_{0} \chi_{0})\; e^{\frac{1}{2} \overleftarrow{\partial} \wedge P} \nn \\ &\stackrel{{\bf CPT}}\longrightarrow& (\varphi^{_{{\bf CPT}}}_{0} \chi^{_{{\bf CPT}}}_{0})\; e^{\frac{1}{2} \overleftarrow{\partial} \wedge P}.
\eea
\subsection{{\bf CPT} in  Non-Abelian Gauge Theories}
The standard model, a non-abelian gauge theory, is ${\bf CPT}$ invariant, but it is not invariant under ${\bf C}$, ${\bf P}$, ${\bf T}$ or products of any two of them. So we focus on discussing just ${\bf CPT}$ for its $S$-matrix when $\theta^{\mu \nu} \neq 0$. The discussion here can be easily adapted to any other non-abelian gauge theory.

\subsubsection{Matter fields coupled to gauge fields}
The interaction representation $S$-matrix is
\bea
{S}^{^{M, G}}_{\theta} = \text{T exp} \; \Big[{-i\int d^{4}x \; {\cal H}^{^{M, G}}_{I
\theta}(x)}\Big]
\eea
where ${\cal H}^{^{M, G}}_{I \theta}$ is the interaction Hamiltonian density for matter
fields (including also matter-gauge field couplings). Under ${\bf CPT}$,
\bea
{\cal H}^{^{M, G}}_{I \theta}(x) \stackrel{{\bf CPT}}\longrightarrow {\cal H}^{^{M, G}}_{I
\theta}(-x)e^{\overleftarrow{\partial} \wedge P}
\eea
where $\overleftarrow{\partial}$ has components $\frac{\overleftarrow{\partial}}{\partial x_{\mu}}$. We write ${\cal H}^{^{M, G}}_{I \theta}$ as
\bea
{\cal H}^{^{M, G}}_{I \theta} = {\cal H}^{^{M, G}}_{I 0} \; e^{\frac{1}{2}
\overleftarrow{\partial} \wedge P}.
\label{eq:matter}
\eea
Thus we can write the interaction Hamiltonian density after ${\bf CPT}$ transformation in
terms of the untwisted interaction Hamiltonian density:
\bea
{\cal H}^{^{M, G}}_{I \theta}(x) \; \; \stackrel{{\bf CPT}}\longrightarrow&& {\cal H}^{^{M, G}}_{I
\theta}(-x)\; e^{\overleftarrow{\partial} \wedge P} \nn \\ &=& {\cal
H}^{^{M, G}}_{I0}(-x)\; e^{-\frac{1}{2}\overleftarrow{\partial} \wedge
P}\; e^{\overleftarrow{\partial} \wedge P}\nn \\
&=&{\cal H}^{^{M, G}}_{I0}(-x)\; e^{\frac{1}{2}\overleftarrow{\partial} \wedge P}.
\eea

Hence under ${\bf CPT}$,
{\small\bea
&&{S}^{^{M, G}}_{\theta} = \text{T exp} \; \Big[-i\int d^{4}x \; {\cal H}^{^{M, G}}_{I
0}(x) \; e^{\frac{1}{2}\overleftarrow{\partial}\wedge P}\Big] \rightarrow \text{T exp} \; \Big[i\int d^{4}x \; {\cal H}^{^{M, G}}_{I
0}(x) \; e^{-\frac{1}{2}\overleftarrow{\partial}\wedge P}\Big]\nn\\
&&~~~~ = ({S}^{^{M, G}}_{-\theta})^{-1}. \nn
\eea}

But it has been shown elsewhere that ${S}^{^{M, G}}_{\theta}$ is independent of $\theta$ \cite{uv-ir}. Hence also ${S}^{^{M, G}}_{\theta}$ is independent of $\theta$.

Therefore a quantum field theory with no pure gauge interaction is ${\bf CPT}$ ``invariant" on ${\calA}_{\theta}({\mathbb R}^{3+1})$. In particular quantum electrodynamics ($QED$) preserves ${\bf CPT}$.

\subsubsection{Pure Gauge Fields}
The interaction Hamiltonian density for pure gauge fields is independent of $\theta^{\mu \nu}$ in the approach of \cite{bal-sasha-babar}:
\bea
{\cal H}_{I \theta}^{^G} = {\cal H}_{I 0}^{^G}\; .
\eea

Hence also the $S$ becomes $\theta$-independent,
\bea
{S}^{^G}_{\theta} = {S}^{^G}_{0},
\eea

and ${\bf CPT}$ holds as a good ``symmetry" of the theory.

\subsubsection{Matter and Gauge Fields}

All interactions of matter and gauge fields can be fully discussed by writing the $S$-operator as\bea
{{\bf S}}^{^{M,G}}_{\theta} = \text{T exp} \; \Big[{-i\int d^{4}x \; {\cal
H}_{I
\theta}(x)}\Big],
\eea
\bea
{\cal H}_{I \theta} = {\cal H}^{^{M, G}}_{I \theta}+{\cal H}^{^G}_{I \theta},
\eea
where
\bea
{\cal H}^{^{M, G}}_{I \theta}={\cal H}^{^{M, G}}_{I 0} \; e^{\frac{1}{2}
\overleftarrow{\partial} \wedge P}\nn
\eea
and
\bea
{\cal H}^{^G}_{I \theta}={\cal H}^{^G}_{I 0}\; .\nn
\eea

In $QED$, ${\cal H}^{^G}_{I \theta}=0$. Thus the $S$-operator ${\bf S}^{^{QED}}_{\theta}$ is
the same as for the $\theta^{\mu \nu} =0$. That is,
\bea
{\bf S}^{^{QED}}_{\theta}={\bf S}^{^{QED}}_{0}.
\eea

Hence ${\bf C}$, ${\bf P}$, ${\bf T}$ and ${\bf CPT}$ are good ``symmetries" for $QED$ on the GM plane.

For a non-abelian gauge theory with non-singlet matter fields, ${\cal H}^{^G}_{I \theta}={\cal H}^{^G}_{I 0} \neq 0$ so that if ${\bf S}^{^{M, G}}_{\theta}$ is the $S$-matrix of the theory,
\bea
{\bf S}^{^{M, G}}_{\theta} \neq {\bf S}^{^{M, G}}_{0}.
\eea

The $S$-operator ${\bf S}^{^{M,G}}_{\theta}$ depends only on $\theta^{0i}$ in a
non-abelian theory, that is,  ${\bf S}^{^{M,G}}_{\theta} =
{\bf S}^{^{M,G}}_{\theta}|_{\theta^{ij}=0}$. Applying ${\bf C}$, ${\bf P}$ and ${\bf T}$ on ${\bf S}^{^{M,G}}_{\theta}$ we can see that ${\bf C}$ and ${\bf T}$ do not affect $\theta^{0i}$ while ${\bf P}$ changes its sign. Thus a non-zero $\theta^{0i}$ contributes to ${\bf P}$ and ${\bf CPT}$ violation. For further analysis see \cite{BPQ}.

\subsection{On Feynman Graphs}

This section uses the results of \cite{bal-sasha-babar} and \cite{bal-sasha-queiroz} where Feynman rules are fully developed and field theories are analyzed further.

\begin{figure}
\centerline{\epsfig{figure=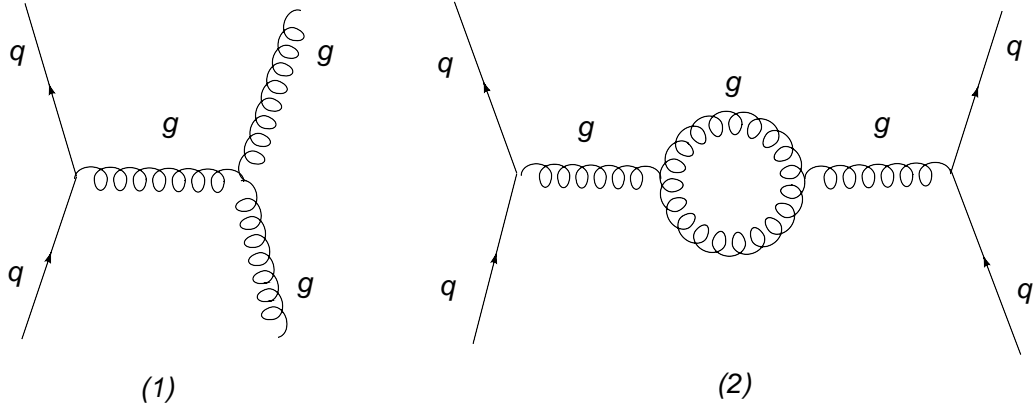}}
\caption{${\bf CPT}$ violating processes on GM plane. (1) shows quark-gluon scattering with a three-gluon vertex. (2) shows a gluon-loop contribution to quark-quark scattering.}
\label{cpt}
\end{figure}

In non-abelian gauge theories, ${\cal H}^{^{G}}_{I \theta}={\cal H}^{^{G}}_{I 0}$ is not zero as gauge fields have self-interactions. The preceding discussions show that the effects of $\theta^{\mu \nu}$ can show up only in Feynman diagrams which are sensitive to products of ${\cal H}^{^{M, G}}_{I \theta}$'s with ${\cal H}^{^{G}}_{I 0}$'s. Fig. (\ref{cpt}) shows two such diagrams.

As an example, consider the first diagram in Fig. (\ref{cpt}) To lowest order, it depends on $\theta^{0i}$.

We can substitute eqn. (\ref{eq:matter}) for ${\cal H}^{^{M, G}}_{I \theta}$ and integrate over ${\bf x}$. That gives,
\bea
{\bf S}^{(2)} =-\frac{1}{2} \int d^{4}x d^{4}y \;  \textrm{T} \Big({\cal H}_{I 0}^{^{M, G}}(x)\; e^{\frac{1}{2} \overleftarrow{\partial}_{0} \theta^{0i} P_{i}}{\cal H}_{I 0}^{^G}(y)\Big)\nn
\eea
where $\overleftarrow{\partial}_{0}$ acts {\it only} on ${\cal H}_{I 0}^{^{M, G}}(x)$ (and not on the step functions in time entering in the definition of $\textrm{T}$.)

Now $P_{i}$, being components of spatial momentum, commutes with
\bea
\int d^{3}y \; {\cal H}^{^{G}}_{I 0}(y)\nn
\eea
and hence for computing the matrix element defining the process ({\it 1}) in Fig. (\ref{cpt}), we can substitute $\overrightarrow{P}_{\textrm{in}}$ for $\overrightarrow{P}$,  $\overrightarrow{P}_{\textrm{in}}$ being the total incident spatial momentum:
\bea
{\bf S}^{(2)} =-\frac{1}{2} \int d^{4}x d^{4}y \;  \textrm{T} \Big({\cal H}_{I 0}^{^{M, G}}(x)\; e^{\frac{1}{2} \overleftarrow{\partial}_{0} \theta^{0i} P^{\textrm{in}}_{i}}{\cal H}_{I 0}^{^G}(y)\Big).
\eea

Thus ${\bf S}^{(2)}$ depends on $\theta^{0i}$ unless
\bea
\theta^{0i}P^{\textrm{in}}_{i} = 0.
\eea

This will happen in the center-of-mass system or more generally if \\
~~$\overrightarrow{\theta^{0}} = $($\theta^{01}$, $\theta^{02}$, $\theta^{03}$) is perpendicular to $\overrightarrow{P}^{\textrm{in}}$.

Under ${\bf P}$ and ${\bf CPT}$, $\theta^{0i} \rightarrow -\theta^{0i}$. This shows clearly that in a general frame, $\theta^{0i}$ contributes to ${\bf P}$ violation and causes ${\bf CPT}$ violation.

The dependence of $S^{(2)}$ on the incident total spatial momentum shows that the scattering matrix is not Lorentz invariant. This noninvariance is caused by the nonlocality of the interaction Hamiltonian density: if we evaluate it at two spacelike separated points, the resultant operators do not commute. Such a violation of causality can lead to Lorentz-noninvariant $S$-operators \cite{bal-sasha-babar}.

The reasoning that reduced $e^{\frac{1}{2}\overleftarrow{\partial} \wedge P}$ to $e^{\frac{1}{2}\overleftarrow{\partial}_{0} \theta^{0i} P^{\textrm{in}}_{i}}$ is valid to all such factors in an arbitrary order in the perturbation expansion of the $S$-matrix and for arbitrary processes, $\overrightarrow{P}^{\textrm{in}}$ being the total incident spatial momentum. As $\theta^{\mu \nu}$ occur only in such factors, this leads to an interesting conclusion: if scattering happens in the
center-of-mass frame, or any frame where $\theta^{0i}P^{\textrm{in}}_{i} = 0$, then the
$\theta$-dependence goes away from the $S$-matrix. That is, $P$ and $CPT$ remain intact if $\theta^{0i}P^{\textrm{in}}_{i} = 0$. The theory becomes $P$ and $CPT$ violating in all other frames.

Terms with products of ${\cal H}^{^{M, G}}_{I \theta}$ and ${\cal H}^{^G}_{I \theta}$ are
$\theta$-dependent and they violate ${\bf CPT}$. Electro-weak and $QCD$ processes will thus
acquire dependence on $\theta$. This is the case when a diagram involves products of
${\cal H}^{^{M, G}}_{I \theta}$ and ${\cal H}^{^G}_{I \theta}$. For example
quark-gluon and quark-quark scattering on the GM plane become $\theta$-dependent ${\bf CPT}$
violating processes (See Fig. (\ref{cpt})).

These effects can be tested experimentally.

\begin{center}
Summary of Chapter \ref{cmb1}
\end{center}

\begin{enumerate}
\item Tiny (small scale) nonuniformities (inhomogeneities and anisotropies) in the CMB radiation suggest the existence of temperature fluctuations (ie. nonequilibrium) in the early universe just before photon-baryon decoupling. These are reflected in the distribution of large scale structures such as galaxy clusters in the universe today.
\item There are problems in the standard model of cosmology: The theory of inflation attempts to explain the high causal connectedness or correlation in the CMB radiation (High isotropy of CMB implies that radiation from two opposite points in the sky must have been in causal contact before decoupling. Decoupling happened in the ``far past'', too close to the big bang singularity, and so such causal contact is not possible with the homogeneous and isotropic metric of standard big bang cosmology), flatness or small curvature of the present universe, absence of primordial or early phase transition byproducts such as monopoles and cosmic strings and the origin of tiny nonuniformities in the highly (large-scale) uniform CMB radiation. A scalar field (inflaton) could have caused a fast expansion of the early universe thus neutralizing accausal, curvature and phase transition byproduct effects and quantum corrections to its dynamics would be responsible for tiny nonuniformities in the CMB radiation.

    Other cosmological problems susceptible to noncommutativity include dark matter (associated with inconsistencies involving excesses in the observed motion of galaxies and clusters), dark energy (associated with observed red-shifts which suggest an accelerated expansion of the universe) and the fact that only four spacetime dimensions are observed even though physical theories predict more than four dimensions for spacetime.

\item Quantum theory predicts a noncommutative structure for spacetime at small scales. Therefore noncommutativity of spacetime will contribute to the tiny nonuniformities of the CMB radiation through it naturally expected affects on the quantum dynamics (taking place precisely at such small scales) of the inflaton.

\item During inflation, metric  fluctuations are negligible compared to inflaton fluctuations. However, at the end of inflation the quantum fluctuations of the inflaton become a \emph{source} of fluctuations in the metric of spacetime as well as of radiation and matter. The power spectrum or Fourier transform of the (metric) two-point correlation amplitude or potential will depend on the spacetime noncommutativity parameter. Using nonequilibrium dynamics one can find the fluctuations in temperature, and corresponding temperature correlations, induced by the metric fluctuations. These temperature fluctuations will then show up in the CMB radiation.
\item One gets a noncommutativity dependent power spectrum, noncommutativity-induced causality violation and a non-Gaussian probability distribution.

\end{enumerate}
\chapter{CMB Power Spectrum and Anisotropy} \label{cmb1}

\vspace{0.5cm} Modern cosmology has now emerged  as a testing ground
for theories beyond the standard model of particle physics. In this
paper, we consider quantum fluctuations of the inflaton scalar field
on certain noncommutative spacetimes and look for noncommutative
corrections in the cosmic microwave background (CMB) radiation.
Inhomogeneities in the distribution of large scale structure and
anisotropies in the CMB radiation can carry traces of
noncommutativity of the early universe. We show that its power
spectrum becomes direction-dependent when spacetime is
noncommutative. (The effects due to noncommutativity can be observed
experimentally in the distribution of large scale structure of
matter as well.) Furthermore, we have shown that the probability
distribution determining the temperature fluctuations is not
Gaussian for  noncommutative spacetimes.

\section{INTRODUCTION}\label{sec:intro}
The CMB radiation shows  how the universe was like when it was only
$400, 000$ years old. If photons and baryons were in equilibrium
before they decoupled from each other, then the CMB radiation we
observe today should have a black body spectrum indicating a smooth
early universe. But in 1992, the Cosmic Background Explorer (COBE)
satellite detected anisotropies in the CMB radiation, which led to
the conclusion that the early universe was not smooth: There were
small perturbations in the photon-baryon fluid.

The theory of inflation  was introduced \cite{guth, Linde, Albrecht}
to resolve the fine tuning problems associated with the standard Big
Bang cosmology. An important property of inflation is that it can
generate irregularities in the universe, which may lead to the
formation of structure. Inflation is assumed to be driven by a
classical scalar field that accelerates the observed universe
towards a perfect homogeneous state. But we live in a quantum world
where perfect homogeneity is never attained. The classical scalar
field has quantum fluctuations around it and these fluctuations act
as seeds for the primordial perturbations over the smooth universe.
Thus according to these ideas, the early universe had
inhomogeneities and we observe them today in the distribution of
large scale structure and anisotropies in the CMB radiation.

Physics at Planck  scale could be radically different. It is the
regime of string theory and quantum gravity. Inflation stretches a
region of Planck size into cosmological scales. So, at the end of
inflation, physics at Planck region should leave its signature on
the cosmological scales too.

There are indications  both from quantum gravity and string theory
that spacetime is noncommutative with a length scale of the order of
Planck length. In this paper we explore the consequences of such
noncommutativity for CMB radiation in the light of  recent
developments in the field of noncommutative quantum field theories
relating to deformed Poincar\'e symmetry.

The early universe and CMB in the noncommutative framework have been
addressed in many places \cite{Greene, Lizzi, Brandenberger1, Huang,
Brandenberger2, queiroz1, Fatollahi2, Fatollahi1}. In \cite{Greene},
the noncommutative parameter $\theta_{\mu \nu} = -\theta_{\nu \mu}
=\textrm{constants}$ with $\theta_{0i} =0$, ($\mu, \nu = 0, 1, 2,
3$, with $0$ denoting time direction), characterizing the Moyal
plane is scale dependent, while \cite{Brandenberger1,
Brandenberger2, Huang} have considered noncommutativity based on
stringy space-time uncertainty relations. Our approach differs from
these authors since our quantum fields obey twisted statistics, as
implied by the deformed Poincar\'e symmetry in quantum theories.

We organize the paper as follows: In section II, we discuss how
noncommutativity breaks the usual Lorentz invariance and indicate
how this breaking can be interpreted as invariance under a  deformed
Poincar\'e symmetry. In section III, we write down an expression for
a scalar quantum field in the noncommutative framework and show how
its two-point function is modified. We review the theory of
cosmological perturbations and (direction-independent) power
spectrum for $\theta_{\mu \nu}=0$ in section IV. In section V, we
derive the power spectrum for the noncommutative Groenewold-Moyal
plane ${\cal A}_{\theta}$ and show that it is direction-dependent
and breaks statistical isotropy. In section VI, we compute the
angular correlations using this power spectrum and show that there
are nontrivial ${\cal O}( \theta^{2})$ corrections to the CMB
temperature fluctuations.  Next, in section VII, we discuss the
modifications of the $n$-point functions for any $n$ brought about
by a non-zero $\theta^{\mu \nu}$ and show in particular that the
underlying probability distribution is not Gaussian. The paper
concludes with section VIII.

\section{Noncommutative Spacetime and Deformed Poincar\'e Symmetry}
At energy scales close to the Planck scale, the quantum nature of
spacetime is expected to become important. Arguments based on
Heisenberg's uncertainty principle and Einstein's theory of
classical gravity suggest that spacetime has a noncommutative
structure at such length scales \cite{doplicher}. We can model such
spacetime noncommutativity by the commutation relations
\cite{ConnesC, Madore, Landi, Bondia} \bea \label{eq:nonSpaceTime}
[\widehat{x}_{\mu}, \widehat{x}_{\nu}] = i \theta_{\mu \nu} \eea
where $\theta_{\mu \nu} = - \theta_{\nu \mu}$ are constants and
$\widehat{x}_{\mu}$ are the coordinate functions of the chosen
coordinate system: \bea \widehat{x}_{\mu}(x) = x_{\mu}. \eea

The above relations depend on choice of  coordinates. The
commutation relations given in eqn. (\ref{eq:nonSpaceTime}) only
hold in special coordinate systems and will look quite complicated
in other coordinate systems. Therefore,  it is important to know in which coordinate system the above simple form for the
commutation relations holds. For cosmological applications, it is
natural to assume that eqn. (\ref{eq:nonSpaceTime}) holds in a
comoving frame, the coordinates in which galaxies are freely
falling. Not only does it make the analysis and comparison with the
observation easier, but also make the time coordinate  the proper
time for us (neglecting the small local accelerations).

The  relations (\ref{eq:nonSpaceTime}) are  not invariant under
naive Lorentz transformations either. But they are invariant under a
deformed Lorentz Symmetry \cite{chaichian}, in which the coproduct
on the Lorentz group  is deformed while the group structure is kept
intact, as we briefly explain below.

The Lie algebra ${\cal P}$ of the Poincar\'e group has generators (basis) $M_{\alpha \beta}$ and $P_{\mu}$. The subalgebra of infinitesimal generators $P_{\mu}$ is abelian and we can make use of this fact to construct a twist element ${\cal F}_{\theta}$ of the underlying quantum group theory \cite{drinfeld, chaichian2, chari}. Using this twist element, the coproduct of the universal enveloping algebra ${\cal U}({\cal P})$ of the Poincar\'e algebra can be deformed in such a way that it is compatible with the above commutation relations.

The coproduct $\Delta_{0}$ appropriate for $\theta_{\mu \nu} =0$ is a symmetric map from ${\cal U}({\cal P})$ to ${\cal U}({\cal P}) \otimes {\cal U}({\cal P})$. It defines the action of ${\cal P}$ on the tensor product of representations. In the case of the generators $X$ of ${\cal P}$, this standard coproduct is
\bea
\Delta_{0}(X) = 1 \otimes X + X \otimes 1.
\eea

The twist element is
\bea
{\cal F}_{\theta} = \textrm{exp}(-\frac{i}{2}\theta^{\alpha \beta}P_{\alpha} \otimes P_{\beta}), ~~~P_{\alpha} = -i \partial_{\alpha}.
\eea
(The Minkowski metric with signature ($-, +, +, +$) is used to raise and lower the indices.)

In the presence of the twist, the coproduct $\Delta_{0}$ is modified to $\Delta_{\theta}$ where
\bea
\Delta_{\theta} = {\cal F}_{\theta}^{-1} \Delta_{0} {\cal F}_{\theta}.
\eea

It is easy to see that the coproduct for translation generators are not deformed,
\bea
\Delta_{\theta}(P_{\alpha}) = \Delta_{0}(P_{\alpha})
\eea
while the coproduct for Lorentz generators are deformed:
\bea
\Delta_{\theta}(M_{\mu \nu}) &=& 1 \otimes M_{\mu \nu} + M_{\mu \nu} \otimes 1 - \frac{1}{2}\Big[(P\cdot \theta)_{\mu}\otimes P_{\nu}-P_{\nu}\otimes (P\cdot \theta)_{\mu} - (\mu \leftrightarrow \nu) \Big], \nn \\
(P \cdot \theta)_{\lambda} &=& P_{\rho}\theta^{\rho}_{\lambda}.
\eea

The algebra ${\cal A}_{0}$ of functions on the Minkowski space ${\cal M}^{4}$ is commutative with the commutative multiplication $m_{0}$:
\bea
m_{0} (f \otimes g)(x) = f(x)g(x).
\eea

The Poincar\'e algebra acts on ${\cal A}_{0}$ in a well-known way
\bea
P_{\mu}f(x) = -i \partial_{\mu}f(x), \; \; \;  M_{\mu \nu}\; f(x) = -i(x_{\mu}\partial_{\nu} - x_{\nu}\partial_{\mu})f(x).
\eea

It acts on tensor products $f \otimes g$ using the coproduct $\Delta_{0}(X)$.

This commutative multiplication is changed in the Groenewold-Moyal algebra ${\cal A}_{\theta}$ to $m_{\theta}$:
\bea
m_{\theta} (f\otimes g)(x) = m_{0}\Big[\textrm{e}^{-\frac{i}{2}\theta^{\alpha \beta}P_{\alpha}\otimes P_{\beta}} \; f \otimes g\Big](x) = (f \star g)(x).
\eea

Equation (\ref{eq:nonSpaceTime}) is a consequence of this $\star$-multiplication:
\bea
[\widehat{x}_{\mu}, \widehat{x}_{\nu}]_{\star} &=& m_{\theta} \; (\widehat{x}_{\mu} \otimes \widehat{x}_{\nu} - \widehat{x}_{\nu} \otimes \widehat{x}_{\mu}) = i \theta_{\mu \nu}.
\eea

The Poincar\'e algebra acts on functions $f \in {\cal A}_{\theta}$ in the usual way while it acts on tensor products $f \otimes g \in {\cal A}_{\theta} \otimes {\cal A}_{\theta}$ using the coproduct $\Delta_{\theta}(X)$ \cite{chaichian, Aschieri}.

Quantum field theories can be constructed on the noncommutative
spacetime ${\cal A}_{\theta}$ by replacing ordinary multiplication
between the fields by $\star$-multiplication and deforming
statistics as we discuss below \cite{bal, uv-ir, bal-statuv-ir,
bal-sasha-babar}. These theories are invariant under the deformed
Poincar\'e action \cite{chaichian, Aschieri, bal-statuv-ir,
bal-sasha-babar} under which $\theta_{\mu\nu}$ is invariant. It is
thus possible to observe $\theta_{\mu\nu}$ without violating
deformed Poincar\'e symmetry. But of course they are not invariant
under the standard undeformed action of the Poincar\'e group as
shown for example by the observability of $\theta_{\mu\nu}$.
\section{Quantum Fields in Noncommutative Spacetime}
It can be shown immediately that the action of the deformed coproduct is not compatible with standard statistics \cite{bal-statuv-ir}. Thus for $\theta^{\mu \nu} =0$, we have the axiom in quantum theory that the statistics operator $\tau_{0}$ defined by
\bea
\tau_{0} \; (\phi \otimes \chi) = \chi \otimes \phi
\eea
is superselected. In particular, the Lorentz group action must and {\it does} commute with the statistics operator,
\bea
\tau_{0} \Delta_{0}(\Lambda) = \Delta_{0}(\Lambda)\tau_{0},
\eea
where $\Lambda \in {\cal P}^{\uparrow}_{+}$, the connected component of the Poincar\'e group.

Also all the states in a given superselection sector are eigenstates of $\tau_{0}$ with the same eigenvalue. Given an element $\phi \otimes \chi$ of the tensor product, the physical Hilbert spaces can be constructed from the elements
\bea
\Big(\frac{1\pm \tau_{0}}{2}\Big) (\phi \otimes \chi).
\eea

Now since $\tau_{0} {\cal F}_{\theta} = {\cal F}^{-1}_{\theta} \tau_{0}$, we have that
\bea
\tau_{0} \Delta_{\theta}(\Lambda) \neq \Delta_{\theta}(\Lambda) \tau_{0}
\eea
showing that the use of the usual statistics operator is not compatible with the deformed coproduct.

But the new statistics operator
\bea
\tau_{\theta} \equiv {\cal F}^{-1}_{\theta} \tau_{0} {\cal F}_{\theta}, \; \; \; \tau_{\theta}^{2} = 1 \otimes 1
\eea
does commute with the deformed coproduct.

The two-particle state $|p, q\rangle_{S_{\theta}, A_{\theta}}$ for bosons and fermions obeying deformed statistics is constructed as follows:
\bea
|p, q\rangle_{S_{\theta}, A_{\theta}} &=& |p\rangle \otimes_{_{S_{\theta}, A_{\theta}}} |q\rangle =\Big(\frac{1 \pm \tau_{\theta}}{2}\Big) (|p\rangle \otimes |q\rangle)\nn \\
&=& \frac{1}{2}\Big(|p\rangle \otimes |q\rangle \pm \textrm{e}^{-i p_{\mu}\theta^{\mu \nu}q_{\nu}}|q\rangle \otimes |p\rangle\Big).
\eea

Exchanging $p$ and $q$ in the above, one finds
\bea
\label{eq:pq-qp}
|p, q\rangle_{S_{\theta}, A_{\theta}} = \pm \; \textrm{e}^{-i p_{\mu}\theta^{\mu \nu}q_{\nu}}|q, p\rangle_{S_{\theta}, A_{\theta}}.
\eea

In Fock space, the above two-particle state is constructed from a second-quantized field $\varphi_{\theta}$ according to
\bea
\frac{1}{2}\langle0|\varphi_{\theta}(x_{1})\varphi_{\theta}(x_{2}) a_{\bf q}^{\dagger}a_{\bf p}^{\dagger}|0\rangle &=& \Big(\frac{1\pm \tau_{\theta}}{2}\Big) (e_{p} \otimes e_{q})(x_{1}, x_{2})\nn \\
&=& (e_{p} \otimes_{S_{\theta}, A_{\theta}} e_{q})(x_{1}, x_{2})\nn \\
&=& \langle x_{1}, x_{2}|p, q\rangle_{S_{\theta}, A_{\theta}}
\eea
where $\varphi_{0}$ is a boson(fermion) field associated with $|p, q\rangle_{S_{0}}$ ($|p, q\rangle_{A_{0}}$).

On using eqn. (\ref{eq:pq-qp}), this leads to the commutation relation
\bea
a_{\bf p}^{\dagger}a_{\bf q}^{\dagger} =  \pm \; \textrm{e}^{i p_{\mu}\theta^{\mu \nu}q_{\nu}}\; a_{\bf q}^{\dagger}a_{\bf p}^{\dagger}.
\eea

Let $P_{\mu}$ be the Fock space momentum operator. (It is the representation of the translation generator introduced previously. We use the same symbol for both.) Then the operators $a_{\bf p}$ , $a_{\bf p}^{\dagger}$ can be written as follows:
\bea
\label{eq:cp-ap}
a_{\bf p} = c_{\bf p} \; \textrm{e}^{-\frac{i}{2}p_{\mu}\theta^{\mu \nu}P_{\nu}}, \; \; \; a_{\bf p}^{\dagger} = c_{\bf p}^{\dagger} \; \textrm{e}^{\frac{i}{2}p_{\mu}\theta^{\mu \nu}P_{\nu}}\; ,
\eea
$c_{\bf p}$'s being $\theta^{\mu \nu}=0$ annihilation operators.

The map from $c_{\bf p}, c_{\bf p}^{\dagger}$ to $a_{\bf p}, a_{\bf p}^{\dagger}$ in eqn. (\ref{eq:cp-ap}) is known as the ``dressing transformation" \cite{Grosse, Faddeev-Zamolodchikov}.

In the noncommutative case, a free spin-zero quantum scalar field of mass $m$ has the mode expansion
\bea
\label{eq:modeExp}
\varphi_{\theta}(x) =\int \frac{d^{3} p}{(2\pi)^{3}} \; (a_{\bf p}\; \textrm{e}_{p}(x) + a_{\bf p}^{\dagger} \; \textrm{e}_{-p}(x))
\eea
where
\bea
\textrm{e}_{p}(x) = \textrm{e}^{-i\; p\cdot x}, \; \; p \cdot x = p_{0}x_{0} - {\bf p}\cdot {\bf x}, \; \; \; p_{0} = \sqrt{{\bf p}^{2} + m^{2}} > 0. \nn
\eea

The deformed quantum field $\varphi_{\theta}$ differs form the undeformed quantum field $\varphi_{0}$ in two ways: $i$.) $\textrm{e}_{p}$ belongs to the noncommutative algebra of ${\cal M}^{4}$ and $ii$.) $a_{\bf p}$ is deformed by statistics. The deformed statistics can be accounted for by writing \cite{gauge-gravity}
\bea
\label{eq:twistedfield}
\varphi_{\theta} = \varphi_{0}\; \textrm{e}^{\frac{1}{2}\overleftarrow{\partial} \wedge P}
\eea
where
\bea
\overleftarrow{\partial} \wedge P \equiv \overleftarrow{\partial}_{\mu}\theta^{\mu \nu}P_{\nu}.
\eea

It is easy to write down the $n$-point correlation function for the deformed quantum field $\varphi_{\theta}(x)$ in terms of the undeformed field $\varphi_{0}(x)$:
\bea
&&\langle 0| \varphi_{\theta}(x_1) \varphi_{\theta}(x_2) \cdots \varphi_{\theta}(x_n)|0 \rangle \nn\\
&&~~~~~=\langle 0|\varphi_0 (x_1) \varphi_0(x_2) \cdots \varphi_0 (x_n)|0\rangle \;
\textrm{e}^{(-{i\over2}\sum^n_{J=2}\sum^{J-1}_{I=1}\ola{\del}_{x_{I}} \wedge \ola{\del}_{x_{J}})}.\nn
\eea

On using
\bea
\varphi_{\theta}(x)=\varphi_{\theta}({\bf x}, t)= \int \frac{d^{3} k}{(2\pi)^{3}} \; \Phi_{\theta}({\bf k}, t) \; \textrm{e}^{i {\bf k} \cdot {\bf x}},
\eea
we find for the vacuum expectation values, in momentum space
{\footnotesize
\bea
&&\langle 0| \Phi_{\theta}({\bf k}_1, t_1) \Phi_{\theta}({\bf k}_2, t_2) \cdots \Phi_{\theta}({\bf k}_n, t_n)|0\rangle=\textrm{e}^{({i\over2}\sum_{J>I}{\bf k}_I \wedge
{\bf k}_J)}\times \nn\\
&&\langle 0| \Phi_0({\bf k}_1, t_1 + {\vec{\theta}^{0} \cdot {\bf k}_2+\vec{\theta}^{0}\cdot {\bf k}_3+
\cdots+\vec{\theta}^{0}\cdot {\bf k}_n \over 2})\Phi_0({\bf k}_2,
t_2+{-\vec{\theta}^{0}\cdot {\bf k}_1+\vec{\theta}^{0}\cdot {\bf k}_3+\cdots+\vec{\theta}^{0}\cdot {\bf k}_n \over 2})\nn\\
&&~~ \cdots \Phi_0({\bf k}_n,
t_n+{-\vec{\theta}^{0}\cdot {\bf k}_1-\vec{\theta}^{0}\cdot {\bf k}_2-\cdots-\vec{\theta}^{0}\cdot {\bf k}_{n-1} \over 2})|0\rangle\nn\\
\eea
}
where
\bea
\label{eq:vecTheta}
\vec{\theta}^{0} = (\theta^{01}, \theta^{02}, \theta^{03}).
\eea

Since the underlying Friedmann-Lema\^itre-Robertson-Walker (FLRW) spacetime has spatial translational invariance,
\bea
{\bf k}_1 + {\bf k}_2 + \cdots + {\bf k}_n = 0,\nn
\eea
the $n$-point correlation function in momentum space becomes
{\footnotesize
\bea
\label{eq:nthetacorr}
&&\langle 0| \Phi_{\theta}({\bf k}_1, t_1) \Phi_{\theta}({\bf k}_2, t_2) \cdots \Phi_{\theta}({\bf k}_n, t_n)|0\rangle\nn\\
&&~~~~= \textrm{e}^{({i\over2}\sum_{J>I}{\bf k}_I \wedge
{\bf k}_J)}\langle 0| \Phi_0({\bf k}_1, t_1 - {\vec{\theta}^{0}\cdot {\bf k}_1 \over 2})\Phi_0({\bf k}_2, t_2
- \vec{\theta}^{0}\cdot {\bf k}_1-{\vec{\theta}^{0}\cdot {\bf k}_2\over2})\nn \\
&&\cdots \Phi_0({\bf k}_n, t_n - \vec{\theta}^{0}\cdot {\bf k}_1 -
\vec{\theta}^{0}\cdot {\bf k}_2 - \cdots -{\vec{\theta}^{0}\cdot {\bf k}_{n-1}}-{\vec{\theta}^{0}\cdot {\bf k}_n\over2})|0\rangle.
\eea
}
In particular, the two-point correlation function is
\bea
\label{eq:2thetacorr}
\langle 0| \Phi_{\theta}({\bf k}_1, t_1) \Phi_{\theta}({\bf k}_2, t_2)|0\rangle = \langle 0| \Phi_0({\bf k}_1, t_1-{\vec{\theta}^{0}\cdot {\bf k}_1 \over 2}) \Phi_0({\bf k}_2, t_2 - {\vec{\theta}^{0}\cdot {\bf k}_1 \over 2})|0\rangle,\nn\\
\eea
since it vanishes unless ${\bf k}_{1} + {\bf k}_{2} =0$ and hence $\textrm{e}^{(\frac{i}{2}\sum_{J>I}\; k_{I} \wedge k_{J})}=1$.

We emphasize that eqns. (27), (29) and (30) come from eqn. (20) which implies
eqns. (21), (23) and (25). They are exclusively due to deformed statistics. The $*$-product is still mandatory when taking products of $\varphi_{\theta}$ evaluated at the same point.

In standard Hopf algebra theory, the exchange operation is to be
performed using the $\mathcal{R}$-matrix times the flip operator
$\sigma$ \cite {mack1, mack2}. It is easy to check that $\mathcal{R}\sigma$ acts as
identity on any pair of factors in eqns. (27) and (29).

One can also explicitly show that the $n$-point functions are
invariant under the twisted Poincar\'e group while those of the
conventional theory are not. Hence the requirement of twisted
Poincar\'e invariance fixes the structure of $n$-point functions. These points are discussed further in \cite{bal-statuv-ir}.

It is interesting to note that the two-point correlation function is nonlocal in time in the noncommutative frame work. Also note the following: Assuming that $\theta^{\mu \nu}$ is non-degenerate, we can write it as
\bea
&&\theta^{\mu \nu} = \alpha \; \epsilon_{ab}\;  e^{\mu}_{a}\; e^{\nu}_{b} + \beta \; \epsilon_{ab}\; f^{\mu}_{a}\; f^{\nu}_{b},\nn \\
&&\alpha, \beta \neq 0, \; \; \epsilon_{ab} = -\epsilon_{ba},\; \; a, b = 1, 2\nn
\eea
where $e_{a}, e_{b}, f_{a}, f_{b}$ are orthonormal real vectors. Thus $\theta^{\mu \nu}$ defines two distinguished two-planes in ${\cal M}^{4}$, namely those spanned by $e_{a}$ and by $f_{a}$. For simplicity we have assumed that one of these planes contains the time direction, say $e_{1} : e^{\mu}_{1} = \delta^{\mu}_{0}$. The $\theta^{0i}$ part then can be regarded as defining a spatial direction $\vec{\theta}^{0}$ as given by eqn. (\ref{eq:vecTheta}).

We will make use of the modified two-point correlation functions given by eqn. (\ref{eq:2thetacorr}) when we define the power spectrum for inflaton field perturbations in the noncommutative frame work.

\section{Cosmological Perturbations and (Direction-Independent) Power Spectrum for $\theta^{\mu \nu}=0$}

In this section we briefly review how fluctuations in the inflaton field cause inhomogeneities in the distribution of matter and radiation following \cite{dodelson}.

The scalar field $\phi$ driving inflation can be split into a zeroth order homogeneous part and a first order perturbation:
\bea
\phi({\bf x}, t) = \phi^{(0)}(t)+\delta \phi({\bf x}, t)
\eea

The energy-momentum tensor for $\phi$ is
\bea
{\cal T}^{\alpha}_{\; \; \; \beta} = g^{\alpha \nu}\frac{\partial \phi}{\partial x^{\nu}}\frac{\partial \phi}{\partial x^{\beta}} - g^{\alpha}_{\; \; \; \beta}\Big[\frac{1}{2}g^{\mu \nu}\frac{\partial \phi}{\partial x^{\mu}}\frac{\partial \phi}{\partial x^{\nu}} + V(\phi)\Big]
\eea

We assume a spatially flat, homogeneous and isotropic (FLRW) background with the metric
\bea
ds^{2} = dt^{2} - {a}^{2}(t)d{\bf x}^{2}
\eea
where ${a}$ is the cosmological scale factor, and nonvanishing $\Gamma$'s
\bea
\Gamma^{0}_{\; \; ij} = \delta_{ij}{a}^{2}H\; \; \;  \textrm{and}\; \; \;  \Gamma^{i}_{\; \; 0j}=\Gamma^{i}_{\; \; j0}=\delta^{i}_{j}H \nn
\eea
where $H$ is the Hubble parameter.

In conformal time $\eta$ where $d \eta =\frac{dt}{{a}(t)}, -\infty < \eta < 0$, the metric becomes
\bea
ds^{2} = a^{2}(\eta)(d\eta^{2} - d{\bf x}^{2}),
\eea
where $a$ is the cosmological scale factor now regarded as a function of conformal time. Using this metric we write the equation for the zeroth order part of $\phi$ \cite{dodelson},
\bea
\ddot{\phi}^{(0)} + 2aH \dot{\phi}^{(0)} + a^{2}V'\phi^{(0)} = 0,
\eea
where overdots denote derivatives with respect to conformal time $\eta$ and $V'$ is the derivative of $V$ with respect to the field $\phi^{(0)}$. Notice that in conformal time $\eta$ we have $\frac{d a(\eta)}{d \eta} = a^{2}(\eta) H$ while in cosmic time $t$ we have $\frac{d {a}(t)}{d t}= {a} H$.

The equation for $\delta \phi$ can be obtained from the first order perturbation of the energy-momentum tensor conservation equation:
\bea
{\cal T}^{\mu}_{\; \; \; \nu;\; \mu} = \frac{\partial {\cal T}^{\mu}_{\; \; \; \nu}}{\partial x^{\mu}} + \Gamma^{\mu}_{\; \; \; \alpha \mu}{\cal T}^{\alpha}_{\; \; \; \nu} - \Gamma^{\alpha}_{\; \; \; \nu \mu}{\cal T}^{\mu}_{\; \; \; \alpha} =0.
\eea

The perturbed part of the energy-momentum tensor $\delta T^{\mu}_{\; \; \; \nu}$ satisfies the following conservation equation in momentum space \cite{dodelson}:
\bea
\frac{\partial \delta T^{0}_{\; \; \; 0}}{\partial t} + ik_{i}\delta T^{i}_{\; \; \; 0} +3H \delta T^{0}_{\; \; \; 0}-H \delta T^{i}_{\; \; \; i} =0,
\eea
where
\bea
T^{\mu \nu}({\bf k}, t)= \int d^{3}x~{\cal T}^{\mu \nu}({\bf x}, t)~\textrm{e}^{-i{\bf k}\cdot {\bf x}}.
\eea

Let $\phi({\bf x}, t) = \int \frac{d^{3}k}{(2\pi)^{3}}~\tilde{\phi}({\bf k}, t)~\textrm{e}^{i{\bf k}\cdot {\bf x}}$. Writing down the perturbations to the energy-momentum tensor in terms of $\tilde{\phi}({\bf k}, t)$,
\bea
\delta T^{i}_{\; \; \; 0} &=& \frac{ik_{i}}{a^{3}} \dot{\tilde{\phi}}^{(0)} \delta \tilde{\phi} , \nn \\
\delta T^{0}_{\; \; \; 0} &=& \frac{-\dot{\tilde{\phi}}^{(0)} \dot{\delta \tilde{\phi}}}{a^{2}} - V'(\tilde{\phi}^{(0)}) \delta \tilde{\phi} , \nn \\
\delta T^{i}_{\; \; \; j} &=& \delta_{ij}\Big(\frac{\dot{\tilde{\phi}}^{(0)} \dot{\delta \tilde{\phi}}}{a^{2}} - V'(\tilde{\phi}^{(0)}) \delta \tilde{\phi} \Big) ,\nn
\eea
the conservation equation becomes
\bea
\ddot{\delta \tilde{\phi}} + 2aH \dot{\delta \tilde{\phi}} + k^{2} \delta \tilde{\phi} = 0.
\eea

Eliminating the middle Hubble damping term by a change of variable \\
~~$\zeta ({\bf k}, \eta) = a(\eta) \delta \tilde{\phi}({\bf k}, \eta)$, the above equation becomes
\bea
\label{eq:harmonic}
\ddot{\zeta}({\bf k}, \eta)+ \omega_{k}^{2}(\eta) \zeta({\bf k}, \eta) = 0, \; \; \; \omega_{k}^{2}(\eta) \equiv \Big(k^{2} -\frac{\ddot{a}(\eta)}{a(\eta)}\Big).
\eea

The mode functions $u$ associated with the quantum operator $\hat{\zeta}$ satisfy
\bea
\label{eq:ukEqn}
\ddot{u}({\bf k}, \eta)+\Big(k^{2} -\frac{\ddot{a}(\eta)}{a(\eta)}\Big)u({\bf k}, \eta)=0
\eea
with the initial conditions $u({\bf k}, \eta_{i}) = \frac{1}{\sqrt{2\omega_{k}(\eta_{i})}}$ and $\dot{u}({\bf k}, \eta_{i})=i \sqrt{\omega_{k}(\eta_{i})}$. Notice that these initial conditions have meaning only when $\omega_{k}(\eta_{i}) > 0$.

We can immediately write down the quantum operator associated with the variable $\zeta$,
\bea
\hat{\zeta}({\bf k}, \eta) = u({\bf k}, \eta)\hat{a}_{\bf k}+u^{*}({\bf k}, \eta)\hat{a}^{\dagger}_{\bf k},
\eea
with the bosonic commutation relations $[\hat{a}_{\bf k}, \hat{a}_{{\bf k}'}] = [\hat{a}_{\bf k}^{\dagger}, \hat{a}_{{\bf k}'}^{\dagger}]=0$ and \\
$[\hat{a}_{\bf k}, \hat{a}_{{\bf k}'}^{\dagger}] = (2 \pi)^{3} \delta^{3}({\bf k} - {\bf k}')$.

During inflation we have scale factor $a(\eta) \simeq -(\eta H)^{-1}$. Thus eqn. (\ref{eq:ukEqn}) takes the form \cite{dodelson}
\bea
\label{eq:evolution}
\ddot{u}+\Big(k^{2} -\frac{2}{\eta^{2}}\Big)u=0.
\eea

When the perturbation modes are well within the horizon, $k|\eta| \gg 1$, one can obtain a properly normalized solution $u({\bf k}, \eta)$ from the conditions imposed on it at very early times during inflation. Such a solution is \cite{dodelson, Mukhanov}
\bea
\label{eq:properSol}
u({\bf k}, \eta)=\frac{1}{\sqrt{2k}} \Big(1-\frac{i}{k \eta}\Big)~\textrm{e}^{-ik(\eta-\eta_{i})}.
\eea

The variances involving $\hat{\zeta}$ and $\hat{\zeta}^{\dagger}$ are
\bea
\label{eq:zeta-corr}
\langle 0|\hat{\zeta}({\bf k}, \eta) \hat{\zeta}({\bf k}', \eta)|0\rangle &=& 0, \nn \\
\langle 0|\hat{\zeta}^{\dagger}({\bf k}, \eta) \hat{\zeta}^{\dagger}({\bf k}', \eta)|0\rangle &=& 0, \nn \\
\langle 0|\hat{\zeta}^{\dagger}({\bf k}, \eta) \hat{\zeta}({\bf k}', \eta)|0\rangle &=& (2\pi)^{3} |u({\bf k}, \eta)|^{2} \delta^{3}({\bf k} - {\bf k}')\nn \\
&\equiv& (2\pi)^{3} P_{\zeta}({\bf k}, \eta) \delta^{3}({\bf k} - {\bf k}')
\eea
where $P_{\zeta}$ is the power spectrum of $\hat{\zeta}$. Eqn. (\ref{eq:zeta-corr}) can be treated as a general definition of power spectrum.

In the case when spacetime is commutative ($\theta^{\mu \nu} =0$), the power spectrum in eqn. (\ref{eq:zeta-corr}) is
\bea
\label{eq:zeta-corr-commu}
\langle 0|\hat{\zeta}^{\dagger}({\bf k}, \eta) \hat{\zeta}({\bf k}', \eta)|0\rangle = (2\pi)^{3} P_{\zeta}(k, \eta) \delta^{3}({\bf k} - {\bf k}').
\eea

The Dirac delta function in eqns. (\ref{eq:zeta-corr}) and (\ref{eq:zeta-corr-commu}) shows that perturbations with different wave numbers are uncoupled as a consequence of the translational invariance of the underlying spacetime. Rotational invariance of the underlying (commutative) spacetime constraints the power spectrum $P_{\zeta}(k, \eta)$ to depend only on the magnitude of ${\bf k}$.

Towards the end of inflation, $k|\eta |$ ($-\infty < \eta <0$) becomes very small. In that case the small argument limit of eqn. (\ref{eq:properSol}),
\bea
\lim_{k|\eta| \rightarrow 0}~~u({\bf k}, \eta)=\frac{1}{\sqrt{2k}}~\frac{-i}{k \eta}~\textrm{e}^{-ik(\eta-\eta_{i})},
\eea
gives the power spectrum $P_{\zeta}(k, \eta) =|u({\bf k}, \eta)|^{2}$. On using $\zeta({\bf k}, \eta)=a(\eta)\delta\tilde{\phi}({\bf k}, \eta)$, we write the power spectrum $P_{\delta \tilde{\phi}}$ for the scalar field perturbations \cite{dodelson}:
\bea
\label{eq:metric-power}
P_{\delta \tilde{\phi}}(k, \eta) =\frac{|u({\bf k}, \eta)|^{2}}{a(\eta)^{2}} =\frac{1}{2k^{3}}\frac{1}{a(\eta)^{2}\eta^{2}}.
\eea
In terms of the Hubble parameter $H$ during inflation ($H \simeq -\frac{1}{a(\eta) \eta}$), the power spectrum becomes
\bea
\label{eq:powerDeltaPhi}
P_{\delta \tilde{\phi}}(k, \eta) = \frac{1}{2k^{3}} H^{2}.
\eea

We are interested in the post-inflation power spectrum for the scalar metric perturbations since they couple to matter and radiation and give rise to inhomogeneities and anisotropies in their respective distributions which we observe. This spectrum comes from the inflaton field since the inflaton field perturbations get transferred to the scalar part of the metric.

We write the perturbed metric in the longitudinal gauge \cite{mukhanov},
\bea
ds^{2} = a^{2}(\eta)\Big[(1+2 \chi({\bf x}, \eta)) d\eta^{2} - (1-2\Psi({\bf x}, \eta))\gamma^{ij}({\bf x}, \eta)dx_{i}dx_{j}\Big],
\eea
where $\chi$ and $\Psi$ are two physical metric degrees of freedom describing the scalar metric perturbations and $\gamma^{ij}$ is the metric of the unperturbed spatial hypersurfaces.

In our model, as in the case of most simple cosmological models, in the absence of anisotropic stress ($\delta T^{i}_{\; \; j} = 0$ for $i \neq j$), the two scalar metric degrees of freedom $\chi$ and $\Psi$ coincide upto a sign:
\bea
\label{eq:psi-phi}
\Psi = -\chi.
\eea

The remaining metric perturbation $\Psi$ can be expressed in terms of the inflaton field fluctuation $\delta \tilde{\phi}$ at horizon crossing \cite{dodelson},
\bea
\tilde{\Psi} \Big|_{\textrm{\tiny{post inflation}}} = \frac{2}{3}aH \frac{\delta \tilde{\phi}}{\dot{\tilde{\phi}}^{(0)}}\Big|_{\textrm{\tiny{horizon crossing}}}
\eea
where $\tilde{\Psi}$ is the Fourier coefficient of $\Psi$.

On using the general definition of power spectrum as in eqn. (\ref{eq:zeta-corr-commu}), the power spectra for $P_{\tilde{\Psi}}$ and $P_{\delta \tilde{\phi}}$ can be connected when a mode $k$ crosses the horizon, i.e. when $a(\eta)H =k$, say for $\eta = \eta_{0}$:
\bea
\label{PhiHorCross}
P_{\tilde{\Psi}}({\bf k}, \eta) = \frac{4}{9} \Big(\frac{a(\eta)H}{\dot{\tilde{\phi}}^{(0)}}\Big)^{2} P_{\delta \tilde{\phi}}\Big|_{a(\eta_{0})H=k}.
\eea

From eqn. (\ref{eq:powerDeltaPhi}), eqn. (\ref{eq:psi-phi}) and
using \bea \label{eq:conversion}
{aH}/{\dot{\tilde{\phi}}^{(0)}}=\sqrt{4 \pi G / \epsilon} \eea at
horizon crossing, where $G$ is Newton's gravitational constant and
$\epsilon$ is the slow-roll parameter in the single field inflation
model \cite{dodelson}, we have the power spectrum (defined as in
eqn. (\ref{eq:zeta-corr-commu})) for the scalar metric perturbation
at horizon crossing, \bea \label{eq:PpsiPphi} P_{\tilde{\Psi}}(k,
\eta(t)) = P_{\Phi_{0}}(k, \eta(t)) = \frac{16 \pi G}{9 \epsilon}
\frac{H^{2}}{2k^{3}}\Big|_{a(\eta_{0})H =k}, \eea Here we wrote
$\Phi_{0}$ for $\tilde{\chi}$.

Note that the Hubble parameter $H$ is (nearly) constant during inflation and also it is the same in conformal time $\eta$ and cosmic time $t$. Since the time dependence of the power spectrum is through the Hubble parameter in eqn. (\ref{eq:PpsiPphi}), we have
\bea
\label{eq:ConstInTime}
P_{\Phi_{0}} (k, \eta(t))= P_{\Phi_{0}} (k, t) \equiv P_{\Phi_{0}} (k) =\textrm{constant in time}.
\eea

The power spectrum in eqn. (\ref{eq:PpsiPphi}) is for commutative spacetime and it depends on the magnitude of ${\bf k}$ and not on its direction. In the next section, we will show that the power spectrum becomes direction-dependent when we make spacetime noncommutative.

\section{Direction-Dependent Power Spectrum}
The two-point function in  noncommutative spacetime, using eqn.
(\ref{eq:2thetacorr}), takes the form \bea \label{eq:modified2pt}
\langle 0| \Phi_{\theta}({\bf k}, \eta) \Phi_{\theta}({\bf k}',
\eta)|0\rangle = \langle 0| \Phi_0({\bf k}, \eta^{-}) \Phi_0({\bf
k}', \eta^{-})|0\rangle~, \eea where $\eta^{-} =\eta( t -
{\vec{\theta}^{0}\cdot {\bf k} \over 2})$.

In the commutative case, the reality  of the two-point correlation
function (since the density fields $\Phi_{0}$ are real) is obtained
by imposing the condition \bea \langle \Phi_{0}({\bf k},
\eta)\Phi_{0}({\bf k}', \eta)\rangle{}^{*} = \langle \Phi_{0}(-{\bf
k}, \eta)\Phi_{0}(-{\bf k}', \eta)\rangle. \eea

But this condition is not correct  when the fields are deformed.
That is because even if $\Phi_{\theta}$ is self-adjoint,
$\Phi_{\theta}({\bf x}, t) \Phi_{\theta}({\bf x}', t') \neq
\Phi_{\theta}({\bf x}', t') \Phi_{\theta}({\bf x}, t)$ for
space-like separations. A simple and natural modification (denoted
by subscript $M$) of the correlation function that ensures reality
involves ``symmetrization" of the product of $\varphi_{\theta}$'s or
keeping its self-adjoint part. That involves replacing the product
of $\phi_\theta$'s by half its anti-commutator,
 \bea
\frac{1}{2}[\varphi_{\theta}({\bf x}, \eta), \varphi_{\theta}({\bf
y}, \eta)]_{+} = \frac{1}{2}\Big(\varphi_{\theta}({\bf x}, \eta)
\varphi_{\theta}({\bf y}, \eta)+ \varphi_{\theta}({\bf y}, \eta)
\varphi_{\theta}({\bf x}, \eta)\Big) .\eea (We emphasize that this
procedure for ensuring reality is a matter of choice)

 For the Fourier
modes $\Phi_\theta$, this procedure gives : \bea
\label{eq:modified2pt2} \langle \Phi_{\theta}({\bf k},
\eta)\Phi_{\theta}({\bf k}', \eta)\rangle_{M} =
\frac{1}{2}\Big(\langle \Phi_{\theta}({\bf k},
\eta)\Phi_{\theta}({\bf k}', \eta)\rangle + \langle
\Phi_{\theta}(-{\bf k}, \eta)\Phi_{\theta}(-{\bf k}',
\eta)\rangle{}^{*}\Big) \eea

After the modification of the correlation function, the power spectrum for scalar metric perturbation takes the form
\bea
\langle \Phi_{\theta}({\bf k}, \eta)\Phi_{\theta}({\bf k}', \eta)\rangle_{M} = (2\pi)^{3} P_{\Phi_{\theta}}({\bf k}, \eta) \delta^{3} ({\bf k}+{\bf k}').
\eea

Using eqns. (\ref{eq:metric-power}), (\ref{PhiHorCross}), (\ref{eq:modified2pt}) and (\ref{eq:modified2pt2}) we write down the modified power spectrum:
\bea
\label{eq:noncommu-power-spectrum}
P_{\Phi_{\theta}}({\bf k}, \eta) = \frac{1}{2}\Big[\frac{4}{9}\Big(\frac{a(\eta)H}{\dot{\tilde{\phi}}^{(0)}}\Big)^{2}\frac{1}{a(\eta)^{2}}\Big(|u({\bf k}, \eta^{-})|^{2} + |u(-{\bf k}, \eta^{+})|^{2} \Big)\Big].
\eea
where $\eta^{\pm} =\eta( t \pm {\vec{\theta}^{0}\cdot {\bf k} \over 2})$. Notice that here the argument of the scale factor $a(\eta)$ is not shifted, since it is not deformed by noncommutativity.

It is easy to show that
\bea
\label{eq:modified-sol}
u({\bf k}, \eta^{\pm})=\frac{\textrm{e}^{-ik\eta^{\pm}}}{\sqrt{2k}} \Big(1-\frac{i}{k \eta^{\pm}}\Big)
\eea
are also solutions of eqn. (\ref{eq:evolution}).

Thus on using eqn. (\ref{eq:conversion}) and the limit $k\eta^{\pm} \rightarrow 0$ of eqn. (\ref{eq:modified-sol}), the modified power spectrum is found to be
\bea
\label{eq:modifiedps}
P_{\Phi_{\theta}}({\bf k}, \eta) &=& \frac{1}{2}\Big[\frac{16 \pi G}{9 \epsilon}\frac{1}{a(\eta)^{2}}\Big(|u({\bf k}, \eta^{-})|^{2} + |u(-{\bf k}, \eta^{+})|^{2} \Big)\Big]\nn \\
&=& \frac{1}{2}\Big[\frac{16 \pi G}{9 \epsilon}\frac{1}{a(\eta)^{2}}\Big(\frac{1}{2k^{3}(\eta^{-})^{2}} + \frac{1}{2k^{3}(\eta^{+})^{2}} \Big)\Big]\nn \\
&=& \frac{8 \pi G}{9 \epsilon}\frac{1}{2k^{3}a(\eta)^{2}}\Big(\frac{1}{(\eta^{-})^{2}} + \frac{1}{(\eta^{+})^{2}} \Big).
\eea

Assuming that the Hubble parameter $H$ is nearly a constant during inflation, the conformal time \cite{dodelson}
\bea
\label{eq:EtaH}
\eta(t) \simeq \frac{-1}{Ha_{0}}~\textrm{e}^{-Ht}.
\eea
gives an expression for $\eta^{\pm}$:
\bea
\label{eq:etapm}
\eta^{\pm} = \eta(t)~\textrm{e}^{\mp \frac{1}{2}H\vec{\theta}^{0}\cdot {\bf k}}.
\eea

On using eqn. (\ref{eq:etapm}) in eqn. (\ref{eq:modifiedps}) we can easily write down an analytic expression for the modified primordial power spectrum at horizon crossing,
\bea
\label{eq:modifiedps2}
P_{\Phi_{\theta}}({\bf k}) = P_{\Phi_{0}}(k) \; \textrm{cosh}(H \vec{\theta}^{0}\cdot {\bf k})
\eea
where $P_{\Phi_{0}}(k)$ is given by eqn. (\ref{eq:PpsiPphi}). Note that the modified power spectrum also respects the ${\bf k} \rightarrow -{\bf k}$ parity symmetry.

This power spectrum depends on both the magnitude and direction of ${\bf k}$ and clearly breaks rotational invariance. In the next section we will connect this power spectrum to the two-point temperature correlations in the sky and obtain an expression for the amount of deviation from statistical isotropy due to noncommutativity.

\section{Signature of Noncommutativity in the CMB Radiation}

We are interested in quantifying the effects of noncommutative scalar perturbations on the cosmic microwave background fluctuations. We assume homogeneity of temperature fluctuations observed in the sky. Hence it is a function of a unit vector giving the direction in the sky and can be expanded in spherical harmonics:
\bea
{\Delta T(\hat{n}) \over T} = \sum_{l m} a_{l m} Y_{l m}(\hat{n}),
\eea
Here $\hat{n}$ is the direction of incoming photons.

The coefficients of spherical harmonics contain all the information encoded in the temperature fluctuations. For $\theta^{\mu \nu}=0$, they can be connected to the primordial scalar metric perturbations $\Phi_{0}$,
\bea
\label{eq:alm}
a_{lm} = 4 \pi (-i)^{l} \; \int \frac{d^{3}k}{(2 \pi)^{3}} \; \Delta_{l}(k) \Phi_{0}({\bf k}, \eta)Y_{lm}^{*}(\hat{k}),
\eea
where $\Delta_{l}(k)$ are called {\it transfer functions}. They describe the evolutions of scalar metric perturbations $\Phi_{0}$ from horizon crossing epoch to a time well into the radiation dominated epoch.

The two-point temperature correlation function can be expanded in spherical harmonics:
\bea
\label{eq:TwoPZero}
\langle {\Delta T(\hat{n}) \over T} {\Delta T(\hat{n}') \over T} \rangle = \sum_{lml'm'} \langle a_{lm} a^{*}_{l'm'}\rangle Y^{*}_{lm}(\hat{n}) Y_{l'm'}(\hat{n}').
\eea

The variance of $a_{lm}$'s is nonzero. For $\theta^{\mu \nu}=0$, we have
\bea
\langle a_{lm} a^{*}_{l'm'}\rangle = C_{l} \delta_{ll'} \delta_{mm'}.
\eea

Using eqn. (\ref{eq:zeta-corr-commu}) and eqn. (\ref{eq:alm}), we can derive the expression for $C_{l}$'s for $\theta^{\mu \nu}=0$:
{\small\bea
&&\langle a_{lm} a^{*}_{l'm'}\rangle \nn\\
&&~~= 16 \pi^{2}(-i)^{l-l'}\; \int \frac{d^{3}k}{(2\pi)^{3}}\frac{d^{3}k'}{(2\pi)^{3}} \; \Delta_{l}(k)\Delta_{l'}(k') \langle \Phi_{0}({\bf k}, \eta)\Phi^{*}_{0}({\bf k}', \eta)\rangle \; Y^{*}_{lm}(\hat{k})Y_{l'm'}(\hat{k}')\nn \\
&&~~= 16 \pi^{2}(-i)^{l-l'} \int \frac{d^{3}k}{(2 \pi)^{3}} \; \Delta_{l}(k)\Delta_{l'}(k) P_{\Phi_{0}}(k) \; Y^{*}_{lm}(\hat{k})Y_{l'm'}(\hat{k})\nn \\
&&~~= \frac{2}{\pi}\int dk \; k^{2} \; (\Delta_{l}(k))^{2} \; P_{\Phi_{0}}(k) \; \delta_{ll'}\delta_{mm'} = C_{l}\; \delta_{ll'} \delta_{mm'},
\eea}
where $P_{\Phi_{0}}(k)$ is given by eqn. (\ref{eq:PpsiPphi}).

When the fields are noncommutative, the two-point temperature correlation function clearly depends on $\theta^{\mu \nu}$. We can still write the two-point temperature correlation as in eqn. (\ref{eq:TwoPZero}):
\bea
\langle {\Delta T(\hat{n}) \over T} {\Delta T(\hat{n}') \over T} \rangle_{_\theta} = \sum_{lml'm'} \langle a_{lm} a^{*}_{l'm'}\rangle_{_\theta} Y_{lm}(\hat{n}) Y^{*}_{l'm'}(\hat{n}').
\eea

This gives
{\small\bea
\label{ThetaAlmEqn}
&&\langle a_{lm} a^{*}_{l'm'}\rangle_{_\theta}\nn\\
&&~~=16 \pi^{2}(-i)^{l-l'} \int \frac{d^{3}k}{(2\pi)^{3}}\frac{d^{3}k'}{(2\pi)^{3}}\Delta_{l}(k)\Delta_{l'}(k') \langle \Phi_{\theta}({\bf k}, \eta)\Phi_{\theta}^{\dagger}({\bf k}', \eta)\rangle_{M} Y^{*}_{lm}(\hat{k})Y_{l'm'}(\hat{k}').\nn\\
\eea}

The two-point correlation function in eqn. (\ref{ThetaAlmEqn}) is calculated during the horizon crossing of the mode {\bf k}. Once a mode crosses the horizon, it becomes independent of time, so that we can rewrite the two-point function as
\bea
\langle \Phi_{\theta}({\bf k}, \eta)\Phi_{\theta}^{\dagger}({\bf k}', \eta)\rangle_{M} = (2\pi)^{3} P_{\Phi_{\theta}}({\bf k}) \delta^{3} ({\bf k} - {\bf k}')
\eea
where $P_{\Phi_{\theta}}({\bf k})$ is given by eqn. (\ref{eq:modifiedps2}).

Thus we write the noncommutative angular correlation function as follows:
\bea
\label{ThetaAlm}
\langle a_{lm} a^{*}_{l'm'}\rangle_{_\theta}&=& 16 \pi^{2}(-i)^{l-l'}\int \frac{d^{3}k}{(2 \pi)^{3}} \;\Delta_{l}(k)\Delta_{l'}(k) P_{\Phi_{\theta}}({\bf k}) \; Y^{*}_{lm}(\hat{k})Y_{l'm'}(\hat{k}).\nn\\
\eea

The regime in which the transfer functions act is well above the noncommutative length scale, so that it is perfectly legitimate to assume that the transfer functions are the same as in the commutative case.

Assuming that the $\vec{\theta^{0}}$ is along the $z$-axis, we have the expansion
\bea
\label{eq:ExrThetaKH}
\textrm{e}^{\pm \vec{H\theta^{0}}\cdot{\bf k}} = \sum_{l=0}^{\infty} i^{l} \sqrt{4\pi(2l+1)} j_{l}(\mp i\theta k H)Y_{l0}(\textrm{cos}\vartheta)
\eea
where $\vec{\theta^{0}}\cdot{\bf k}=\theta k \; \textrm{cos}\vartheta$ and $j_{l}$ is the spherical Bessel function.

On using eqn. (\ref{eq:ExrThetaKH}) and the identities $j_{l}(-z) = (-1)^{l}j_{l}(z)$ and \\
~~$j_{l}(iz) = i^{l}\; i_{l}(z)$, where $i_{l}$ is the modified spherical Bessel function, we can write eqn. (\ref{eq:modifiedps2}) as
\bea
\label{PowerThetaExp}
P_{\Phi_{\theta}} ({\bf k}) = P_{\Phi_{0}} (k)\sum_{l=0, \; l: \textrm{\tiny{even}}}^{\infty}\sqrt{4\pi(2l+1)}\; i_{l}(\theta k H)\; Y_{l0}(\textrm{cos}\vartheta).
\eea

Using eqns. (\ref{ThetaAlm}) and (\ref{PowerThetaExp}),  we rewrite eqn. (\ref{ThetaAlm}) as,
{\footnotesize\bea
\label{eq:ThetaAngular}
\langle a_{lm} a^{*}_{l'm'}\rangle_{_\theta} &=& \frac{2}{\pi}\int d k \sum_{l''=0, \; l'': \textrm{\tiny{even}}}^{\infty}  (i)^{l-l'} (-1)^{m}(2l''+1)  \; k^{2} \Delta_{l}(k)\Delta_{l'}(k) P_{\Phi_{0}} (k) i_{l''}(\theta kH)\nn \\
&& \times  \sqrt{(2l+1)(2l'+1)} \left( \begin{array}{ccc}
l & l' & l'' \\
0 & 0 & 0 \end{array} \right)\left( \begin{array}{ccc}
l & l' & l'' \\
-m & m' & 0 \end{array} \right),
\eea}
the Wigner's 3-j symbols in eqn. (\ref{eq:ThetaAngular}) being related to the integrals of spherical harmonics:
\bea
&&\int d\Omega_{k}~Y_{l,-m}(\hat{k}) Y_{l'm'}(\hat{k}) Y_{l''0}(\hat{k})\nn\\
&&~~~~= \sqrt{(2l+1)(2l'+1)(2l''+1)/4\pi} \left( \begin{array}{ccc}
l & l' & l'' \\
0 & 0 & 0 \end{array} \right)\left( \begin{array}{ccc}
l & l' & l'' \\
-m & m' & 0 \end{array} \right).\nn\\
\eea

We can also get a simplified form of eqn. (\ref{eq:ThetaAngular}) by expanding the modified power spectrum in eqn. (\ref{eq:modifiedps2}) in powers of $\theta$ up to the leading order:
\bea
\label{eq:ThetaExp}
P_{\Phi_{\theta}} ({\bf k}) \simeq P_{\Phi_{0}} (k) \Big[1 + \frac{H^{2}}{2} (\vec{\theta}^{0}\cdot {\bf k})^{2}\Big].
\eea
A modified power spectrum of this form has been considered in \cite{Ackerman}, where the rotational invariance is broken by introducing a (small) nonzero vector. In our case, the vector that breaks rotational invariance is $\vec{\theta^{0}}$ and it emerges naturally in the framework of field theories on the noncommutative Groenewold-Moyal spacetime. We have also an exact expression for $P_{\Phi_{\theta}}(\bf k)$ in eqn. (\ref{eq:modifiedps2}).

Work is in progress to find a best fit for the data available and thereby to determine the length scale of noncommutativity.

The direction-dependent primordial power spectrum discussed in \cite{Ackerman} is considered in a model independent way in \cite{Kamionkowski} to compute minimum-variance estimators for the coefficients of direction-dependence. A test for the existence of a preferred direction in the primordial perturbations using full-sky CMB maps is performed in a model independent way in \cite{Christian}. Imprints of cosmic microwave background anisotropies from a non-standard spinor field driven inflation is considered in \cite{Mota1}. Anisotropic dark energy equation of state can also give rise to a preferred direction in the universe \cite{Mota2}.

\section{Non-causality and Noncommutative Fluctuations}

In the noncommutative frame work, the  expression for the two-point correlation function for the field $\varphi_\theta$ contains real and imaginary parts. We identified the real part with the observed temperature correlations which are real. This gave us the modified power spectrum
\bea
P_{\Phi_{\theta}}({\bf k}) = P_{\Phi_{0}}(k) \; \textrm{cosh}(H
\vec{\theta}^{0}\cdot {\bf k}). \eea

In this section we discuss the imaginary part of the  two-point
correlation function for the field $\varphi_\theta$. In position
space, the imaginary part of the two-point correlation function is
obtained from the ``anti-symmetrization" of the fields for a
space-like separation: \bea \label{re:non-causality}
\frac{1}{2}[\varphi_{\theta}({\bf x}, \eta), \varphi_{\theta}({\bf
y}, \eta)]_{-} = \frac{1}{2}\Big(\varphi_{\theta}({\bf x}, \eta)
\varphi_{\theta}({\bf y}, \eta)- \varphi_{\theta}({\bf y}, \eta)
\varphi_{\theta}({\bf x}, \eta)\Big). \eea

The commutator of deformed fields, in general, is nonvanishing for
space-like separations. This  type of non-causality is an inherent
property of noncommutative field theories constructed on the
Groenewold-Moyal spacetime \cite{Bal-locality}.

To study this non-causality, we consider two smeared fields localized at ${\bf x}_{1}$ and ${\bf x}_{2}$. (The expression for non-causality diverges for conventional choices for $P_{\Phi_{0}}$ if we do not smear the fields. See after eqn. (\ref{non-causal}).) We write down smeared fields at ${\bf x}_{1}$ and ${\bf x}_{2}$.
\bea
&&\varphi(\alpha, {\bf x}_{1}) = \Big(\frac{\alpha}{\pi}\Big)^{3/2}\int d^{3}x~\varphi_{\theta}({\bf x})~e^{-\alpha({\bf x} - {\bf x}_{1})^{2}}, \\
&&\varphi(\alpha, {\bf x}_{2}) = \Big(\frac{\alpha}{\pi}\Big)^{3/2}\int d^{3}x~\varphi_{\theta}({\bf x})~e^{-\alpha({\bf x} - {\bf x}_{2})^{2}},
\eea
where $\alpha$ determines the amount of smearing of the fields. We have
\bea
\lim_{\alpha  \rightarrow \infty}\Big(\frac{\alpha}{\pi}\Big)^{3/2}\int d^{3}x~\varphi_{\theta}({\bf x})~e^{-\alpha({\bf x} - {\bf x}_{1})^{2}}=\varphi_{\theta}({\bf x}_{1}).
\eea
The scale $\alpha$ can
be thought of as  the width of a wave packet which is a measure of
the size of the spacetime region over which an experiment is performed.

We can now write down the uncertainty relation for the fields $\varphi(\alpha, {\bf x}_{1})$ and $\varphi(\alpha, {\bf x}_{2})$ coming from eqn. (\ref{re:non-causality}):
\bea
\label{eq:noncausal}
\Delta \varphi(\alpha, {\bf x}_{1}) \Delta \varphi(\alpha, {\bf x}_{2}) \geq \frac{1}{2} \Big| \langle 0 |[\varphi(\alpha, {\bf x}_{1}), \varphi(\alpha, {\bf x}_{2})]|0\rangle \Big|
\eea

{\it This equation is an expression for the violation of causality due to noncommutativity.}

Notice that, in momentum space, we can rewrite the commutator in terms of the primordial power spectrum $P_{\Phi_{0}}(k)$ at horizon crossing using the discussion following eqn. (\ref{eq:modified2pt2}):
{\footnotesize
\bea
\label{eq:commutator}
\frac{1}{2}\langle0|[\Phi_{\theta}({\bf k}, \eta), \Phi_{\theta}({\bf k}', \eta)]_{-}|0\rangle \Big|_{\textrm{horizon crossing}} = (2\pi)^{3}P_{\Phi_{0}}(k) \; \textrm{sinh}(H \vec{\theta}^{0}\cdot {\bf k})~\delta^{3}({\bf k}+{\bf k}')\nn\\
\eea
}

We can calculate the right hand side of eqn. (\ref{eq:noncausal})
{\footnotesize
\bea
&&\langle 0 |[\varphi(\alpha, {\bf x}_{1}), \varphi(\alpha, {\bf x}_{2})]|0\rangle \nn\\
&&~~=\Big(\frac{\alpha}{\pi}\Big)^{3}\int d^{3}x d^{3}y~\langle 0 |[\varphi_{\theta}({\bf x}), \varphi_{\theta}({\bf y})]|0\rangle~e^{-\alpha({\bf x} - {\bf x}_{1})^{2}}e^{-\alpha({\bf y} - {\bf x}_{2})^{2}}\nn \\
&&~~=\Big(\frac{\alpha}{\pi}\Big)^{3}\int d^{3}x d^{3}y \frac{d^{3}k}{(2\pi)^{3}} \frac{d^{3}q}{(2\pi)^{3}}~\langle 0 |[\Phi_{\theta}({\bf k}), \Phi_{\theta}({\bf q})]|0\rangle~e^{-i{\bf k}\cdot{\bf x}-i{\bf q}\cdot{\bf y}}e^{-\alpha[({\bf x} - {\bf x}_{1})^{2}+({\bf y} - {\bf x}_{2})^{2}]}\nn \\
&&~~=\frac{2}{(2\pi)^{3}}\Big(\frac{\alpha}{\pi}\Big)^{3}\int d^{3}x d^{3}y~d^{3}k d^{3}q~P_{\Phi_{0}}(k)~\textrm{sinh}(H\vec{\theta}^{0}\cdot {\bf k})~\delta^{3}({\bf k} + {\bf q})\times\nn\\
&&~~~~~~~~~~~~e^{-i{\bf k}\cdot{\bf x}-i{\bf q}\cdot{\bf y}}e^{-\alpha[({\bf x} - {\bf x}_{1})^{2}+({\bf y} - {\bf x}_{2})^{2}]}\nn \\
&&~~=\frac{2}{(2\pi)^{3}}\Big(\frac{\alpha}{\pi}\Big)^{3}\int d^{3}x d^{3}y d^{3}k~P_{\Phi_{0}}(k)\; \textrm{sinh}(H\vec{\theta}^{0}\cdot {\bf k})~e^{-i{\bf k}\cdot({\bf x}-{\bf y})}e^{-\alpha[({\bf x} - {\bf x}_{1})^{2}+({\bf y} - {\bf x}_{2})^{2}]}\nn \\
&&~~=\frac{2}{(2\pi)^{3}}\Big(\frac{\alpha}{\pi}\Big)^{3}\int d^{3}k~P_{\Phi_{0}}(k)~\textrm{sinh}(H\vec{\theta}^{0}\cdot {\bf k})~\int d{\bf x} d{\bf y}e^{-i{\bf k}\cdot({\bf x}-{\bf y})}e^{-\alpha[({\bf x} - {\bf x}_{1})^{2}+({\bf y} - {\bf x}_{2})^{2}]}\nn \\
&&~~=\frac{2}{(2\pi)^{3}}\int
d^{3}k~P_{\Phi_{0}}(k)~\textrm{sinh}(H\vec{\theta}^{0}\cdot {\bf
k})~e^{-\frac{{\bf k}^{2}}{2 \alpha} -i{\bf k}\cdot({\bf x}_{1}-{\bf
x}_{2})}. \label{non-causal-commu} \eea
}
This gives for eqn. (\ref{eq:noncausal}),
\bea &&\Delta \varphi(\alpha, {\bf x}_{1}) \Delta
\varphi(\alpha, {\bf x}_{2}) \geq \Big|\frac{1}{(2\pi)^{3}}\int
d^{3}k~P_{\Phi_{0}}(k)~\textrm{sinh}(H\vec{\theta}^{0}\cdot {\bf
k})~e^{-\frac{{\bf k}^{2}}{2 \alpha} -i{\bf k}\cdot({\bf x}_{1}-{\bf
x}_{2})}\Big| \label{non-causal} \nn\\\eea
The right hand side of eqn.
(\ref{non-causal}) is divergent for  conventional asymptotic
behaviours of $P_{\Phi_{0}}$ (such as $P_{\Phi_{0}}$ vanishing for
large $k$ no faster than some inverse power of $k$) when $\alpha
\rightarrow \infty$ and thus the Gaussian width becomes zero. This
is the reason for introducing smeared fields.

Notice that the amount of causality violation given in eqn. (\ref{non-causal}) is direction-dependent.

The uncertainty relation given in eqn. (\ref{non-causal}) is purely due to spacetime noncommutativity as it vanishes for the case $\theta^{\mu \nu} =0$. It is an expression of causality violation.

\section{Non-Gaussianity from noncommutativity}
In this section, we briefly explain how $n$-point correlation functions become non-Gaussian when the fields are noncommutative, assuming that they are Gaussian in their commutative limits.

Consider a noncommutative field $\varphi_{\theta}({\bf x}, t)$. Its first moment is obviously zero:
\bea
\langle \varphi_{\theta}({\bf x}, t)\rangle = \langle \varphi_{0}({\bf x}, t)\rangle =0.\nn
\eea

The information about noncommutativity is contained in the higher moments of $\varphi_{\theta}$. We show that the $n$-point functions cannot be written as sums of products of two-point functions. That proves that the underlying probability distribution is non-Gaussian.

The $n$-point correlation function is
\bea
C_{n}(x_{1}, x_{2}, \cdots, x_{n}) = \langle \varphi_{\theta}({\bf x}_{1},
t_{1}) \cdots \varphi_{\theta}({\bf x}_{n}, t_{n}) \rangle
\eea

Since $\varphi_{0}$ is assumed to be Gaussian and $\varphi_{\theta}$ is given in terms of $\varphi_{0}$ by eqn. (\ref{eq:twistedfield}), all the odd moments of $\varphi_{\theta}$ vanish.

But the even moments of $\varphi_{\theta}$ need not vanish and do not split into sums of products of its two-point functions in a familiar way.

Non-Gaussianity cannot be seen at the level of two-point functions. Consider the two-point function $C_{2}$. We write this in momentum space in terms of $\Phi_{0}$:
{\footnotesize
\bea
C_{2}=\langle \Phi_{\theta}({\bf k}_1,t_1)\Phi_{\theta}({\bf k}_2,t_2) \rangle = e^{-{i\over2}({\bf k}_2\wedge{\bf k}_1)} \Big\langle \Phi_0({\bf k}_1,t_1+{\vec{\theta}^{0}\cdot {\bf k}_2 \over 2})\Phi_0({\bf k}_2,t_2-{\vec{\theta}^{0}\cdot {\bf k}_1 \over 2}) \Big\rangle.\nn\\
\eea
}
where ${\bf k}_i\wedge{\bf k}_j \equiv k_{i}\theta^{ij}k_{j}$.

Making use of the translation invariance ${\bf k}_1+{\bf k}_2=0$, the above equation becomes
{\footnotesize
\bea
\langle \Phi_{\theta}({\bf k}_1,t_1)\Phi_{\theta}({\bf k}_2,t_2) \rangle &=& \Big\langle \Phi_0({\bf k}_1,t_1-{\vec{\theta}^{0}\cdot {\bf k}_1 \over 2})\Phi_0({\bf k}_2,t_2-\vec{\theta}^{0}\cdot {\bf k}_1-{\vec{\theta}^{0}\cdot {\bf k}_2 \over 2}) \Big\rangle.\nn\\
\eea
}
Non-Gaussianity can be seen in all the $n$-point functions for $n \geq 4$  and even $n$. Still they can all be written in terms of correlation functions of $\Phi_{0}$. For example, let us consider the four-point function $C_{4}$:
{\footnotesize
\bea
&&C_{4}={\langle}\Phi_{\theta}({\bf k}_1,t_1)\Phi_{\theta}({\bf k}_2,t_2)\Phi_{\theta}({\bf k}_3,t_3)\Phi_{\theta}({\bf k}_4,t_4){\rangle}\nn\\
&&=e^{-{i\over2}({\bf k}_3\wedge{\bf k}_2+{\bf k}_3\wedge{\bf k}_1+{\bf k}_2\wedge{\bf k}_1)}\Big{\langle} \Phi_0({\bf k}_1,t_1-{\vec{\theta}^{0}\cdot {\bf k}_1 \over 2})\Phi_0({\bf k}_2, t_2-\vec{\theta}^{0}\cdot {\bf k}_1 - {\vec{\theta}^{0}\cdot {\bf k}_2 \over 2})\times\nn\\
&&\Phi_0({\bf k}_3, t_3 - \vec{\theta}^{0}\cdot {\bf k}_1 - \vec{\theta}^{0}\cdot {\bf k}_2 - {\vec{\theta}^{0}\cdot {\bf k}_3 \over 2})\Phi_0({\bf k}_4, t_4 - \vec{\theta}^{0}\cdot {\bf k}_1 - \vec{\theta}^{0}\cdot {\bf k}_2 - \vec{\theta}^{0}\cdot {\bf k}_3 - {\vec{\theta}^{0}\cdot {\bf k}_4 \over 2})\Big{\rangle}\nn
\eea
}
Here we have used translational invariance, which implies that ${\bf k}_1+{\bf k}_2+{\bf k}_3+{\bf k}_4=0$. Using this equation once more to eliminate ${\bf k}_{4}$, we find
{\footnotesize
\bea
&&C_{4}=e^{-{i\over2}({\bf k}_3\wedge{\bf k}_2+{\bf k}_3\wedge{\bf k}_1+{\bf
k}_2\wedge{\bf k}_1)}\Big{\langle}\Phi_0({\bf k}_1, t_1-{\vec{\theta}^{0}\cdot {\bf k}_1 \over 2})\Phi_0({\bf k}_2, t_2 - \vec{\theta}^{0}\cdot {\bf k}_1 - {\vec{\theta}^{0}\cdot {\bf k}_2 \over 2}) \times \nn \\
&&\times~\Phi_0({\bf k}_3, t_3 - \vec{\theta}^{0}\cdot {\bf k}_1 - \vec{\theta}^{0}\cdot {\bf k}_2 - {\vec{\theta}^{0}\cdot {\bf k}_3 \over 2})\Phi_0({\bf
k}_4, t_4 - {\vec{\theta}^{0}\cdot {\bf k}_1 + \vec{\theta}^{0}\cdot {\bf k}_2 + \vec{\theta}^{0}\cdot {\bf k}_3 \over 2})\Big{\rangle}\nn
\eea
}

Assuming Gaussianity for the field $\Phi_{0}$ and denoting $\Phi_{0}({\bf k}_{i}, t_{i})$ by $\Phi_{0}^{(i)}$, we have,
{\footnotesize
\bea
\langle \Phi^{(1)}_{0}\Phi^{(2)}_{0}\cdots \Phi^{(i)}_{0}\Phi^{(i+1)}_{0}\cdots\Phi^{(n)}_{0}\rangle &=& \langle \Phi^{(1)}_{0}\Phi^{(2)}_{0}\rangle \langle \Phi^{(3)}_{0}\Phi^{(4)}_{0}\rangle
\cdots \langle \Phi^{(i)}_{0}\Phi^{(i+1)}_{0}\rangle
\cdots \langle \Phi^{(n-1)}_{0}\Phi^{(n)}_{0}\rangle\nn \\
&&+~\textrm{permutations}~~(\textrm{for}~n~\textrm{even})
\eea
}
and
\bea
\langle \Phi^{(1)}_{0}\Phi^{(2)}_{0}\cdots \Phi^{(i)}_{0}\Phi^{(i+1)}_{0}\cdots\Phi^{(n)}_{0}\rangle  = 0~~(\textrm{for}~n~\textrm{odd}).
\eea

Therefore $C_{4}$ is
{\footnotesize
\bea
&&{\langle}\Phi_{\theta}({\bf k}_1, t_1)\Phi_{\theta}({\bf k}_2, t_2)\Phi_{\theta}({\bf k}_3, t_3)\Phi_{\theta}({\bf
k}_4, t_4){\rangle}\nn\\
&&~~=e^{-{i\over2}({\bf k}_3\wedge{\bf k}_2+{\bf k}_3\wedge{\bf
k}_1+{\bf k}_2\wedge{\bf k}_1)}\Big(\Big{\langle}\Phi_0({\bf k}_1, t_1-{\vec{\theta}^{0}\cdot {\bf k}_1\over2})\Phi_0({\bf k}_2,
t_2-\vec{\theta}^{0}\cdot {\bf k}_1-{\vec{\theta}^{0}\cdot {\bf k}_2\over2})\Big{\rangle}\nn\\&&~~\Big{\langle}\Phi_0({\bf k}_3, t_3-\vec{\theta}^{0}\cdot {\bf k}_1-\vec{\theta}^{0}\cdot {\bf k}_2-{\vec{\theta}^{0}\cdot {\bf k}_3\over2})\Phi_0({\bf k}_4, t_4 - {\vec{\theta}^{0}\cdot {\bf k}_1 + \vec{\theta}^{0}\cdot {\bf k}_2 + \vec{\theta}^{0}\cdot {\bf k}_3\over2})\Big{\rangle}\times\nn\\
&&~~+\Big{\langle}\Phi_0({\bf k}_1, t_1-{\vec{\theta}^{0}\cdot {\bf k}_1\over2})\Phi_0({\bf k}_3,
t_3-\vec{\theta}^{0}\cdot {\bf k}_1-\vec{\theta}^{0}\cdot {\bf k}_2-{\vec{\theta}^{0}\cdot {\bf k}_3\over2})\Big{\rangle}\times\nn\\
&&~~\Big{\langle}\Phi_0({\bf k}_2,
t_2-\vec{\theta}^{0}\cdot {\bf k}_1-{\vec{\theta}^{0}\cdot {\bf k}_2\over2})\Phi_0({\bf k}_4,
t_4-{\vec{\theta}^{0}\cdot {\bf k}_1+\vec{\theta}^{0}\cdot {\bf k}_2 + \vec{\theta}^{0}\cdot {\bf k}_3\over2})\Big{\rangle}\nn\\
&&~~+\Big{\langle}\Phi_0({\bf k}_1, t_1-{\vec{\theta}^{0}\cdot {\bf k}_1\over2})\Phi_0({\bf k}_4,
t_4-{\vec{\theta}^{0}\cdot {\bf k}_1+\vec{\theta}^{0}\cdot {\bf k}_2+\vec{\theta}^{0}\cdot {\bf k}_3\over2})\Big{\rangle}\times\nn\\
&&~~\Big{\langle}\Phi_0({\bf k}_2,
t_2-\vec{\theta}^{0}\cdot {\bf k}_1-{\vec{\theta}^{0}\cdot {\bf k}_2\over2})\Phi_0({\bf k}_3,
t_3-\vec{\theta}^{0}\cdot {\bf k}_1-\vec{\theta}^{0}\cdot {\bf k}_2-{\vec{\theta}^{0}\cdot {\bf k}_3\over2})\Big{\rangle}\Big).
\eea
}

Using spatial translational invariance for each two-point function, we have
{\footnotesize
\bea
&&{\langle}\Phi_{\theta}({\bf k}_1, t_1)\Phi_{\theta}({\bf k}_2, t_2)\Phi_{\theta}({\bf k}_3, t_3)\Phi_{\theta}({\bf k}_4, t_4){\rangle} \nn\\
&&~~= \Big[\Big{\langle}\Phi_0({\bf k}_1, t_1-{\vec{\theta}^{0}\cdot {\bf k}_1\over2})\Phi_0({\bf k}_2, t_2-{\vec{\theta}^{0}\cdot {\bf k}_1\over2})\Big{\rangle}\Big{\langle}\Phi_0({\bf k}_3, t_3-{\vec{\theta}^{0}\cdot {\bf k}_3\over2})\Phi_0({\bf k}_4, t_4-{\vec{\theta}^{0}\cdot {\bf k}_3\over2})\Big{\rangle}\Big]\nn\\
&&~~+ \textrm{e}^{-i{\bf k}_2\wedge{\bf k}_1}\Big[\Big{\langle}\Phi_0({\bf k}_1, t_1-{\vec{\theta}^{0}\cdot {\bf k}_1\over2})\Phi_0({\bf k}_3, t_3-\vec{\theta}^{0}\cdot {\bf k}_2-{\vec{\theta}^{0}\cdot {\bf k}_1\over2})\Big{\rangle}\nn \\
&&~~\Big{\langle}\Phi_0({\bf k}_2, t_2-\vec{\theta}^{0}\cdot {\bf k}_1-{\vec{\theta}^{0}\cdot {\bf k}_2\over2})\Phi_0({\bf k}_4, t_4-{\vec{\theta}^{0}\cdot {\bf k}_2\over2})\Big{\rangle}\Big]\nn\\
&&~~+\Big[\Big{\langle}\Phi_0({\bf k}_1, t_1-{\vec{\theta}^{0}\cdot {\bf k}_1\over2})\Phi_0({\bf
k}_4, t_4 - {\vec{\theta}^{0}\cdot {\bf k}_1\over2})\Big{\rangle}\times\nn \\
&&~~\Big{\langle}\Phi_0({\bf k}_2, t_2 - \vec{\theta}^{0}\cdot {\bf k}_1 - {\vec{\theta}^{0}\cdot {\bf k}_2\over2})\Phi_0({\bf
k}_3, t_3 - \vec{\theta}^{0}\cdot {\bf k}_1 - {\vec{\theta}^{0}\cdot {\bf k}_2\over2})\Big{\rangle}\Big].
\eea
}
Notice that the second term has a non-trivial phase which depends on the
spatial momenta ${\bf k}_{1}$ and ${\bf k}_{2}$ and the noncommutative parameter
$\theta$. As $C_{4}$ cannot be written as sums of products of $C_{2}$'s in a standard way, we see that the noncommutative probability distribution is non-Gaussian. Also it should be noted that we still cannot achieve Gaussianity of $n$-point functions even if we modify them by imposing the reality condition as we did for the two-point case.

Non-Gaussianity affects the CMB distribution and also the large scale structure (the large scale distribution of matter in the universe). We have not considered the latter. An upper bound to the amount of non-Gaussianity coming from noncommutativity can be set by extracting the four-point function from the data.
\section{Conclusions: Chapter \ref{cmb1}}
In this chapter, we have shown that the introduction of spacetime noncommutativity gives rise to  nontrivial contributions to the CMB temperature fluctuations. The two-point correlation function in momentum space, called the power spectrum, becomes direction-dependent. Thus spacetime noncommutativity breaks the rotational invariance of the CMB spectrum. That is, CMB radiation becomes statistically anisotropic. This can be measured experimentally to set bounds on the noncommutative parameter. The next chapter (see \cite{numerical}) presents numerical fits to the available CMB data to put bounds on $\theta$.

We have also shown that the probability distribution governing correlations of fields on the Groenewold-Moyal algebra ${\cal A}_{\theta}$ are non-Gaussian. This affects the correlation functions of temperature fluctuations. By measuring the amount of non-Gaussianity from the four-point correlation function data for temperature fluctuations, we can thus set further limits on $\theta$.

We have also discussed the signals of non-causality of non-commutative field theories in the temperature fluctuations of the CMB spectrum. It will be very interesting to test the data for such signals.
\begin{center}
Summary of Chapter \ref{cmb2}
\end{center}

\begin{itemize}
\item The noncommutativity parameter is not constrained by WMAP data, however ACBAR and CBI data restrict the lower bound of its energy scale to be around $10$ TeV
\item Upper bound for the noncommutativity parameter: $\sqrt{\theta} <  1.36 \times 10^{-19}\textrm{m}$. This corresponds to a ~$10$ TeV~ lower bound for the energy scale.
\item Amount of non-causality coming from spacetime noncommutativity for the fields of primordial scalar perturbations that are space-like separated
    \bea
&&\Delta \varphi(\alpha, {\bf x}_{1}) \Delta
\varphi(\alpha, {\bf x}_{2}) \geq \Big|\frac{1}{(2\pi)^{3}}\int
d^{3}k~P_{\Phi_{0}}(k)~\textrm{sinh}(H\vec{\theta}^{0}\cdot {\bf
k})~e^{-\frac{{\bf k}^{2}}{2 \alpha} -i{\bf k}\cdot({\bf x}_{1}-{\bf
x}_{2})}\Big|.\nn
\eea
\end{itemize}
\chapter{Constraint from the CMB,  Causality}\label{cmb2}

We try to constrain the noncommutativity length scale of the theoretical model given in \cite{cmbpaper} using the observational data from ACBAR, CBI and five year WMAP. The noncommutativity parameter is not constrained by WMAP data, however ACBAR and CBI data restrict the lower bound of its energy scale to be around $10$ TeV. We also derive an expression for the amount of non-causality coming from spacetime noncommutativity for the fields of primordial scalar perturbations that are space-like separated. The amount of causality violation for these field fluctuations are direction dependent.


\section{Introduction}
In 1992, the Cosmic Background Explorer (COBE) satellite detected anisotropies in the CMB radiation, which led to the conclusion that the early universe was not smooth: there were small density perturbations in the photon-baryon fluid before they decoupled from each other. Quantum corrections to the inflaton field generate perturbations in the metric and these perturbations could have been carried over to the photon-baryon fluid as density perturbations.  We then observe them today in the distribution of large scale structure and anisotropies in the CMB radiation.

Inflation \cite{Starobinsky79, Starobinsky82, guth, Linde, Albrecht} stretches a region of Planck size into cosmological scales. So, at the end of inflation, physics at the Planck scale can leave its signature on cosmological scales too. Physics at the Planck scale is better described by models of quantum gravity or string theory. There are indications from considerations of either quantum gravity or string theory that spacetime is noncommutative with a length scale of the order of Planck length. CMB radiation, which consists of photons from the last scattering surface of the early universe can carry the signature of spacetime noncommutativity. With these ideas in mind, in this paper, we look for a constraint on the noncommutativity length scale from the WMAP5 \cite{WMAP1, WMAP2, WMAP3}, ACBAR \cite{ACBAR1, ACBAR2, ACBAR3} and CBI \cite{CBI1, CBI2, CBI3, CBI4, CBI5} observational data.

In a noncommutative spacetime, the commutator of quantum fields at space-like separations does not in general vanish, leading to violation of causality. This type of violation of causality in the context of the fields for the primordial scalar perturbations is also discussed in this paper. It is shown that the expression for the amount of causality violation is direction-dependent.

In \cite{Sachin}, it was shown that causality violation coming from noncommutative spacetimes leads to violation of Lorentz invariance in certain scattering amplitudes. Measurements of these violations would be another way to put limits on the amount of spacetime noncommutativity.

This paper is a sequel to an earlier work \cite{cmbpaper}. The latter explains the theoretical basis of the formulae used in this paper. In \cite{queiroz1} another approach of noncommutative inflation is considered based on target space noncommutativity of fields \cite{queiroz1}.


\section{Likelihood Analysis for Noncomm. CMB}

The CMBEasy \cite{Doran} program calculates CMB power spectra based on a set of parameters and a cosmological model. It works by calculating the transfer functions $\Delta_{l}$ for multipole $l$ for scalar perturbations at the present conformal time $\eta_{0}$ as \cite{Seljak}
\bea
\label{eq:Delta}
\Delta_l(k, \eta=\eta_0) = \int_{0}^{\eta_{0}} d\eta~S(k,\eta) j_l[k(\eta_0 - \eta)],
\eea
where $S$ is a known ``source" term and $j_{l}$ is the spherical Bessel function. (Here ``scalar perturbations" mean the scalar part of the primordial metric fluctuations. Primordial metric fluctuations can be decomposed into scalar, vector and second rank tensor fluctuations according to their transformation properties under spatial rotations \cite{brandenberger}. They evolve independently in a linear theory. Scalar perturbations are most important as they couple to matter inhomogeneities. Vector perturbations are not important as they decay away in an expanding background cosmology. Tensor perturbations are less important than scalar ones, they do not couple to matter inhomogeneities at linear order. In the following discussion we denote the amplitudes of scalar and tensor perturbations by $A_s$ and $A_T$ respectively.) The lower limit of the time integral in eq. (\ref{eq:Delta}) is taken as a time well into the radiation dominance epoch. Eq. (\ref{eq:Delta}) shows that for each mode ${\bf k}$, the source term should be integrated over time $\eta$.

The transfer functions for scalar perturbations are then integrated over ${\bf k}$ to obtain the power spectrum for multipole moment $l$,
\bea
C^{(0)}_{l} = (4\pi^2) \int dk~k^2 P_{\Phi_{0}}(k) |\Delta_l(k,\eta=\eta_0)|^2,
\eea
where $P_{\Phi_{0}}$ is the initial power spectrum of scalar perturbations (cf. Ref. \cite{cmbpaper}.), taken to be $P_{\Phi_{0}}(k) = A_{s}k^{-3+(n_s-1)}$ with a spectral index $n_{s}$.

The coordinate functions $\widehat{x}_{\mu}$ on the noncommutative Moyal plane obey the commutation relations
\bea
[\widehat{x}_{\mu}, \widehat{x}_{\nu}] = i \theta_{\mu \nu},~~\theta_{\mu \nu} = -\theta_{\nu \mu} = \textrm{const}.
\eea
We set $\vec{\theta}^{0} \equiv (\theta^{01}, \theta^{02}, \theta^{03})$ to be in the third direction. In that case, $\vec{\theta}^{0} = \theta~\widehat{\theta}^{0}$ where the unit vector $\widehat{\theta}^{0}$ is $(0, 0, 1)$.

We now write down eq. (79) of \cite{cmbpaper},
\bea
\label{eq:ThetaAngular}
\langle a_{lm} a^{*}_{l'm'}\rangle_{_\theta} &=& \frac{2}{\pi}\int d k \sum_{l''=0, \; l'': \textrm{\tiny{even}}}^{\infty}  i^{l-l'} (-1)^{m}(2l''+1)k^{2} \Delta_{l}(k)\Delta_{l'}(k)P_{\Phi_{0}} (k) i_{l''}(\theta kH)\nn \\
&& \times  \sqrt{(2l+1)(2l'+1)}\left( \begin{array}{ccc}
l & l' & l'' \\
0 & 0 & 0 \end{array} \right)\left( \begin{array}{ccc}
l & l' & l'' \\
-m & m' & 0 \end{array} \right),
\eea
where $i_l$ is the modified spherical Bessel function and $H$ is the Hubble parameter during inflation. In the limit when $\theta = 0$  eq. (\ref{eq:ThetaAngular}) leads to the usual $C_l$'s \cite{dodelson}:
\bea
C_l=\frac{1}{2l+1} \sum_{m} \langle a_{lm} a^*_{lm}\rangle_0 = (4\pi^2) \int dk~k^2 P_{\Phi_{0}}(k) |\Delta_l(k,\eta=\eta_0)|^2.
\eea

Our goal is to compare theory with the observational data from WMAP5, ACBAR and CBI. These data sets are only available for the diagonal terms $l=l'$ of eq. (\ref{eq:ThetaAngular}), and for the average over $m$ for each $l$, so we consider only this case. Taking the average over $m$ of eq. (\ref{eq:ThetaAngular}), for $lm = l'm'$ the sum collapses to
{\footnotesize
\bea
\label{eq:i-zero}
C^{(\theta)}_l &\equiv& \frac{1}{2l+1} \sum_{m} \langle a_{lm} a^*_{lm}\rangle_\theta =\int dk\, k^2 P_{\Phi_{0}}(k) |\Delta_l(k,\eta=\eta_0)|^2 i_0(\theta k H),\\
C^{(0)}_l &=& C_l .
\eea
}

The CMBEasy integrator was modified to include the additional $i_0$ code and the Monte Carlo Markov-chain (MCMC) facility of the program was used to find best-fit values for $\theta H$ along with the other parameters of the standard $\Lambda$CDM cosmology.

In the first run the parameters were fit using a joint likelihood derived from the WMAP5, ACBAR and CBI data.  The outcome of this analysis was inconclusive, as the resulting value was unphysically large.  This result can be understood by examining the WMAP5 data alone and considering a $\chi^2$ goodness-of-fit test, using

\bea
\label{eq:chi-square}
\chi^2 = \sum_l \left(\frac{C^{(\theta)}_{l} - C_{l,data}}{\sigma_l}\right)^2,
\eea
where $C_{l,data}$ is the power spectrum and $\sigma_l$ is the standard deviation for each $l$ as reported by WMAP observation.

\begin{figure}
\includegraphics[height=9cm]{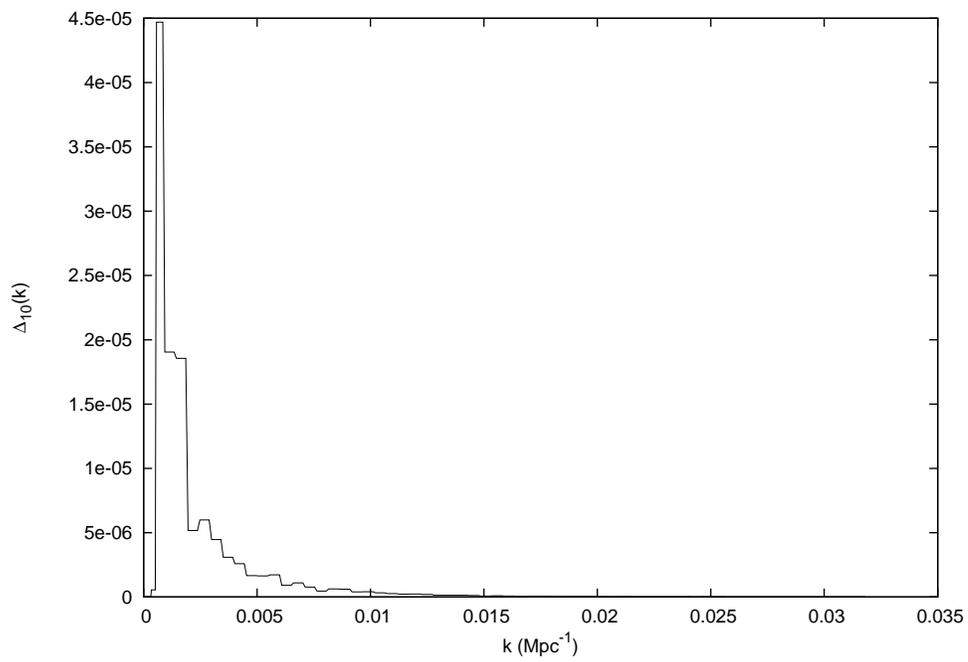}
\caption{Transfer function $\Delta_{l}$ for $l=10$ as a function of $k$. It peaks around $k=0.001$ Mpc$^{-1}$.} \label{fig:1}
\end{figure}

\begin{figure}
\includegraphics[height=9cm]{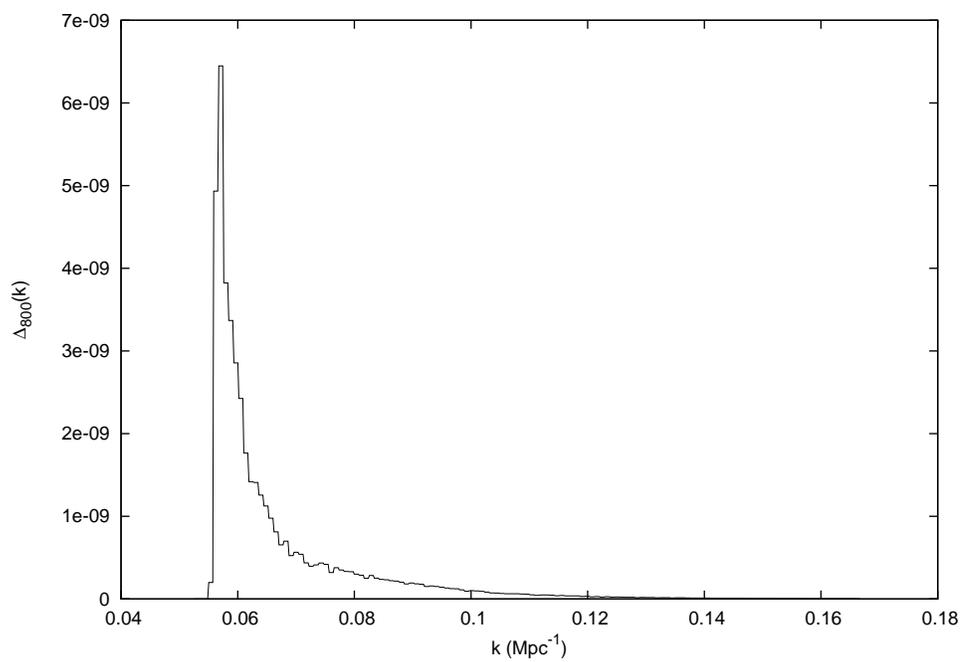}
\caption{Transfer function $\Delta_{l}$ for $l=800$ as a function of $k$. It peaks around $k=0.06$ Mpc$^{-1}$.} \label{fig:2}
\end{figure}

We expect noncommutativity to have a negligible effect on most of the parameters of the standard $\Lambda$CDM cosmology. We therefore consider the effect on the CMB power spectrum of varying only the new parameter $H\theta$. To determine its effect, we consider the shape of the transfer functions $\Delta_l(k)$ as calculated by CMBEasy. The graphs of two such functions are shown in Figs. \ref{fig:1} and \ref{fig:2}. As can be seen, these functions drop off rapidly with $k$, but extend to higher $k$ with increasing $l$. (For example, in Fig. \ref{fig:1}, the transfer function for $l=10$, $\Delta_{10}$, peaks around $k=0.001$ Mpc$^{-1}$ while in Fig. \ref{fig:2}, the transfer function for $l=800$, $\Delta_{800}$, peaks around $k=0.06$ Mpc$^{-1}$.)
As $i_0$ is a monotonically increasing function of $k$ starting at $i_{0}(0)=1$, this means that transfer functions of higher multipoles will feel the effect of noncommutivity first.

\begin{figure}
\includegraphics[height=9cm]{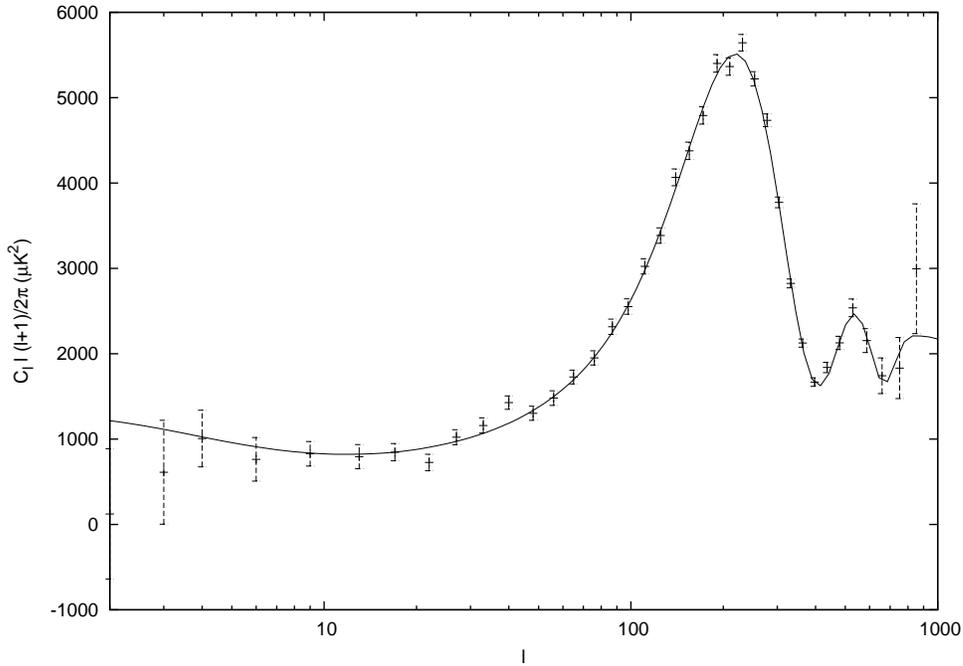}
\caption{CMB power spectrum of $\Lambda$CDM model (solid curve) compared to the WMAP data (points with error bars).} \label{fig:3}
\end{figure}

The spectrum from the WMAP observation is shown in Fig. \ref{fig:3}.  Note in particular that the last data point, corresponding to $l=839$ falls significantly above the theoretical curve. This means that $\chi^2$ can be lowered by a significant amount by using an unphysical value of $H\theta$ to fit this last point, so long as doing so does not also raise adjacent points too far outside their error bars.  Performing the calculation shows that is indeed what happens. We therefore conclude that the WMAP data do not constrain $H\theta$.

\begin{figure}
\includegraphics[height=9cm]{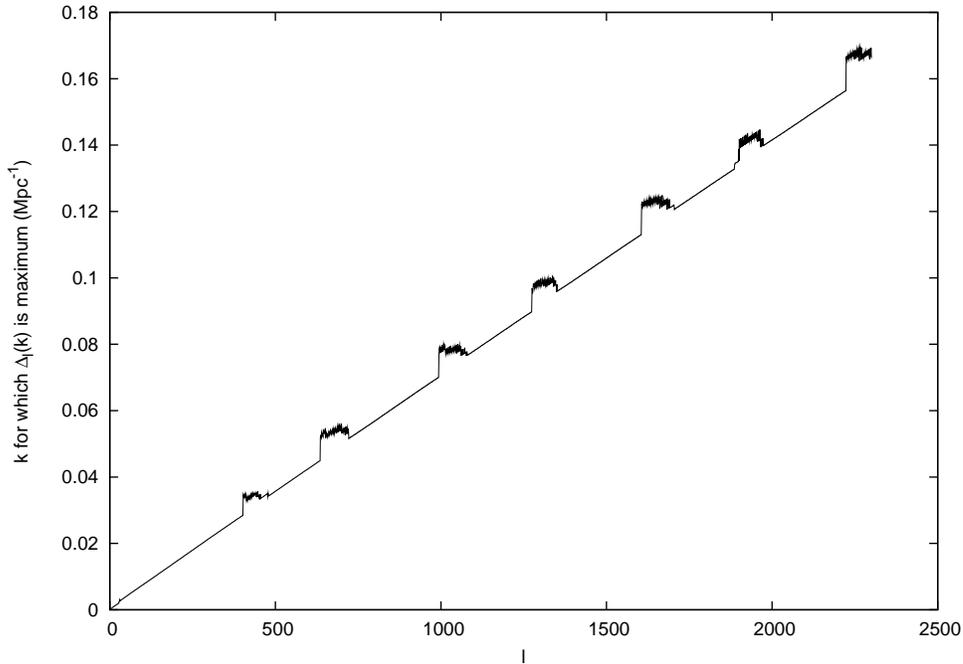}
\caption{The values of $k$ which maximize $\Delta_l(k)$, as a function of $l$} \label{fig:4}
\end{figure}

Fig. \ref{fig:4} shows the values of $k$ which maximize $\Delta_l(k)$, as a function of $l$, which in turn gives a rough estimate of the region over which the transfer functions contribute the most to the integral in eq. (4), and hence the region over which changes in $i_0(H \theta k)$
will most change the corresponding $C_l$. Thus to improve the bound on $H\theta$, we need data at higher $l$ ($l > 839$). In addition, tighter error bars at these higher $l$ will, of course, also help constrain the new parameter.

Based on this analysis we performed a second run of CMBEasy excluding the WMAP data.  This run resulted in a smaller, but still unphysically large, value of $H\theta$.  To see why this happens, we again consider the effect of varying only the new parameter $H\theta$ and examine the behavior of $\chi^2$.

\begin{figure}
\includegraphics[height=9cm]{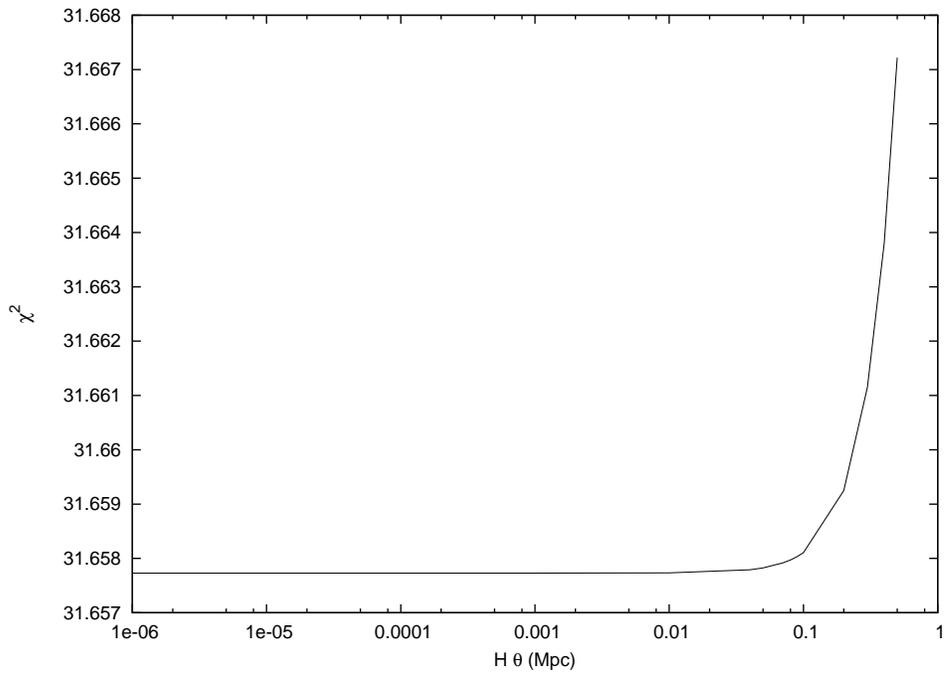}
\caption{$\chi^2$ versus $H\theta$ for ACBAR data} \label{fig:5}
\end{figure}

ACBAR and CBI are CMB data on small-scales (ACBAR and CBI give CMB power spectrum for multipoles up to $l=2985$ and $l=3500$ respectively) and hence may be better suited to determination of $H\theta$.  A plot of $\chi^2$ versus $H\theta$ for ACBAR+CBI data is shown in Fig. \ref{fig:5}.
The plateau between $H \theta = 0$ Mpc and $H \theta = 0.01$ Mpc is not physical, it results from limited numerical precision. Therefore, likelihoods calculated in this range only restrict $H\theta < 0.01$ Mpc and hence cannot indicate whether the best fit is at $H \theta = 0$ Mpc or some small non-zero value.

However, it is possible to put a constrant on the energy scale of spacetime noncommutativity from $H\theta <0.01$ Mpc. We discuss this below.

We can use the ACBAR+WMAP3 constraint on the amplitude of scalar power spectrum $A_{s} \simeq 2.15 \times 10^{-9}$ and the slow-roll parameter $\epsilon < 0.043$ \cite{ACBAR1} to find the Hubble parameter during inflation. The expression for the amplitude of the scalar power spectrum
\bea
A_{s} = \frac{1}{\pi \epsilon}\Big(\frac{H}{M_{p}}\Big)^{2},
\eea
where $M_{p}$ is the Planck mass, gives an upper limit on Hubble parameter:
\bea
H < 1.704 \times 10^{-5} M_{p}.
\eea

On using this upper limit for $H$ in the relation $H \theta <0.01~\textrm{Mpc}$, we have $\theta < 1.84 \times 10^{-9}m^{2}$.

We are interested to know the noncommutativity parameter at the end of inflation. That is, we should know the value of the cosmological scale factor $a$ when inflation ended. Most of the single field slow-roll inflation models work at an energy scale of $10^{15}$ GeV or larger \cite{dodelson}. Assuming that the reheating temperature of the universe was close to the GUT energy scale ($10^{16}$ GeV), we have for the scale factor at the end of inflation the value $a\simeq10^{-29}$ \cite{dodelson}. Thus we have for the noncommutativity parameter, $\sqrt{\theta} < (1.84~a \times 10^{-9})^{1/2} = 1.36 \times 10^{-19}\textrm{m}$. This corresponds to a lower bound for the energy scale of $10$ TeV.


\section{Non-causality from Noncommutative Fluctuations}

In the noncommutative frame work, the expression for the two-point correlation function for the field $\varphi_\theta$ for the scalar metric perturbations contains hermitian and anti-hermitian parts \cite{cmbpaper}. Taking the hermitian part, we obtained the modified power spectrum
\bea
P_{\Phi_{\theta}}({\bf k}) = P_{\Phi_{0}}(k) \; \textrm{cosh}(H\vec{\theta}^{0}\cdot {\bf k}),
\eea
where $P_{\Phi_{0}}(k)$ is the power spectrum for the scalar metric perturbations in the commutative case (as discussed in \cite{cmbpaper}), $H$ is the Hubble parameter during inflation. The constant spatial vector $\vec{\theta}^{0}$ is a measure of noncommutativity. The parameter $\theta$ is related to $\vec{\theta}^{0}$ by $\vec{\theta}^{0}=\theta \hat{z}$ if we choose the $z$-axis in the direction of $\vec{\theta}^{0}$, $\hat{z}$ being a unit vector. Also,
\bea
\Phi_{\theta}({\bf k}, t)= \int d^{3}x~\varphi_\theta({\bf x}, t)~\textrm{e}^{-i{\bf k}\cdot {\bf x}}.
\eea
This modified power spectrum was used to calculate the CMB angular power spectrum for the two-point temperature correlations.

In this section \footnote{This section is based on the work of four of us with Sang Jo. It has been described in \cite{cmbpaper}, but not published.}, we discuss the imaginary part of the two-point correlation function for the field $\varphi_\theta$. In position space, the imaginary part of the two-point correlation function is
obtained from the ``anti-symmetrization" (taking the anti-hermitian part) of the product of fields for a space-like separation:
\bea
\label{re:non-causality}
\frac{1}{2}[\varphi_{\theta}({\bf x}, \eta), \varphi_{\theta}({\bf
y}, \eta)]_{-} = \frac{1}{2}\Big(\varphi_{\theta}({\bf x}, \eta)
\varphi_{\theta}({\bf y}, \eta)- \varphi_{\theta}({\bf y}, \eta)
\varphi_{\theta}({\bf x}, \eta)\Big).
\eea
The commutator of deformed fields, in general, is nonvanishing for space-like separations. This  type of non-causality is an inherent property of noncommutative field theories constructed on the
Groenewold-Moyal spacetime \cite{Sachin}.

To study this non-causality, we consider two smeared fields localized at ${\bf x}_{1}$ and ${\bf x}_{2}$. (The expression for non-causality diverges for conventional choices for $P_{\Phi_{0}}$ if we do not smear the fields. See after eq. (\ref{non-causal}).) We write down smeared fields at ${\bf x}_{1}$ and ${\bf x}_{2}$.
\bea
&&\varphi(\alpha, {\bf x}_{1}) = \Big(\frac{\alpha}{\pi}\Big)^{3/2}\int d^{3}x~\varphi_{\theta}({\bf x})~e^{-\alpha({\bf x} - {\bf x}_{1})^{2}}, \\
&&\varphi(\alpha, {\bf x}_{2}) = \Big(\frac{\alpha}{\pi}\Big)^{3/2}\int d^{3}x~\varphi_{\theta}({\bf x})~e^{-\alpha({\bf x} - {\bf x}_{2})^{2}},
\eea
where $\alpha$ determines the amount of smearing of the fields. We have
\bea
\lim_{\alpha  \rightarrow \infty}\Big(\frac{\alpha}{\pi}\Big)^{3/2}\int d^{3}x~\varphi_{\theta}({\bf x})~e^{-\alpha({\bf x} - {\bf x}_{1})^{2}}=\varphi_{\theta}({\bf x}_{1}).
\eea
The scale $1/\sqrt{\alpha}$ can be thought of as  the width of a wave packet which is a measure of
the size of the spacetime region over which an experiment is performed.

We can now write down the uncertainty relation for the fields $\varphi(\alpha, {\bf x}_{1})$ and $\varphi(\alpha, {\bf x}_{2})$ coming from eq. (\ref{re:non-causality}):
\bea
\label{eq:noncausal}
\Delta \varphi(\alpha, {\bf x}_{1}) \Delta \varphi(\alpha, {\bf x}_{2}) \geq \frac{1}{2} \Big| \langle 0 |[\varphi(\alpha, {\bf x}_{1}), \varphi(\alpha, {\bf x}_{2})]|0\rangle \Big|
\eea

{\it This equation is an expression for the violation of causality due to noncommutativity.}

We can connect the power spectrum for the field $\Phi_0$ at horizon crossing with the commutator of the fields given in eq. (\ref{re:non-causality}):
{\footnotesize
\bea
\label{eq:commutator}
\frac{1}{2}\langle0|[\Phi_{\theta}({\bf k}, \eta), \Phi_{\theta}({\bf k}', \eta)]_{-}|0\rangle \Big|_{\textrm{horizon crossing}} = (2\pi)^{3}P_{\Phi_{0}}(k)~\textrm{sinh}(H \vec{\theta}^{0}\cdot {\bf k})~\delta^{3}({\bf k}+{\bf k}').\nn\\
\eea
}
Here we followed the same derivation given in \cite{cmbpaper}, using a commutator for the fields to start with, instead of an anticommutator of the fields, to obtain the above result.

The right hand side of eq. (\ref{eq:noncausal}) can be calculated as follows:
{\footnotesize
\bea
&&\langle 0 |[\varphi(\alpha, {\bf x}_{1}), \varphi(\alpha, {\bf x}_{2})]|0\rangle =\Big(\frac{\alpha}{\pi}\Big)^{3}\int d^{3}x d^{3}y~\langle 0 |[\varphi_{\theta}({\bf x}), \varphi_{\theta}({\bf y})]|0\rangle~e^{-\alpha({\bf x} - {\bf x}_{1})^{2}}e^{-\alpha({\bf y} - {\bf x}_{2})^{2}}\nn \\
&&~~=\Big(\frac{\alpha}{\pi}\Big)^{3}\int d^{3}x d^{3}y \frac{d^{3}k}{(2\pi)^{3}} \frac{d^{3}q}{(2\pi)^{3}}~\langle 0 |[\Phi_{\theta}({\bf k}), \Phi_{\theta}({\bf q})]|0\rangle~e^{-i{\bf k}\cdot{\bf x}-i{\bf q}\cdot{\bf y}}e^{-\alpha[({\bf x} - {\bf x}_{1})^{2}+({\bf y} - {\bf x}_{2})^{2}]}\nn \\
&&~~=\frac{2}{(2\pi)^{3}}\Big(\frac{\alpha}{\pi}\Big)^{3}\int d^{3}x d^{3}y d^{3}k~P_{\Phi_{0}}(k)\; \textrm{sinh}(H\vec{\theta}^{0}\cdot {\bf k})~e^{-i{\bf k}\cdot({\bf x}-{\bf y})}e^{-\alpha[({\bf x} - {\bf x}_{1})^{2}+({\bf y} - {\bf x}_{2})^{2}]}\nn \\
&&~~=\frac{2}{(2\pi)^{3}}\int
d^{3}k~P_{\Phi_{0}}(k)~\textrm{sinh}(H\vec{\theta}^{0}\cdot {\bf
k})~e^{-\frac{{\bf k}^{2}}{2 \alpha} -i{\bf k}\cdot({\bf x}_{1}-{\bf
x}_{2})}. \label{non-causal-commu}
\eea
}

This gives for eq. (\ref{eq:noncausal}),
\bea
&&\Delta \varphi(\alpha, {\bf x}_{1}) \Delta
\varphi(\alpha, {\bf x}_{2}) \geq \Big|\frac{1}{(2\pi)^{3}}\int
d^{3}k~P_{\Phi_{0}}(k)~\textrm{sinh}(H\vec{\theta}^{0}\cdot {\bf
k})~e^{-\frac{{\bf k}^{2}}{2 \alpha} -i{\bf k}\cdot({\bf x}_{1}-{\bf
x}_{2})}\Big|.
\label{non-causal}\nn\\
\eea
The right hand side of eq. (\ref{non-causal}) is divergent for  conventional asymptotic
behaviours of $P_{\Phi_{0}}$ (such as $P_{\Phi_{0}}$ vanishing for
large $k$ no faster than some inverse power of $k$) when $\alpha
\rightarrow \infty$ and thus the Gaussian width becomes zero. This
is the reason for introducing smeared fields.

Notice that the amount of causality violation given in eq. (\ref{non-causal}) is direction-dependent.

The uncertainty relation given in eq. (\ref{non-causal}) is purely due to spacetime noncommutativity as it vanishes for the case $\theta^{\mu \nu} =0$. It is an expression of causality violation.
\begin{figure}
\includegraphics[height=9cm]{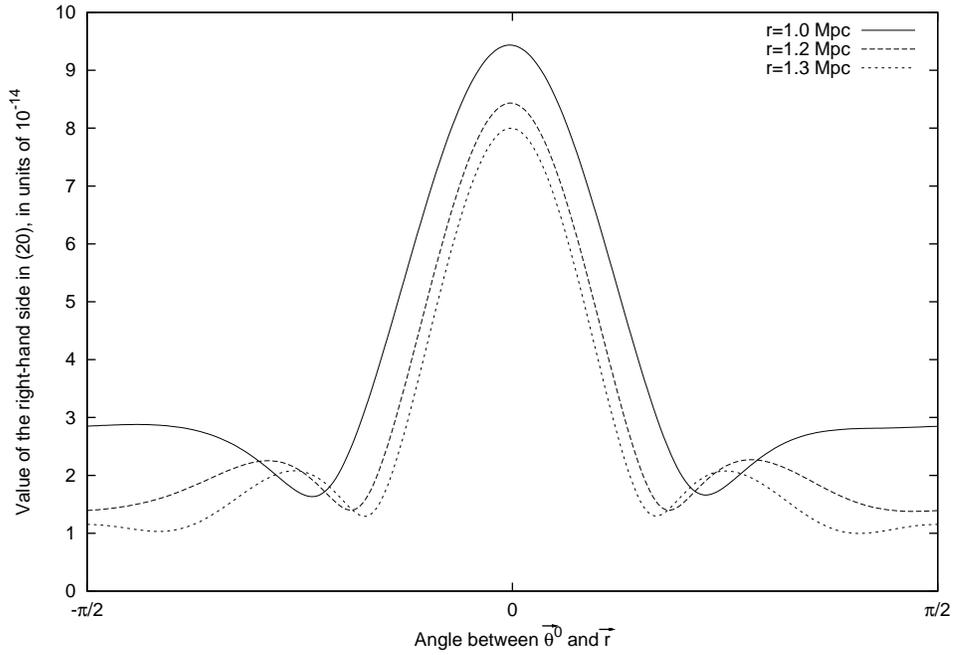}
\caption{The amount of causality violation with respect to the relative orientation between the vectors $\vec{\theta}^{0}$ and ${\bf r}={\bf x}_{1}-{\bf x}_{2}$. It is maximum when the angle between the two vectors is zero. Notice that the minima do not occur when the two vectors are orthogonal to each other. This plot is generated using the Cuba integrator \cite{cuba}.} \label{fig:6}
\end{figure}

This amount of causality violation may be expressed in terms of the CMB temperature fluctuation $\Delta T/T$. We have the relation connecting the temperature fluctuation we observe today and the primordial scalar perturbation $\Phi_{\theta}$,
\bea
\label{eq:deltaToverT}
{\Delta T(\hat{n}, \eta_{0}) \over T} &=& \sum_{l m} a_{l m}(\eta_{0}) Y_{l m}(\hat{n}),\nn \\
a_{lm}(\eta_{0}) &=& 4 \pi (-i)^{l}~\int \frac{d^{3}k}{(2 \pi)^{3}}~\Delta_{l}(k, \eta_{0}) \Phi_{\theta}({\bf k})Y_{lm}^{*}(\hat{k}),
\eea
where $\hat{n}$ is the direction of incoming photons and the transfer functions $\Delta_{l}$ take the primordial field perturbations to the present time $\eta_{0}$. We can rewrite the commutator of the fields in terms of temperature fluctuations $\Delta T/T$ using eq. (\ref{eq:deltaToverT}), but the corresponding correlator differs from the one for the CMB temperature anisotropy. It is not encoded in the two-point temperature correlation functions which as we have seen are given by the correlators of the anti-commutator of the fields.

In Fig. \ref{fig:6}, we show the dependence of the amount of non-causality on the relative orientation of the vectors $\vec{\theta}^{0}$ and ${\bf r}={\bf x}_{1}-{\bf x}_{2}$. The amount of causality violation is maximum when the two vectors are aligned.

\section{Conclusions: Chapter \ref{cmb2}}
The power spectrum becomes direction dependent in the presence of spacetime noncommutativity, indicating a preferred direction in the universe. We tried a best-fit of the theoretical model in \cite{cmbpaper} with the WMAP data and saw that to improve the bound on $H\theta$, we need data at higher $l$. (The last data point for WMAP is at $l=839$.) We therefore conclude that the WMAP data do not constrain $H\theta$. We also see that tighter error bars at these higher $l$ will also help constrain the noncommutativity parameter. The small-scale CMB data like ACBAR and CBI give the CMB power spectrum for larger multipoles and hence may be better suited for the determination of $H\theta$. ACBAR+CBI data only restrict $H\theta$ to $H\theta < 0.01$ Mpc and do not indicate whether the best fit is at $H \theta = 0$ Mpc or some small non-zero value. However, this restriction corresponds to a lower bound for the energy of $\theta$ of around $10$ TeV.

Further work is needed before rejecting the initial hypothesis that the other parameters of the $\Lambda$CDM cosmology are unaffected by noncommutivity. It requires performing a full MCMC study of all seven parameters.

Also, we have shown the existence and direction-dependence of non-causality coming from spacetime noncommutativity for the fields describing the primordial scalar perturbations when they are space-like separated. We see that the amount of causality violation is maximum when the two vectors, $\vec{\theta}^{0}$ and ${\bf r}={\bf x}_{1}-{\bf x}_{2}$, are aligned. Here ${\bf r}$ is the relative spatial coordinate of the fields at spatial locations ${\bf x}_{1}$ and ${\bf x}_{2}$.
\begin{center}
Summary of Chapter \ref{fdt}
\end{center}

\begin{itemize}
\item Deformed Lorentz invariance leads to noncausal correlations which ``correspond'' to corrections $\delta\chi_\theta$ to susceptibility $\chi$ in linear response theory.

\item Linear response theory involves determination of the linear dependence \\
~~$\lang\delta q\rang\approx \chi f$ of the expectation value $\lang\delta q\rang$ of the change $\delta q$ in a dynamical variable or coordinate $q$ of a physical system when the Hamiltonian $H$ of the system is perturbed ~$H\ra H+qf$~ by applying a weak external force $f$ to the system.
\item There are acausal corrections $\delta\chi_\theta$ to  susceptibility $\chi$ due to spacetime noncommutativity. For input with a single frequency $\omega_0$ the momentum dependence $\widetilde{\delta\chi_\theta}(\vec{k},\omega)$ of the corrections $\delta\chi_\theta$ to the output due to noncommutativity display zeroes and oscillations which are potential experimental signals for noncommutativity.

\end{itemize}

\chapter{Finite Temperature Field Theory}\label{fdt}

In this paper, we initiate the study of finite temperature quantum field theories (QFT's) on the Moyal plane. Such theories violate causality which influences the properties of these theories. In particular, causality influences the fluctuation-dissipation theorem: as we show, a disturbance in a spacetime region $M_1$ creates a response in a spacetime region $M_2$ spacelike with respect to $M_1$ ($M_1\times M_2$). The relativistic Kubo formula with and without noncommutativity is discussed in detail, and the modified properties of relaxation time and the dependence of mean square fluctuations on time are derived. In particular, the Sinha-Sorkin result \cite{sorkin-sinha} on the logarithmic time dependence of the mean square fluctuations is discussed in our context.

We derive an exact formula for the noncommutative susceptibility in terms of the susceptibility for the corresponding commutative case. It shows that noncommutative corrections in the four-momentum space have remarkable periodicity properties as a function of the four-momentum $k$. They have direction dependence as well and vanish for certain directions of the spatial momentum. These are striking observable signals for noncommutativity.

The Lehmann representation is also generalized to any value of the noncommutativity parameter $\theta^{\mu\nu}$ and finite temperatures.

\section{INTRODUCTION}

The Moyal plane is the algebra $\A_\theta(\mathbb{R}^d)$ of functions on $\mathbb{R}^d$ with the $\ast$-product given by
\bea
\label{moyal1}&&(f\ast g)(x)=f(x)e^{{i\over 2}\ola{\del}_\mu\theta^{\mu\nu}\ora{\del}_\nu}g(x)\eqv f(x)e^{{i\over 2}\ola{\del}\wedge\ora{\del}}g(x),~~f,g\in \A_\theta(\mathbb{R}^d),\nn\\
&&\theta_{\mu\nu}=-\theta_{\nu\mu}=\txt{constant}.
\eea
If $\hat{x}_\mu$ are coordinate functions, $\hat{x}_\mu(x)=x_\mu$, then (\ref{moyal1}) implies that
\bea
[\hat{x}_\mu,\hat{x}_\nu]=i\theta_{\mu\nu}.
\eea
Thus $\A_\theta(\mathbb{R}^d)$ is a deformation of $\A_0(\mathbb{R}^d)$ \cite{qft-us}.

There is an action of a Poincar$\acute{\txt{e}}$-Hopf algebra with a "twisted" coproduct on $\A_\theta(\mathbb{R}^d)$. Its physical implication is that QFT's can be formulated on $\A_\theta(\mathbb{R}^d)$ compatibly with the Poincar$\acute{\txt{e}}$ invariance of Wightman functions \cite{drinfeld,qft-us}. There is also a map of untwisted to twisted fields corresponding to $\theta_{\mu\nu}=0$ and $\theta_{\mu\nu}\neq 0$ (``the dressing transformation''  \cite{Grosse,Faddeev-Zamolodchikov}). For matter fields, if these are $\vphi_0$ and $\vphi_\theta$,

\bea
&&\vphi_\theta(x)=\vphi_0(x)e^{{1\over 2}\ola{\del}_\mu\theta^{\mu\nu}P_\nu}\eqv \vphi_0(x)e^{{1\over 2}\ola{\del}\wedge P},\\
&&P_\mu=\txt{Total momentum operator}.
\eea
While there is no twist factor $e^{{1\over 2}\ola{\del}\wedge P}$ for gauge fields, the gauge field interactions of a matter current with a gauge field are twisted as well:
\bea
\H^\theta_I(x)=\H^0_I(x)e^{{1\over 2}\ola{\del}\wedge P},
\eea
where $\H^0_I$ can be the standard interaction $J^{0\mu}A_\mu$ of an untwisted matter current to the untwisted gauge field $A_\mu$.

The twisted fields $\vphi_\theta$ and $\H^\theta_I$ are not causal (local). Thus even if $\vphi_0$ and $\H^0_I$ are causal fields,

\bea
&&[\vphi_0(x),\vphi_0(y)]=0,\\
&&[\H^0_I(x),\H^0_I(y)]=0,\\
&&[\H^0_I(x),\vphi_0(y)]=0,~~x\times y
\eea
($x\times y$ means that $x$ and $y$ are relatively spacelike),
that is not the case for the corresponding twisted fields. For example,

\bea
&&[\vphi_\theta(x),\H^\theta_I(y)]=e^{-{i\over 2}{\del\over\del x^\mu}\theta^{\mu\nu}{\del\over\del y^\nu}}\vphi_0(x)\H^0_I(y)-e^{-{i\over 2}{\del\over\del y^\mu}\theta^{\mu\nu}{\del\over\del x^\nu}}\H^0_I(y)\vphi_0(x)\neq 0,\nn\\
&&~~~~ x\times y.
\eea

Thus acausality leads to correlation between events in spacelike regions. The study of these correlations at finite temperatures  at the level of linear response theory (Kubo formula) is the central focus of this paper. We will also formulate the Lehmann representation for relativistic fields at finite temperature for $\theta_{\mu\nu}\neq 0$. It is possible that some of our results for $\theta_{\mu\nu}=0$ and $\theta_{\mu\nu}\neq 0$ are known \cite{fradkin}.

In section 3, we review the standard linear response theory \cite{fradkin} and the striking work of Sinha and Sorkin \cite{sorkin-sinha}.
We also discuss the linear response theory for relativistic QFT's at finite temperature for $\theta_{\mu\nu}=0$. It leads to a natural lower bound on relaxation time, a modification of the result ``$(\Delta r)^2\approx \txt{constant}\times\Delta t$'' of Einstein and its generalization ``$(\Delta r)^2\approx \txt{constant}\times \log\Delta t$'' to the `` quantum regime'' by Sinha and Sorkin \cite{sorkin-sinha}.

Section 4 contains the linear response theory for the twisted QFT's for $\theta_{\mu\nu}\neq 0$. A striking result we find is the existence of correlations between spacelike events: A disturbance in a spacetime region $M_2$ evokes a fluctuation in a spacetime region $M_1$ spacelike with respect to $M_2$ $(M_1\times M_2)$. Noncommutative corrections in four-momentum space also have striking periodicity properties and zeros as a function of the four-momentum $k$. They are also direction-dependent and vanish in certain directions of the spatial momentum $\vec{k}$. All these results are discussed in this section.

The results of this section have a bearing on the homogeneity problem in cosmology. It is a problem in causal theories \cite{trodden-vachaspati}. The noncommutative theories are not causal and hence can contribute to its resolution.

In section 5, we derive the finite temperature Lehmann representation for \\
~~$\theta_{\mu\nu}=0$ and generalize it to $\theta_{\mu\nu}\neq 0$. The Lehmann representation is known to be useful for the investigation of QFT's. The concluding remarks are in section 6.

\section{Review of standard theory: Sinha-Sorkin results}
Let $H_0$ be the Hamiltonian of a system in equilibrium at temperature $T$. It is described by the Gibbs state $\omega_\beta$ which gives for the mean value $\omega_\beta(A)$ of an observable $A$,
\bea
&&\omega_\beta(A)={{\Tr}~e^{-\beta H_0}A\over {\Tr}~e^{-\beta H_0}}.
\eea

We assume that $H_0$ has no explicit time dependence, otherwise it is arbitrary and can describe an interacting system.

We now perturb the system by an interaction $H'(t)$ so that the Hamiltonian becomes
\bea
&&H(t)=H_0+H'(t).
\eea

When $H'$ is treated as a perturbation, the change $\omega_\beta(\delta A(t))$ in the expectation value of an observable $A(t)$ in the Heisenberg  picture at time $t$ is
\bea
\omega_\beta(\delta A(t))=\omega_\beta(U^{-1}_I(t)A~U_I(t))-\omega_\beta(A),
\eea
where
\bea
&&U_I(t)=\T e^{-{i\over\hbar}\int_{-\infty}^t d\tau H_I(\tau)}\\
&&H_I(\tau)=e^{{i\over\hbar} H_0\tau}H'(\tau)e^{-{i\over\hbar} H_0\tau}.
\eea
Hence to leading order,
\bea
\omega_\beta(\delta A(t))&=&-{i\over\hbar}\int_{-\infty}^t d\tau~ \omega_\beta([A,H_I(\tau)])\\
         &=&-{i\over\hbar}\int_{-\infty}^\infty d\tau~\theta(t-\tau)\omega_\beta([A,H_I(\tau)]).
\eea
The linear response theory is based on this formula. It is completely general and applies equally well to quantum mechanics and QFT's. But in the latter case, the spatial dependence of the observable should also be specified.

For illustration of known results, we now specialize to quantum mechanics with one degree of freedom and to a dynamical variable $A(t)=x(t)=x(t)^\dagger$ and $H'(t)=x(t)f(t)$ where $f$ is a weak external force. Then,
\bea
&&\omega_\beta(\delta x(t))=-{i\over\hbar}\int^\infty_{-\infty}d\tau~ \theta(t-\tau)\omega_\beta([x(t),x(\tau)])~f(\tau)\\
&&~~~~=\int^\infty_{-\infty}~\chi(t-\tau)~f(\tau),
\eea
where $\chi$ is the susceptibility:
\bea
\chi(t)=-{i\over\hbar}\theta(t)\omega_\beta([x(t),x(0)]).
\eea
We have the following expressions:
\bea
&&W(t)=\omega_\beta(x(t)x(0))=S(t)+iA(t),\\
&&S(t)={1\over2}\omega_\beta(\{x(t),x(0)\}),~~A(t)=-{i\over2}\omega_\beta([x(t),x(0)]),\nn\\
&&\chi(t)={2\over\hbar}~\theta(t)A(t).
\eea

The significant properties of these correlation functions are as follows:
\begin{enumerate}
\item Unitarity:
\bea
&&H_0^\dagger=H_0,~~x(t)^\dagger=x(t)~~\Ra~~\overline{S(t)}=S(t),~~\overline{A(t)}=A(t).\nn
\eea

\item Time translation invariance:
\bea
&&S(-t)=S(t),~~A(-t)=-A(t)~~\Ra~~\overline{W(t)}=W(-t)\nn
\eea
from time independence of $H_0$.

\item The KMS condition: (with $\hbar=1$.)
\bea
W(-t-i\beta)=W(t).
\eea
\end{enumerate}

Denoting the Fourier transform of these functions, including $\chi$, by a tilde ~$\widetilde{}$~, as for instance
\bea
\widetilde{W}(\omega)=\int dt~e^{i\omega t}W(t),
\eea
one finds
\bea
&&\widetilde{W}(\omega)=e^{\beta\omega}\widetilde{W}(-\omega),\\
&&\txt{Im}\widetilde{\chi}(\omega)=-{1\over 2}(1-e^{-\beta\omega})\widetilde{W}(\omega),\\
\label{symmetric2corrft}&&\widetilde{S}(\omega)=-\coth{\beta\omega\over 2}\txt{Im}\widetilde{\chi}(\omega).
\eea

The important aspect of these relations is that the dissipative part $\txt{Im}\widetilde{\chi}$ of the (Fourier transform of) susceptibility $\chi$ completely determines all the two point correlations, and hence also the real part $\txt{Re}\widetilde{\chi}$ of $\widetilde{\chi}$.

$\txt{Re}\widetilde{\chi}$ can also be determined from $\txt{Im}\widetilde{\chi}$ by the Kramers-Kronig relation \cite{fradkin}.

 Following an argument, presented in \cite{sorkin-sinha}, which exploits the properties of the Heaviside function $\theta$, we can write
\bea
&&\txt{Im}\widetilde{\chi}(\omega)
=-{i\over 2}\widetilde{\chi}'(\omega),\nn\\
\eea
where
\bea
&&\chi'(t):=\txt{sgn}(t)~\chi(|t|),\nn\\
&&\txt{sgn}(t)=\theta(t)-\theta(-t).
\eea
Therefore, (\ref{symmetric2corrft}) becomes
\bea
\label{symmetric3corrft}&&\widetilde{S}(\omega)={i\over 2}\coth{\beta\omega\over 2}~\widetilde{\chi}'(\omega).
\eea
The Fourier transform of (\ref{symmetric3corrft}) gives
\bea
\label{symmetric2corr}S(t)={1\over 2\beta}P\int^\infty_{-\infty}dt'~\txt{sgn}(t'-t)~\chi(|t'-t|)\coth{\pi t'\over\beta},
\eea

where $P$ denotes the principal value of $\coth$.
$\txt{Re}\widetilde{\chi}$ does not contribute to (\ref{symmetric2corr}).

This equation has important physics. In time $\Delta t$, the operator changes by $\Delta x(t)=x(t+\Delta t)-x(t)$. With $t=0$, the square displacement due to equilibrium fluctuations is thus
\bea
\omega_\beta(\Delta x(0)~^2)=2[S(0)-S(\Delta t)]
\eea
so that we obtain the Sinha-Sorkin formula
{\footnotesize
\bea
&&{1\over 2}\omega_\beta(\Delta x(0)~^2)\nn\\
&&~~~~={i\over 2\beta}P\int _0^\infty dt'\chi(t')[2\coth (\Omega t')-\coth(\Omega(t'+\Delta t))-\coth(\Omega(t'-\Delta t))],~~\Omega={\pi\over\beta}.\nn\\
\eea
}
Sinha and Sorkin \cite{sorkin-sinha} have analyzed this equation for the (realistic) ansatz
\bea
\label{ansatz}\chi(t)=\mu[1-e^{-{t\over\tau}}]\theta(t)~\sr{t\gg \tau}{\ral}~ \mu ~\theta(t-\tau),
\eea
where $\tau$ is the relaxation time.

In that case,
\bea
\label{sinha.sorkin.rl}{1\over 2}\omega_\beta(\Delta x(0)~^2)={\mu\hbar\over \pi}\ln {[\sinh(\Omega |\Delta t-\tau|)\sinh(\Omega |\Delta t+\tau|)]^{1\over 2}\over \sinh(\Omega\tau)},
\eea
where we have restored $\hbar$.

Sinha and Sorkin \cite{sorkin-sinha} observed that (\ref{sinha.sorkin.rl}) gives Einstein's relation in the classical regime:
\bea
\label{clregime}\beta\hbar \ll\tau\ll\Delta t: ~~{1\over 2}\omega_\beta(\Delta x(0)~^2)\approx {\mu\over\beta}\Delta t.
\eea
But in addition they found a \emph{logarithmic} dependence of $\Delta t$ in the "quantum" regime:
\bea
\label{qregime}\tau\ll \Delta t\ll \beta\hbar:~~{1\over 2}\omega_\beta(\Delta x(0)~^2)={\mu\hbar\over \pi}\ln{\Delta t\over\tau}.
\eea
They have emphasized that this behavior can be tested experimentally.

They also discuss a regime between the classical and quantum extremes which interpolates (\ref{clregime}) and (\ref{qregime}).

\section{Quantum Fields on Commutative Spacetime}

Hereafter, we set $\hbar=c=1$.

We now specialize to QFT's for $\theta_{\mu\nu}=0$. For simplicity, we take
\bea
H'(t)=e\int d^3y~ N_0(y)\vphi_0(y),
\eea
where $N_0(y)$ is the number density of a charged spinor field $\psi_0$,
\bea
N_0(y)=\psi_0^\dagger(y)\psi_0(y).
\eea
$\vphi_0$ is the externally imposed scalar potential and the subscript denotes that $\theta_{\mu\nu}=0$ for these fields. Again for simplicity, we choose $A$ as well to be the number density at a spacetime point $x$. Then
\bea
&&\omega_\beta(\delta N_0(x))=-{ie\over\hbar}\int d^4y~\theta(x_0-y_0)\omega_\beta([N_0(x),N_0(y)])\vphi_0(y).
\eea
The natural definition of susceptibility in this case is
\bea
\chi_\beta(x,y)=-i{e\over\hbar}\theta(x_0-y_0)\omega_\beta([N_0(x),N_0(y)]).
\eea

With this definition,
\bea
\omega_\beta(\delta N_0(x))=\int d^4y~ \chi_\beta(x,y)\vphi_0(y).
\eea

We will now analyze this formula.

\begin{flushleft}{\textbf{The Kubo formulae}}\end{flushleft}


The susceptibility $\chi_\beta$ is related to the Wightman function
\bea
W^\beta_0(x,y)={i\over\hbar}\omega_\beta(N_0(x)N_0(y))
\eea
and the autocorrelation and commutator functions
\bea
&&W^\beta_0(x,y)=S^\beta_0(x,y)+iA^\beta_0(x,y),\nn\\
&&S^\beta_0(x,y)={1\over 2\hbar}\omega_\beta (N_0(x)N_0(y)+N_0(y)N_0(x)),\nn\\
&&A^\beta_0(x,y)={-i\over 2\hbar}\omega_\beta([N_0(x),N_0(y)]),\nn\\
&&\chi_\beta(x,y)=2e\theta(x_0-y_0)A^\beta_0(x,y).
\eea

There are more nontrivial conditions coming from the KMS condition which we now discuss.

By assumption, $H_0$ commutes with spacetime translations and rotations as dictated by the Poincar$\acute{\txt{e}}$ algebra. So $\omega_\beta$ enjoys these symmetries and \\
$W^\beta_0(x,y),~S^\beta_0(x,y),~A^\beta_0(x,y)$ depend only on $x_0-y_0$ and $(\vec{x}-\vec{y})^2$. Hence they are even in $\vec{x}-\vec{y}$:
\bea
W^\beta_0(x_0,\vec{x}_0~;~y_0,\vec{y})&=&W^\beta_0(x_0,\vec{y}_0~;~y_0,\vec{x})~~\txt{etc}.\\
 &=& \hat{W}^\beta_0(x_0-y_0~;~(\vec{x}_0-\vec{y})^2).
\eea

As $\hat{W}^\beta_0(x_0-y_0~;~(\vec{x}_0-\vec{y})^2)$ can contain terms with $\theta(x_0-y_0)$, we cannot always claim that it is even in $x_0-y_0$ as well. The same goes for $S^\beta_0$ and $A^\beta_0$.

\subsubsection{Spacelike Disturbances}
If $x$ and $y$ are relatively spacelike,
~$[N_0(x),N_0(y)]=0$~ because of causality (locality).

So if $\vphi_0=0$ outside the spacetime region $D_2$ and we observe the fluctuation in a spacetime region $D_1$ spacelike with respect to $D_2$, then the fluctuation vanishes:
\bea
\omega_\beta(\delta N_0(x))=0~~ \txt{if}~~ x\in D_2,~~ \txt{Supp}\vphi_0=D_2,~~ D_1\times D_2.
\eea
Here Supp denotes the support of the function $\vphi_0$ (it is zero in the complement of the support).

Thus we easily recover the prediction of causality for $\theta_{\mu\nu}=0$ \cite{fradkin}.

\subsubsection{Timelike Disturbances}
In this case, the point of observation $x$ is causally linked to the spacetime region $D_2$. Hence $[N_0(x),N_0(y)]$ need not vanish if $x\in D_1$.

We can model the analysis of this case to the one in Section 2 if $H_0$ is the time translation generator of the Poincar$\acute{\txt{e}}$ group for $\vphi_0=0$. We assume that to be the case.

Following section 2, we now introduce the correlator
\bea
\label{wightmann2fxn}W_0^\beta(x,y)=\omega_\beta(N_0(x)N_0(y)).
\eea
By relativistic invariance, $W_0^\beta$ depends only on $(\vec{x}-\vec{y})^2$. Since $\theta(x_0-y_0)$ is Lorentz invariant when $x-y$ is timelike, it can also depend on $\theta(x_0-y_0)$. Thus $W_0^\beta$ depends on $(\vec{x}-\vec{y})^2$ and $x_0-y_0$ and we can rewrite (\ref{wightmann2fxn}) as
\bea
W_0^\beta((\vec{x}-\vec{y})^2,x_0-y_0)=\omega_\beta(N_0(x)N_0(y)).
\eea
We can thus focus on
\bea
\overline{W}_0^\beta(\vec{x}^2,x_0)=\omega_\beta(N_0(x)N_0(y)).
\eea

It is important that \emph{it is even in} $\vec{x}$.
We cannot say that about $x_0$ because of the potential presence of $\theta(x_0)$.

Now
\bea
&&\overline{W}_0^\beta(\vec{x}^2,x_0)=\omega_\beta(N_0(0)N_0(x))=W_0^\beta(\vec{x}^2,-x_0).
\eea
The presence of $\vec{x}$ thus does not affect the symmetry properties in $x_0$. That is the case also with regard to the KMS condition. We write all these conditions explicitly now: write
\bea
W_0^\beta(\vec{x}^2,x_0)=S_0^\beta(\vec{x}^2,x_0)+iA_0^\beta(\vec{x}^2,x_0),
\eea
where
\bea
&&S_0^\beta(\vec{x}^2,x_0)={1\over 2}\omega_\beta(N_0(x)N_0(0)+N_0(0)N_0(x)),\nn\\
&&A_0^\beta(\vec{x}^2,x_0)=-{i\over 2}\omega_\beta([N_0(x),N_0(0)]).
\eea
Then
\bea
&&\chi_\beta(\vec{x}^2,x_0)=2e\theta(x_0)A_0^\beta(\vec{x}^2,x_0),
\eea
where we have written the susceptibility as a function of $\vec{x}^2$ and $x_0$.
Then as before
\begin{enumerate}
\item $S_0^\beta$ and  $A_0^\beta$ are real functions:
\bea
\overline{S}_0^\beta(\vec{x}^2,x_0)=S_0^\beta(\vec{x}^2,x_0),~~ \overline{A}_0^\beta(\vec{x}^2,x_0)=A_0^\beta(\vec{x}^2,x_0).
\eea
\item $S_0^\beta$ is even in $x_0$ and $A_0^\beta$ is odd in $x_0$:
\bea
{S}_0^\beta(\vec{x}^2,-x_0)=S_0^\beta(\vec{x}^2,x_0),~~ {A}_0^\beta(\vec{x}^2,-x_0)=-A_0^\beta(\vec{x}^2,x_0).
\eea
\item We have the KMS condition
\bea
W_0^\beta(\vec{x}^2,-x_0-i\beta)=W_0^\beta(\vec{x}^2,x_0),
\eea
where we have set the speed of light $c$ equal to $1$.
\end{enumerate}
[We will rewrite ~$\chi_\beta,~\widetilde{\chi}_\beta$~ as ~$\chi^\beta_0,~\widetilde{\chi}^\beta_0$~ to emphasize that they correspond to $\theta_{\mu\nu}=0$.]
Thus from the Fourier transforms distinguished by tildes, as in
\bea
\widetilde{W}_0^\beta(\vec{x}^2,\omega)=\int dx_0~e^{i\omega x_0}W_0^\beta(\vec{x}^2,x_0),
\eea
we get
\bea
&&\widetilde{W}_0^\beta(\vec{x}^2,\omega)=e^{\beta\omega}\widetilde{W}_0^\beta(\vec{x}^2,-\omega),\\
\label{ImSusc1}&&\txt{Im}\widetilde{\chi}^\beta_0(\vec{x}^2,\omega)=-{e\over 2}(1-e^{\beta\omega})\widetilde{W}_0^\beta(\vec{x}^2,-\omega),\\
\label{symmcorr1}&&e\widetilde{S}^\beta_0(\vec{x}^2,\omega)=-\coth{\beta\omega\over 2}\txt{Im}\widetilde{\chi}^\beta_0(\vec{x}^2,\omega)
\eea
 Now following an argument analogous to the one that yielded (\ref{symmetric3corrft}), we are able to write
\bea
\txt{Im}\widetilde{\chi}_0^\beta(\vec{x}^2,\omega)
&=&-{i\over 2}\widetilde{\chi}'{}^\beta_0(\vec{x}^2,\omega),\nn\\
\eea
where
\bea
\chi'{}^\beta_0(\vec{x}^2,x_0)&:=&\txt{sgn}(x_0,\vec{x})~\chi_0^\beta(\vec{x}^2,|x_0|),\nn\\
\txt{sgn}(x_0,\vec{x})&=&\theta(x_0-|\vec{x}|)-\theta(-x_0-|\vec{x}|).\nn\\
\eea
Therefore, (\ref{symmcorr1}) becomes
\bea
\label{symmcorr2}&&e\widetilde{S}^\beta_0(\vec{x}^2,\omega)=-\coth{\beta\omega\over 2}\txt{Im}\widetilde{\chi}^\beta_0(\vec{x}^2,\omega)={i\over 2}\coth{\beta\omega\over 2}\widetilde{\chi}'{}^\beta_0(\vec{x}^2,\omega).
\eea
The Fourier transform of (\ref{symmcorr2}) gives

\bea
e{S}_0^\beta(\vec{x}^2,x_0)={1\over 2\beta}P\int dx'_0~\txt{sgn}(x'_0-x_0,\vec{x})\chi_0^\beta(\vec{x}^2,|x'_0-x_0|)\coth{\pi x'_0\over\beta}.
\eea

The expression for the mean square equilibrium fluctuation $\omega_\beta(\Delta N_0^2)(\vec{x}^2,0)$ follows as before:
{\footnotesize
\bea
&&{1\over2}\omega_\beta (\Delta N_0^2)((\vec{x}-\vec{y})^2,0)={1\over2}\omega_\beta( (N_0(\vec{x},x_0+\Delta x_0)-N_0(\vec{y},x_0))^2)\nn\\
&&~~~~=e(~S^\beta_0(\vec{0}^2,0)-S^\beta_0((\vec{x}-\vec{y})^2,\Delta x_0)~)= {1\over 2\beta}\{~2\int_{|\vec{0}|}^\infty dx'_0~\chi^\beta_0(\vec{0}^2,|x'_0|)~\coth{\pi x'_0\over \beta}\nn\\
&&~~~~-\int_{|\vec{x}-\vec{y}|}^\infty dx'_0~\chi^\beta_0((\vec{x}-\vec{y})^2,|x'_0|)(\coth{\pi(x'_0+\Delta x_0)\over \beta}+\coth{\pi(x'_0-\Delta x_0)\over \beta})~\}\label{fluctuation1}\nn\\
\eea
}
So nothing much has changed until this point except for the additional dependence of correlations on $\vec{x}^2$.

An ansatz like (\ref{ansatz}) for susceptibility is no longer appropriate now. That is because if
\bea
x_0^2<\vec{x}^2,
\eea
then as we saw $\chi_0^\beta(\vec{x}^2,x_0)$ is zero by causality.

Thus the relaxation time $\tau$ in units of $c$ has the lower bound $|\vec{x}|$:
\bea
\tau \geqslant |\vec{x}|.
\eea
$\tau$ is a function of $\vec{x}^2$, and we write $\tau(\vec{x}^2)$. Then the generalization of the ansatz (\ref{ansatz}) is
\bea
\chi^\beta_0(\vec{x}^2,x_0)=\mu[1-e^{-{x_0-|\vec{x}|\over\tau(\vec{x}^2)}}]\theta(x_0-|\vec{x}|)~\sr{x_0-|\vec{x}|\gg \tau}{\ral}~ \mu ~\theta(x_0-|\vec{x}|-\tau(\vec{x}^2)).
\eea
This lets us evaluate the mean square fluctuation of number density
{\footnotesize
\bea
&&\label{m2}{1\over 2}\omega_\beta(\Delta N_0^2)((\vec{x}-\vec{y})^2,0)={\mu\hbar\over \pi} \ln {[\sinh\Omega|\Delta x_0-\tau((\vec{x}-\vec{y})^2)|~\sinh\Omega|\Delta x_0+\tau((\vec{x}-\vec{y})^2)|]^{1\over 2}\over \sinh\Omega \tau(0) },\nn\\
\eea
}
where ~~$\Omega={\pi\over \hbar\beta}$.

Following Sinha and Sorkin \cite{sorkin-sinha}, we assume that
\bea
\label{bounds}\Delta x_0 \gg \tau(\vec{x}^2)\geqslant |\vec{x}|.
\eea
There are thus four time scales:
\bea
\beta\hbar,~~|\vec{x}|,~~\tau(\vec{x}^2),~~\Delta x_0,
\eea
where we have restored $\hbar$. With the assumption (\ref{bounds}), we have four possibilities to consider:
\begin{enumerate}
\item $\beta\hbar \ll |\vec{x}|\ll \tau(\vec{x}^2) \ll \Delta x_0$,

\item $ |\vec{x}|\ll \beta\hbar\ll \tau(\vec{x}^2) \ll \Delta x_0$,

\item $ |\vec{x}|\ll \tau(\vec{x}^2)\ll  \beta\hbar\ll \Delta x_0$,

\item $ |\vec{x}|\ll \tau(\vec{x}^2)\ll \Delta x_0 \ll \beta\hbar$.

\end{enumerate}

Case 1: \emph{The classical Regime}

Case 1 is the "classical" limit. We get back Einstein's result in this case:

\bea
&&{1\over 2}\omega_\beta(\Delta N_0^2)((\vec{x}-\vec{y})^2,0)\nn\\
&&~~~~={\mu\over \beta}(\Delta x_0-\tau(0))=\mu kT(\Delta x_0-\tau(0)).
\eea

Cases 2 and 3 interpolate the classical regime and the extreme quantum regime of case 4. So let us first consider Case 4.

Case 4: \emph{The Extreme Quantum Regime}

This is the  new regime where Sinha and Sorkin \cite{sorkin-sinha} found a logarithmic dependence on time $\Delta t$ of mean square fluctuations. It is now changed significantly.

\bea
{1\over 2}\omega_\beta(\Delta N_0^2)((\vec{x}-\vec{y})^2,0)={\mu\hbar\over\pi} \ln(~{\Delta x_0\over \tau(0)}[1-({\tau((\vec{x}-\vec{y})^2)\over \Delta x_0})^2]^{1\over 2}~).
\eea

As for the cases 2 and 3, our results are as follows:

\emph{Case 2}: The same as \emph{Case 1}.
\bea
{1\over 2}\omega_\beta(\Delta N_0^2)((\vec{x}-\vec{y})^2,0)={\mu\over \beta}(\Delta x_0-\tau(0)).
\eea

\emph{Case 3}:
\bea
{1\over 2}\omega_\beta(\Delta N_0^2)((\vec{x}-\vec{y})^2,0)={\mu\over\beta}\Delta x_0+{\mu\hbar\over\pi}\ln{\hbar\beta\over 2\pi\tau(0)}.
\eea

\section{Quantum Fields on the Moyal Plane}
 For the Moyal plane, we must use the twisted fields and interactions as explained in the Introduction. That leads to the following expression for $\delta N_\theta$:
 \bea
 \delta N_\theta(x)=-i\int_{-\infty}^\infty dx'_0~\theta(x_0-x'_0)\omega_\beta([N_\theta(x),H_I^\theta(x'_0)]),
 \eea
where
\bea
N_\theta=N_0e^{{1\over 2}\ola{\del}\wedge P},~~H_I(x_0)=e\int d^3 x~\H_I^0(x)e^{{1\over 2}\ola{\del}\wedge P},
\eea
$\H_I^0$ being the interaction Hamiltonian density in the interaction representation.

Note that $e^{{1\over 2}\ola{\del}\wedge P}$ reduces to $e^{{1\over 2}\ola{\del}_0\theta^{0i}P_i}$ on integration over $d^3x$. But we will not use this simplification yet.

We shall first discuss the dependence on $\theta$ of two-point correlators.

Let us first examine the twisted Wightman function:
\bea
&&W_\theta^\beta(x,y)=\omega_\beta(N_\theta(x)N_\theta(y))\nn\\
\label{ncwightmanfxn1}&&~~~~=e^{-{i\over 2}{\del\over\del x^\mu}\theta^{\mu\nu}{\del\over\del y^\nu}}\omega_\beta(N_0(x)N_0(y)e^{-{i\over 2}({\ola{\del}\over\del x^\mu} +{\ola{\del}\over\del y^\mu})\theta^{\mu\nu}P_\nu}).
\eea

We can write this as an integral (and sum) over states with total momentum $p$ such as
\bea
\label{ncwightmanfxn2}\langle p,...|e^{-\beta P_0}N_0(x)N_0(y)e^{-{i\over 2}({\ola{\del}\over\del x^\mu} +{\ola{\del}\over\del y^\mu})\theta^{\mu\nu}P_\nu} |p,...\rangle,
\eea
where the dots indicate that there will in general be many states contributing to a state of given total momentum $p$. We can write (\ref{ncwightmanfxn2}) as
\bea
\langle p,...|e^{-\beta P_0}N_0(x)N_0(y)e^{-{i\over 2}\ola{ad}P_\mu\theta^{\mu\nu}P_\nu} |p,...\rangle,
\eea
where $ad P_\mu A=[P_\mu,A]$. for any operator $A$. But
\bea
\langle p,...|[P_\mu,A] |p,...\rangle=0
\eea
for any $A$. Consequently (\ref{ncwightmanfxn2}) is
\bea
W_\theta^\beta(x,y)=e^{-{i\over 2}{{\del}\over\del x^\mu}\theta^{\mu\nu}{{\del}\over\del y^\nu}}W_0^\beta(x,y).
\eea
But now we can write $W_0^\beta(x,y)$ as we wrote it earlier:
\bea
W_0^\beta(x,y)\ra W_0^\beta((\vec{x}-\vec{y})^2,x_0-y_0).
\eea
It depends on $x-y$. Hence in the exponential,
\bea
{{\del}\over\del x^\mu}\theta^{\mu\nu}{{\del}\over\del y^\nu}=-{{\del}\over\del x^\mu}\theta^{\mu\nu}{{\del}\over\del x^\nu}=0.
\eea
Similarly,
\bea
&&S_\theta^\beta(x,y)={1\over 2}\omega_\theta(N_\theta(x)N_\theta(y)+N_\theta(y)N_\theta(x) )=S_0^\beta((\vec{x}-\vec{y})^2,x_0-y_0),\nn\\
&&A_\theta^\beta(x,y)=-{i\over 2}\omega_\theta([N_\theta(x),N_\theta(y)])=A_0^\beta((\vec{x}-\vec{y})^2,x_0-y_0)
\eea
and they have the properties listed earlier.

But we cannot conclude that $\delta N_\theta$ is independent of $\theta^{\mu\nu}$ as well. Specializing to
\bea
\H_I^0=N_0\vphi_0,
\eea
we find
\bea
&&\delta N_\theta(x)=\delta N_\theta{}^1(x)-\delta N_\theta{}^2(x),\\
&&\delta N_\theta{}^1(x)=-i\int d^4x'~\theta(x_0-x_0')e^{-{i\over 2}{{\del}\over\del x^\mu}\theta^{\mu\nu}{{\del}\over\del x'{}^\nu}}\omega_\beta(N_0(x)\H_I^0(x') e^{-{i\over 2}({\ola{\del}\over\del x^\mu}+{\ola{\del}\over\del x'{}^\mu})\theta^{\mu\nu}P_\nu})\nn\\
\eea
with a similar expression for $\delta N_\theta^2(x)$.
The last exponential can be replaced by 1 as before. Also, integration over $\vec{x}'$ reduces
~$e^{-{i\over 2}{{\del}\over\del x^\mu}\theta^{\mu\nu}{{\del}\over\del x'{}^\nu}}$~ to ~$e^{-{i\over 2}{{\del}\over\del x^i}\theta^{i0}{{\del}\over\del x'{}^0}},$
\bea
e^{-{i\over 2}{{\del}\over\del x^\mu}\theta^{\mu\nu}{{\del}\over\del x'{}^\nu}} \ra e^{-{i\over 2}{{\del}\over\del x^i}\theta^{i0}{{\del}\over\del x'{}^0}}.
\eea
Thus
\bea
\delta N_\theta^1=-i\int d^4x'~\theta(x_0-x_0')e^{-{i\over 2}{{\del}\over\del x^i}\theta^{i0}{{\del}\over\del x'{}^0}}\omega_\beta(N_0(x)N_0(x'))\vphi_0(x')
\eea
and similarly
\bea
\delta N_\theta^2=-i\int d^4x'~\theta(x_0-x_0')e^{{i\over 2}{{\del}\over\del x^i}\theta^{i0}{{\del}\over\del x'{}^0}}\omega_\beta(N_0(x')N_0(x))\vphi_0(x').
\eea

We now discuss the two cases where $x$ is space- and time-like with respect to supp $\vphi_0$.

\begin{flushleft}\emph{$x$ spacelike with respect to Supp $\vphi_0$}:\end{flushleft}

This is the case where we anticipate qualitatively new results.

 While calculating $\delta N^1_\theta(x')-\delta N^2_\theta(x'),$ we cannot set

 \bea
 N_0(x)N_0(x')=N_0(x')N_0(x)~~~~  (\txt{from causality})
 \eea
because the exponentials in the integrand translate the arguments $x$ and $x'$, and can bring them to timelike separations. With this in mind, we can write
\bea
&&\delta N_\theta(x)=-i\int d^4x'~\theta(x_0-x'_0)\cos[{1\over 2}{\del\over\del x^i}\theta^{i0}{\del\over\del x^0{}'}] \omega_{\beta}([N_0(x),N_0(x')])\vphi_0(x')\nn\\
&&~~-\int d^4x'~\theta(x_0-x'_0)\sin[{1\over 2}{\del\over\del x^i}\theta^{i0}{\del\over\del x^0{}'}] \omega_{\beta}(N_0(x)N_0(x')+N_0(x')N_0(x))\vphi_0(x').\nn\\
\eea
We can replace~ $\cos({1\over 2}{\del\over\del x^i}\theta^{i0}{\del\over\del x^0{}'})$~ by ~$\cos({1\over 2}{\del\over\del x^i}\theta^{i0}{\del\over\del x^0{}'})-1=2\sin^2({1\over 4}{\del\over\del x^i}\theta^{i0}{\del\over\del x^0{}'})$~ as the extra term contributes $0$ by causality. This shows that this term is $O((\theta^{i0})^2)$.
Finally,
{\footnotesize
\bea
&&\delta N_\theta(x)=-\int d^4x'~\theta(x_0-x'_0)\sin[{1\over 2}{\del\over\del x^i}\theta^{i0}{\del\over\del x^0{}'}] \omega_{\beta}(N_0(x)N_0(x')+N_0(x')N_0(x))\vphi_0(x') \nn\\
&&~~+2i\int d^4x'~\theta(x_0-x'_0)\sin^2[{1\over 4}{\del\over\del x^i}\theta^{i0}{\del\over\del x^0{}'}] \omega_{\beta}([N_0(x),N_0(x')])\vphi_0(x').
\eea
}

This shows clearly that there is an acausal fluctuation in $\delta N_\theta(x)$ when $\vphi_0$ (the ``chemical potential'') is fluctuated in a region $D_2$ spacelike with respect to $x$.

But it occurs only when time-space noncommutativity $(\theta^{0i})$ is non-zero.

We will come back to this term after also briefly looking at the case where $x$ is not spacelike with respect to $D_2$.

\begin{flushleft}\emph{$x$ is not spacelike with respect to Supp $\vphi_0$}\end{flushleft}

The only change as compared to the spacelike case is that we must restore the extra term, which contributed $0$ in the spacelike case, but does not do that now.

 We can simplify notation by defining $\Delta N_\theta(x)$ for any $x$ as follows:
 {\footnotesize
 \bea
&&\Delta N_\theta(x)=-\int d^4x'~\theta(x_0-x'_0)\sin[{1\over 2}{\del\over\del x^i}\theta^{i0}{\del\over\del x^0{}'}] \omega_{\beta}(N_0(x)N_0(x')+N_0(x')N_0(x))\vphi_0(x')\nn \\
&&~~+2i\int d^4x'~\theta(x_0-x'_0)\sin^2[{1\over 4}{\del\over\del x^i}\theta^{i0}{\del\over\del x^0{}'}] \omega_{\beta}([N_0(x),N_0(x')])\vphi_0(x').
\eea
}
Then

a) If $x~\times$~Supp~$\vphi_0$,
\bea
\delta N_\theta(x)=\Delta N_\theta(x).
\eea

b) If $x$ is not spacelike with respect to Supp $\vphi_0$,
\bea
\delta N_\theta(x)=i\int d^4x'~\theta(x_0-x'_0) \omega_{\beta}([N_0(x),N_0(x')])\vphi_0(x') + \Delta N_\theta(x).
\eea

\subsection{An exact expression for susceptibility}

We want to write
\bea
\delta N_\theta(x)=\int d^4x'~\chi_\theta(x,x')\vphi_0(x'),
\eea
where $\chi_\theta$ is the deformed susceptibility.

We will succeed in doing that by deriving an exact expression for the Fourier transform
\bea
\widetilde{\chi}_\theta(k)=\int d^4x~e^{ikx}\chi_\theta(x),~~ kx=k_0x_0-\vec{k}\cdot\vec{x},
\eea
in terms of $\widetilde{\chi}_0(k)$. The corrections to $\widetilde{\chi}_0(k)$ have remarkable zeros and direction dependence which we will soon point out.

We can write
\bea
\delta N_\theta(x)=\delta N_0(x)+\Delta N_\theta(x),
\eea
where
\bea
\delta N_0(x)=\int d^4x'~\chi_0(x-x')\vphi_0(x')
\eea
and
\bea
&& \Delta N_\theta(x)=\Delta N^1_\theta(x)-\Delta N^2_\theta(x),\nn\\
&&\Delta N_\theta^{(1)}(x)=-2\int d^4x'~\theta(x_0-x'_0)\sin({1\over 2}{\del\over\del x^i}\theta^{i0}{\del\over\del x'_0})S^\beta_0(x-x')\vphi_0(x')\nn\\
\label{suscept1}&&~~~~:=\int d^4x'~\chi_\theta^{(1)}(x-x')\vphi_0(x'),\\
&&\Delta N_\theta^{(2)}(x)=-4\int d^4x'~\theta(x_0-x'_0)\sin^2({1\over 4}{\del\over\del x^i}\theta^{i0}{\del\over\del x'_0})A^\beta_0(x-x')\vphi_0(x')\nn\\
\label{suscept2}&&~~~~:=\int d^4x'~\chi_\theta^{(2)}(x-x')\vphi_0(x').\\
\eea
In (\ref{suscept1}) and (\ref{suscept2}), ~${\del\over\del x'_0}=({\del\over\del x'_0})_1+({\del\over\del x'_0})_2, $~ where the first differentiates just $S^\beta_0$ and the second differentiates just $\vphi_0$.

On partially integrating the second derivative, it cancels the first derivative acting on $S^\beta_0$ leaving a derivative ${\del\over\del x'_0}$ acting on $\theta(x_0-x'_0)$. So finally
\bea
\chi^{(1)}_\theta(x)=2S^\beta_0(x)\sin({1\over 2}{\ola{\del}\over\del x^i}\theta^{i0}{\ora{\del}\over\del x_0})\theta(x_0)
\eea
and similarly,
\bea
\chi^{(2)}_\theta(x)=-4A^\beta_0(x)\sin^2({1\over 4}{\ola{\del}\over\del x^i}\theta^{i0}{\ora{\del}\over\del x_0})\theta(x_0).
\eea

Let us Fourier transform these expressions setting
\bea
&&\widetilde{\chi}^{(1)}_\theta(k)=\int d^4x~e^{ikx}\chi^{(1)}_\theta(x),\\
&&\widetilde{\chi}^{(2)}_\theta(k)=\int d^4x~e^{ikx}\chi^{(2)}_\theta(x)
\eea
and similarly for $\widetilde{S}(k),~\widetilde{A}(k)$. Then
\bea
&&\widetilde{\chi}^{(1)}_\theta(k)={1\over\pi}\int dx_0~\theta(x_0)[\int dq_0~e^{i(k_0-q_0)x_0}\sin{k_i\theta^{i0}(k_0-q_0)\over 2}~\widetilde{S}(\vec{k},q_0) ],\nn\\
&&\widetilde{\chi}^{(2)}_\theta(k)=-{2\over\pi}\int dx_0~\theta(x_0)[\int dq_0~e^{i(k_0-q_0)x_0}\sin^2{k_i\theta^{i0}(k_0-q_0)\over 4}~\widetilde{A}(\vec{k},q_0) ].\nn\\
\eea
Here we can write $\widetilde{S}$ and $\widetilde{A}$ in terms of $\txt{Im}\widetilde{\chi}_0$:
\bea
&&\widetilde{S}(\vec{k},k_0)=-\coth{\beta k_0\over 2}~\txt{Im}\widetilde{\chi}_0(\vec{k},k_0),\\
&&\widetilde{A}(\vec{k},k_0)=i\txt{Im}\widetilde{\chi}_0(\vec{k},k_0).
\eea

Finally for the twisted susceptibility $\chi_\theta'$,
\bea
\chi_\theta=\chi_0+\chi_\theta^{(1)}+\chi_\theta^{(2)},
\eea
where we have \emph{exact} expressions for $\chi_\theta^{(j)}$ in terms of $\txt{Im}\chi_0$.

\subsection{Zeros and Oscillations in $\widetilde{\chi}_\theta^{(j)}$}

A generic $\txt{Im}\widetilde{\chi}_0$ is the \emph{superposition} of terms with $\delta$-function supports at frequencies $\omega$, that is, of terms
\bea
\label{zerosupp}\delta(k_0-\omega)\txt{Im}\widetilde{\chi}^R_0(\vec{k},\omega)
\eea
($R$ standing for ``reduced'').

We now focus on a single frequency $\omega$, that is, the case where $\txt{Im}\widetilde{\chi}_0(\vec{k},k_0)$ equals (\ref{zerosupp}). Then
\bea
&&\widetilde{\chi}_\theta^{1}(k)=-{i\over\pi}\coth{\beta\omega\over 2}~{1\over k_0-\omega}~\sin{k_i\theta^{i0}(k_0-\omega)\over 2}~\txt{Im}\widetilde{\chi}^R_0(\vec{k},\omega)\\
&&\widetilde{\chi}_\theta^{2}(k)={2\over\pi}~{1\over k_0-\omega}~\sin^2{k_i\theta^{i0}(k_0-\omega)\over 4}~\txt{Im}\widetilde{\chi}^R_0(\vec{k},\omega).
\eea

These corrections have striking zeros and oscillations which would be characteristic signals for noncommutativity. Thus,

a)
\bea
\label{zeros1}\widetilde{\chi}^{(1)}_\theta(k)=\widetilde{\chi}^{(2)}_\theta(k)=0~~~\txt{if}~~~{k_i\theta^{i0}(k_0-\omega)\over 2}=2n\pi,~~n\in \mathbb{Z}.
\eea
$\widetilde{\chi}_\theta^{(1)}$ actually vanishes at all $n\pi$.

b) Regarding the oscillations, they are from the $\sin$ and $\sin^2$ terms. The sine repeats if its argument is changed by
\bea
\label{zeros2}2n\pi
\eea
 while the $\sin^2$ term does so if its argument is changed by
 \bea
 \label{zeros3}n\pi
 \eea
$(n\in \mathbb{Z})$.
These are multiplying backgrounds with no particular oscillatory behavior.

Both $a)$ and $b)$ are characteristic features of the Moyal Plane and in principle accessible to experiments. We emphasize that that both these effects are direction-dependent.

These features may have applications to the homogeneity problem in cosmology \cite{trodden-vachaspati}.

\section{Finite temperature Lehmann representation}
 The Lehmann representation in QFT expresses the two-point vacuum correlation functions of a fully interacting theory in terms of their free field values.  It is exact and captures the properties emerging from the spectrum of $P_\mu$ and Poincar$\acute{\txt{e}}$ invariance in a useful manner.

  We have seen in Section 4 that all the two-point correlations at finite temperature for $\theta^{\mu\nu}\neq 0$ can be expressed in terms of the corresponding expressions for $\theta^{\mu\nu}= 0$.
  In this section, we treat the $\theta^{\mu\nu}=0$ case in detail which then also covers the $\theta^{\mu\nu}\neq 0$ case.

  First we state some notation. The single particle states are normalized according to
  \bea
  \label{normalzn}\langle k'|k\rangle=2|k_0|\delta^3(k'-k),~~k_0=(\vec{k}^2+m^2)^{1\over 2},
  \eea
  where $m$ is the particle mass. The scalar product of $n$-particle states such as $|k_1,...,k_n\rangle$ then follows, (with appropriate symmetrization factors which we will not display here or below).
  We will also not display degeneracy indices such as those from color: their treatment is easy. For a similar reason, we consider spin $0$ fields.

  For the normalization (\ref{normalzn}), the volume form $dV_n$ for the $n$-particle state is a product of factors

 ${d^3k_j\over 2|k_{j0}|}$:
 \bea
 dV_n=\prod_{j=1}^nd\mu_j,~~d\mu_j={d^3k_j\over 2|k_{0j}|},~~|k_{j0}|=\sqrt{\vec{k}_j^2+m_j^2}.
  \eea

Now consider
\bea
W_0^\beta(x)=\omega_\beta(\vphi_0(x)\vphi_0(x')),
\eea
where $\vphi_0$ is a scalar field for $\theta^{\mu\nu}=0$ and $H$ is the total time-translation generator of the Poincar$\acute{\txt{e}}$ group. Its spacetime translation invariance implies that
\bea
\omega_\beta(\vphi_0(x)\vphi_0(x'))=\omega_\beta(\vphi_0(x-x')\vphi_0(0)).
\eea

We assume as usual that
\bea
\label{vacuumcondn}\langle 0| \vphi_0(x)| 0\rangle=0.
\eea

We can write
\bea
&&W_0^\beta(x)={\langle 0| e^{-\beta H}\vphi_0(x)\vphi_0(0)|0 \rangle +\omega_\beta(\vphi_0(x)|0 \rangle\langle 0|\vphi_0(0))\over Z(\beta)}+\widehat{W}_0^\beta(x),\nn\\
&&Z(\beta):={\Tr}e^{-\beta H}.
\eea

We shall see that the vacuum contributions are separated out in the first two terms and that vacuum intermediate states do not contribute to $\widehat{W}_0^\beta$.

We now consider the three terms separately.

\bea 1)~~~~{1\over Z(\beta)} \langle 0| e^{-\beta H}\vphi_0(x)\vphi_0(0)|0 \rangle={1\over Z(\beta)}W_0^0(x)\eqv {1\over Z(\beta)}W(x).
\eea
Here $W(x)$ is the zero-temperature Wightman function with its standard spectral representation:
\bea
W(x)=\int dM^2~\rho(M^2)\Delta_+(x,M^2),~~\Delta_+(x,M^2)=\int d^4p~\delta(p^2-M^2)\theta(p_0)e^{ipx}.\nn\\
\eea

{\footnotesize\bea
2)~\omega_\beta(\vphi_0(x)|0\rangle\langle 0|\vphi_0(0))={1\over Z(\beta)}\sum_{n\geqslant 1}\int dV_n~\langle k_1,...,k_n|e^{-\beta H}\vphi_0(x)|0\rangle\langle 0|\vphi_0(0)|k_1,...,k_n\rangle\nn\\
\eea}
where the $n=0$ term has been omitted in the sum as it contributes $0$ by (\ref{vacuumcondn}).

Using
\bea
\vphi_0(x)=e^{iPx}\vphi_0(0)e^{-iPx},
\eea
where $P_\mu$ generates translations $(P_0=H)$, we find
\bea
&&\omega_\beta(\vphi_0(x)|0\rangle\langle 0|\vphi_0(0))={1\over Z(\beta)}\int d^4 k~\theta(k_0)e^{-\beta k_0+ikx}\rho(k^2),\\
&&\rho(k^2)=\sum_n\int \prod_{j=1}^n~\delta(k^2_j-m^2_j)\theta(k_{j0})\delta^4(\sum k_j-k)~|\langle k_1,...,k_n|\vphi_0(0)|0\rangle|^2,\nn\\
\eea
$\rho$ being the zero-temperature spectral function.

Thus
\bea
&&\omega_\beta(\vphi_0(x)|0\rangle\langle 0|\vphi_0(0))={1\over Z(\beta)}\int dM^2~\rho(M^2)\Delta_+(x,M^2;\beta),\\
&&\Delta_+(x,M^2;\beta)=\int d^4k~\theta(k_0)\delta(k^2-M^2)e^{-\beta k_0+ikx}.
\eea
For $\beta=0$, $\Delta_+(x,M^2;0)$ is the free field zero-temperature Wightman function. It vanishes when $\beta\ra \infty$.
{\footnotesize
\bea
3) \widehat{W}_0^\beta(x)={1\over Z(\beta)}\sum_{n,m\geqslant 1}\int dV_n dV_m~\langle k_1,...,k_n |e^{-\beta H}\vphi_0(x)|q_1,...,q_m\rangle~\langle q_1,...,q_n |\vphi_0(0)|k_1,...,k_m\rangle.\nn
\eea
}

The vacuum contributions ($n$~and /or $m=0$) have already been considered and need not be included here.

Elementary manipulations like those above show that
\bea
&&\widehat{W}_0^\beta(x)={1\over Z(\beta)}\int d^4K d^4Q~\theta(K_0)\theta(Q_0)e^{-\beta K_0+i(K-Q)x}\times\nn\\
&&~~~~\{~\sum_{n,m\geqslant1}\int\prod_{j=1}^nd^4k\theta(k_{j0})\delta(k^2_j-m^2_j)\prod_{j=1}^md^4q\theta(q_{j0})\delta(q^2_j-m^2_j)\times\nn\\
&&~~~~\delta^4(\sum k_j-K)\delta^4(\sum q_j-Q)~|\langle k_1,...,k_n|\vphi_0(0)|q_1,...,q_m\rangle|^2\}.\nn\\
\eea
The term in braces, by relativistic invariance, depends only on $K^2,~Q^2$ and \\
~~$(K+Q)^2$.
As $K_\mu,~Q_\mu$ are timelike with $K_0,Q_0>0$, we have, as in scattering theory,

\bea
(K+Q)^2\geqslant (\sqrt{K^2}+\sqrt{Q^2})^2.
\eea
Call the terms in braces as $\rho(K^2,Q^2,(K+Q)^2)$. Then
{\footnotesize
\bea
&&\widehat{W}_0^\beta(x)={1\over Z(\beta)}\int dM^2dN^2dR^2~\rho(M^2,N^2,R^2)\times\nn\\
&&~~\{~\int d^4K~\theta(K_0)\delta(K^2-M^2) \int d^4Q~\theta(Q_0)\delta(Q^2-N^2)~\delta((K+M)^2-R^2)e^{-\beta K_0+i(K-Q)x} ~\}.\nn\\
\eea
}
The term in braces here is the elementary function appropriate for $\widehat{W}_\theta^\beta$.

The full spectral representation for $W_\theta^\beta$ is obtained by adding those of its terms given above.

\section{Conclusions: Chapter \ref{fdt}}A major result of this chapter is the derivation of acausal and noncommutative effects in finite temperature QFT's. They are new and are expected to have applications for instance in the homogeneity problem in cosmology.

We have also treated the finite temperature Lehmann representation on the commutative and Moyal planes in detail. This representation succinctly expresses the spectral and positivity properties of the underlying QFT's in a transparent manner and are thus expected to be useful.
\chapter{Conclusions} \label{ch:conclus}
We have given a brief review of quantum theory as well as an introduction to quantum field theory in noncommutative spacetime. The concept of deformed Lorentz invariance in noncommutative spacetime led to the following effects which may be susceptible to experimental tests.

\begin{enumerate}

\item Deformed statistics of quantum fields whose consequences include

1) modification of the statistical interparticle force and hence degeneracy pressure which determines the fate of galactic nuclei after fuel burning seizes,

2) the possibility of observing Pauli forbidden transitions,

3) observation of Lorentz, P, PT, CP, CPT and causality violations.

\item  The presence of noncommutativity dependent temperature fluctuations in the CMB radiation, through
a noncomutativity dependent post inflation power spectrum; giving an estimated upper bound for the noncommutativity parameter and a corresponding lower bound for the energy scale.

\item Encounter with noncommutativity-induced causality violation and a non-Gaussian probability distribution during cosmological inflation.

\item Noncommutativity induces noncausal, and potentially periodic, corrections to the susceptibility in linear response theory.

\end{enumerate}


To summarize we have investigated, in the context of quantum field theory, the scope of applicability of a new concept of Lorentz invariance. This new concept is a deformation of the usual concept of Lorentz invariance motivated by the form of invariance in Moyal's treatment of quantum mechanics. The investigations were based on certain available theoretical models and experimental data. Results of these investigations can point to alternative and hopefully simpler solutions to both expected and observed physical phenomena whose experimental energies fall within the range of validity of the noncommutativity models.


\appendix
%
%

\chapter{Some physical concepts}

\section{Motion of an electron in constant magnetic field}\label{electron-motion}
When an electron moves in a constant magnetic field the coordinates of the center of its circular motion (ie. guiding center)  become noncommutative when the system is quantized canonically. The Lagrangian and equations of motion
\bea
&&L={m\vec{v}^2\over 2}-e\vec{v}\cdot\vec{A},~~\vec{A}=-{1\over 2}~\vec{x}\times \vec{B},\nn\\
&&m{d \vec{v}\over
dt}=e\vec{v}\times \vec{B},~~\vec{v}={d\vec{x}\over dt}\nn
\eea

have the solution
\bea
&&\vec{x}(t)=\vec{x}_0+{m\over
eB}~{\hat{\vec{B}}\times\vec{v}_0}+\hat{\vec{B}}~(\hat{\vec{B}}\cdot{\vec{v}_0})(t-t_0)\nn\\
&&~~~~-{\hat{\vec{B}}\times \vec{v}_0}~{\cos(\om (t-t_0))\over\om}-{\hat{\vec{B}}\times (\hat{\vec{B}}\times \vec{v}_0)}~{\sin(\om(t-t_0))\over\om},\nn\\
 &&\om={e|\vec{B}|\over m}={eB\over m}.\nn
\eea
The position of the center of circular motion is
\bea
 &&\vec{x}_c(t)=\vec{x}_0+{m\over
eB}~{\hat{\vec{B}}\times\vec{v}_0}+\hat{\vec{B}}~(\hat{\vec{B}}\cdot{\vec{v}_0})(t-t_0).\nn
\eea
and the canonical momentum is
\bea
&&\vec{p}={\del L\over\del\vec{v}}=m\vec{v}-e\vec{A}.\nn
\eea
One gets the canonical commutation relations
\bea
&&[x^i(t),x^j(t)]=0=[p^i(t),p^j(t)]~~\forall t,\nn\\
&& [x^i(t),p^j(t)]=[x^i(t),mv^j(t)-eA^j(x(t))]=[x^i(t),mv^j(t)]=-i\hbar\delta^{ij}~\forall t,\nn
\eea
from which one can verify that
\bea
&& [v^i(t),v^j(t)]=i{e\hbar B\over m^2}\vep^{ik j}\hat{B}^k~~\forall t,\nn\\
&&~~\Ra~~ [x^i_c(t),x^j_c(t)]=i\theta^{ij}=i{\hbar\over eB}\vep^{ikj}\hat{B}^k~~\forall t.
\eea
Here $\theta^{ij}={\hbar\over eB}\vep^{ikj}\hat{B}^k$ is not invertible as a $3\times 3$ matrix as $\det_{3\times 3}\theta=0$. However if we arrange the system such that $\hat{\vec{B}}\cdot{\vec{v}_0}=0$,~~say with $B^k=B\delta ^{zk},~~v_0^k=v^x_0\delta^{xk}+v_0^y\delta^{yk}$, then the motion stays in the $x-y$ plane and $\theta^{ij}={\hbar\over eB}\vep^{i3j}$ is now invertible as a $2\times 2$ matrix.

\section{Symmetries and the least action principle}
  \subsection{Use of symmetries}
A major reason for the use of symmetries to analyze physical systems stems from the fact that the kinematics and/or dynamics of a physical system can be cast in terms of nonanalytic and/or analytic (differential or integral) constraints or equations which may also be derivable from a least action principle. The symmetry group of the action or Lagrangian is a subgroup of the symmetry group of the equations. The key observation that the solution space of the equations is invariant under the symmetry group of the equations implies that the complete space of solutions can be generated from only a few simple solutions. Moreover, most of the physically relevant information about the solution space of the equations is contained in their symmetry group. In particular, one expects that each independent solution of the equations has a simple correspondence with an irreducible representation of the symmetry group. Therefore instead of trying to solve the equations directly, one could rather consider the problem of finding the irreducible representations of the symmetry group. The group theoretic analysis is most useful for interacting physical systems where the interactions lead to coupled nonlinear equations for which even the simplest solution can be difficult to find. One may postulate that whenever two separate systems couple, one or more of the variables involved should be modified or extended such that their individual symmetry groups become either 1) independent symmetry groups of the coupled system or 2) subgroups of a larger symmetry group of the coupled system or 3) identified; that is, merged together into a larger unifying symmetry group.
 Another major reason for the use of symmetries is that they identify physically observable quantities, such as interaction amplitudes or potentials, as those that can survive the symmetry transformation. Together with an action principle, the symmetries also provide conservation laws and conserved quantities (Noether's theorem) which help simplify the analysis of interactions.
\subsection{Analogy and least action principle}
 Biological systems, their developments and the interactions among them may be characterized by the way they respond to the variety of (certain) natural changes in their supporting environments (``external'' changes) and  also to a variety of changes in their most basic or defining configurations (``internal'' changes) in these environments. Similarly, mathematical structures, operations on them and the relations between them can be characterized by the way they respond to a variety of special maps or transformations among their supporting spaces which are the spaces on which they are defined or configured and also to a variety of special maps or transformations among the spaces consisting of the structures and classes of structures themselves. Many mathematical models for (elementary) physical systems (their configurations and interactions in space and time or simply spacetime) can be based on a least action principle for a composite or derived mathematical structure on spacetime called the action functional. The action functional is a configuration-dependent variable that is written as a sum total of a Lagrangian over the domain (the region of spacetime in which the system can be variously configured) of the physical system. The Lagrangian is a quantity written in terms of spacetime variables and spacetime-dependent configuration variables for the physical system. A classical physical system is then characterized by its symmetries; those transformations or changes in spacetime variables and/or configuration variables and Lagrangian that do not alter the outcome of (or the equations of motion resulting from) the least action principle. The least action principle asserts that within a given spacetime domain, supporting all possible configurations of the physical system, the actual configuration of the physical system is the one for which the action functional is minimum. The domain of the system in spacetime may either be a collection of points (eg. the system is a set of ''events''), a one dimensional path (eg. the system is a mechanical ''particle'') or a hypersurface in general (eg. the system is either an extended classical object or a quantum event). In quantum theory, it turns out that one needs to average quantities over the configuration space domain of the physical system with a probability density function given by the exponential of the classical action. The exponential form of the probability distribution is due to the correspondence between the additive nature of the classical action and the multiplicative nature of the joint probability distribution for a collection of noninteracting systems.

 \section{Renormalizability}
 NB: Here the term ``classical'' is synonymous to ``low energies'' meanwhile the term ``quantum'' is synonymous to ``all possible energies''.

 In quantum theory, the probability amplitude for the evolution of a physical system from an initial quantum configuration (or a set of possible initial quantum configurations) to a final quantum configuration (or a set of possible final quantum configurations) may be defined or postulated in terms of certain functionals known as Green's functions. For a noninteracting theory these probability amplitudes are finite. The introduction of interactions leads to initial/final quantum configuration-dependent quantum corrections to the probability amplitudes. Some of these corrections contain purely divergent parts. The finite parts of the divergent corrections can be isolated with the help of a regularization procedure. In some cases the remaining purely divergent parts can be eliminated by simple redefinitions of the parameters in the classical action and hence a few additional parameters to be determined experimentally.

 This observation therefore suggests that whenever there are interactions one has corresponding initial/final state dependent quantum corrections to the experimentally measured values of the parameters found in the classical action as well. The elimination of the purely divergent parts of the corrections is known as renormalization and theories in which the simple parameter redefinitions are sufficient to eliminate all possible divergences are said to be renormalizable. Nonrenormalizable theories are known as effective (as opposed to fundamental) theories since due to divergences they can be valid only for a restricted range of initial/final quantum configurations. Effective theories are expected to arise as consequences of fundamental theories. There are several possible regularization procedures resulting in different renormalized values for the same quantity. A theory may have more than one symmetry and when none of the possible regularization procedures can preserve all the symmetries then anomalies, which may be presented as a failure of the conservation law of Noether currents, arise rendering the theory nonrenormalizable in some cases. Anomalies signal a possibility of incompleteness of the theory that may be for example due to a failure to take into account extra degrees of freedom (ie. a missing piece of the configuration space of the system) posing as topological nontriviality of spacetime and/or configuration space, or considering too many degrees of freedom such as the case where a reducible space rather than an irreducible one is used.

 A consequence of renormalization is that requiring nondependence of the Green's functions and the measurable or measured coupling constant and/or mass on the regularization parameter implies a first order differential relationship between the Green's functions and the measured coupling constant and/or mass. The solution to this differential relationship indicates a scaling behavior for the Greens's function as the measured coupling constant and/or mass is varied through a single real parameter that may be thought of as a parameter for the group of all possible renormalization schemes. The fixed points of this variation may indicate possible phase transitions which are marked changes in the behavior of the Green's functions as the initial/final quantum configurations of the system are varied. This is because a change in renormalization scheme causes a change in the renormalized or measured coupling constant and/or mass (which in turn depend only on the initial/final quantum configurations of the system) and so may be regarded as being equivalent to a change in the initial/final quantum configurations of the system. The scaling behavior together with symmetry properties of the Greens function give a qualitative description of the Green's function, and hence of the quantum configurations of the system, especially near the critical or fixed points.

\section{Rules for writing probability amplitudes of physical processes}
A sample Lagrangian is that of QED
\bea
\L={1\over 4}(\del_\mu A_\nu-\del_\nu A_\mu)(\del^\mu A^\nu-\del^\nu A^\mu)+i\bar{\psi} \gamma^\mu(\del_\mu-iqA_\mu)\psi.
\eea

\begin{enumerate}
\item Sketch all possible \emph{connected} Feynman diagram(s) of the process and indicate momenta.
\item Each external (initial/final) line (``half of a propagator'') represents the \emph{normalized} Fourier coefficient of the classical field; which includes polarization/spin ``vectors'' or directions.
\item Each internal line represent a full time-ordered (ie. Feynman) propagator.
\item Each vertex represents one or more of the following: coupling constants, momentum vectors, spin matrices, representation matrices/tensors, ~etc, as they appear in the (Fourier transformed) Lagrangian. When written out, a vertex with external lines has the form of the Fourier transform of a current from the Lagrangian.
\item Conserve overall momentum,~ conserve momentum at each vertex. This may be done either directly or by including delta functions.
\item For each loop, integrate over the residual momentum that remains after momentum conservation has been applied to all vertices surrounding the loop.
\item Trace over $\gamma$-matrices in a \emph{purely fermion} loop.
\item Divide the amplitude of each diagram by its ``symmetry factor'' which represents how many times the given diagram has been over counted as compared to those other diagrams that lack the symmetries of the given diagram.
\item Add together the contributions from each diagram to get the total amplitude of the process.
\end{enumerate}

\chapter{Quantization}
\section{Canonical quantization, deformation quantization and noncommutative geometry}\label{cano-quant}
The form of the classical action in the Lagrangian and Hamiltonian pictures is
\bea
&&S[q]=\int d\ld ~L(\ld,q,\dot{q})=\int p_idq^i-\int d\ld~ H(\ld,q,p),~~p={\del L\over\del\dot q},\nn\\
&&L(\ld,q,\dot{q})=\sum_i\L_i(\ld,q,\dot{q}),~~H(\ld,q,p)=\sum_i\H_i(\ld,q,p).
\eea
One can identify the canonical $1$-form
\bea
&&A_I({\ld},x)dx^I\eqv{p}_i dq^i+d{\al}({\ld},q,{p}),~~x^I=(x^0_i,x^1_i)=(q_i({\ld}),{p}_i({\ld})),\nn\\
\label{spotential}&&A({\ld},x)=(A^0_i(x,{\ld}),A^1_i(x,{\ld}))=({\del {\al}\over\del q^i},{p}_i({\ld})+{\del {\al}\over\del {p}^i})
 \eea
from the Legendre transformation $H({\ld},d{\ld},q,{p})={p}_idq^i-L({\ld},d{\ld},q,dq)|_{{p}^i={\del L\over\del dq^i}}$.

A canonical transformation is any symmetry of the Lagrangian $L$ ($L$ changes by at most a total derivative) which is also a symmetry of the Hamiltonian $L$ ($H$ changes by at most a total derivative) and since $H=A-L$ it means that a canonical transformation is any symmetry of $L$ for which $A$ changes by at most a total differential ($\delta A=d\beta$ ~or equivalently~ $\delta \Omega=\delta dA=d\delta A=0$).

 Therefore a canonical transformation is a transformation on phase space \\
 ~$T^\ast(\M)\simeq \{q,p\},~~\M\simeq \{q\}$~ that preserves the exterior differential \\
 ~$\Omega=dA,~A\in T^\ast(T^\ast(\M))/\M$. The relationship between the canonical $2$-form \\
 $\Omega=d A=\del_IA_Jdx^I\wedge dx^J$ and $A=A_Idx^I$ is analogous to the relationship between the electromagnetic $2$-form and its $1$-form. The infinitesimal transformation of any 2-form $\Omega$
{\footnotesize
 \bea
 &&\delta\Omega=\delta(\Omega_{IJ}dx^I\wedge dx^J)=\delta\Omega_{IJ} dx^I\wedge dx^J+\Omega_{IJ}\delta dx^I\wedge dx^J+\Omega_{IJ}dx^I \wedge\delta dx^J\nn\\
 &&~~=\delta x^K\del_K\Omega_{IJ} dx^I\wedge dx^J+\Omega_{KJ}\del_I\delta x^K dx^I\wedge dx^J+\Omega_{IK} dx^I\del_J\delta x^K\wedge dx^J\nn\\
 &&~~=(\del_K\Omega_{IJ}+\del_J\Omega_{KI}+\del_I\Omega_{JK}) \delta x^Kdx^I\wedge dx^J-[\del_I(\Omega_{JK}\delta x^K)-\del_J(\Omega_{IK}\delta x^K)]\times\nn\\
&&~~~~dx^I\wedge dx^J\nn
\eea
can be written in the general form
\bea
 &&\delta \Omega=\pounds_{{\delta x}}\Omega=i_{{\delta x}}d\Omega+d(i_{{\delta x}}\Omega).
 \eea
}
Similarly the infinitesimal transformation of any 1-form $A$ is given by
 \bea
 &&\delta A=\pounds_{{\delta x}}A=i_{{\delta x}}dA+d(i_{{\delta x}}A)\ra i_{{\delta x}}\Omega+d(i_{{\delta x}}A).
 \eea

Therefore a canonical transformation $\delta A=d\beta$ is given by

\bea
&&i_{{\delta x}}\Omega=-df~~(\txt{ie.}~~\delta x^I=\Omega^{IK}\del_Kf\eqv\{x^I,f\}~),
\eea
 where ~~$\Omega_{IK}\Omega^{KJ}=\delta_I^J$ and the Poisson bracket ~$\{f,g\}=\Omega^{IJ}\del_If\del_Jg$~ can be infered.
This produces the desired symmetry conditions
\bea
&&\delta A=i_{{\delta x}}\Omega+d(i_{{\delta x}}A)=d(-f+i_{{\delta x}}A),\nn\\
&&\delta \Omega=i_{{\delta x}}d\Omega+d(i_{{\delta x}}\Omega)=0.
\eea

The vector field~
\bea
\xi_f=\xi_f^I\del_I\eqv\delta x^I\del_I=-{\Omega^{IJ}}\del_If\del_J
 \eea
associated with the canonical transformations is known as a Hamiltonian vector field. The following relations hold following the Jacobi identity for the Poisson bracket:
\bea
&&[\pounds_{\xi_f},\pounds_{\xi_g}]\psi=\pounds_{[\xi_f,\xi_g]}\psi=\pounds_{\xi_{\{f,g\}}}\psi,
\eea
where $\psi\in {\mathbb{C}^N\over T^\ast(\M)}\eqv \F(T^\ast(\M))$.

\textbf{Remarks:}
\begin{itemize}
\item One has the Liouville  measure ~$d\mu=\Omega^{\wedge D}=\Omega\wedge\Omega^{\wedge(D-1)}=\sqrt{\det\Omega}~d^{2D}x$~ on $T^\ast(\M)$.

\item Canonical transformations (or canonical invariants rather) provide a way to derive quantization conditions. If only canonical path deformations are allowed then
\bea
&&\label{BScondition2}\delta\oint_CA= \oint_C\delta A=0~~~~\Ra~~~~\oint_CA=const.\eqv K~~\forall C,\nn\\
&&[~\txt{Also verify using}~~\oint_C A=\int_{C^0}dA=\int_{C^0}\Omega ~],
\eea
which reproduces the Bohr-Sommerfeld quantization condition (\ref{BScondition1}) when $K$ takes on integer values. However $K\in \mathbb{R}$ in general and therefore one can have continuous as well as discrete values for the spectra of quantum mechanical observables. A generalization of the canonical invariant $\oint_C A$ to a situation where $A$ is noncommutative (eg. an $A$ that contains the nonabelian gauge potential) is the path ordered loop integral (known as Wilson loop)
\bea
&&\Tr~Pe^{\oint_C A}=e^{K},~~{\del K\over\del C}=0,
\eea
which is a gauge\footnote{Gauge transformations are examples of canonical transformations.} invariant and $P$ denotes path ordering.

\item  One learns that a canonical transformation is generated by a function on $T^\ast(\M)\simeq\{(q,{p})\}$, where ~$\M\simeq\{q\}~\&~T(\M)\simeq\{(q,dq)\}$, through a Poison bracket constructed from $\Omega$. Note that $A\in {T^\ast(T^\ast(\M))\over\M},~\Omega\in {T^\ast(T^\ast(\M))\wedge T^\ast(T^\ast(\M))\over \M}$. In particular, the generator or generating function associated to time translations $\delta x^I=\delta t{dx^I\over dt}$ is the Hamiltonian $H$. Conversly, to every function is associated a canonical transformation whose generator is the function.

\item Canonical quantization is a parallel or correspondence where canonical transformations are mapped to unitary linear operators (or unitary transformations) of the set of operators $O(\H)$ on a Hilbert space $\H$; the classical observables or the generating functions of the canonical transformations are mapped to hermitian or antihermitian linear operators which are generators of the unitary transformations on $\H$ and the Poisson bracket $\{,\}$ is mapped to the commutator $[,]$ in $O(\H)$. Thus canonical transformations are to the symplectic 2-form $\Omega$ as unitary transformations are to the inner product $\langle~|~\rangle$ of the Hilbert space.

 One can construct a quantum  Hilbert space $\H_S=(\F(\M),\lang|\rang)$ from the pointwise product algebra $(\F(\M),\txt{pt$\cdot$wise})$ of the space of complex functions $\F(\M)$ on $\M\simeq\{q\}$ with an inner product given by $\lang f|g\rang=\int d\mu(q)~\overline{f(q)}~g(q)$. On $\H_S$ the commutation relation $[\hat{q},\hat{p}]=i\hbar$ implies that \\ $\hat{q}=\mu_q,~\hat{p}=-i\hbar\del_q=-i\hbar{\del\over\del q}$ which is known as Schrodinger representation where the position operators $\hat{q}$ act as multiplication operators
 \bea
 &&\mu_q:\H_S\ra\H_S,~\mu_q\xi(q)=q\xi(q).
 \eea
 The quantum operators ~$Q(T^\ast(\M))\subset O(\H_S)$~ are then given by
 \bea
 &&Q(T^\ast(\M))=\{Q(f),~f\in \F(T^\ast(\M)) \},\nn\\
 &&Q:\F(T^\ast(\M))\ra O(\H_S),~f(q,p)\mapsto Q(f)(q,p)= f(q,-i\hbar\del_q).\nn
 \eea
 Since the points $x^I=(q^i,p^i)$ of $T^\ast(\M)$ on which the commutative algebra of (generating) functions $\F(T^\ast(\M))$ is defined act like linear functionals $\delta_x:f\ra \delta_x(f)=\int_{T^\ast(\M)} dy~\delta(y-x)~f(y)=f(x)$ on the space of functions $\F(T^\ast(\M))$, the role of these points may be played by linear functionals $\O^\ast(\H)=\{\chi:O(\H)\ra \mathbb{C},~~\chi(a+b)=\chi(a)+\chi(b)\}$ on the algebra of linear operators $O(\H)$ on the Hilbert space $\H$.

\item Deformation quantization is an alternative method of quantization that arises because the algebra of operators $O(\H)$ on the quantum mechanical Hilbert space $\H$ can be shown to be equivalent to a noncommutative $\ast$-product function algebra ~$(\F(T^\ast(\M)),\ast\txt{-wise})$, the commutator in which reduces to the Poisson bracket of the classical function algebra \\ $(\F(T^\ast(\M)),\txt{pt$\cdot$wise})$ in a certain limit.
 That is, the Poisson algebra can be obtained from a noncommutative deformation $(\F(T^\ast(\M)),\ast\txt{-wise})$ of the commutative function algebra $(\F(T^\ast(\M)),\txt{pt$\cdot$wise})$:
 \bea
 &&(\F(T^\ast(\M)),\txt{pt$\cdot$wise})\ra (\F(T^\ast(\M)),\ast\txt{-wise}),\nn\\
 &&~~~~~~(fg)(q,{p})=f(q,{p})g(q,{p})\mapsto (f\ast g)(q,{p})=f(q,{p})e^{\ola{\del}_I\Omega^{IJ}\ora{\del}_J}g(q,{p}).\nn\\
 \eea
 Here one may again construct a quantum  Hilbert space $\H_M=(\F(T^\ast(\M)),\lang|\rang)$ from the $\ast$-product product algebra $(\F(T^\ast(\M)),\ast\txt{-wise})$ of the space of complex functions $\F(T^\ast(\M))$ on $T^\ast(\M)\simeq\{(q,p)\}$ with an inner product given by $\lang f|g\rang=\int d\mu(q,p)~\overline{f(q,p)}\ast g(q,p)$. On $\H_M$ the commutation relation $[\hat{q},\hat{p}]=i\hbar$ implies that both $\hat{q}=\mu_q,~\hat{p}=\mu_p$ act (reducibly) as multiplication operators
 \bea
 &&\mu_q,\mu_p:\H_M\ra\H_M,~\mu_q\xi(q,p)=q\ast\xi(q,p)=(q+{i\over 2}\del_p)\xi(q,p),\nn\\
 &&\mu_p\xi(q,p)=p\ast\xi(q,p)=(p-{i\over 2}\del_q)\xi(q,p),\nn\\
 &&{df(q_t,p_t)\over dt}=(H\ast f-f\ast H)(q_t,p_t),
 \eea
 which is however only left multiplication but we however have both Left and right independent multiplication operators $\mu^{L,R}_q,\mu^{L,R}_p$.~ Due to the simple nature of the algebra~
 $\mu^c={1\over 2}(\mu^L_{(q,p)}+\mu^R_{(q,p)})$~ gives a commutative coordinate representation $\mu^c=\mu_{(q_c,p_c)}$ that is insensitive to the $\ast$-product.

 Deformation quantization provides an example of noncommutative geometry since any $C^\ast$-algebra can be realized as an algebra of operators $O(\H)$ on a Hilbert space $\H$ and noncommutative geometry involves the representation of an arbitrary $\ast$-algebra $\A$ as a noncommutative algebra of functions on its dual $\A^\ast=\{\chi:\A\ra \mathbb{C}^N,~~\chi(a+b)=\chi(a)+\chi(b)\}$. That is
 \bea
 R:\A\ra R(\A)=\{\td{a}:\A^\ast\ra \mathbb{C}^N,~~\td{a}(\chi)=\chi(a),~(\td{a}\ast \td{b})(\chi)=\chi(ab)\}.\nn\\
 \eea
\item Thus quantizing a given classical system involves the representation theory of the algebra(s) and symmetry group(s) of the classical system.
\end{itemize}
\subsection{Star products and regularization}
The star product construction is a trick one may use, whenever convenient, to find characteristic representations \\ $\pi:\A\ra\pi(\A)\simeq\F(X)|_{fg\ra f\ast g}\simeq \{\td{a}:\{\pi\}\simeq X\ra \mathbb{C}^N,~\pi\mapsto \td{a}(\pi)=\pi(a)\}$ of a given algebra $\A$ by modifying the product on the algebra of functions \\
$\F(X)=\mathbb{C}^N/X$ on some topological space $X\simeq \{\pi\}$. The characteristic representation of the algebra product on the function space is known as a star product:
\bea
&&\pi(ab)=(\td{a}\ast\td{b})(\pi).
\eea
An example is given by the group algebra $G^\ast$ of a group $G$.
\bea
&&G^\ast=\txt{Span}\{\hat{f}=L(f)=\sum_{g\in G}f(g)~g,~~f\in\F(G)\},~~\F(G)=\{f:G\ra \mathbb{C}\},\nn\\
&&L(f)L(h)=L(f\ast h),\nn\\
&&(f\ast h)(g)=\sum_{u\in G}f(u)~h(u^{-1}g)= \sum_{u\in G}f(gu^{-1})~h(u) \neq (h\ast f)(g).\nn
\eea

As another example let $\A$ be the Moyal-Weyl algebra;
\bea
&&\A=\{W(f)=\hat{f}\},\nn\\
&&\hat{f}=\sum_p\td{f}_pe_p(\hat{x})=\sum_xf(x)\sum_pe_p(\hat{x}-x)=\sum_xf(x)\hat{\delta}_x,\nn\\
&&\hat{\delta}_x=\sum_y\delta(x-y)\hat{\delta}_y\eqv W(\delta_x),~~~~x,p\in \mathbb{R}^{d+1}
\eea
generated by the linear operators $\{\hat{x}^\mu\}$;
\bea
\label{mplane2}&&[\hat{x}^\mu,\hat{x}^\nu]=i\theta^{\mu\nu}.
\eea
The major point here is to be able to invert (in an unambiguous way) the series expansion
\bea
\label{weylseries}\hat{f}=\sum_xf(x)\hat{\delta}_x.
\eea
 This is possible if a unique linear functional $\phi\eqv \sum_{\hat{x}}$ (an analog of the integral) can be found such that $\phi(\hat{\delta}_x\hat{\delta}_y)\sim\delta_{xy}$. To find this functional, consider the
generators of real translations $\{\hat{\del}^\mu\}$ on this algebra given by
\bea
&&[\hat{x}^\mu,\hat{\del}_\nu]=-\delta^\mu{}_\nu,~~[\hat{\del}_\mu,\hat{\del}_\nu]=0,\nn\\
&&(~\txt{compare with}~~\hat{y}_\nu=-i\theta_{\nu\al}\hat{x}^\al=-i\theta^{-1}{}^{\nu\al}\hat{x}^\al,~~[\hat{y}_\mu,\hat{y}_\nu]=i\theta_{\mu\nu}\nn\\
&&~~~~~~~~~~~~[\hat{x}^\mu,\hat{y}_\nu]=-\delta^\mu_\nu~).
\eea
That such $\hat{\del}_\mu$'s exist may be seen by representing the algebra as an algebra of differential operators on a function space $\F(\mathbb{R}^{d+1})=\F(\{x\}),~~\hat{x}^\mu=x^\mu+{i\over 2}\theta^{\mu\nu}\del_\nu$. More simply, $\hat{\del}_\mu=-i\theta^{-1}{}^{\mu\nu}\ad_{\hat{x}^\nu}\eqv-i\theta_{\mu\nu}\ad_{\hat{x}^\nu}=\ad_{\hat{y}_\mu}$ where the algebra (\ref{mplane2}) implies that
\bea
[\ad_{\hat{x}^\mu},\hat{x}^\nu]\hat{f}=\ad_{\hat{x}^\mu}(\hat{x}^\nu\hat{f})-\hat{x}^\nu\ad_{\hat{x}^\mu}(\hat{f}) = i\theta^{\mu\nu}\hat{f}
\eea
and one easily sees that $[\ad_{\hat{x}^\mu},\ad_{\hat{x}^\nu}]=\ad_{[\hat{x}^\mu,\hat{x}^\nu]}=\ad_{i\theta^{\mu\nu}}=0 $.

Then
\bea
&&\hat{\delta}_x=\sum_pe_p(\hat{x}-x)=\sum_p e^{ip(\hat{x}-x)}=e^{-ix\hat{\del}}~\sum_pe_p(\hat{x})~e^{ix\hat{\del}},\nn\\
&&[\hat{\del}_\mu,\hat{\delta}_x]=-\del_\mu\hat{\delta}_x ~~\Ra~~ \del_\mu{\Tr}\hat{\delta}_x=0~~\txt{as}~~{\Tr}(AB)={\Tr}(BA)
\eea
which means that ${\Tr}\hat{\delta}_x=c=const.$ and the normalization ${\Tr}\hat{\delta}_x=1$ gives
\bea
&&{\Tr}(\hat{\delta}_x\hat{\delta}_y)=\delta(x-y)
\eea
since
\bea
&&\hat{\delta}_x\hat{\delta}_y=\sum_{pp'}e^{i(p+p')\hat{x}}e^{{i\over 2}p\wedge p'}e^{-ipx-ip'y}\nn\\
\eea
and\footnote{Moreover one can simplify further to obtain
\bea
&&\hat{\delta}_x\hat{\delta}_y=\sum_{pp'}e^{i(p+p')\hat{x}}e^{{i\over 2}p\wedge p'}e^{-ipx-ip'y}={e^{-ix({\theta\over 2})^{-1}y} \over \det ({\theta\over 2})^{-1}}~\sum_z~\hat{\delta}_z~e^{iz({\theta\over 2})^{-1}(x-y)}\nn\\
&&~~~~=e^{-ix({\theta\over 2})^{-1}y}~\sum_k~\hat{\delta}_{{\theta\over 2}k}~e^{-ik(x-y)}\eqv \Gamma_{xy}{}^z\hat{\delta}_z,~~\sum_z \Gamma_{xy}{}^z=\delta(x-y),\nn\\
&&\Tr (\hat{\delta}_x\hat{\delta}_y\hat{\delta}_z )=\Gamma_{xy}{}^z={1\over\det({\theta\over 2})^{-1}}~e^{-ix({\theta\over 2})^{-1}y-ix({\theta\over 2})^{-1}z-iy({\theta\over 2})^{-1}z}\nn\\
&&~~~~~~~~={1\over\det({\theta\over 2})^{-1}}~e^{-i(x-z)({\theta\over 2})^{-1}(y+z)}\eqv \txt{Cycl}_{xyz}(~{1\over\det({\theta\over 2})^{-1}}~e^{-i(x-z)({\theta\over 2})^{-1}(y+z)}~),\nn\\
&&\Tr (\hat{\delta}_x\hat{\delta}_y\hat{\delta}_z\hat{\delta}_w )= \Gamma_{xy}{}^{\al}\Gamma_{\al z}{}^w=\Gamma_{xy}{}^w~\delta(x+y-z-w)=\Gamma_{xy}{}^{-z}~\delta(x+y-z-w),\nn\\
&&\Tr (\hat{\delta}_{x_1}\hat{\delta}_{x_2}...\hat{\delta}_{x_{n-1}}\hat{\delta}_{x_n} )= \Gamma_{x_1x_2}{}^{z_1}\Gamma_{z_1 x_3}{}^{z_2}\Gamma_{z_2 x_4}{}^{z_3} ...\Gamma_{z_{k-1}x_{k+1}}{}^{z_k} \Gamma_{z_{k}x_{k+2}}{}^{z_{k+1}}\nn\\
&&~~~~...\Gamma_{z_{n-5}x_{n-3}}{}^{z_{n-4}}\Gamma_{z_{n-4}x_{n-2}}{}^{z_{n-3}}\Gamma_{z_{n-3}x_{n-1}}{}^{z_{n-2}}\Gamma_{z_{n-2}x_n}{}^{z_{n-1}}\Tr(\hat{\delta}_{z_{n-1}}) \nn\\
&&~~~~= \Gamma_{x_1x_2}{}^{z_1}\Gamma_{z_1 x_3}{}^{z_2}\Gamma_{z_2 x_4}{}^{z_3} ...\Gamma_{z_{k-1}x_{k+1}}{}^{z_k} \Gamma_{z_{k}x_{k+2}}{}^{z_{k+1}}\nn\\
&&~~~~...\Gamma_{z_{n-5}x_{n-3}}{}^{z_{n-4}}\Gamma_{z_{n-4}x_{n-2}}{}^{z_{n-3}}\Gamma_{z_{n-3}x_{n-1}}{}^{x_n}\nn\\
&&~~~~=\Tr (\hat{\delta}_{x_1}\hat{\delta}_{x_2}...\hat{\delta}_{x_{n-4}}\hat{\delta}_{x_{n-3}}\hat{\delta}_{x_n+x_{n-1}-x_{n-2}} )~\Gamma_{(x_n+x_{n-1}-x_{n-2})~x_{n-2}}{}^{x_n}.
\eea
The appearance of delta functions indicates that \emph{\textbf{Moyal noncommutativity is not strong enough to be able to regularize all possible $n$-point correlations. $(2k+1)$-point correlations are fully regularized but $2k$-point correlations are only partially regularized.}}.
}
\bea
&&e^{ia\hat{x}}= \sum_\al\delta(a-\al)e^{i\al\hat{x}}=\sum_\al \sum_ze^{i(a-\al)z}e^{i\al\hat{x}}=\sum_z~e^{iaz}~\sum_\al e^{i\al\hat{x}-i\al z}~\nn\\
&&~~~~=\sum_z\hat{\delta}_ze^{iaz}~~\Ra~~\Tr ~e^{ia\hat{x}}=\sum_z~\Tr\hat{\delta}_z~e^{iaz}=\sum_z e^{iaz}=\delta(a).\nn\\
\eea
Thus the trace ~$\Tr\eqv \sum_{\hat{x}}$~ provides a means to invert the series (\ref{weylseries}).

We can therefore define the linear functionals $A^\ast\simeq\{\delta_x,~x\in \mathbb{R}^D\}\simeq \mathbb{R}^{D}$ by
\bea
&&\delta_x:\A\ra \mathbb{C},~\delta_x(\hat{f})={\Tr}( \hat{f}\hat{\delta}_x)=(\Tr\hat{\delta}_x\circ W)(f)=f(x).\nn\\
\eea
Since $W(f_1)W(f_2)...W(f_n)=W(f_1\ast f_2\ast...\ast f_n)$ one has
\bea
&&\delta_x(\hat{f}_1\hat{f}_2...\hat{f}_n)= {\Tr}(\hat{\delta}_x \hat{f}_1\hat{f}_2...\hat{f}_n)={\Tr}(\hat{\delta}_x W(f_1)W(f_2)...W(f_n))\nn\\
&&~~~~=({\Tr}\hat{\delta}_x\circ W) (f_1\ast f_2\ast...\ast f_n)=(f_1\ast f_2\ast...\ast f_n)(x),\nn\\ &&(f\ast g)(x)=f(x)e^{{i\over 2}\ola{\del}_\mu\theta^{\mu\nu}\ora{\del}_\nu}g(x),~~~~[\hat{\del}_\mu,W(f)]=W(\del_\mu f).
\eea

For noncommutativity of the form \\
$[\hat{x}^\mu,\hat{x}^\nu]=iC^{\mu\nu}{}_\al \hat{x}^\al,~\txt{Cyclic}_{\mu\nu\al}(~C^{\mu\nu}{}_\rho C^{\al\rho}{}_\ld~)=0$, for the purpose of inverting the series (\ref{weylseries}) one may define a conjugate \\
$\hat{\delta}_x^B=\sum_p A(p)~e^{iB(p)\hat{x}-ipx}$
to $\hat{\delta}_x=\sum_p e^{ip\hat{x}-ipx}$ such that $\Tr (\hat{\delta}_x\hat{\delta}_y^B)=\delta(x-y)$. Again one assumes that $\hat{\del}_\mu$ can be found such that
\bea
&&[\hat{x}^\mu,\hat{\del}_\nu]=-\delta^\mu{}_\nu,~~[\hat{\del}_\mu,\hat{\del}_\nu]=0.
\eea
That such $\hat{\del}_\mu$'s exist may be seen by representing the algebra as an algebra of differential operators on a function space $\F(\mathbb{R}^{d+1})=\F(\{x\})$.
\bea
&&\hat{x}^\mu\ra  x^\nu{E}^\mu{}_\nu(i\del),~~\del={\del\over\del x},\nn\\
\label{lie-rep}&&E^\al{}_\nu(i\del)~\del^\nu E^\bi{}_\mu(i\del)-E^\bi{}_\nu(i\del)~\del^\nu E^\al{}_\mu(i\del)=C^{\al\bi}{}_\nu~ E^\nu{}_\mu(i\del), \\
&&eg.~~{E}^\nu{}_\mu(i\del)=({F(i\del)\over 1-e^{-F(i\del)}})^\nu{}_\mu=(\int_0^1dt~e^{-tF(i\del)})^{-1}{}^\nu{}_\mu,~~F^\mu{}_\nu(i\del)=C^{\mu\al}{}_\nu ~i\del_\al,\nn\\
&&~~\txt{as well as}~~ E^\nu{}_\mu(i\del)=F^\nu{}_\mu(i\del)~~ (\txt{due to the Jacobi id}),
\eea
where the interchange ~$i\del\lra x$~ in any given representation ~$\hat{x}\ra f(x,i\del)$~ produces another representation $f(i\del,x)$. In this case $\ad_{\hat{x}^\mu}$'s are derivations but they rotate the coordinates rather than translate them as was the case with $[\hat{x}^\mu,\hat{x}^\nu]=i\theta^{\mu\nu}$. The $\hat{\del}_\mu$'s may be represented
\footnote{The action of generators can be seen by making an infinitesimal variation and using the Hausdorf-Campbell formula:
\bea
&& f({\hat{x}}+\delta {\hat{x}})\approx f({\hat{x}})+\delta {\hat{x}}^i\int_0^1 ds~e^{s~ \ola{\txt{ad}}_{\hat{x}^j}\ora{\del}_j}~\del_i f({\hat{x}}),\nn\\
&&\del_i f(s~ \ola{\txt{ad}_{\hat{x}}}+{\hat{x}})={{\del f(y)\over\del y^i}}|_{y^j=s~ \ola{\txt{ad}_{\hat{x}^j}}+{\hat{x}^j}},\nn\\
&&\txt{ad}_{\hat{x}}=(\txt{ad}_{\hat{x}^i})=(\txt{ad}_{\hat{x}^1},\txt{ad}_{\hat{x}^2},...,\txt{ad}_{\hat{x}^D}),\nn\\
&&\txt{ad}_a(b)=[a,b]=-[b,a]=-(b)\ola{\txt{ad}_a},\nn\\
&&\delta\hat{x}^j=d\hat{x}^i~\hat{J}_i{}^j=d\hat{x}^i~J_i{}^j(\hat{x}),~~d\hat{x}^i\ra 0,
\eea
where its is assumed that $\hat{f}=f(\hat{x})$ can be expanded in the specific form
\bea
&&f({\hat{x}})=\sum_{p\in \mathbb{C}^D}f_1(p)~f_2(p\cdot {\hat{x}}),~~p\cdot {\hat{x}}=p_i ~\hat{x}^i.
\eea
Rotations and translations are isometries of $g_{ij}=\delta_{ij}$ and are given by
\bea
&&\del_iJ_{mj}(x)+\del_jJ_{mi}(x)=0,\nn\\
&&J_{ij}(x)=\delta_{ij}~~\txt{for translations}~~\&~~\nn\\
&&J_{mi}(x)=(c_{mij}-c_{mji})x^j~~\txt{for rotations},\nn
\eea
where the $c$'s are constants.
}
by operators $\hat{\del}_\mu=Q_\mu(\del)$ ~such that~
\bea
&&dQ_\mu(\del)=E^{-1}{}^\nu{}_\mu(i\del)~d\del_\nu~\sr{eg.}{=}~(\int_0^1dt~e^{-tF(i\del)})^\nu{}_\mu~d\del_\nu\nn
\eea
and their action is as follows:
\bea
&&Q_\mu(\del)~T_J(x,\del)={\del \over\del u^\mu}T_J(x,\del)={\del X^K\over\del u^\mu}\{ {\del\over\del X^K}T_J(x,\del)+\Gamma_{K J}{}^L(\del)~ T_L(x,\del)\},\nn\\
&&u^\mu=x^\nu{E}^\mu{}_\nu(i\del),~~X^I=(x^\mu,\del_\mu),\nn\\
&& \Gamma_{KJ}{}^L= {\del u^\rho\over\del X^K}{\del\over\del X^J}{\del X^L\over \del u^\rho}= {\del u^\rho\over\del X^J}{\del\over\del X^K}{\del X^L\over \del u^\rho} =-{\del^2u^\rho\over\del X^K\del X^J}{\del X^L\over \del u^\rho}  ,\nn
\eea
where we only need its restriction on $x$-functions, $T_J(x,\del)\ra f_\mu(x)$.

If $e^{i\al\hat{x}}e^{i\beta\hat{x}}=e^{iK(\al,\beta)\hat{x}}$~ then $\Tr (\hat{\delta}_x\hat{\delta}_y^B)=\delta(x-y)$~ requires the functions $A,B$ to satisfy
\footnote{ Note that
\bea
&&\hat{\delta}_x\hat{\delta}_y^B=\sum_{pp'} A(p)~e^{ip\hat{x}}e^{iB(p')\hat{x}}e^{-ipx-ip'y}\nn\\
&&~~~~=\sum_{pp'} A(p)~e^{iK(p,B(p'))\hat{x}}~e^{-ipx-ip'y}=\sum_{pp'}\sum_z \hat{\delta}_z~A(p)~e^{iK(p,B(p'))z}~e^{-ipx-ip'y}\nn\\
&&~~~~=\Gamma_{xy}{}^z(B)~\hat{\delta}_z,~~\Gamma_{xy}{}^z(B)=\sum_{pp'}A(p)~e^{iK(p,B(p'))z}~e^{-ipx-ip'y},\nn\\
&&\Tr \hat{\delta}_x=1~~\txt{requires the existence  of}~~[\hat{x}^\mu,\hat{\del}_\nu]=-\delta^\mu{}_\nu,~[\hat{\del}_\mu,\hat{\del}_\nu]=0.\nn
\eea
 \emph{\textbf{Lie algebra type noncommutativity may be strong enough to regularize all possible $n$-point correlations unlike Moyal noncommutativity}}.
 }
\bea
&& K(p,B(-p))=0,~~A(p)=\det~\del_{p'} K(p,B(p'))|_{p'=-p}.\\
&&(~\txt{Therefore}~~\hat{\delta}_x^B=\sum_p A(p)~e^{iB(p)\hat{x}-ipx}=\sum_p e^{ip\hat{x}-iB^{-1}(p)x}\nn\\
&&~~~~~~~~~~~~=\sum_z\hat{\delta}_z~\sum_p e^{ipz-iB^{-1}(p)x}~).
\eea

That is to say that $p+p'=0$ is a solution of $K(p,B(p'))=0$ and the factor $\det\del_{p'} K(p,B(p'))|_{p'=-p}$  coming from $\delta(K(p,B(p')))$ at $p+p'=0$ needs to be canceled by the amplitude $A(p)$.
The uniqueness of the inversion here depends upon the solutions $\hat{\del}_\mu$ and $B$ of the relations
\bea
&&[\hat{x}^\mu,\hat{\del}_\nu]=-\delta^\mu{}_\nu,~[\hat{\del}_\mu,\hat{\del}_\nu]=0,~~~~K(p,B(-p))=0.
\eea
Finally, with $W(f_1)W(f_2)...W(f_n)=W(f_1\ast f_2\ast...\ast f_n)$ one defines
\bea
&&\delta_x(\hat{f}_1\hat{f}_2...\hat{f}_n)= {\Tr}(\hat{\delta}^B_x \hat{f}_1\hat{f}_2...\hat{f}_n)={\Tr}(\hat{\delta}^B_x W(f_1)W(f_2)...W(f_n))\nn\\
&&~~~~=({\Tr}\hat{\delta}^B_x\circ W) (f_1\ast f_2\ast...\ast f_n)=(f_1\ast f_2\ast...\ast f_n)(x),\nn\\
&&[\hat{\del}_\mu,W(f)]=W(\del_\mu f).
\eea
Note that one now also has the equivalent mirror algebra
\bea
&&\A^B=\{W^B(f)=\hat{f}^B=\sum_xf(x)\hat{\delta}_x^B\},\nn\\
&&\hat{f}^B=\sum_xf(x)\hat{\delta}_x^B=\sum_p\td{f}(p)~A(p)~e^{iB(p)\hat{x}}.
\eea
 The corresponding set of linear functionals is
\bea
A^\ast_B\simeq\{\delta^B_x=\Tr\circ\mu_{\hat{\delta}_x},~x\in \mathbb{R}^D\}\simeq \mathbb{R}^{D}~~~~\forall B.
\eea

\section{The quantum field}
A point particle's instance-wise trajectory $\Gamma:~]0,1[\subset \mathbb{R}~\ra \mathbb{R}^{d+1},~\tau\mapsto\gamma^\mu(\tau)$ in spacetime $\mathbb{R}^{d+1}$ may be regarded as a field or collection \\
$\{c_\tau=(\gamma(\tau),\rho_\tau(\mathbb{R}^{d+1});~\tau\in ]0,1[~\}$ of point-like spacetime distributions, with each instance $\gamma(\tau)$ represented by its localization or density or support  function \\
$\rho_\tau(y)=\delta(y-\gamma(\tau))$ in spacetime $\mathbb{R}^{d+1}$.
For the value of any property $P$ (eg. position, velocity, energy, momentum, etc) of the point particle that depends only on instances $\gamma(\tau)$ of its trajectory $\Gamma$ one then has the decomposition
\bea
&&P=P(\gamma(\tau))=\sum_{y\in \mathbb{R}^{d+1}}P(y)~\delta(y-\gamma(\tau))\eqv \sum_yp_y(\gamma(\tau))
\eea
where $\delta(y-\gamma(\tau))$ represents the density or support of the particle at the instant $\gamma(\tau)$ meanwhile $p_y(\gamma(\tau))=P(y)\delta(y-\gamma(\tau))\eqv P(\gamma(\tau))\delta(y-\gamma(\tau))$ is the (probability of) presence/influence, at the instance $\tau$, of the property $P$ at/on a generic point $y\in \mathbb{R}^{d+1}$.

Now consider a wave packet (generically a field) $\psi:\D\subset \mathbb{R}^{d+1}\ra \mathbb{C},~x\mapsto\psi(x)$ which describes the energy ( presence or existence ) distribution or concentration of a large collection of particles. Just as the instance-wise trajectory $\Gamma$ of the point particle  was decomposed into point-like (or spacetime $\delta$-) distributions according to its instances $\{\gamma(\tau);~\tau\in ]0,1[~\}$ one can also decompose the wave packet into spacetime modes, which are spacetime $\delta$-distributions, ($\delta$-modes)  as
\bea
&&\psi(x)=\sum_y\psi(y)~\delta(y-x)=\sum_y\psi(x)~\delta(y-x)\eqv\sum_{y}\psi_y~\delta_y(x)=\sum_{y}\psi_x~\delta_y(x).\nn
\eea
where $\psi_x$ is the amplitude of the space-time mode that is $\delta$-localized at $x$.

The $\delta$ decomposition is done in analogy (and should be interpreted similarly) to the plane-wave (ie. Fourier) or exponential (e) decomposition
\bea
&&\psi(x)=\sum_k\wt{\psi}_k~e_k(x)\eqv \sum_k\wt{\psi}(k)~e^{ikx}
\eea
where $\wt{\psi}_k$ is the amplitude of the energy-momentum mode that is $e$-localized at $x$.

In principle one has an arbitrarily large number of possible types of decompositions (or transforms). The basic idea is to describe the interaction of two systems (wave packet, point particles, fields, etc) in terms of the interactions/correlations of their individual modes.

Of course one also has an $e$-decomposition of the instance-wise property $P$
of the point-like particle:
\bea
P(\gamma(\tau))=\sum_k\wt{P}_k~e_k(\gamma(\tau))=\sum_k\wt{P}_k~e^{ik\gamma(\tau)}.
\eea

The quantum field $\Psi:\D\subset \mathbb{R}^{d+1}\ra U(\mathbb{C}),~x\mapsto \Phi(x)$ is an operator-valued wave packet
\bea
\Psi(x)=\sum_k \wt{\Psi}_k~e_k(x)=\sum_y \Psi_y~\delta_y(x)=...
\eea
In a noncommutative spacetime $\mathbb{R}_\theta^{d+1}$ with coordinates ~$\hat{x}^\mu,~[\hat{x}^\mu,\hat{x}^\nu]\neq 0$~ these decompositions may be written analogously as
\bea
&&\Psi(\hat{x})=\sum_k \wt{\Psi}_k\ot \hat{e}_k=\sum_y \Psi_y\ot\hat{\delta}_y=...,\nn\\
&&\hat{e}_k=e^{ik\hat{x}},~~\hat{\delta}_y=\sum_{k}e^{ik(\hat{x}-y)}.
\eea
When $\hat{x}$ is commutative, we have the two-point correlation duality
\bea
&&\lang\hat{e}_{k_1}...\hat{e}_{k_m}\hat{e}^\ast_{k'_1}...\hat{e}^\ast_{k'_n}\rang_{NC}=\sum_{\hat{x}}\hat{e}_{k_1}...\hat{e}_{k_m}\hat{e}^\ast_{k'_1}...\hat{e}^\ast_{k'_n}=\delta(\sum k-\sum k'),\nn\\
&&\lang\hat{\delta}_{y_1}...\hat{\delta}_{y_m}\hat{\delta}^\ast_{y'_1}...\hat{\delta}^\ast_{y'_n}\rang_{NC}=\sum_{\hat{x}}\hat{\delta}_{y_1}...\hat{\delta}_{y_m}\hat{\delta}^\ast_{y'_1}...\hat{\delta}^\ast_{y'_n}
=\prod_{i=1}^{m-1}\delta( y_i-y_{i+1})~\prod_{j=1}^{n-1}\delta( y'_j-y'_{j+1})\nn\eea
where the former is an expression of momentum conservation.
The purpose of noncommutativity(NC) is to spread out all the delta functions in the latter expression, ie. to make the spacetime $\delta$-distribution nonsingular ( although this does not happen for $2n+1$-point functions in the case of Moyal noncommutativity $[[\hat{x}^\mu,\hat{x}^\nu],\hat{x}^\al]=0$. Moyal NC also maintains translational invariance/momentum conservation expressed by the former correlation expression but breaks rotational invariance and hence any angular momentum conservation). In Lie algebra type NC ~$[\hat{x}^\mu,\hat{x}^\nu]=C^{\mu\nu}{}_\al\hat{x}^\al$~ full smearing may be achieved (Here both rotational and translational invariance, and hence angular momentum and momentum conservation, are broken).

\section{The algebra of quantum fields}

Consider the algebra of free causal/accausal real (or hermitian) quantum fields $\A=\{\phi\}$
\bea
&&[\phi(x),\phi(y)]=i\Delta(x,y)=-i\oint_{C}{d^4x\over (2\pi)^4}{e^{-ikx}\over k^2-m^2},\nn\\
&&[\phi_x,\phi_y]=i\Theta_{xy},~~\Theta_{xy}=-\oint_{C}{d^4x\over (2\pi)^4}{e^{-ikx}\over k^2-m^2},
\eea
where the integral in $k_0$ is an integral along any closed contour $C$ in the complex $k_0$ plane that encloses all two poles of the integrand that are located at \\
$k_0=\pm\sqrt{\vec{k}^2+m^2}$.
Regarding $\theta^{\mu\nu}$ as a 2-point correlation function in the directions of spacetime, then one can employ the star product technique to calculate correlation functions of quantum fields:
\bea
&&g(\mu_1,\mu_2,...)={\Tr}(\hat{x}^{\mu_1}\hat{x}^{\mu_2}...~\hat{\delta}_x)=x^{\mu_1}\ast x^{\mu_2}\ast...,\nn\\
&&W(x_1,x_2,...)={\Tr}(\phi_{x_1}\phi_{x_2}...~\hat{\delta}_\vphi)\eqv {\Tr}_\vphi(\phi_{x_1}\phi_{x_2}...)=\vphi_{x_1}\ast_{\Delta} \vphi_{x_2}\ast_{\Delta}...,\nn\\
&&\hat{\delta}_\vphi=\int D\Pi~e^{i\sum_y\Pi_y(\phi_y-\vphi_y)}.\nn\\
&&G(x_1,x_2,..)={\Tr}(T(\phi_{x_1}\phi_{x_2}..)T\hat{\delta}_\vphi)=T({\Tr}(\phi_{x_1}\phi_{x_2}..\hat{\delta}_\vphi))=T(\vphi_{x_1}\ast_{\Delta} \vphi_{x_2}\ast_{\Delta}..),\nn\\
&&T\hat{\delta}_\vphi=\int D\Pi~Te^{i\sum_y\Pi_y(\phi_y-\vphi_y)},\nn\\
&&A(x_1,x_2,...)={\Tr}(T(e^{iS_E[\phi]}\phi_{x_1}\phi_{x_2}... \hat{\delta}_\vphi),~~\txt{(amplitude of a dynamical process)},\nn\\
&&\ast_\Delta\eqv e^{-{i\over 2}\sum_{xy}{\ola{\delta}\over\delta\vphi_x}\Theta_{xy}{\ora{\delta}\over\delta\vphi_y} },\nn\\
&&G(x,y)=\vphi_x\vphi_y+\txt{sign}(x_0-y_0)~\Theta_{xy}\nn\\
&&~~~~=\vphi_x\vphi_y+\txt{sign}(x_0-y_0)~\theta((x_0-y_0)^2-(\vec{x}-\vec{y})^2)~\Theta_{xy} ,\nn\\
&&T(\phi_x\phi_y)={1\over 2}(\phi_x\phi_y+\phi_y\phi_x)+\txt{sign}(x_0-y_0)~[\phi_x,\phi_y],\nn\\
&&\theta((x_0-y_0)^2-(\vec{x}-\vec{y})^2)~\Theta_{xy}=\Theta_{xy},\nn\\
\eea
where $T$ denotes time ordering.
With this analogy, $\theta^{\mu\nu}$ may be interpreted as the probability amplitude or potential that a straight line trajectory into the $\mu$ direction will spontaneously turn in the $\nu$ direction.

If the spacetime on which the quantum field is defined is also noncommutative as the Moyal plane then the algebra of the free quantum fields
\bea
&&[\phi_x,\phi_y]=0,~~\txt{whenever}~~\Theta_{xy}=0
\eea
becomes
\bea
&&\phi_x\phi_y=e^{i\theta^{\mu\nu}{\del\over\del y^\mu}{\del\over\del x^\nu}}\phi_y\phi_x,~~\txt{whenever}~~\Theta_{xy}=0,\nn\\
&&\phi_x=\phi^0_x~e^{{1\over 2}{\ola{\del}}_\mu\theta^{\mu\nu}P_\nu},~~[\phi^0_x,\phi^0_y]=0~~\txt{whenever}~~\Theta_{xy}=0.
\eea
Thus the $\ast$ to be used in the Green's functions and process amplitudes is a composition of two $\ast$'s
\bea
&&\ast=\ast_\Delta\circ\ast_\theta= e^{-{i\over 2}\sum_{yz}{\ola{\delta}\over\delta\vphi_y}\Theta_{yz}{\ora{\delta}\over\delta\vphi_z}}\circ e^{-{i\over 2}\ola{\delta}_\mu\theta^{\mu\nu}\ora{\delta}_\nu }=e^{-{i\over 2}\sum_{yz}{\ola{\delta}\over\delta\vphi_y}\Theta_{yz}{\ora{\delta}\over\delta\vphi_z}-{i\over 2}\ola{\delta}_\mu\theta^{\mu\nu}\ora{\delta}_\nu} ,\nn\\
&&G(x_1,x_2,...)=({\Tr}_\vphi\circ{\Tr}_\theta)(T\phi_{\hat{x}_1}\phi_{\hat{x}_2}...)={\Tr}_\vphi({\Tr}_\theta(T\phi_{\hat{x}_1}\phi_{\hat{x}_2}...)) \nn\\
&&=T(\vphi_{x_1}\ast \vphi_{x_2}\ast...) .
\eea
The two $\ast$'s commute (ie. $\ast_\Delta\circ\ast_\theta=\ast_\theta\circ\ast_\Delta$) since they act on different spaces (spacetime and the internal space of the quantum fields). Therefore to consider the $\hat{x}$'s dynamical one may simply add a suitable term $\Gamma[\X(\hat{x})]$ to the action ~$S[\phi]$ ($
S[\phi]\ra S[\phi]+\Gamma[\X]$)~
 which describes the dynamics
\footnote{ Time evolution in terms of the Hamiltonian is given by
\bea
&&i\del_0\phi_x=[H,\phi_x]~~\Ra~~i\del_0\phi'_x=[H'-H_0,\phi'_x],~~\del_0H_0=0,\nn\\
&&H'=e^{-ix_0H_0}H e^{ix_0H_0},~~\phi'_x=e^{-ix_0H_0}\phi_x e^{ix_0H_0}~~(~\phi_x=e^{ix_0H_0}\phi'_x e^{-ix_0H_0}~),\nn\\
&&\Ra~(\txt{solution})~\phi'_x=C_{\vec{x}}+\overline{T}~e^{-i\int_{-\infty}^{x_0}dt~(H'-H_0)}\hat{\vphi}_{\vec{x}}Te^{i\int_{-\infty}^{x_0}(H'-H_0)}=C_{\vec{x}}+\delta\phi'_x,\nn\\
&&~~~~\del_0C_{\vec{x}}=0,~~[H'-H_0,C_{\vec{x}}]=0,~~\del_0\hat{\vphi}_{\vec{x}}=0,~~[H'-H_0,\hat{\vphi}_{\vec{x}}]\neq 0,\nn\\
&&\Ra~~\phi_x=e^{iH_0x_0}C_{\vec{x}}e^{-iH_0x_0}+\overline{T}~e^{-i\int_{-\infty}^{x_0}dt~(H-H_0)}e^{iH_0x_0}\hat{\vphi}_{\vec{x}}e^{-iH_0x_0}Te^{-i\int_{-\infty}^{x_0}(H-H_0)}\nn\\
&&~~~~~~~~=C_x+\overline{T}~e^{-i\int_{-\infty}^{x_0}dt~(H-H_0)}\hat{\vphi}_xTe^{i\int_{-\infty}^{x_0}(H-H_0)}=C_{x}+\delta\phi_x\nn\\
&&~~~~~~~~=C_x+{\S}^\dg~T({\S}\hat{\vphi}_x),\\
&&T(\delta\phi_x\delta\phi_y...)=T(\overline{T}~e^{-i\int_{-\infty}^{x_0}(H-H_0)}\hat{\vphi}_xTe^{i\int_{-\infty}^{x_0}(H-H_0)}\overline{T}e^{-i\int_{-\infty}^{y_0}(H-H_0)}\hat{\vphi}_yTe^{i\int_{-\infty}^{y_0}(H-H_0)}...)\nn\\
&&~~~~={\S}^\dg~T({\S}\hat{\vphi}_x\hat{\vphi}_y...),~~~~~~H-H_0= \int d^{D-1}x~({\del(\L-\L_0)\over\del\del_0\phi}\del_0\phi-(\L-\L_0)).
\eea
Here
\bea
&&T e^{i\int _{t_1}^{t_2}H}= e^{idt_2H(t_2)}...e^{idt_1H(t_1)},~~~~(T e^{i\int _{t_1}^{t_2}H})^{-1}=e^{-idt_1H(t_1)}...e^{-idt_2H(t_2)}=\overline{T} e^{-i\int _{t_1}^{t_2}H}.\nn
\eea
\bea
&&\overline{T}~e^{-i\int_{-\infty}^{x_0}H_I}=e^{-idt_{-\infty}{H_I}(-\infty)}...e^{-idx_0{H_I}(x_0)}\nn\\
&&~~=e^{-idt_{-\infty}{H_I}(-\infty)}...e^{-idx_0{H_I}(x_0)}~e^{-idx_0{H_I}(x_0)}...e^{-idt_{\infty}{H_I}(\infty)}~e^{idt_{\infty}{H_I}(\infty)}...e^{idx_0{H_I}(x_0)}\nn\\
&&~~=\overline{T} e^{-i\int_{-\infty}^\infty dt {H_I}}~T e^{i\int_{x_0}^\infty dt {H_I}}={\S}^\dg~T e^{i\int_{x_0}^\infty dt {H_I}}.
\eea

If $C_x=C_{\vec{x}}$, ie. all commuting, and $\L_I$ contains no time derivatives then
\bea
&&H_I=H-H_0= -\int d^{D-1}x~\L_I,\nn\\
&&S_I[\phi]=\int dt~H_I=S_I[C+U^{-1}\hat{\vphi}U]=S_I[C+\hat{\vphi}]~~\txt{as}~~\txt{ $S_I[\phi]$ is local},\nn\\
\eea
where (if it exists) $\hat{\vphi}$ may be chosen such that $[\hat{\vphi}_x,\hat{\vphi}_y]$ commutes with all quantum operators, a property that does not hold for $\phi$ in general.
}
of $\phi$

For example one can consider
\bea
&&[\hat{x}_\mu(u),\hat{x}_\nu(v)]=i\theta_{\mu\nu}(u,v),~~~~\ast_\theta=e^{-{i\over 2}\sum_{uv}{\ola{\delta}\over\delta x^\mu(u)}\theta^{\mu\nu}(u,v){\ora{\delta}\over\delta x^\nu(v)}},\nn\\
&&\Gamma[\hat{x}]=\sum_{u}\al_\mu(u) ~\gamma^\mu(u,\hat{x}(u),\del_u \hat{x}(u),...),\nn\\
&&
\eea
The Moyal coordinate $\hat{x}$ is seen to have been evolved from a general dynamical noncommutative coordinate $\X$ to the form \\ $[\hat{x}_\mu(u),\hat{x}_\nu(v)]=i\theta_{\mu\nu}(u,v)$ by $\Gamma[\X]=\Gamma[c+U^{-1}\hat{x}U]$ in the same way that $\phi$ is evolved into $\hat{\vphi}$ by $S[\phi]=S[C+U^{-1}\hat{\vphi}U]$. Therefore the dynamical quantum theory of an ``interacting'' membrane embedded in spacetime is a theory of noncommutative spacetime.

Here one may say that the field $\phi$ propagates in a dynamical (ie. curved) spacetime (whose metric is induced by the classical path $\hat{x}(u)$ defined by ~\\ ${\delta\Gamma[\hat{x}]\over\delta\hat{x}^\mu(u)}=0$)
\bea
g_{\mu\nu}=h_{ab}{\del u^a\over\del x^\mu}{\del u^b\over\del x^\nu}=(h^{ab}{\del x^\mu\over\del u^a}{\del x^\nu\over\del u^b})^{-1},
\eea
where for the special case of two parameters (ie. ``string'') one can set $h_{ab}=\eta_{ab}$ and one would then say that the field $\phi$ is propagating in a ``stringy spacetime''.
\bea
&&S[\phi]=\int Dx~\L[\hat{x},\phi[\hat{x}],\del_{\hat{x}}\phi[\hat{x}],...]=\int(\prod_u d^{d+1}x(u))~\L[\hat{x},\phi[\hat{x}],\del_{\hat{x}}\phi[\hat{x}],...]\nn
\eea
where any sum $\sum_\mu$ has been replaced by $\sum_\mu\sum_u.$ One may also combine the $\vphi$ and $x$ spaces thus combining the two products:
\bea
&&\vphi_i=(\vphi_y,x_\mu(u)),~~~~\Theta_{ij}=\left(
                \begin{array}{cc}
                  \Theta_{yz} & 0 \\
                  0 & \theta_{\mu\nu}(u,v) \\
                \end{array}
              \right).\nn\\
&&{\delta\over\delta\vphi_i}={\delta\psi_j(\vphi)\over\delta\vphi_i}({\delta\over\delta\psi_j}+\Gamma_j).
\eea

\subsection{Operator product ordering and physical correlations}
\begin{itemize}
\item Linear transforms, such as the Fourier transform of functions, enable information to be processed (encoded/stored/transported/decoded) deterministically.
In general, functions of the noncommuting variable $\phi$ may be analyzed by defining Fourier transforms (now however depending on the order in the operator products) in analogy to commutative variables.
In particular for the description of natural processes or phenomena one can define a time-ordered Fourier transform required by their transitive past-future time direction; recall that $\phi$ can be expressed as a time ordered function of $\hat{\psi}=C+\hat{\vphi}$. \emph{Any natural process or phenomenon may be regarded as a sequence of localized spacetime ``events'' that is well ordered in time (non-relativistic sense) or proper time (relativistic sense) or any other suitable parameter.} This natural time ordering is trivial in a commutative theory but nontrivial in a noncommutative theory; it provides a starting point for defining kinematic variables by eliminating the inherent operator ordering ambiguity in the noncommutative theory.
\bea
&&f_\O[\hat{\psi}]=\int DJ~\td{f}[J]~\O(\prod_xe^{-iJ_x\hat{\psi}_x}).\nn\\
&&f_T[\hat{\psi}]=\int DJ~\td{f}[J]~T(\prod_xe^{-iJ_x\hat{\psi}_x}),\nn\\
&&T (e^{-iJ_x\hat{\psi}_x}e^{-iJ_y\hat{\psi}_y})~{}^2=\theta(x_0-y_0)~e^{-iJ_x\hat{\psi}_x}e^{-iJ_y\hat{\psi}_y}+\theta(y_0-x_0)~e^{-iJ_y\hat{\psi}_y}e^{-iJ_x\hat{\psi}_x}\nn\\
&&=e^{-iJ_x\hat{\psi}_x-iJ_y\hat{\psi}_y}~e^{-{1\over 2}J_x\txt{sign}(x_0-y_0)i\Theta_{xy}J_y}=e^{-iJ_x\hat{\psi}_x}e^{-iJ_y\hat{\psi}_y}~e^{{1\over 2}J_x(1-\txt{sign}(x_0-y_0))i\Theta_{xy}J_y}\nn\\
&&~~=e^{-iJ_x\hat{\psi}_x}e^{-iJ_y\hat{\psi}_y}~e^{J_x\theta(y_0-x_0)i\Theta_{xy}J_y}=e^{-iJ_y\hat{\psi}_y}e^{-iJ_x\hat{\psi}_x}~e^{-J_x\theta(x_0-y_0)i\Theta_{xy}J_y} ,\nn\\
&&T(A_1A_2...A_n)=\prod_{i<j}T_{ij}~(A_1A_2...A_n),~~T(\al A)=\al T(A)~~\forall \al\in \mathbb{C}.\nn\\
&&~~\Ra~~f_T[\hat{\psi}]=\int DJ~\td{f}[J]~e^{-i\sum_xJ_x\hat{\psi}_x}~e^{-{1\over 2}\sum_{xy}J_x{1\over 2}\txt{sign}(x_0-y_0)i\Theta_{xy}J_y} \nn\\
\label{ordering1}&&~~~~~~~~~~~~~~~~~=e^{{1\over 2}\sum_{xy}{1\over 2}\txt{sign}(x_0-y_0)i\Theta_{xy}{\delta\over\delta\hat{\psi}_x}{\delta\over\delta\hat{\psi}_y}}~f_W[\hat{\psi}],
\eea
where $\O$ denotes general ordering, $T$ is time ordering and $W$ is symmetric (Weyl) ordering.

\subsection{From Weyl or symmetric ordering to normal or classical ordering}
 One can further write Weyl ordering in terms of Normal ordering by the decomposition
    \bea
    &&\hat{\psi}_x=\hat{\psi}^{+}_x+\hat{\psi}^{-}_x,~~[\hat{\psi}^{+}_x,\hat{\psi}^{+}_y]=0,~~[\hat{\psi}^{-}_x,\hat{\psi}^{-}_y]=0 ,\nn\\
    &&f_W[\psi]=\int DJ\td{f}[J]~W(\prod_xe^{-iJ_x\hat{\psi}_x})=\int DJ\td{f}[J]~e^{-i\sum_xJ_x\hat{\psi}_x}\nn\\
    &&~~~~=\int DJ\td{f}[J]~e^{-i\sum_xJ_x(\hat{\psi}^{+}_x+\hat{\psi}^{-}_x)}\nn\\
    &&~~~~=\int DJ\td{f}[J]~e^{-i\sum_xJ_x\hat{\psi}^{+}_x-i\sum_yJ_y\hat{\psi}^{-}_y}\nn\\
    &&~~~~=\int DJ\td{f}[J]~e^{{1\over 2}\sum_{xy}J_xJ_y[\hat{\psi}^{+}_x,\hat{\psi}^{-}_y]}~e^{-i\sum_xJ_x\hat{\psi}^{-}_x}e^{-i\sum_yJ_y\hat{\psi}^{+}_y}\nn\\
    &&~~~~=\int DJ\td{f}[J]~e^{{1\over 2}\sum_{xy}J_xJ_y[\hat{\psi}^{+}_x,\hat{\psi}^{-}_y]}~N(e^{-i\sum_xJ_x\hat{\psi}_x})\nn\\
    &&~~~~=e^{-{1\over 2}\sum_{xy}[\hat{\psi}^{+}_x,\hat{\psi}^{-}_y]{\delta\over\delta\hat{\psi}_x}{\delta\over\delta\hat{\psi}_y}}\int DJ\td{f}[J]~N(e^{-i\sum_xJ_x\hat{\psi}_x})\nn\\
    &&~~~~=e^{-{1\over 2}\sum_{xy}[\hat{\psi}^{+}_x,\hat{\psi}^{-}_y]{\delta\over\delta\hat{\psi}_x}{\delta\over\delta\hat{\psi}_y}}f_N[\hat{\psi}].\nn\\
    &&f_T[\hat{\psi}]=e^{{1\over 2}\sum_{xy}{1\over 2}\txt{sign}(x_0-y_0)i\Theta_{xy}{\delta\over\delta\hat{\psi}_x}{\delta\over\delta\hat{\psi}_y}}~f_W[\hat{\psi}]\nn\\
    &&~~~~=e^{{1\over 2}\sum_{xy}({1\over 2}\txt{sign}(x_0-y_0)i\Theta_{xy}-[\hat{\psi}^{+}_x,\hat{\psi}^{-}_y]){\delta\over\delta\hat{\psi}_x}{\delta\over\delta\hat{\psi}_y}}~f_N[\hat{\psi}]\nn\\
    &&~~~~=e^{{1\over 2}\sum_{xy}({1\over 2}\txt{sign}(x_0-y_0)[\hat{\psi}_x,\hat{\psi}_y]-[\hat{\psi}^{+}_x,\hat{\psi}^{-}_y]){\delta\over\delta\hat{\psi}_x}{\delta\over\delta\hat{\psi}_y}}~f_N[\hat{\psi}]\nn\\
    &&~~~~=e^{{1\over 2}\sum_{xy}\Delta_F(x,y){\delta\over\delta\hat{\psi}_x}{\delta\over\delta\hat{\psi}_y}}~f_N[\hat{\psi}].\nn\\
    &&\Delta_F(x,y)={1\over 2}\txt{sign}(x_0-y_0)[\hat{\psi}_x,\hat{\psi}_y]-{1\over 2}([\hat{\psi}^{+}_x,\hat{\psi}^{-}_y]+[\hat{\psi}^{+}_y,\hat{\psi}^{-}_x] )\nn\\
    &&~~~~\eqv {1\over 2}\txt{sign}(x_0-y_0)([\hat{\psi}^{+}_x,\hat{\psi}^{-}_y]-[\hat{\psi}^{+}_y,\hat{\psi}^{-}_x] )-{1\over 2}([\hat{\psi}^{+}_x,\hat{\psi}^{-}_y]+[\hat{\psi}^{+}_y,\hat{\psi}^{-}_x] )\nn\\
    &&~~~~=\theta(x_0-y_0)~[\hat{\psi}^{-}_x,\hat{\psi}^{+}_y]+\theta(y_0-x_0)~[\hat{\psi}^{-}_y,\hat{\psi}^{+}_x].\nn\\
    &&T(f_T[\hat{\psi}]g_T[\hat{\psi}])=(fg)_T[\hat{\psi}].
    \eea
 Therefore in the coherent (ie. classical) state defined by
 \bea
 &&\hat{\psi}^{+}_x|\psi\rangle=\psi^{+}_x|\psi\rangle=|\psi\rangle\psi^{+}_x,\nn\\
 &&\langle\psi|f_T[\hat{\psi}]|\psi\rangle=\langle\psi|e^{{1\over 2}\sum_{xy}\Delta_F(x,y){\delta\over\delta\hat{\psi}_x}{\delta\over\delta\hat{\psi}_y}}f_N[\hat{\psi}]|\psi\rangle\nn\\
 &&~~~~=\langle\psi|\psi\rangle~e^{{1\over 2}\sum_{xy}\Delta_F(x,y){\delta\over\delta\psi_x}{\delta\over\delta\psi_y}} f[\psi]\nn\\
 &&~~~~=\langle\psi|\psi\rangle\int D\psi'~f[\psi']~{1\over\sqrt{\det\Delta_F}}e^{{1\over 2}\sum_{xy}(\psi_x-\psi_x')\Delta_F^{-1}(x,y)(\psi_y-\psi_y')}\nn\\
 &&~~~~=\langle\psi|\psi\rangle\int D\psi'~f[\psi+\psi']~{1\over\sqrt{\det\Delta_F}}e^{{1\over 2}\sum_{xy}\psi_x'\Delta_F^{-1}(x,y)\psi_y'},
 \eea
 where $\psi=0$ corresponds to a possible vacuum state (~ground state or local minimum energy configuration: \\
 ${\delta E[\psi,\dot{\psi}]\over\delta\psi_x}=0,{\delta E[\psi,\dot{\psi}]\over\delta\dot{\psi}_x}=0,~~E[\psi,\dot{\psi}]=\int d^{D-1}x~({\del\L\over\del\del_0\psi}\del_0\psi-\L) $~) and $\psi$ may be identified with $\langle\phi\rangle=C+\langle\hat{\vphi}\rangle$. In the case of more than one independent vacua $\{\psi_i\}$ the vacuum amplitude has a matrix structure
 \bea
 &&\langle\psi_i|f_T[\hat{\psi}]|\psi_j\rangle=\langle\psi_i|e^{{1\over 2}\sum_{xy}\Delta_F(x,y){\delta\over\delta\psi_x}{\delta\over\delta\psi_y}}f_N[\hat{\psi}]|\psi_j\rangle\nn\\
 &&~~~~=\langle\psi_i|\psi_j\rangle~e^{{1\over 2}\sum_{xy}\Delta_F(x,y){\delta\over\delta\psi_x}{\delta\over\delta\psi_y}} f[\psi]|_{\psi=\psi_i^{+}+\psi_j^{-}}.
 \eea

 One notes that the eigenfunctionals (which describe stationary or ``elementary'' processes) of the operator $\sum_{xy}\Delta_F(x,y){\delta\over\delta\psi_x}{\delta\over\delta\psi_y}$ are individually ``unaffected'' by the quantization. One may also consider expanding any given process in terms of these stationary processes.

Since $S[C+U^{-1}\hat{\vphi}U]=S[\phi]$ describes the background (as opposed to ``particles'' or excitations) dynamics ${\delta S[\phi]\over\delta \phi_x}=0$ of the quantum field $\phi$, one can define a corresponding action $\Gamma[\psi]$ that describes the background dynamics ${\delta \Gamma[\psi]\over\delta\psi_x}=0$ of $\psi$ (ie. the same dynamics in the classical picture) by
\bea
&& e^{-i\Gamma[\psi]}=e^{{1\over 2}\sum_{xy}\Delta_F(x,y){\delta\over\delta\psi_x}{\delta\over\delta\psi_y}}~ e^{-iS[\psi]},
\eea
which is given by the choice \\
$f[\psi]=e^{-iS[\psi]}$; that is $\hat{\psi}_{x_1}=\hat{\psi}_{x_2}=...=\hat{\psi}_{x_n}=const$.

\item Thus n-point scattering amplitudes correspond to the choice
\bea
&&f[\psi]=e^{-iS_I[\psi]}\psi_{x_1}\psi_{x_2}...\psi_{x_n},\nn\\
&&G(x_1,...,x_n;\psi)={f_{eff}[\psi]\over e^{{1\over 2}\sum_{xy}\Delta_{xy}{\delta\over\delta\psi_x}{\delta\over\delta\psi_y}}~e^{-iS_I[\psi]}}\nn\\
&&~~~~={e^{{1\over 2}\sum_{xy}\Delta_{xy}{\delta\over\delta\psi_x}{\delta\over\delta\psi_y}}~(~e^{-iS_I[\psi]}\psi_{x_1}\psi_{x_2}...\psi_{x_n}~)\over e^{{1\over 2}\sum_{xy}\Delta_{xy}{\delta\over\delta\psi_x}{\delta\over\delta\psi_y}}~e^{-iS_I[\psi]}}.\nn
\eea

In the presence of spacetime noncommmutativity an appropriate choice would be
\bea
&&f[\psi]=e^{-iS_{\ast I}[\psi]}\psi_{x_1}\ast\psi_{x_2}\ast...\ast\psi_{x_n}.
\eea

\item In momentum space one can choose
\bea
&&f[\psi]=e^{-iS_I[\psi]}\td{\psi}_{k_1}\td{\psi}_{k_2}...\td{\psi}_{k_n}~\delta(\sum_ik_i),\nn\\
\eea
where $\td{\psi}_{k}=\int d^4x~\psi_x~e^{ikx}$ and in the presence of Moyal noncommutativity one can choose
\bea
&&f[\psi]=e^{-iS_{\ast I}[\psi]}\td{\psi}_{k_1}\td{\psi}_{k_2}...\td{\psi}_{k_n}~e^{-{i\over 2}\sum_{i<j}k^\mu_i\theta_{\mu\nu}k^\nu_j}~\delta(\sum_ik_i).\nn\\
&&(\widetilde{\psi^n_x})_q=\sum_{q_1,q_2,...,q_n}\delta(q_1+...+q_n-q)~\widetilde{\psi}_{q_1}...\widetilde{\psi}_{q_n}.
\eea

\textbf{\textbf{Remarks:
}}

\item The definition of the functional includes the definition of pointwise products and therefore the use of alternative ordering to break Weyl symmetric ordering in functionals generalizes the implementation/or detection of noncommutativity.

\item To involve fermions one may directly extend the field $\phi$ to become a superfield; ie. $\phi_x\ra \Phi_{s},~~s_M=(x_\mu,\vartheta_i)=(x_\mu,\theta_\sigma,\bar{\theta}_\sigma),~~\vartheta_i\vartheta_j=-\vartheta_j\vartheta_i $.
    \bea
    &&\Phi_{s}=\phi_x+\psi_x\theta+\bar{\psi}_x\bar{\theta}+\theta A\bar{\theta}+F\theta^2+\bar{F}\bar{\theta}^2+\chi_x\theta\bar{\theta}^2+\bar{\chi}_x\bar{\theta}\theta^2+D\theta^2\bar{\theta}^2.\nn\\
    &&\mu=0,1,...,D-1,~~i=1,2,...,2^{D\over 2},~~\sigma=1,2,...,{1\over 2}2^{D\over 2}=1,2,...,2^{{D\over 2}-1}.\nn
    \eea
 A superspace extension of $g_x^{\mu\nu}\nabla_\mu\nabla_\nu \phi_x=0$ would be
 \bea
 g_s^{MN}\nabla_M\nabla_N \Phi_s=0.
 \eea

\item If the semiclassical quantization procedure is applied to Moyal spacetime
\bea
&&[\hat{x}_\mu,\hat{x}_\nu]=i\theta_{\mu\nu}=-i\theta_{\nu\mu},
\eea
one would obtain
{\footnotesize\bea
&&\langle x|f_{\O}(\hat{x})|x\rangle=\langle x|x\rangle~e^{{i\over 2}\sum_{\mu\nu}\Delta_{\mu\nu}\del_\mu\del_\nu}~f(x)=\int d^Dx'~f(x')~{e^{{1\over 2}(x_\mu-x'_\mu)\Delta^{-1}_{\mu\nu}(x_\nu-x'_\nu)}\over\sqrt{\det\Delta}},\nn\\
&&\Delta_{\mu\nu}=i\O_{\mu\nu}\theta_{\mu\nu}+i\N_{\mu\nu},~~\O_{\mu\nu}=-\O_{\nu\mu},~~\N_{\mu\nu}=\N_{\nu\mu}=\N_{\mu\nu}(\theta),\nn
\eea}
where $\O,\N$ are yet to be determined physical constants. If Lorentz invariance is required then $\Delta_{\mu\nu}=\ld~\eta^{\mu\nu}$.

By analogy we may write $\hat{x}_\mu=\hat{x}^+_\mu+\hat{x}^-_\mu,~~[\hat{x}^+_\mu,\hat{x}^+_\nu]=0,~~[\hat{x}^-_\mu,\hat{x}^-_\nu]=0$; then
\bea
&&i\theta_{\mu\nu}=[\hat{x}_\mu,\hat{x}_\nu]=[\hat{x}^+_\mu,\hat{x}^-_\nu]+[\hat{x}^-_\mu,\hat{x}^+_\nu]=[\hat{x}^-_\mu,\hat{x}^+_\nu]-[\hat{x}^-_\nu,\hat{x}^+_\mu],\nn\\
&&\O_{\mu\nu}={1\over 2}\txt{sign}(\mu-\nu),~~i\N_{\mu\nu}={1\over 2}([\hat{x}^-_\mu,\hat{x}^+_\nu]+[\hat{x}^-_\nu,\hat{x}^+_\mu]),\nn\\
&&\Delta_{\mu\nu}=\theta(\mu-\nu)~[\hat{x}^-_\mu,\hat{x}^+_\nu]+\theta(\nu-\mu)~[\hat{x}^-_\nu,\hat{x}^+_\mu].
\eea

However, for a particle moving in spacetime (or equivalently a particle-like spacetime), the coordinates $\X_\mu(\tau)=c_\mu+U^{-1}\hat{x}_\mu(\tau) U$ already possess the natural ordering wrt the parameter $\tau$,
\bea
&&[\hat{x}_\mu(\tau),\hat{x}_\nu(\tau')]=i\theta_{\mu\nu}(\tau,\tau'),\nn\\
&&[\hat{x}_M,\hat{x}_N]=i\theta_{MN},~~M=(\tau,\mu),~~N=(\tau',\nu).\nn\\
&&\X_M=c_\mu+U^{-1}\hat{x}_M U.\nn\\
&&d\tau^2=\eta_{\mu\nu}dx^\mu(\tau) dx^\nu(\tau).
\eea
Recall that \emph{Any natural process or phenomenon may be regarded as a sequence of localized spacetime ``events'' that is well ordered in time (non-relativistic sense) or proper time (relativistic sense) or any other suitable parameter.}

One can therefore write
\bea
&&\Delta_{MN}=\theta(\tau-\tau')~[\hat{x}^-_M,\hat{x}^+_N]+\theta(\tau'-\tau)~[\hat{x}^-_N,\hat{x}^+_M].
\eea
Similarly for a string-like space-time
\bea
&&[\hat{x}_\mu(\tau,\sigma),\hat{x}_\nu(\tau',\sigma')]=i\theta_{\mu\nu}(\tau,\sigma,\tau',\sigma'),\nn\\
&&[\hat{x}_M,\hat{x}_N]=i\theta_{MN},~~M=(\tau,\sigma,\mu),~~N=(\tau',\sigma',\nu),\nn\\
&&d\tau^2=\sum_\sigma\eta_{\mu\nu}dx^\mu_\sigma(\tau) dx^\nu_\sigma(\tau),
\eea
 one can write
\bea
&&\Delta_{MN}=\theta(\tau-\tau')~[\hat{x}^-_M,\hat{x}^+_N]+\theta(\tau'-\tau)~[\hat{x}^-_N,\hat{x}^+_M].
\eea
and for a brane-like spacetime one has
\bea
&&[\hat{x}_M,\hat{x}_N]=i\theta_{MN},~~M=(u_a,\mu),~~N=(u'_a,\nu),\nn\\
&&\Delta_{MN}=\theta(u_0-u'_0)~[\hat{x}^-_M,\hat{x}^+_N]+\theta(u'_0-u_0)~[\hat{x}^-_N,\hat{x}^+_M].
\eea
\end{itemize}

\section{Hamilton-Jacobi theory}
Here is an example of how solutions of differential equations may be found using knowledge of their symmetries.

 Consider the configurations or canonical ``flow'' parameter $\ld$ of (\ref{spotential}) and replace $q$ by $x\in \mathbb{R}^{d+1}$, then the simplectic potential and $\ld$-flow (with generating function $\J$) equations are given by
\bea
&&A(x,p)=p_\mu dx^\mu=p^0dt-p_idx^i,~~~~\J=\J(x,p)\nn\\
&&{dx^\mu\over d\ld}={\del \J\over\del p^\mu},~~{dp^\mu\over d\ld}=-{\del \J\over\del x^\mu},~~{d\J\over d\ld}=0,
\eea
where the Poisson bracket is given by $\{f,g\}_1={\del f\over \del p^\mu}{\del g\over \del x^\mu}-{\del g\over \del p^\mu}{\del f\over \del x^\mu}$.

Choosing the flow parameter $\ld$ to be timelike ($\J\ra H$), ie. ${dt\over d\ld}=1={\del H\over\del p^0}$, implies that $\J=H=p^0+h(t,x^i,p^i)=p^0(t)+h(t,x^i(t),p^i(t))$ and the equations of motion become
\bea
&&{dx^i\over dt}={\del H\over\del p^i}={\del h\over\del p^i},~~{dp^i\over dt}=-{\del H\over\del x^i}=-{\del h\over\del x^i},~~{dp^0\over dt}=-{\del H\over\del t}=-{\del h\over\del t}.\nn
\eea
Thus for any given $F=F(x,p)=F(t,x^i(t),p^i(t))$,
\bea
&&{d F\over dt}=\{H,F\}_1={\del H\over \del p^\mu}{\del F\over \del x^\mu}-{\del F\over \del p^\mu}{\del H\over \del x^\mu}={\del F\over\del t}+{\del F\over \del p^i}{\del h\over \del x^i}-{\del h\over \del p^i}{\del F\over \del x^i}\nn\\
&&~~~~= {\del F\over\del t}+\{h,F\},~~~~~~\{f,g\}={\del f\over \del p^i}{\del g\over \del x^i}-{\del g\over \del p^i}{\del f\over \del x^i}.
\eea

Note that $H=p^0+h$ defines a hypersurface since $H$ is a constant of motion.
Recall that a canonical transformation $(t,x^i,p^i,p^0)\ra (\td{t},\td{x}^i,\td{p}^i,\td{p}^0)$ is one where
\bea
&&\td{A}(\td{x},\td{p})=\td{p}^0d\td{t}-\td{p}_id\td{x}^i\eqv A(x,p)+ dW=p^0dt-p_idx^i+dW,\nn\\
&&~~~~\td{H}=\td{p}^0+\td{h}=\td{p}^0(\td{t})+\td{h}(\td{t},\td{x}^i(\td{t}),\td{p}^i(\td{t})),\nn\\
&&{d F\over d\td{t}}={\del F\over\del \td{t}}+\{\td{h},F\},~~F=F(\td{t},\td{x}^i(\td{t}),\td{p}^i(\td{t})),\nn\\
&&{d \td{x}^i\over d\td{t}}={\del \td{h}\over\del \td{p}^i},~~{d \td{p}^i\over d\td{t}}=-{\del \td{h}\over\del \td{x}^i},~~~~~~\{f,g\}={\del f\over \del \td{p}^i}{\del g\over \del \td{x}^i}-{\del g\over \del \td{p}^i}{\del f\over \del \td{x}^i}.
\eea
That is, the transformation preserves the symplectic form  (and hence the Poisson bracket) while changing the symplectic potential by a total differential.

It may be possible to choose the function $W$ such that\\ $\td{t}=t~~(\txt{synchronization}),~~\td{H}=0$~~$(~\td{p}^0=-\td{h}~),~~~~\td{p}^0-p^0=-h$ in which case $\td{x}^i,\td{p}^i$ become (arbitrary) constants of motion given by
\bea
&&\td{p}^i=-{\del W\over\del \td{x}^i},~~{p}^i=-{\del W\over\del {x}^i},~~0={\del W\over\del \td{p}^i},\nn\\
&&\td{p}^0-p^0={\del W\over\del t}=-h(t,x^i,{\del W\over\del x^i}).\nn\\
&&W=W(t,x^i,\td{x}^i).
\eea
If~ ${\del h\over\del t}=0~~(ie.~~h(t,x,p)=h(x,p))$~ then we may write
 \bea
&&W(t,x^i,\td{x}^i)=S(x^i,\td{x}^i,\td{p}^i)-\al(\td{x}^i,\td{p}^i) t,~~~~\al(\td{x}^i,\td{p}^i)=h(x^i,{\del S\over\del x^i}),\nn\\
&&\td{p}^j=-{\del S(x^i(t),\td{x}^i,\td{p}^i)\over\del\td{x}^j}+{\del\al(\td{x}^i,\td{p}^i)\over\del \td{x}^j} t,\nn\eea
where the constants depend on initial conditions thus
\bea
&&\al=\al(\td{x}^i,\td{p}^i)=h(x^i(0),p^i(0))\eqv h_0,\nn\\
&&\td{x}^j=\td{x}^j(x^i(0),p^i(0)),~~~~\td{p}^j=\td{p}^j(x^i(0),p^i(0)).
\eea
The choice of how the $\td{x}^i$ and $\td{p}^i$ depend on $x^i(0)$ and $p^i(0)$ is a matter of convenience and one can for example choose $\td{x}^i=x^i(0),~~\td{p}^i=p^i(0)$. In this case\\ $W(t,x^i,\td{x}^i)=S(x^i,x^i(0),p^i(0))-h(x^i(0),p^i(0)) t$ where we need to remember that
\bea
&&{\del W(t,x^i,\td{x}^i)\over \del \td{p}^i}={\del S(x^i,\td{x}^i,\td{p}^i)\over\del\td{p}^i}-{\del \al(\td{x}^i,\td{p}^i)\over\del \td{p}^i} t=0\nn\\
&&\Ra~~{\del W(t,x^i,x^i(0))\over \del p^i(0)}={\del S(x^i,x^i(0),p^i(0))\over\del p^i(0)}-{\del h(x^i(0),p^i(0))\over\del p^i(0)} t=0.\nn\\
\eea

For the case $h=\vec{p}^2+V(\vec{x})$, one has
\bea
&&\al(\td{x}^i,\td{p}^i)=h(x^i(0),p^i(0))=({\del S\over\del \vec{x}})^2+V(\vec{x}),\nn\\
&&\vec{p}^2=({\del S\over\del \vec{x}})^2=({\del S\over\del r})^2+{1\over r^2\sin^2\theta}({\del S\over\del \theta})^2+{1\over r^2}({\del S\over\del \vphi})^2=p_r^2+{1\over r^2\sin^2\theta}p_\theta^2+{1\over r^2}p_\vphi^2\nn\\
\eea so that
\bea
&&\vec{p}={\del S_u(x^i,\td{x}^i,\td{p}^i)\over\del \vec{x}}=\vec{u}(x^i,\td{x}^i,\td{p}^i)\sqrt{\al(\td{x}^i,\td{p}^i)-V(\vec{x})},~~\vec{u}^2=1,\nn\\
&&S_u(x^i,\td{x}^i,\td{p}^i)=\int_{x^i(0)}^{x^i}d\vec{y}\cdot\vec{u}(y^i,\td{x}^i,\td{p}^i) \sqrt{\al(\td{x}^i,\td{p}^i)-V(\vec{y})}+S_u(x^i(0),\td{x}^i,\td{p}^i),\nn\\
\eea
where a convenient choice of integration path depends on the choice of $\vec{u}$ whose set is as large as $SO(d)$ for $x^i\in \mathbb{R}^d$. Similarly for the case $h=\sqrt{\vec{p}^2+m^2}+V(\vec{x})$ one obtains
{\footnotesize\bea
&&S_u(x^i,\td{x}^i,\td{p}^i)=\int_{x^i(0)}^{x^i}d\vec{y}\cdot\vec{u}(y^i,\td{x}^i,\td{p}^i) \sqrt{[\al(\td{x}^i,\td{p}^i)-V(\vec{y})]^2-m^2}+S_u(x^i(0),\td{x}^i,\td{p}^i),\nn\\
&&W_u(t,x^i,\td{x}^i)=\int_{x^i(0)}^{x^i}d\vec{y}\cdot\vec{u}(y^i,\td{x}^i,\td{p}^i) \sqrt{[\al(\td{x}^i,\td{p}^i)-V(\vec{y})]^2-m^2}+S_u(x^i(0),\td{x}^i,\td{p}^i)\nn\\
&&~~~~-\al(\td{x}^i,\td{p}^i)t.\nn
\eea}
If we set $\td{p}=t_0=-{\del W\over\del\al}$ then we obtain
\bea
\int_{x^i(0)}^{x^i}d\vec{y}\cdot\vec{u}(y^i,\td{x}^i,\td{p}^i) {\al(\td{x}^i,\td{p}^i)-V(\vec{y})\over\sqrt{[\al(\td{x}^i,\td{p}^i)-V(\vec{y})]^2-m^2}}=t-t_0.
\eea
For bounded motion in a central potential where
\bea
&&h=\sqrt{\vec{p}^2+m^2}+V(r)= \sqrt{p_r^2+{L^2\over r^2}+m^2}+V(r)=\vep,\nn\\
&&L^{ij}={1\over\sqrt{2}}(x^ip^j-x^jp^i),~~{dL^{ij}\over dt}=\{h,L^{ij}\}=0,\nn\\
&&L^2=L_{ij}L^{ij}=\vec{x}^2\vec{p}^2-(\vec{x}\cdot\vec{p})^2=r^2(\vec{p}^2-p^2_r),~~p_r={\vec{x}\over r}\cdot \vec{p},
\eea
 the solutions to\\ $p_r=-{\del S(r,\vep,L)\over\del r}=-\sqrt{[\vep-V(r)]^2-m^2-{L^2\over r^2}}=0$ give the extreme bounds $r_-,r_+,...$ of the orbit and thus with $t_0=-{\del W\over\del\vep}$
\bea
&&\int_{r(t_0)}^{r(t)}dr' {\vep-V(r')\over\sqrt{[\vep-V(r')]^2-m^2-{L^2\over r'{}^2}}}=t-t_0\nn\\
&&\Ra \int_{r_-}^{r_+}dr {\vep-V(r)\over\sqrt{[\vep-V(r)]^2-m^2-{L^2\over r^2}}}=T.
\eea
gives a measure of the period(s) $T$ of the motion.

For the simple time dependent case $h=p^2+xt$, one can write \\
$W(t,x)=a(t)+xb(t)+x^2c(t)$ and then solve for the t dependent coefficients using \\
${\del W\over\del t}=-h(t,x^i,{\del W\over\del x^i})$.

\section{(Orbital) angular momentum and spherical functions}
Consider the angular momentum operator given by $L_{ij}={1\over\sqrt{2}}(x_i\del_j-x_j\del_i) $, then
{\small\bea
&&L^2=r^2\vec{\del}^2-(D-2)\vec{x}\cdot\vec{\del}-(\vec{x}\cdot\vec{\del})^2\nn\\
&&~~~~=r^2\vec{\del}^2-(D-2)r{\del\over\del r}-r{\del\over\del r}r{\del\over\del r}=r^2\vec{\del}^2-({D}-1)r{\del\over\del r}-r^2{\del^2\over\del r^2}\nn\\
&&\Ra~~\vec{\del}^2={\del^2\over\del r^2}+{{D}-1\over r}{\del\over\del r}+{L^2\over r^2}={1\over r^{{D}-1}}{\del\over\del r}r^{{D}-1}{\del\over\del r}+{L^2\over r^2}.\nn\\
&&\vec{\del}^2\psi=({1\over r^{{D}-1}}{\del\over\del r}r^{{D}-1}{\del\over\del r}+{L{}^2\over r^2})\psi={1\over r^{{D}-1\over 2}}({\del^2\over\del r^2}+{L^2-(1-D)(3-D)/4 \over r^2})(r^{{D}-1\over 2}\psi),\nn\\
&&{1\over f}{d\over dx}f{d\over dx}\psi={1\over f^{1\over 2}}\{~{d^2\over dx^2}+f^{-{1\over 2}}(f(f^{-{1\over 2}})')'~\}(f^{1\over 2}\psi)={1\over f^{1\over 2}}\{~{d^2\over dx^2}+{f'{}^2-2ff''\over 4f^2}~\}(f^{1\over 2}\psi)\nn\\
&&~~\eqv {1\over f^{1\over 2}}\{~{d^2\over dx^2}+\bi(f)~\}(f^{1\over 2}\psi),\nn\\
\label{lap-sym-trans}&&{1\over g}{d\over dx}g{d\over dx}\psi={\sqrt{f\over g}}~\{~{1\over f}{d\over dx}f{d\over dx}+\bi(g)-\bi(f)~\}~({\sqrt{g\over f}}~\psi).
\eea}
Therefore, a function satisfying \\
$\vec{\del}^2G(r)=\delta^{D}(\vec{r})$~ is
~$G(r)=r^{\al},~~\al=3-{D},~~L^2=-\al(\al+{D})=-3(3-{D}),~~{D}\neq 2$.

\subsection{$\del^2$ in a minimally coupled system?}

In arbitrary orthogonal coordinates $\vec{x}=\vec{x}(u^a)$, the divergence operator is given by
\bea
&&\del_iV^i =\frac{1}{\sqrt{h}}\del_a(\sqrt{h}V^a)=
\frac{1}{\sqrt{h}}\del_a(\frac{\sqrt{h}}{h^2_a}e^i_aV^i)\eqv \frac{1}{\sqrt{h}}\del_a(\frac{\sqrt{h}}{h^2_a}V_a),\nn\\
&&h=\prod_a h_a,~~ h_a={|\vec{e}_a|},~~\vec{e}_a=e^i_a={\del x^i\over\del u^a}.
\eea
For a $U(1)$ gauge covariant divergence $(\del_i+A_i)V^i$ one may simply make the replacement \\
$\sqrt{h}\ra U\sqrt{h} $ so that
\bea
(\del_i+A_i)V^i=\frac{1}{U\sqrt{h}}\del_a(\frac{U\sqrt{h}}{h^2_a}V_a),
\eea
where $\del_iU=A_iU$.
Writing $P_i={\del_i}+A_i,~~J_{ij}=x_iP_j-x_jP_i$,
\bea
&&[P_i,x_j]=\delta_{ij},~~[P_i,P_j]=\del_iA_j-\del_jA_i,~~[x^2,J_{ij}]=0,~~[P^2,J_{ij}]\neq 0\nn\\
&&\vec{P}^2={D-2\over r^2}\vec{x}\cdot \vec{P}+{1\over r^2}(\vec{x}\cdot \vec{P})^2+{J^2\over r^2}\nn\\
&&~~~~={\del^2\over\del r^2}+{D-1+2rA_r\over r}{\del\over\del r}+{(D-2)rA_r+r\del_r(rA_r)+r^2A_r^2+J^2\over r^2}\nn\\
&&~~~~= {1\over f(r)}{\del\over\del r}f(r){\del\over\del r}+{(D-2)rA_r+r\del_r(rA_r)+r^2A_r^2+J^2\over r^2},\nn\\
&&f(r)=e^{\int^rdr'{D-1+2r'A_{r'}\over r'}},~~rA_r=\vec{x}\cdot\vec{A}(\vec{x}),~~\del_r={\del\over\del r}.
\eea
The condition ~$[x^2,J_{ij}]=0$~ is essential for the independence of the angular part from the radial parts.
Notice that for the case where $\vec{x}\cdot\vec{A}=const$, such as \\
$A^i=\al{x^i\over r^2}+{1\over 2}B^{[ij]}(\vec{x})~x^j=-\al\del_i({1\over r})+{1\over 2}B^{[ij]}(\vec{x})~x^j$, the expression for $P^2$ simplifies drastically and is almost equivalent to $\del^2$ except that the angular momentum $J_{ij}$ (and hence $J^2$) is slightly modified.

In the case $A_i={1\over 2}B_{ij}x^j,~~B_{ij}=-B_{ji}=const$ one also has $\del_iA^i=0$ and so expanding $P^2$ implies that $\del^2$ can be written as
\bea
\del^2={1\over r^{D-1}}{d\over dr}r^{D-1}{d\over dr}+{J^2\over r^2}-{1\over 2}B_{ij}J^{ij}+{1\over 4}B_{ki}B_{kj}x_ix_j.
\eea

\subsection{Spherical eigenfunctions}
In spherical coordinates (with $\tht_0=0$)
{\footnotesize\bea
&&x_0=y=r\cos\tht_0\sin\tht_1\sin\tht_2\sin\tht_3\sin\tht_4\sin\tht_5...\sin\tht_{{D}-1},\\
&&x_1=x=~~~~~~~r\cos\tht_1\sin\tht_2\sin\tht_3\sin\tht_4\sin\tht_5...\sin\tht_{{D}-1}\\
&&x_2=z=~~~~~~~~~~~~~~~r\cos\tht_2\sin\tht_3\sin\tht_4\sin\tht_5...\sin\tht_{{D}-1}\\
&&x_3=~~~~~~~~~~~~~~~~~~~~~~~~~~~~r\cos\tht_3\sin\tht_4\sin\tht_5...\sin\tht_{{D}-1}\\
&&x_4=~~~~~~~~~~~~~~~~~~~~~~~~~~~~~~~~~~~r\cos\tht_4\sin\tht_5...\sin\tht_{{D}-1}\\
&&...\\
&&x_k=~~~~~~~~~~~~~~~~~~~~~~~~~~~~~~~~r\cos\tht_k\sin\tht_{k+1}...\sin\tht_{{D}-1}\\
&&...\\
&&x_{{D}-2}=~~~~~~~~~~~~~~~~~~~~~~~~~~~~~~~~~~~~~~~r\cos\tht_{{D}-2}\sin\tht_{{D}-1}\\
&&x_{{D}-1}=~~~~~~~~~~~~~~~~~~~~~~~~~~~~~~~~~~~~~~~r\cos\tht_{{D}-1}
\eea}
the Laplacian is given by
{\tiny
\bea
&&\vec{\del}^2={1\over r^{{D}-1}}{\del\over\del r}r^{{D}-1}{\del\over\del r}+{1\over r^2}\sum^{{D}-1}_{k=1}{1\over \sin^2\theta_{k+1}\sin^2\theta_{k+2}...\sin^2\theta_{{D}-1}}{1\over\sin^{k-1}\theta_k }{\del\over\del\theta_k}\sin^{k-1}\theta_k{\del\over\del\theta_k}\nn\\
&&={1\over r^{{D}-1}}{\del\over\del r}r^{{D}-1}{\del\over\del r}+{1\over r^2}\{{1\over \sin^2\theta_{2}...\sin^2\theta_{{D}-1}}{\del^2\over\del\theta_1^2}+{1\over \sin^2\theta_{3}...\sin^2\theta_{{D}-1}}{1\over\sin\theta_2 }{\del\over\del\theta_2}\sin\theta_2{\del\over\del\theta_2}\nn\\
&&+{1\over \sin^2\theta_{4}...\sin^2\theta_{{D}-1}}{1\over\sin^{2}\theta_3 }{\del\over\del\theta_3}\sin^{2}\theta_3{\del\over\del\theta_3}+{1\over \sin^2\theta_{5}...\sin^2\theta_{{D}-1}}{1\over\sin^{3}\theta_4 }{\del\over\del\theta_4}\sin^{3}\theta_4{\del\over\del\theta_4}\nn\\
&&+...+\nn\\
&&{1\over \sin^2\theta_{{D}-1}}{1\over\sin^{{D}-3}\theta_{{D}-2} }{\del\over\del\theta_{{D}-2}}\sin^{{D}-3}\theta_{{D}-2}{\del\over\del\theta_{{D}-2}}+{1\over\sin^{{D}-2}\theta_{{D}-1} }{\del\over\del\theta_{{D}-1}}\sin^{{D}-2}\theta_{{D}-1}{\del\over\del\theta_{{D}-1}}\}\nn\\
\eea
}
Now any operator $Q$ that has a complete set of eigenfunctions \\
$\Q=\{\psi,~Q\psi=\ld \psi,~~\ld\in \mathbb{C}\}$  on a space $X$ (ie. every element in $\F(X)$ can be expressed as a linear combination of the elements of $\Q,~~\F(X)=\txt{Span}\Q$) can be defined in $X$ and/or $\F(X)$ only up to similarity transformations since the spectrum of $PQP^{-1}$ is isomorphic to the spectrum of $Q$. The Laplacian $\del^2$ is such an operator.
One can make the identification
\bea
&&L^2=\del^2_\Omega=\sum^{{D}-1}_{k=1}{1\over \sin^2\theta_{k+1}\sin^2\theta_{k+2}...\sin^2\theta_{{D}-1}}{1\over\sin^{k-1}\theta_k }{\del\over\del\theta_k}\sin^{k-1}\theta_k{\del\over\del\theta_k}\nn\\
\eea

The solution in a spherically symmetric potential $f(r)$ may be separated as follows
{\footnotesize
\bea
&&(\del^2+f(r))\psi(\vec{x})=0,\nn\\
&&\psi(\vec{x})=\sum_{m_1...m_{{D}-2}{\td{L}}}\al_{m_1...m_{{D}-2}{\td{L}}}~\Theta^{m_1}_1(\theta_1)~\Theta^{m_1m_2}_2(\theta_2)~...~\Theta^{m_{k-1}m_k}_k(\theta_k)~...~\Theta^{m_{{D}-2}{\td{L}}}_{{D}-1}(\theta_{{D}-1})\times\nn\\
&&~~~~~R^{\td{L}}(r)\eqv \sum_{m_1...m_{{D}-2}{\td{L}}}\al_{m_1...m_{{D}-2}{\td{L}}}~Y^{m_1...m_{{D}-2}{\td{L}}}(\theta_1,...,\theta_{D-1})~R^{\td{L}}(r),\nn\\
&&{\del^2\over\del\theta_1^2}\Theta^{m_1}_1(\theta_1)=m_1^2\Theta^{m_1}_1(\theta_1),\nn\\
&&({m_1^2\over\sin^2\theta_2}+{1\over\sin\theta_2}{\del\over\del\theta_2}\sin\theta_2{\del\over\del\theta_2})\Theta^{m_1m_2}_2(\theta_2)=m^2_2\Theta^{m_1m_2}_2(\theta_2),\nn\\
&&~~~~....\nn\\
&&({m_{k-1}^2\over\sin^2\theta_k}+{1\over\sin^{k-1}\theta_k}{\del\over\del\theta_k}\sin^{k-1}\theta_k{\del\over\del\theta_k})\Theta^{m_{k-1}m_k}_k(\theta_k)=m^2_k\Theta^{m_{k-1}m_k}_k(\theta_k),\nn\\
&&~~~~....\nn\\
&&({m_{{D}-2}^2\over\sin^2\theta_{{D}-1}}+{1\over\sin^{{D}-2}\theta_{{D}-1}}{\del\over\del\theta_{{D}-1}}\sin^{{D}-2}\theta_{{D}-1}{\del\over\del\theta_{{D}-1}})\Theta^{m_{{D}-2}{\td{L}}}_{{D}-2}(\theta_{{D}-1})\nn\\
&&~~~~={\td{L}}^2\Theta^{m_{{D}-2}{\td{L}}}_{{D}-1}(\theta_{{D}-1}),\nn\\
&&({1\over r^{{D}-1}}{\del\over\del r}r^{{D}-1}{\del\over\del r}+{{\td{L}}^2\over r^2}+f(r))R^{\td{L}}(r)=0.
\eea
}
Writing ${s_k}=\cos\theta_k$ the $k$th equation is

\bea
&&\{~{1\over (1-{s_k}^2)^{k\over 2}}{d\over d{s_k}}(1-{s_k}^2)^{k\over 2}{d\over d{s_k}}+{m_{k-1}^2-(1-{s_k}^2)m^2_k\over (1-{s_k}^2)^2}~\}~\Theta^{m_{k-1}m_k}_k({s_k})=0,\nn
\eea
\bea
&&\label{nlegendre}{1\over (1-{s_k}^2)^{k\over 4}}\{{d^2\over d{s_k}^2}+{4k-k^2+m_{k-1}^2-(1-{s_k}^2)(2k-k^2+m^2_k)\over (1-{s_k}^2)^2}\}\times\nn\\
&&~~~~[(1-{s_k}^2)^{k\over 4}\Theta^{m_{k-1}m_k}_k]=0.
\eea
The Legendre equation
\bea
&&((1-x^2)y')'+(l(l+1)-{m^2\over 1-x^2})y=0,\nn\\
&&{1\over (1-x^2)^{1\over 2}}({d^2\over dx^2}+{4-m^2+l(l+1)(1-x^2)\over (1-x^2)^2})((1-x^2)^{1\over 2}y)=0\nn\\
\eea
has solutions
\bea
y_{ml}(x)={(-1)^m\over 2^ll!}(1-x^2)^{m\over 2}{d^{l+m}\over dx^{l+m}}(x^2-1)^l,~~~~-l\leq m\leq l.
\eea
To get the solution to (\ref{nlegendre}) one should replace ~$m^2$~ with ~$(2-k)^2-m^2_{k-1}$~ and replace ~$l(l+1)$~ with  ~$-(2k-k^2+m_k^2)$.

The eigenvalue equation with $f(r)={\beta\over r}$ may be written as
\bea
&&(\vec{\del}^2+{\beta\over r}-\ld)R=({1\over r^{{D}-1}}{\del\over\del r}r^{{D}-1}{\del\over\del r}+{L{}^2\over r^2}+{\beta\over r}-\ld)R\nn\\
&&~~~~={1\over r^{{D}-1\over 2}}({\del^2\over\del r^2}+{L^2-(1-D)(3-D)/4 \over r^2}+{\beta\over r}-\ld)(r^{{D}-1\over 2}R)\nn\\
&&~~~~={1\over r^{{D}-1\over 2}}({\del^2\over\del r^2}+{\tilde{L}^2\over r^2}+{\beta\over r}-\ld)(r^{{D}-1\over 2}R)  =0,\nn\\
\eea

As $r\ra 0$ the term in $\td{L}^2$ dominates and as $r\ra \infty$ the term in $\ld$ dominates. These suggest
the change of variables \\
$F(r)=r^{D-1\over 2}R(r)=r^{\td{l}+1} e^{-r\sqrt{\ld}}~u(r),~~-\td{L}^2=\td{l}(\td{l}+1)$ that may simplify the equation.
\bea
&&(r^{\td{l}+1} e^{-r\sqrt{\ld}}~u(r))''=[(-\sqrt{\ld}+{\td{l}+1\over r}+{u'\over u})F]'\nn\\
&&~~~~=[(-\sqrt{\ld}+{\td{l}+1\over r}+{u'\over u})^2-{\td{l}+1\over r^2}+{u''\over u}-{u'^2\over u^2}]F\nn\\
&&~~~~=[\ld+{\td{l}(\td{l}+1)\over r^2}-{2(\td{l}+1)\sqrt{\ld}\over r}+(-2\sqrt{\ld}+{2(\td{l}+1)\over r}){u'\over u}+{u''\over u}]F\nn\\
&&\{~{d^2\over dr^2}+(-2\sqrt{\ld}+{2(\td{l}+1)\over r}){d\over dr}+{\beta-2(\td{l}+1)\sqrt{\ld}\over r}~\}u=0\nn\\
&&\{~r{d^2\over dr^2}+(-2r\sqrt{\ld}+2(\td{l}+1)){d\over dr}+\beta-2(\td{l}+1)\sqrt{\ld}~\}u=0\nn\\
\eea
Let the solution be in the form $u=\sum_{k=N}^{\infty}\al_k r^k$, then
\bea
&&\sum_{k=N}^\infty k(k-1)\al_k r^{k-1}-2\sqrt{\ld}\sum_{k=N}^\infty k\al_k r^{k}\nn\\
&&~~~~+2(\td{l}+1)\sum_{k=N}^\infty k\al_k r^{k-1}+({\beta-2(\td{l}+1)\sqrt{\ld}})\sum_{k=N}^\infty \al_k r^{k}=0\nn\\
&&\sum_{k=N}^\infty k(k-1)\al_k r^{k-1}-2\sqrt{\ld}\sum_{k=N}^\infty k\al_k r^{k}\nn\\
&&~~~~+2(\td{l}+1)\sum_{k=N}^\infty k\al_k r^{k-1}+({\beta-2(\td{l}+1)\sqrt{\ld}})\sum_{k=N}^\infty \al_k r^{k}=0\nn\\
&&\sum_{k=N}(k(k-1)+2k(\td{l}+1))\al_k r^{k-1}+\sum_{k=N}(\beta-2\sqrt{\ld}(k+\td{l}+1))\al_k r^k=0,\nn\\
&&(N(N-1)+2N(\td{l}+1))\al_N r^{N-1}\nn\\
&&~~~~+\sum_{k=N-1}(k+1)(k+2(\td{l}+1))\al_{k+1} r^{k}+\sum_{k=N}(\beta-2\sqrt{\ld}(k+\td{l}+1))\al_k r^k=0,\nn\\
\eea
With $N=0$,
\bea
\al_{k+1} =-{\beta-2\sqrt{\ld}(k+\td{l}+1)\over (k+1)(k+2(\td{l}+1))}\al_k\eqv [k]_{\beta\ld\td{l}}~\al_k=[k]_{\beta\ld\td{l}} !~\al_0,
\eea
and the series terminates (eg. when bound states, $|R(r)|<\infty~~ \forall r,~r=|\vec{x}_1-\vec{x}_2|$, are desired) at
\bea
&&\ld\ra\ld_{kl}={\beta^2\over 4(k+\td{l}+1)^2},~~~~E_{12}^{k\td{l}}=-{2m_1m_2\over \hbar^2(m_1+m_2)}({q_1q_2\over 4\pi\vep_0})^2{1\over 4(k_{12}+\td{l}_{12}+1)^2}.\nn\\
&&\ld=-{2m\over\hbar^2}E,~~~~\beta={2m\over\hbar^2}{qq\over 4\pi\vep_0}.\nn\\
&&R(r)=\al_0~r^{1-D\over 2}r^{\td{l}+1}e^{-r\sqrt{\ld}}\sum_{k=0}^\infty [k]_{\beta\ld\td{l}} !~ r^k\nn\\
\label{radhom}&&~~~~\sr{D\ra\infty}{\ral}~r^{D\over 2}e^{-r\sqrt{\ld}}~e^{r\sqrt{\ld}}=r^{D\over 2},\nn\\
&&[k]_{\beta\ld\td{l}}={2\sqrt{\ld}(k+\td{l}+1)-\beta\over (k+1)(k+2(\td{l}+1))}.
\eea
where
\bea
\td{L}^2=L^2-(1-D)(3-D)/4~~\Ra~~\td{l}(\td{l}+1)=l(l+1)-{(D-1)(D-3)\over 4}.\nn
\eea
Thus (\ref{radhom}) implies that radial and angular motion decouple and motion becomes free from any interactions. Therefore systems that can travel in extra dimensions can avoid interactions with those confined to fewer dimensions.

In spectroscopic notation
\bea
E_n=-{2m\over \hbar^2}({qq\over 4\pi\vep_0})^2{1\over 4n^2},~~~~n=k+\td{l}+1~~\Ra~~0\leq \td{l}\leq n-1.
\eea
If a constant weak $B_{ij}$-field,~~$B^2\ll B~$~, is included, then $\ld\ra \ld+{e\over\hbar}j^BB$ where $j^B$ is the angular momentum quantum number satisfying
\bea
&&B_{ij}J_{ij}\psi_{j^B}=j^B\psi_{j^B}.\nn\\
\eea
Therefore
\bea
&&E_{nj^B}={\hbar^2\over 2m}{eB\over 2\hbar}j^B-{2m\over \hbar^2}({qq\over 4\pi\vep_0})^2{1\over 4n^2}= {e\hbar B\over 4m}j^B-{2m\over \hbar^2}({qq\over 4\pi\vep_0})^2{1\over 4n^2}\nn\\
&&n=k+\td{j}+1~~\Ra~~0\leq \td{j}\leq n-1.
\eea
In spectroscopic notation, quantum orbits (known as orbitals) are label are given as
\bea
&&n^{2s+1}l_j,~~~~|l-s|\leq j\leq l+s,~~n=k+j+1.\nn\\
&&l=0,1,2,3,4,5,6~...\nn\\
&&~~~~\eqv S ~(\txt{sharp}), ~P~(\txt{principal}),~D ~(\txt{diffused}),~F ~(\txt{fundamental}),~G,~H,~...\nn\\
\eea
where the possible total angular momentum quantum numbers \\
$\vec{J}^2=(\vec{L}+\vec{S})^2=j(j+1)$ follow from the vector inequality
\bea
(|\vec{L}|-|\vec{S}|)^2\leq (\vec{L}+\vec{S})^2\leq (|\vec{L}|+|\vec{S}|)^2
\eea
and
\bea
&&\vec{L}^2=l(l+1),~~\vec{S}^2=s(s+1).
\eea

One observes that to label representations
 \bea
 &&R:SO(n)\ra GL(N)\ra \O(\F(\mathbb{R}^n)),~~g\vartriangleright|m_1,...,m_{n-2},l>\nn\\
 &&~~~~=\sum_{m'_1,...,m'_{n-2},l'}|m'_1,...,m'_{n-2},l><m'_1,...,m'_{n-2},l'|g|m_1,...,m_{n-2},l>\nn\\
 &&~~~~\mapsto R(g)\vartriangleright Y^{m_1...m_{{n}-2}{\td{L}}}\nn\\
 \eea
for $SO(n)$, one can consider the sequence
$SO(n)\subset SO(n-1)...\subset SO(3)\subset SO(2)$ with their respective quadratics Casimirs $L^2_{(n)},L^2_{(n-1)}...L^2_{(3)},L^2_{(2)}$ all commuting and therefore can serve as labels. These Casimirs may be identified with the numbers $m_1,m_2,...,m_{n-2},\td{L}$ given in the spherically symmetric equation above. \\
Since $L^2_{(k)}\leq L^2_{(k+1)}$, numbers may be assigned such that \\
$-|l_{k+1}|\leq l_{k}\leq |l_{k+1}|,~~k=2,3,...,n$, where $L^2_{(k)}=L^2_{(k)}(l_{k})=l_{k}(l_{k}+1)$. States may be labeled as $|l_{2}l_{3}...l_{n-1}l_{n}\rangle \eqv |m_1,...,m_{n-2},l>$.

\chapter{Variation principle and classic symmetries}

\section{Division of spaces}
If $A,B$ are two spaces, then their quotient is given by the set of isomorphic maps
\bea
A/B=\{f;~f:B\ra A,~~f(a)=f(b)~\Lra~ a=b \}.\nn
\eea
If one wishes the maps to be parallel then the following condition may be included: $f(a)=g(a)~~\Lra~~f=g~~\forall~f,g\in A/B$.

If $A\simeq F\times B$ then one says that $A$ is a bundle of fibers $\{F\}$ (or fiber bundle $\pi:A\ra B$) over $B$ and  $A/B$ is the space of sections of $A$ by $B$. The partial equality ~$\simeq$~ means ``similar to'' and its actual meaning depends on the context.

\subsection{Spectrum of a group algebra}
In the case of groups, if $A=G$ a group and subgroup $F=H\subset G$,~ with an action $\rho:G\times H\ra H,~~(g,h)\ra \rho(g,h)=gh$;~ eg. $G=SO(n+1),~~H=SO(n)$, then $B=G/H\simeq \{gH,~g\in G\}\simeq \{Hg,~g\in G\}$. Thus one has a fiber bundle structure $G\simeq H\times {G\slash H}$.

If a subgroup $H$ is not normal; ie.~ $gH=Hg$ does not hold for at least one $g\in G$, then a normal subgroup $H_G$ may be constructed from it as
\bea
H_G=\{ghg^{-1},~~G\in G,~h\in H\}
\eea
since $\td{g}~gHg^{-1}=\td{g}g H(\td{g}g)^{-1}~\td{g}$. One can also define a commuting element $S_0(s)$ for any element $s\in G$,~ $S_0(s)$ ~being an element of the group algebra
\bea
&&\A_G\simeq \{a=a(\al)=\int_{g\in G} d\mu(g)~\al(g)~g,~~\al:G\ra \mathbb{C}^N\},\nn\\
&&a(\al)a(\beta)=a(\al\ast\beta),\nn\\
&&(\al\ast\beta)(g)=\int_{x\in G} d\mu(x)~\al(gx^{-1})\beta(x)=\int_{x\in G} d\mu(x)~\al(x)\beta(x^{-1}g),\nn
\eea
where $\mu$ is the left-translation invariant measure\footnote{Section \ref{haar-measure}.} on $G$.
{\footnotesize
\bea
S_0(s)=\{h_0=\int_{g\in G} d\mu(g)~gsg^{-1},~~s\in H\},~~~~\int_{g\in G} d\mu(g'g)=\int_{g\in G} d\mu(g)~~\forall g'\in G.\nn\\
\eea
}
The number of unique such elements is equal to the number of conjugacy classes of $G$ since
$S_0(g)=S_0(hgh^{-1})$. That is, the center  $Z(\A_G)$ of $\A_G$ is as large as the set of conjugacy classes $\{[g],~g\in G\}$.
\bea
&&Z(\A_G)= \{S_0([g]),~~g\in G\}\simeq \{[g],~g\in G\}.\nn\\
&&Z(G)=G\cap Z(\A_G)
\eea
and the irreducible representations of $G$ or of $\A_G$ are parametrized by the spectrum\footnote{Section \ref{spectral-theory}} $\sigma(Z(\A_G))$.

 \section{Gauge symmetry and Noether's theorem}
 A $U(1)$ gauge transformation is a continuous local transformation of the electromagnetic potential $A(x)\ra A(x)-{1\over g(x)}d g(x)$ that preserves the Maxwell Lagrangian for electromagnetism. Gauge symmetry may also be defined for an interacting theory, in which case, it may be associated to the conservation of electric charge by Noether's theorem which associates, along the classical path $\delta S=0$, a ``complete'' set of conservation laws and hence a ``complete'' set of conserved charges to any continuous global symmetry of a classical theory. A continuous symmetry of a classical theory is a transformation that changes the differential action or Lagrangian only by an exact form $\delta_{\xi} \L=d\K$~~(a canonical transformation)~ and hence does not change the equations of motion $\delta S=\delta\int\L=0$. The Noether charges $\{Q^a\}$ for a given symmetry give a canonical representation for the generators of the symmetry group. The characteristic values or spectra, which may be referred to as possible physical realizations of the Noether charges $\{Q^a\}$, correspond to the irreducible representations of the symmetry group.

 \section{Symmetry breaking/violation}

Certain internal (non spacetime) symmetries are broken by the observation that ``elementary'' particles come with different masses.
A natural way to characterize this symmetry breaking is through a procedure known as dynamical or ``spontaneous'' (implicit in general) symmetry breaking. In this procedure the physical system around it's ground configuration (lowest energy configuration) is seen to have evolved from a more symmetric system at high energy/temperature configurations (ie. high kinetic energy, referring to a situation where the kinetic terms are dominating in the Lagrangian). As the system evolves to lower energy configurations (a situation where the interaction or potential energy terms are dominating) it has more than one local minimum energy configuration to randomly/spontaneously choose from. The space of all configurations with a given local minimum of energy is known as a vacuum or a vacuum manifold.
The local extrema may be obtained by solving
\bea
&&{\del H\over\del\{\del_0\vphi\}}=0,~~{\del H\over\del\{\vphi\}}=0,\nn\\
&&H=\int d^3x~\H=\int d^3x~({\del\L\over\del\{\del_0\vphi\}}\{\del_0\vphi\}-\L),
\eea
where $\{\vphi\}$ is the collection of all fields involved.

In one case all fields take zero values in the vacuum and in this case the field theory around this vacuum retains the original symmetry and the vacuum is said to be invariant under the symmetry. In the other case, one or more of the fields assume non-zero values in the vacuum and consequently the field theory around the vacuum cannot retain all of the original symmetry and the symmetry is said to have been spontaneously broken by this supposedly spontaneous or random choice of the vacuum.

There is also empirical (explicit in general) symmetry breaking which involves the introduction of noninvariant terms into the Lagrangian in order that theoretical results (eg. calculated interaction amplitudes) agree with experimental results of certain processes observed to violate the symmetry.

\section{Action/on-shell symmetries}
At the level of the action, any two theories with the same number of degrees of freedom (dofs) are equivalent in that they can~be related by an invertible transformation \\
 $\{\vphi_1\}\ra \{\vphi_2\}=g_{12}(\{\vphi_1\}),~~S_1[\{\vphi_1\}]\ra S_2[\{\vphi_2\}]=f_{12}[\vphi_1,S_1[\vphi_1],\delta_{\vphi_1}S_1[\vphi_1],...]$.
\bea
&&S_1[\{\vphi_1\}]=\int d\mu(x)\L_1(x,\{\vphi_1\},\del\{\vphi_1\},...),\nn\\
&&S_2[\{\vphi_2\}]=\int d\mu(x)\L_2(x,\{\vphi_2\},\del\{\vphi_2\},...).
\eea
However the equations of motion
\bea
&&{\delta S_1[\{\vphi_1\}]\over \delta\{\vphi_1\}(x)}=0,~~~~{\delta S_2[\{\vphi_2\}]\over \delta\{\vphi_2\}(x)}=0
\eea
may not be invariant under the transformation (that is the two solution spaces are not isomorphic). Thus the space of all action equivalent theories $T=\{S_i[\{\vphi_i\}]\}$ for a given number of degrees of freedom has an action intertheory symmetry group $\G$ that interconnects the different theories in $T$. The usual symmetry groups of physics are ``fixed points'' of $T$.
\bea
&&\{\vphi_1\}\ra \{\vphi_2\}=g_{12}(\{\vphi_1\}),\nn\\
&&S_1[\{\vphi_1\}]\ra S_2[\{\vphi_2\}]=\al S_1[a\{\vphi_1\}+b]+\beta\nn\\
&&~~\Ra~~{\delta S_2[\{\vphi_2\}]\over \delta\{\vphi_2\}(x)}=\al~{\delta\{\vphi_1\}(y)\over\delta g_{12}(\{\vphi_1\}(x))}{\delta S_1[a\{\vphi_1\}+b]\over \delta\{\vphi_1\}(y)}=0.
\eea
where $\al,\beta,a,b$ are constants.
That is, they map an action to one that is similar to itself and for these special cases, the equations of motion are interelated even though the symmetry group of the equations of motion can be larger than that of the action. The duality symmetries of string theory arise as special cases of $\al,a\neq 1$ and $\beta,b=0$.

\section{Noether's theorem and Ward-Takahashi identities}
For simplicity we consider an action with at most first derivatives but the discussion can be extended to the case with any number of higher derivatives.

Consider a physical system described by a particular configuration (trajectory or path)~ \\
$\phi\in \{f:\D\subseteq\mathbb{R}^{d+1}\ra \E\subseteq\M=\mathbb{C}^N\}$. Assume that the dynamics of the system is determined by a least action principle with an action $S[f]=\int_\D d\mu ~\L(x,f(x),\del f(x))$. That is, among all the possible configurations $f$ marked by any given \emph{bound} \\
~$\del\E=f(\del \D)$ (ie.~$\delta f|_{f\in\del \E}\eqv \delta f(x)|_{x\in \del \D}=0$), the classical physical configurations(s) is (are) the one(s) for which the action is extremized $\delta S[f]|_{f=\phi}=0$. Therefore physically the Euler-Lagrange equations describe the only classically possible dependence(s) of $\phi$ on $x\in\D$ for any given boundary $\del \E=\phi(\del\D)$.
{\footnotesize
\bea
&&\delta S[{f}]= \int_\D d^nx~\{~\del_\mu(\delta{f}~{\del \L\over\del\del_\mu{f}})+\delta{f}~({\del \L\over\del{f}}-\del_\mu~{\del \L\over\del\del_\mu{f}} )~\}\nn\\
&&~~=\int_{x\in\del \D} dS_\mu(u)~\delta{f}(x(u))~{\del \L\over\del\del_\mu{f}}(x(u))+\int_{x\in\D} d^nx~\delta{f}(x)~({\del \L\over\del{f}}(x)-\del_\mu~{\del \L\over\del\del_\mu{f}}(x) ),\nn\\
&&dS_\mu(u)=\vep_{\mu\nu_1...\nu_{D-1}}{\del(x^{\nu_1},...,x^{\nu_{D-1}})\over\del (u^1,...,u^{D-1})}d^{D-1}u.
\eea
}
Therefore $\delta S[f]|_{f=\phi}=0,~~\delta f(x)|_{x\in\del\D}=0$ implies that
\bea
\label{euler-lagrange}{\del \L\over\del\phi}(x)-\del_\mu~{\del \L\over\del\del_\mu\phi}(x)=0~~\forall x\in \D
\eea
or simply
\bea
\label{euler-lagrange2}{\del \L\over\del\phi}-\del_\mu~{\del \L\over\del\del_\mu\phi}=0.
\eea

Now different physical observers describe the behavior of the system with different points of view, which range from the use of different coordinates (labels or parameters) $x\mapsto y$ and/or different integration domains $\D\ra \D'$ to reordering and/or rescaling/translating of the components of the field variable $\phi$ (~$\phi(x)\mapsto \phi'(x')$~) and of the Lagrangian $\L$. According to the \textbf{theory (or principle) of relativity}, these observers should still use the same least action principle and hence the same equations of motion, among other things, to describe the behavior of $\phi$.
That is, $\delta S'[f]_{f=\phi'}=0$ implies that

\bea
&&\label{euler-lagrange}{\del \L'\over\del\phi'}(x',..)-\del'_\mu~{\del \L'\over\del\del'_\mu\phi'}(x',..)=0~~\forall x'\in \D',\nn\\
&&(\del_\mu f)'(x')={\del\over\del x^\mu{}'}~f'(x')\eqv \del_\mu'f'(x').
\eea

One can check that for smooth transformations (ie. smoothly related observers), the difference between $\D$ and $\D'$ can be fully specified through a change in the integration measure of the action.
{\footnotesize
\bea
&&S[\phi,{\D}]=\int_{\D} d^nx~\L(x,\phi(x),\del\phi(x)),\nn\\
&&x\ra x'=x+\delta x,\nn\\
&&\phi(x)\ra \phi'(x')\eqv M(x,\phi(x),\del\phi(x),...)=\phi'(x')-\phi(x')+\phi(x')\eqv \bar{\delta}\phi(x')+\phi(x')\nn\\
&&~~~~\Ra~~~~\bar{\delta}\phi(x)=\phi'(x)-\phi(x)=M(x-\delta x,\phi(x-\delta x),\del\phi(x-\delta x),...)-\phi(x),\nn\\
&&d^nx\ra \det{\del(x+ \delta x)\over\del x}~d^nx = \det(\mathbb{I}+\del\delta x)~d^nx\approx  (1+{\Tr}(\del\delta x))~d^nx=(1+\del_\mu \delta x^\mu)d^nx,\nn\\
&&\delta\L=\bar{\delta}\L+\delta x^\mu~\del_\mu \L+\bar{\delta}\phi ~{\del \L\over\del\phi}+\bar{\delta}\del_\mu\phi ~{\del \L\over\del\del_\mu\phi}=\bar{\delta}\L+ \delta x^\mu~\del_\mu \L+\bar{\delta}\phi ~{\del \L\over\del\phi}+\del_\mu\bar{\delta}\phi ~{\del \L\over\del\del_\mu\phi} ,\nn\\
&&~~~~=\bar{\delta}\L-\del_\mu\delta x^\mu~\L+\del_\mu(\delta x^\mu~\L+\bar{\delta}\phi~{\del \L\over\del\del_\mu\phi})+\bar{\delta}\phi({\del \L\over\del\phi}-\del_\mu{\del \L\over\del\del_\mu\phi} ),\nn\\
&&\delta S[\phi,{\D}]:=S'[\phi',\D']-S[\phi,\D]\nn\\
&&~~~~=\int_{\D'} d^nx'~\L'(x',\phi'(x'),\del'\phi'(x'))-\int_\D d^nx~\L(x,\phi(x),\del\phi(x))\nn\\
&&~~~~={\int_{\D}} (\delta d^nx~\L+d^nx~\delta\L)\approx{\int_{\D}} \{(1+\del_\mu \delta x^\mu)d^nx~\L+d^nx~\delta\L\} \nn\\
&&~~~~= {\int_{\D}} d^nx~\{~\bar{\delta}\L+\del_\mu(\delta x^\mu~\L+\bar{\delta}\phi~{\del \L\over\del\del_\mu\phi})+\bar{\delta}\phi~({\del \L\over\del\phi}-\del_\mu~{\del \L\over\del\del_\mu\phi} )~\}\nn\\
&&~~~~= {\int_{\D}} d^nx~\{\bar{\delta}\L+\del_\mu~J^\mu+\bar{\delta}\phi~ E\}= {\int_{\D}} d^nx~(\bar{\delta}\L+\del_\mu~\al^\mu)=0~~~~\forall \D,\nn\\
&&~~\al^\mu=\delta x^\mu~\L+\bar{\delta}\phi~{\del \L\over\del\del_\mu\phi}+f_1^\mu+f_2^\mu,~~\del_\mu f_1^\mu=0,~~~~\int_{\del\D} dS_\mu f_2^\mu=0.\nn\\
\eea
}
In particular, for domains that can be continuously shrunk to a point, one has that $\bar{\delta}\L+\del_\mu \al^\mu=0$ at every point. Here, $\bar{\delta}$ is the functional variation \\
$\bar{\delta}F(u)=(F'-F)(u)=F'(u)-F(u)$; ie. it is the part of the variation that is not due to the ``visible'' arguments for the function involved.
\bea
&&\bar{\delta}\phi(x)=\phi'(x)-\phi(x),\nn\\
&&\bar{\delta}\L(x,\phi(x),\del\phi(x))=\L'(x,\phi(x),\del\phi(x))-\L(x,\phi(x),\del\phi(x)).
\eea
For any system of observers whose functional forms of the Lagrangian $\L$ can differ only by a total divergence $\del_\mu \beta^\mu$ there is a conserved current $J^\mu=\al^\mu+\beta^\mu$;
\bea
\bar{\delta}\L=\del_\mu \beta^\mu~~\Ra~~\delta S=\int \del_\mu J^\mu=\int\del_\mu(\al^\mu+\beta^\mu)=0.
\eea

In the case that involves a system with a dynamic domain $\D$ such as a smoothly expanding universe, the dynamics of $\D$ can be accounted for by introducing a dynamical metric field $g_{\mu\nu}$ whose dynamics is also determined by the Euler-Lagrange equations.
In a more convenient form for other purposes, one may express the general variation of the action as
\bea
&&\delta S=\int d^Dx\{\td{\delta}\L+\del_\mu(\delta x^\mu\L)\}=0~~\Ra~~\td{\delta}\L+\del_\mu(\delta x^\mu\L)=0,
\eea
where $\td{\delta}x=x\td{\delta},~~ \td{\delta}\del=\del\td{\delta}$.

In ~$D= 1+0$~ dimensions for example,
\bea
&&\int d^Dx~\L(x,\phi(x),\del\phi(x))\ra \int d\ld ~ L(\ld,q(\ld),\dot{q}(\ld)),~~\del_\mu\ra {d\over d\ld},\nn\\
&&\al=\delta\ld~ L+\bar{\delta}q^i(\ld)~{\del L\over\del\dot{q}^i},~~\dot{q}={dq\over d\ld}.\nn\\
&&\bar{\delta}L={d\over d\ld}\beta~~\Ra~~{d\over d\ld}(\al+\beta)=0.
\eea

Analogously in quantum field theory where we have the quantum measure $d\mu_\phi$  involving a sum over all possible $\phi$ configurations (or "paths") in $\D$, an amplitude $G$ for a physical process is given by the expectation value $\langle F(\{\phi\})\rangle$, wrt the quantum measure, of a homogeneous polynomial $F(\{\phi\})$ of the fields. The invariance of $G$ may be expressed as follows:
\bea
&&G=\langle F(\{\phi\})\rangle=\int_{\vphi\in\M/\D} d\mu_\vphi F(\{\vphi\})=\int_{\vphi\in\M/\D} D\vphi~F(\{\vphi\})e^{{i\over\hbar}S[\vphi,\D]}\nn\\
&&~~\ra~ \int_{\vphi\in\M/\D} D'\vphi'~F'(\{\vphi'\})~e^{{i\over\hbar}S'[\vphi',\D']}\nn\\
&&~~~~=\int_{\vphi\in\M/\D} D\vphi~\det({D'\vphi'\over D\vphi})~(F(\{\vphi\})+\delta F(\{\vphi\}))~e^{{i\over\hbar}(S[\vphi,\D]+\delta S[\vphi,\D])}\nn\\
&&~~\approx \int_{\vphi\in\M/\D} D\vphi~ e^{{\tr}{\td{\delta}(\td{\delta}\vphi)\over\td{\delta}(\vphi)}}~(F(\{\vphi\})+\delta F(\{\vphi\}))~e^{{i\over\hbar}(S[\vphi,\D]+\delta S[\vphi,\D])},\nn\\
&&\delta G\approx\int_{\vphi\in\M/\D} D\vphi~[~\delta F(\{\vphi\})+F(\{\vphi\})(~{i\over\hbar}~\delta S[\vphi,\D]+{\tr}{\td{\delta}(\td{\delta}\vphi)\over\td{\delta}(\vphi)}~)~]~e^{{i\over\hbar}S[\vphi,\D]}=0\nn\\
\label{ward-takahashi}&&\langle\delta F(\{\phi\})\rangle+\langle F(\{\phi\})(~{i\over\hbar}~\int_\D \del_\mu J^\mu+{\tr}{\td{\delta}(\td{\delta}\phi)\over\td{\delta}(\phi)})\rangle=0.
\eea
The relation ~(\ref{ward-takahashi})~ is the quantum analog of the classical Noether's theorem and is known as Ward-Takahashi identity.

An expression for the trace ${\tr}$ is
\bea
&&{\tr}{\td{\delta}(\td{\delta}\phi)\over\td{\delta}(\phi)}=\sum_{xy}\sum_{\xi}\xi^\ast(x){\td{\delta}\td{\delta}\phi(y)\over\td{\delta}\phi(x)}\xi(y),~~~~{\td{\delta}\phi(y)\over\td{\delta}\phi(x)}=\delta^n(x-y).\nn\\
&&\sum_\xi \xi^\ast(x)\xi(y)=\delta^n(x-y),~~~~\sum_x\xi'{}^\ast(x)\xi(x)=\delta_{\xi'\xi}.\nn\\
&&\phi(x)=\sum_{\xi}\td{\phi}_\xi~\xi(x).
\eea
For the case of global spacetime translations where $\td{\delta}\phi=-b^\mu\del_\mu\phi$,
\bea
&&{\tr}{\td{\delta}(\td{\delta}\phi)\over\td{\delta}(\phi)}=-\sum_{\xi}\int_\D d^nx~b^\mu~\xi^\ast(x)\del_\mu\xi(x),~~~~\xi|_{\del \D}=0.
\eea
This vanishes if the domain $\D$ is symmetric as we have an odd integrand.

In general,
\bea
&&{\tr}{\td{\delta}(\td{\delta}\phi)\over\td{\delta}(\phi)}={\Tr}(P\cdot{\td{\delta}(\td{\delta}\phi)\over\td{\delta}(\phi)})=\sum_{xy}P(x,y){\td{\delta}(\td{\delta}\phi(x))\over\td{\delta}(\phi(y))} \nn\\
&&P(x,y)=\sum_{\xi\xi'}\xi^\ast(x)g^{\xi\xi'}\xi'(y),~~g^{\xi\xi'}=g^{-1}_{\xi\xi'},~~g_{\xi\xi'}=\sum_x\xi^\ast(x)\xi'(x).\nn
\eea
where $P$ is a projection that may be constructed from a complete set of functions which can span the solution space of the classical trajectory~ given by ${\td{\delta} S[\phi]\over\td{\delta}\phi}=0$.

In order to obtain an analogous situation to the classical case, we need to define
\bea
Z[\C]=\int_{\vphi\in\C(\D)} D\vphi~e^{{i\over\hbar}S[\vphi,\D]},
\eea
where ~$\C(\D)=\{f;~f:\D\subseteq \mathbb{R}^{d+1}\ra\M,~~f_1(\del\D)=f_2(\del \D)~\forall f_1,f_2\}\subset {\M/\D}$~ is the configurations space.

\subsection{Dynamics using differential forms}
In terms of differntial forms the terms of matter and fermion actions are
\bea
&&i\int_\D~d^Dx~\overline{\psi}\gamma^\mu(\del_\mu-ieA_\mu)\psi=i\int_\D\overline{\psi} \gamma\ast(d\psi-ieA\psi) ,~~\gamma=\gamma_\mu dx^\mu,\nn\\
&&{1\over 2}\int_{\D} d^Dx~\del_\mu\phi^\dg\del^\mu\phi={1\over 2}\int_\D d\phi^\dg\ast d\phi,\nn\\
&&{1\over 4}\int_\D F_{\mu\nu}F^{\mu\nu}={1\over 4}\int_\D F\ast F,~~F=dA,~~A=A_\mu dx^\mu.
\eea
Infinitesimal transformations are given by
\bea
&& \delta f=i_{\delta x}df+d(i_{\delta x}f)~~\forall~f=f_{\mu_1...\mu_p}dx^{\mu_1...\mu_p},\nn
\eea
where
\bea
&&i_{\delta x}f=f_{\mu_1...\mu_p}\delta x^{[\mu_1}dx^{\mu_2...\mu_p]}\nn
\eea
and the $\ast$ and $d$ operations are
\bea
&& \ast f=f_{\mu_1...\mu_p}\vep^{\mu_1...\mu_p}{}_{\mu_{p+1}...\mu_D}dx^{\mu_{p+1}...\mu_D}\eqv f^\ast_{\mu_{p+1}...\mu_D}dx^{\mu_{p+1}...\mu_D},\nn\\
&&df=\del_{[\mu}f_{\mu_1...\mu_p]}dx^{\mu\mu_1...\mu_p},~~f_{\mu_1...\mu_p}=f_{\mu_1...\mu_p}(x).\nn
\eea
Example of transformation:
\bea
&&\delta(F\ast F)=2(\delta F)\ast F=2~d(i_{\delta x}F)\ast F\nn\\
&&~~~~=2~d((i_{\delta x}F)\ast F)-2~(i_{\delta x}F)(d\ast F).\nn\\
\eea
The Lagrangian in general is given by
\bea
&& \L=\L(f,\ast f,df,\ast df,d\ast f,\ast d\ast f,...).
\eea

\section{Faddeev-Popov gauge gixing method}
The definition of gauge fields in the classical or low energy action involves irrelevant degrees of freedom (in the form of invariance under gauge transformations) that must be eliminated (through gauge fixing: ie. by imposing any constraint that breaks the gauge symmetry completely) when attempting to obtain physical solutions to the equations of motion resulting from the least action principle. Similarly this elimination has to be done when attempting to quantize (ie. extend to all possible energies) the classical gauge theory since quantization involves summing over contributions from relevant degrees of freedom only.  The Faddeev-Popov gauge fixing method is one method of implementing gauge fixing in quantum theory.

The convenient (i.e. Euclidean) measure $d\mu(\A)$ and action $S[A]$, in the partition function
\bea
&&\label{partfunc1}Z_1[J]= \int DA~e^{-S[A]+JA}\eqv\int d\mu(\A)~e^{-S[A]+JA},\nn\\
&&~~JA=\int d^Dx~\Tr (J_\mu(x)A^\mu(x))=\int d^Dx~J^a_\mu(x)A^{a\mu}(x)),~~\nn
\eea
are invariant under the gauge transformation
\bea
&&A\ra A^g=g^{-1}Ag+g^{-1}dg=A+g^{-1}Dg,~~Dg=dg+[A,g].
\eea
Here $\A$ is the ${\G}$-bundle $\A=\{A\}\simeq\A\slash {\G}~\times {\G}$ of all gauge equivalent potentials $[A]$. This means that
\bea
&&d\mu(\A)=d\mu(\A\slash {\G})~d\mu({\G}),\nn\\
&&\A\slash {\G}=\{[A],~~A\in \A\},~~~~[A]=\{g^{-1}Ag+g^{-1}dg,~g\in {\G}\}.\nn
\eea
Therefore there is over counting in the partition function (\ref{partfunc1}) as it includes integration over the group $\G$ under which the integrand is invariant at $J=0$ which is the most important point in the definition and applications of the partition function to averaging of quantities \\
~$\lang Q(A)\rang={1\over Z[0]}Q({\delta Z[J]\over \delta J}|_{J=0})$~ as well as in evaluating effective actions. One simply needs to divide $Z[0]$ by the volume of the group $\int d\mu(\G)$ in order to remove the redundant factor and so the corrected partition function is
\bea
&&\label{partfunc}Z[0]= \int d\mu(\A\slash {\G})~e^{-S[A]}
\eea
now having the less convenient measure $d\mu(\A\slash {\G})$. The Faddeev-Popov method involves rewriting $Z[0]$ in terms of the more convenient measure $d\mu(\A)$ by choosing a path (a section or gauge fixing condition $G[A]-h=0,~{\del h\over\del A}=0$) other than $g=const$ through the bundle $\{[A]\}\times \G\simeq \{([A],g)\}$. The path should cut through any given fiber $[A]$ only once : ie $G[A^{g'}]\neq G[A^g]$ unless $g=g'$.
We insert the identity
\bea
&& 1=\int_{g\in\G}DG[A^g]~\delta(G[A^g]-h)=\int_{g\in\G}d\mu(\G)~|\det{\delta G[A^g]\over\delta g}|~\delta(G[A^g]-h)\nn
\eea
 into the integral expression for $Z[0]$.
\bea
&&Z[0]= \int d\mu(\A\slash {\G})~e^{-S[A]}\nn\\
&&~~~~=\int d\mu(\A\slash {\G})~DG[A^g]~\delta(G[A^g]-h)~e^{-S[A]}\nn\\
&&~~~~=\int d\mu(\A\slash {\G})~d\mu(\G)~|\det{\delta G[A^g]\over\delta g}|~\delta(G[A^g]-h)~e^{-S[A]}\nn\\
&&~~~~=\int d\mu(\A)~|\det{\delta G[A^g]\over\delta g}|~\delta(G[A^g]-h)~e^{-S[A]}\nn\\
&&~~~~=\int d\mu(\A)~|\det{\delta G[A^g]\over\delta g}|~\delta(G[A^g]-h)~e^{-S[A^g]}\nn\\
&&~~~~=\int d\mu(\A)~|\det{\delta G[A^g]\over\delta g}|_{g=1}~\delta(G[A]-h)~e^{-S[A]}\nn\\
&&~~~~=\int d\mu(\A)~|\det{\delta(G[A^g]-h)\over\delta g}|_{g=1}~\delta(G[A]-h)~e^{-S[A]}\nn\\
&&~~~~=\int d\mu(\A)~|\det{\delta (G[A+g^{-1}Dg]-h)\over\delta g}|_{g=1}~\delta(G[A]-h)~e^{-S[A]}.\nn
\eea
But
\bea
&&G[A+g^{-1}Dg]-h=G[A]-h+{\delta (g^{-1}D_\mu g)\over\delta g}~{\delta G[A]\over\delta A_\mu}\nn\\
&&~~~~+{1\over 2!}{\delta (g^{-1}D_\mu g~g^{-1}D_\nu g)\over\delta g}~{\delta^2 G[A]\over\delta A_\mu \delta A_\nu}+...\nn\\
&&~~~~=G[A]-h+{\delta (g^{-1}D_\mu g)\over\delta g}~{\delta G[A]\over\delta A_\mu}+g^{-1}D_\mu g{\delta (g^{-1}D_\nu g)\over\delta g}~{\delta^2 G[A]\over\delta A_\mu \delta A_\nu}+...\nn\\
\eea
and so only the first derivative term in the expansion can survive in the presence of ~$\delta(G[A]-h)$~ and upon setting $g=1$.
\bea
&&Z[0]=\int d\mu(\A)~|\det {\delta (g^{-1}D_\mu g)\over\delta g}~{\delta G[A]\over\delta A_\mu}|_{g=1}~\delta(G[A]-h)~e^{-S[A]}\nn\\
&&~~~~=\int d\mu(\A)~|\det D_\mu {\delta G[A]\over\delta A_\mu}|~\delta(G[A]-h)~e^{-S[A]}\nn\\
&&~~~~=\int d\mu(\A)DcD\overline{c}~\delta(G[A]-h)~e^{-S[A]-\Tr(\overline{c}D_\mu( {\delta G[A]\over\delta A_\mu})c)}\nn\\
&&~~~~=\int d\mu(\A)DcD\overline{c}~\delta(G[A]-h)~e^{-S[A]-\overline{c}^aD_\mu({\delta G^a[A]\over\delta A^b_\mu})c^b}~~\forall h,\nn\\
&&c^ac^b=-c^bc^a,~~c^a\overline{c}^b=-\overline{c}^bc^a,~~\overline{c}^a\overline{c}^b=-\overline{c}^b\overline{c}^a
\eea
where integration over spacetime is understood.

Since $h$ is arbitrary we can use equivalently
\bea
Z[0]=\int d\mu(\A)DcD\overline{c} Dh~F[h]~\delta(G[A]-h)~e^{-S[A]-\overline{c}^aD_\mu({\delta G^a[A]\over\delta A^b_\mu})c^b}.
\eea
In particular $F[h]=e^{-{1\over 2\al} \Tr h^2}=e^{-{1\over 2\al} h^ah^a}$ gives
\bea
&&Z[0]=\int d\mu(\A)DcD\overline{c}~e^{-S[A]-{1\over 2\al}G^a[A]G^a[A]-\overline{c}^aD_\mu({\delta G^a[A]\over\delta A^b_\mu})c^b},\nn\\
&&Z[J]=\int d\mu(\A)DcD\overline{c}~e^{-S[A]-{1\over 2\al}G^a[A]G^a[A]-\overline{c}^aD_\mu({\delta G^a[A]\over\delta A^b_\mu})c^b+AJ}.
\eea

\chapter{Geometry and Symmetries}
\section{Manifold structure}
A real $D$-dimensional manifold $\M_D$ is a collection of differentiable invertible maps from an arbitrarily given space $\M$ onto $\mathbb{R}^D$. That is $\M_D\simeq{\M\over \mathbb{R}^D}\subset\{\vphi:\mathbb{R}^{D}\ra \M\}$. One may ignore the dimension label $D$ when it is understood and write simply ~$\M\simeq {\M\over \mathbb{R}^D}$.
The function space over $\M$ is $\F(\M)=\mathbb{C}^N/ \M$  and the tangent vector bundle $T(\M)$ and dual tangent vector bundle $T^\ast(\M)$ are given by
\bea
&&T(\M)=\{t:\F(\M)\ra \F(\M),~~t\circ(fg)=t\circ f~g+f~t\circ g,\nn\\
&&~~~~t\circ (f+g)=t\circ f+t\circ g\},\nn\\
&&T^\ast(\M)=\{t^\ast:T(\M)\ra \mathbb{C}^N,~~t^\ast\circ(t_1+t_2)=t^\ast\circ t_1+t^\ast\circ t_2\}.\nn\\
\eea
and their fields (or sections) are  $T(\M)/\M$ and $T^\ast(\M)/\M$ respectively. One can equally construct tensor fields and dual tensor fields which are sections\\~$\T(\M)/\M,~~\T^\ast(\M)/\M$~ of the tensor and dual tensor algebras
\bea
&&\T(\M)=\mathbb{C}^N\oplus \bigoplus_{k=1}^\infty T(\M)^{\otimes k}=\mathbb{C}^N\oplus T(\M)\otimes\bigoplus_{k=0}^\infty T(\M)^{\otimes k},\nn\\
&&\T^\ast(\M)=\mathbb{C}^N\oplus \bigoplus_{k=1}^\infty T^\ast(\M)^{\otimes k}=\mathbb{C}^N\oplus T^\ast(\M)\otimes\bigoplus_{k=0}^\infty T^\ast(\M)^{\otimes k}.\nn\\
\eea

\section{Relativity or Observer Symmetry}
According to a universal observer, the dynamics of a physical system may be described by a ``path''\footnote{A ``path in the universal space $E$'' is a ``configuration in spacetime $X$''.} $\Gamma: X\ra E$; ie. $\Gamma\in E/X=\{\psi:X\ra E\}$, in the universal (experimental, operational or investigational) space $E=\{f:\A\slash X\ra \A\slash X\}$ of space and time (spacetime) $X\simeq \mathbb{R}^D$, where $\A$ is any suitable algebra. However a local or limited observer (sees the path as $\psi:\D\subseteq X\ra E,~~\psi(\D)=\Gamma(X)$) can only access an observer domain $\D$ of spacetime that serves as a parameter space and is different for different local observers although the physical system (ie. its ``path'' in the universal space), and of course the universal space, look the same according to the different observers. We assume that the local observers are careful observers, where a careful observer is one that is aware that he needs to make several observations using as many different frames of reference $\D$ as possible before attempting to make any general conclusions about the behavior of the physical system.

According to the universal observer, it is therefore natural to regard each local observer  $O\in G$ as merely a member of the set of structure preserving transformations $G=\{g:E\slash X\ra E\slash X,~~(g\circ\Gamma)(X)=\Gamma(X)\}\subset E$ on spacetime based systems or paths, where the structure to be preserved is the ``path'' $\Gamma$ of the physical system in the universal space $E$ of $X$. \\ Therefore $\Gamma\in \C=\{c=G\circ f\eqv [f],~~f\in E\}\subset E/G$~ since \\
$g\circ G:=\{gh;~ h\in G\}=G~~\forall g\in G$,~~where $\C$ is the space of all symmetric path configurations.

For simplicity, we will make the restriction $E\simeq \mathbb{C}^{N_1}\times \mathbb{C}^{N_2}\times...\times \mathbb{C}^{N_n}$. Now two local observers $O,O'$ define the path $\Gamma$ as ${\psi}:\D\subseteq X\ra E,~~{\psi}(\D)=\Gamma(X)$~ and  ~${\psi}':\D'\subseteq X\ra E,~~{\psi}'(\D')=\Gamma(X)$~ respectively. \\ Which means that ${\psi}'(\D')={\psi}(\D)$.
From experimenting with local observer relabeling properties of a function, components of a vector field, components of a spinor field, components of a tensor field (infinitesimal polygons), ... in ordinary spaces one finds that respectively,
\bea
&&\al'(u')=\al(u),~~\al:X\ra \mathbb{R},\nn\\
&&\al'(u')=e^{i\theta(u,u',{\del u'\over\del u},...)}\al(u),~~\al:X\ra \mathbb{C},~~\theta(u,u',{\del u'\over\del u},...)\in \mathbb{R},\nn\\
&&\del'_\mu={\del u^\nu\over\del u'{}^\mu}\del_\nu,\nn\\
&&\al'{}^\mu(u')={\del u'{}^\mu\over\del u^\nu}\al^\nu(u),\nn\\
&&\al'{}^a(u')=S^a{}_b(u,u',{\del u'\over\del u})\al^b(u),\nn\\
&&~~~~...
\eea
Therefore ``pointwise'', ~${\psi}'(\D')={\psi}(\D)$~ may be written as
{\footnotesize\bea
&&{\psi}'_{a_1a_2...a_n}(u')=R_{a_1}{}^{b_1}(u,u',{\del u'\over\del u},..)R_{a_2}{}^{b_2}(u,u',{\del u'\over\del u},..)...R_{a_n}{}^{b_n}(u,u',{\del u'\over\del u},..)~{\psi}_{b_1b_2...b_n}(u)\nn\\
&&~~~~\forall ~u'\in \D',~u\in\D, \nn\\
\eea}
a more general form of which being
\bea
&&{\psi}'_{a_1a_2...a_n}(u')=R_{a_1a_2...a_n}{}^{b_1b_2...b_n}(u,u',{\del u'\over\del u},..)~{\psi}_{b_1b_2...b_n}(u)+b_{a_1a_2...a_n}(u,u',{\del u'\over\del u},..).\nn\\
&&~~~~\forall ~u'\in \D',~u\in\D. \nn\\
\eea
and yet a more general form being
\bea
&&{\psi}'_{a_1a_2...a_n}(u')=R_{a_1a_2...a_n}(u,u',\del u',..,{\psi}(u),\del {\psi}(u),..).\nn\\
&&~~~~\forall ~u'\in \D',~u\in\D. \nn\\
\label{observers1}&&{\psi}'(u')=R(u,u',\del u',..,{\psi}(u),\del {\psi}(u),..).
\eea
In general $\psi$ may be expanded as a sum of products of elementary functions $\{e\}$:
\bea
&&\psi(u)=\sum_{k}\td{\psi}_{i_1i_2...i_k}{e}^{i_1}(u){e}^{i_2}(u)...{e}^{i_k}(u).
\eea

Internal symmetries are those for which $u=u'~~\forall u\in\D,~u'\in\D'$. One notes here that the observer domains $\D,\D'$ may be specified through differentiable-invertible maps $\vphi:U\subset \M\ra X,~m\mapsto u$ and $\vphi':U'\subset \M\ra  X,~m'\mapsto u'$ (with $\vphi:U\cap U'\ra \D\subseteq X,~~\vphi':U\cap U'\ra \D'\subseteq X$) so that $u'$ and $u$ correspond to the same point $m=\vphi^{-1}(u)=\vphi'{}^{-1}(u')~$ in the intersection $U\cap U'$ on some abstract space $\M$, in which case any given complete collection $\{U_a\},~\bigcup_aU_a\supseteq\M$ of pre-observer domains is said to define a differentiable manifold $\M$ over $X$ meanwhile any corresponding appropriate choices $\{E(\M)=\{f:\A\slash \M\ra \A\slash \M\}\}$ for the universal space $E$ are fiber bundles $\{\pi:E\ra \M\}$ over $\M$. The (internal) transition relation among the various observers may be expressed thus
\bea
&& u_a=\vphi_a(m)=\vphi_a\circ\vphi_b^{-1}\circ\vphi_b(m)=\vphi_a\circ\vphi_b^{-1}(u_b)\nn\\
&&~~~~=\vphi_{ab}(u_b)=\vphi_{ab}(\vphi_{bc}(u_c))=\vphi_{ab}\circ\vphi_{bc}(u_c)=\vphi_{ac}(u_c).\nn\\
&&\vphi_{ac}\circ\vphi_{cb}=\vphi_{ab}.\\
&&\psi_a(\D_a)=\psi_b(\D_b)~\iff~\psi_a\circ \vphi_a(U_a\cap U_b)=\psi_b\circ\vphi_b(U_a\cap U_b)\nn\\
&&~~~~\iff~ \psi_a=\psi_b\circ \vphi_{ba},~~~~\vphi_{ab}=\vphi_a\circ\vphi_b^{-1}.
\eea

The form of the representation function $R$ in (\ref{observers1}) is determined by consistency with observational facts. For example, in quantum theory $E$ is a noncommutative space [as decided by observations] meanwhile $X$ can also be noncommutative [as decided by observations]; then for the noncommutativity to be physical or observable, the underlying algebraic (eg. commutation) relations, or their functional form equivalently, need to be the same (just as the path $\Gamma(X)$ is) for each local observer and consequently the (local) observer relabeling or reparametrization tensor $R$ needs to take on a form that can support this preservation of the algebraic relations on relabeling. The following section \ref{hopf} is an attempt at such transformations.

If the path $\Gamma$ is defined by a least action principle
\bea
&&S[{\psi},\D]=\int_{\D}d\mu(\D)~\L(u,{\psi}(u),\del {\psi}(u),...),\nn\\
&&{\delta S[{\psi},\D]\over\delta {\psi}(u)}=0,~~~~ {\delta\psi(u)}|_{u\in\del\D}=0,
\eea
then ${\psi}'(\D')={\psi}(\D)$ requires that
\bea
&&S'[{\psi}',\D']=\int_{\D'}d\mu(\D')~\L'(u',{\psi}'(u'),\del {\psi}'(u'),...),\nn\\
&&{\delta S'[{\psi}',\D']\over\delta {\psi}'(u')}=0,~~~~ {\delta\psi'(u')}|_{u'\in\del\D'}=0
\eea
as well.

In the case where $X$ is an algebra $\hat{X}=\{\hat{x}\}$ specified by commutation relations, one may first determine the spectra\\
$\sigma(\hat{x}^\mu)=\{\ld^\mu\in \mathbb{\mathbb{C}};~(\hat{x}^\mu-\ld^\mu 1)^{-1}~\nexists\}~~\forall \mu$. Then the ``functional'' form of the relativistic path $\hat{\psi}:\hat{\D}\subseteq\hat{X}\ra \hat{E},~~\hat{\psi'}(\hat{\D}')=\hat{\psi}(\hat{\D})$  may be expressed as
\bea
&&\hat{\psi}(\hat{u})=\sum_{k}\td{\hat{\psi}}_{i_1...i_k}{e}^{i_1}(\hat{u}){e}^{i_2}(\hat{u})...{e}^{i_k}(\hat{u}),\nn\\
&&e^i(\hat{u})=(e^i\circ\vphi)(\hat{x})=\oint_{D(\sigma(\hat{x}))} (e^i\circ\vphi)(z)~\prod_\mu{dz^\mu\over z^\mu-\hat{x}^\mu},
\eea
where one now has (index) reordering or permutation or braiding symmetry due to the noncommutativity and again this reordering must be consistent with the relation $\hat{\psi'}(\hat{\D}')=\hat{\psi}(\hat{\D})$.

\section{Hopf symmetry transformations}\label{hopf}
The permutation group $\mathcal{S}_n$ and braid group $\mathcal{B}_n$ may be defined as follows:
\bea
&&T_n=\{1,{t}_{12},{t}_{23},...,{t}_{n-1~n}\}\eqv \{1,t_1,t_2,...,t_{n-1}\},\nn\\
&&t_k={t}_{k~k+1}:V^{\otimes N}\ra V^{\otimes N},~v_1\otimes...\otimes v_N\mapsto v_1\otimes...\otimes {t}(v_k\otimes v_{k+1})\otimes...\otimes v_N,\nn\\
&&{t}:V\otimes W\ra W\otimes V,\nn\\
&&B_n=\{b_i\in T_n~;~~b_ib_j=\delta_{ij}b_i^2+b_jb_ib_jb_i^{-1}\delta_{i\pm 1~j}+b_jb_i(1-\delta_{ij}-\delta_{i\pm 1~j}) \}\nn\\
&&~~~~=\{b_i\in T_n~;~~[b_i,b_j]=b_jb_i(b_jb_i^{-1}-1)\delta_{i\pm 1~j}=b_ib_j(1-b_ib_j^{-1})\delta_{j\pm 1~i} \}\nn\\
&&\B_n=\{g_i=u_1^{\circ n_1}\circ u_2^{\circ n_2}\circ...\circ u_n^{\circ n_i},~~n_1+...+n_i=i,~n_r\in \mathbb{Z};\nn\\
&&~~~~~~~~~~(u_1,...,u_n)\in B_n{}^n\eqv B_n\times B_n{}^{n-1}\}.\nn\\
&&S_n=\{s_i\in T_n;~~s_is_j=\delta_{ij}+(s_js_i)^2\delta_{i\pm 1~j}+s_js_i(1-\delta_{ij}-\delta_{i\pm 1~j}) \}\nn\\
&&\S_n=\{g_i=u_1^{\circ n_1}\circ u_2^{\circ n_2}\circ...\circ u_n^{\circ n_i},~~n_1+...+n_i=i,~n_r\in\{0,1\};\nn\\
&&~~~~~~~~~~(u_1,...,u_n)\in S_n{}^n\eqv S_n\times S_n{}^{n-1}\}.\nn\\
\eea

One may summarize the defining properties of a Hopf algebra \\
$H=(A=\{a,b,...\},F,\mu,\Delta,\eta,\vep,S,\tau)\eqv$ (Vector space,~Field,~product,~coproduct,~unit,~counit,~antipode,~braiding) as follows.
\begin{itemize}
\item Unit:
\bea
&&\eta_a:F\ra A,~\ld\mapsto \ld a.~\forall a\in A.\nn\\
&&~~~~\eta:=\eta_{1_A}:F\ra A,~\ld\mapsto \ld 1_A.\nn\\
\eea
\item Product:
\bea
&&\mu:A\otimes A\ra A.\nn\\
\eea
\item Coproduct
\bea
&&\Delta=\Delta^2:A\ra A\otimes A,~~a\mapsto \Delta(a)=\sum_{\al\beta}C_{\al\beta}~\al(a)\otimes \beta(a)\eqv a_{(1)}\otimes a_{(2)}\nn\\
&&~~~~\eqv a_\al\ot a^\al,\nn\\
&&~~~~\Delta(a\otimes b)=a_{(1)}\otimes b_{(1)}\otimes a_{(2)}\otimes b_{(2)}~~(\txt{an alternative}).\nn\\
&&~~~~\Delta\circ\mu=(\mu\otimes\mu)\circ (id\otimes \tau\otimes id)\circ(\Delta\otimes\Delta),~~~~\Delta(ab)=\Delta(a)\Delta(b).\nn\\
&&~~~~\Delta^1=id.\nn\\
&&~~~~\Delta^3:=(id\otimes \Delta)\circ\Delta=(\Delta\otimes id)\circ\Delta,\nn\\
&&~~~~\Delta^3(g)=g_{(1)}\otimes g_{(2)(1)}\otimes g_{(2)(2)}=g_{(1)(1)}\otimes g_{(1)(2)}\otimes g_{(2)}\eqv g_{(1)}\otimes g_{(2)}\otimes g_{(3)}.\nn\\
&&~~~~\Delta^k:A\ra A^{\otimes k}=A\otimes A^{\otimes (k-1)},\nn\\
&&~~~~\Delta^{k+1}=((id\otimes)^{i-1}\Delta(\otimes id)^{k-i})\circ\Delta^{k},~~i=1,2,...,k. \nn\\
&&~~~~\Delta^k(g)=g_{(1)}\otimes..\otimes g_{(i-1)}\otimes g_{(i)(1)}\otimes g_{(i+1)(2)}\otimes g_{(i+2)}\otimes ...\otimes g_{(k-1)} \nn\\
&&~~~~\eqv g_{(1)}\otimes g_{(2)}\otimes ...\otimes g_{(k)}.\nn\\
\eea
\item Counit:
\bea
&& \vep: A\ra F,~a\mapsto \vep(a),~\vep(1_A)=1_F,\nn\\
&&~~~~\vep\circ\mu_A=\mu_F\circ(\vep\ot\vep),~~~~\vep(ab)=\vep(a)\vep(b).\nn\\
&&~~~~(id\otimes\vep)\circ\Delta=(\vep\otimes id)\circ\Delta=id,\nn\\
&&~~~~g_{(1)}\vep(g_{(2)})=\vep(g_{(1)})g_{(2)}=g.\nn\\
&&~~~~((id\otimes)^{i-1}\vep(\otimes id)^{k-i})\circ\Delta^{k}=\Delta^{k-1},~~i=1,2,...,k.\nn\\
&&~~~~ \vep(g_{(i)}) ~g_{(1)}\otimes...\otimes g_{(i-1)}\otimes g_{(i+1)}\otimes...\otimes g_{(k)}=g_{(1)}\otimes...\otimes g_{(k-1)}.\nn\\
\eea
\item Antipode:
{\small\bea
&&S:A\ra A,\nn\\
&&~~~~\mu\circ(id\otimes S)\circ \Delta=\mu\circ(S\otimes id)\circ \Delta=\eta\circ\vep,\nn\\
&&~~~~g_{(1)}S(g_{(2)})=S(g_{(1)})g_{(2)}=\eta(\vep(g))=\vep(g)1_G,\nn\\
&&~\Ra~S(gh)=S(h)S(g),~S(1_G)=1_G,~(S\otimes S)\circ\Delta=\tau\circ\Delta\circ S,~\vep\circ S=\vep.\nn\\
&&~~~~((id\otimes)^{i-1}S(\otimes id)^{k-i})\circ\Delta^{k}\sr{?}{=}\eta\circ\vep~\Delta^{k-1},\nn\\
&&~~~~g_{(1)}\otimes...\otimes g_{(i-1)}\otimes S(g_{(i)})g_{(i+1)}\otimes g_{(i+2)}\otimes...\otimes g_{(k)}\nn\\&&~~~~~~~~\sr{?}{=}\vep(g)~1_G\otimes g_{(1)}\otimes g_{(2)}\otimes ...\otimes g_{(k-2)}.\nn
\eea}
\item Boundary:
\bea
&&\mu^k: A^{\otimes k}\ra A,~~a_1\otimes a_2\otimes...\otimes a_k\mapsto a_1a_2...a_k.\nn\\
&&~~\del_i\eqv \mu_{i~mod_{k+1}(i+1)}=(id\otimes)^{i-1}\mu^2(\otimes id)^{k-i}: A^{\otimes k}\ra A^{\otimes(k-1)},~i=1,2,..,k.\nn\\
&&mod_N(n)=\left\{
             \begin{array}{ll}
               n, & n< N \\
               min(\{n\}), & n= N\\
               mod_N(n-N), & n> N
             \end{array}
           \right\}\nn\\
&&~~~~= n~\theta(N-n)+min(\{n\})~\delta_{nN}+mod_N(n-N)~\theta(n-N),\nn\\
&&\del=\sum^k_{i=1}(-1)^{i-1}\del_i,~~~~\del^2=0.
\eea
\end{itemize}

\subsubsection{Example}
$H=(\A=\{a,b,c,...\},\mu=\mu_\A,\Delta,\tau,\vep,\eta,S),~~\mu,\Delta,\tau,\vep,\eta,S~~$linear. $Z(\A)=\A\cap\A'$. ``First'' define $\Delta$ such that (ie. check that) $(id\ot\Delta)\circ\Delta=(\Delta\ot id)\circ\Delta$. For example, if ~$\pi:H\ra O(\B),~\B=(B,\mu_B)$~ then $\Delta$ will be defined such that $\pi(a)\circ \mu_B=\mu_B\circ \Delta(\pi(a))\eqv \mu_B\circ\Delta(\pi)\circ\Delta(a)$.
\bea
&&\Delta(a)=a_\al\ot a^\al,~~\Delta(ab)=(ab)_\al\ot(ab)^\al,\nn\\
&&\Delta(a)\Delta(b)=a_\al b_\beta\ot a^\al b^\beta,\nn\\
&&(ab)_\al=a_\gamma b_\rho \ld^{\gamma \rho}{}_\al,~~(ab)^\al=\ld^\al{}_{\gamma \rho}a^\gamma b^\rho ,~~ \ld^{\gamma \rho}{}_\al \ld^\al{}_{\gamma' \rho'}= \delta ^\gamma{}_{\gamma'}\delta^\rho{}_{\rho'}\nn\\
&&~~\Ra~~\Delta(ab)=\Delta(a)\Delta(b).\nn\\
&&S(a_\al)=\eta(\vep(a))~(a_\rho a^\rho)^{-1}a_\al ,~~S(a^\al)=a^\al(a_\rho a^\rho)^{-1}\eta(\vep(a)),\nn\\
&&\vep(a_\al)=\delta_{aa_\al},~~\vep(a^\al)=\delta_{aa^\al},\nn\\
&&\vep(ab)=\vep(a)\vep(b),~~S(ab)=S(b)S(a).
\eea

\subsection{Quasi-tringular Hopf algebras and R-matrix}
If ~$\tau \circ\Delta=Q\circ\Delta$,~ then one may write $\T=Q\circ \tau$, where ~$\T\circ\Delta=\Delta$.
\bea
&&\T_{i}\T_{i+1}\T_{i}=\T_{i+1}\T_{i}\T_{i+1}\nn\\
&&\Ra~~ Q_{i~i+1}Q_{i~i+2}Q_{i+1~i+2}=Q_{i+1~i+2}Q_{i~i+2}Q_{i~i+1}.
\eea
For example, if $Q=\ad_R$; ie. $\tau\circ\Delta(h)=R\circ\Delta(h)\circ R^{-1}$ then we also have
\bea
&& R_{i~i+1}R_{i~i+2}R_{i+1~i+2}=R_{i+1~i+2}R_{i~i+2}R_{i~i+1}.
\eea

\subsection{Action}
$H=\{h,g,...\}$ acts on a product algebra $A=A_1\otimes A_2\otimes...\otimes A_k=\{a,b,c,...\}$ through an action $\rho$.
\bea
&&\rho:H\otimes A\ra A,~~(h,a)\mapsto \rho(h,a)=\rho(h)a\eqv \rho_ha\eqv h\rhd a,\nn\\
&&\rho_ha=\Delta(\rho_h)(a_1\otimes a_2\otimes...\otimes a_k)=\rho_{h_{(1)}}a_1\otimes \rho_{h_{(2)}}a_2\otimes ... \otimes \rho_{h_{(k)}}a_k.\nn\\
&&\rho_h(abc...)=\rho_{h_{(1)}}a~\rho_{h_{(2)}}b ~\rho_{h_{(3)}}c~...,~~a,b,c...\in A. \nn\\
&&\txt{e.g.~left action (left ``translation'')}~~\rho_ha=L_ha=\Delta(h)(a_1\otimes a_2\otimes ... \otimes a_k)\nn\\
&&~~~~~~~~=h_{(1)}a_1\otimes h_{(2)}a_2\otimes ... \otimes h_{(k)}a_k.\nn\\
&&\txt{~~~~adjoint action}~\rho_ha=ad_ha:=h_{(1)}aS(h_{(2)})=ad_{h_{(1)}}a_1\otimes ad_{h_{(2)}}a_2\otimes ... \otimes ad_{h_{(k)}}a_k\nn\\
&&~~~~~~~~\Ra~~ad_h(ab)=h_{(1)}abS(h_{(2)})= h_{(1)}a S(h_{(2)})h_{(3)}bS(h_{(4)}) =ad_{h_{(1)}}a~ad_{h_{(2)}}b.\nn\\
\eea

More simply for the adjoint action, one can write
\bea
&&ga= g_{(1)}a S(g_{(2)})~g_{(2)}=ad_{g_{(1)}}a~g_{(2)},\nn\\
&&gab=ad_{g_{(1)}}a~g_{(2)}b=ad_{g_{(1)}}a~ad_{g_{(2)(1)}}b~g_{(2)(2)}= ad_{g_{(1)}}a~ad_{g_{(2)}}b~g_{(3)}\nn\\
&&~~~~=ad_{g_{(1)}}(ab)~g_{(2)}.\nn\\
&&g\psi_1\psi_2...\psi_k=ad_{g_{(1)}}(\psi_1\psi_2...\psi_k)~g_{(2)}=ad_{g_{(1)}}\psi_1~ad_{g_{(2)}}\psi_2~...~ad_{g_{(k)}}\psi_k~g_{(k)},\nn\\
\eea
where each of $\psi_i$'s may be a tensor product as well; ie.
\bea
\psi_i\in \mathcal{T}(A)=F\oplus A\oplus A^{\otimes 2}\oplus A^{\otimes 3}\oplus...\oplus A^{\otimes n}~~~~\forall i.
\eea

\subsection{Duality and integration}
The set of linear functionals ~$H^\ast\eqv A^\ast=\{f,~f: H\ra F,~ a\ra f(a)\eqv (f,a)\}$~  is the dual of $H$. That is, $(,):H^\ast\otimes H\ra F$.  For purposes of (co)homology indicated by the maps $\del_i=\mu_{i~mod_{k+1}(i+1)}:A^{\otimes k}\ra A^{\otimes (k-1)},~~\Delta_i: A^{\otimes k}\ra A^{\otimes (k+1)}$,~ $\mu$ and $\Delta$ are dual to each other. Similarly, $\vep$ and $\eta$ are duals.
\bea
&&(ff',a):= (\mu (f\otimes f'),a)=(f\otimes f',\Delta(a)).\nn\\
&&(\Delta(f),a\otimes b):=(f,\mu(a\otimes b))=(f,ab) .\nn\\
&&(f,\eta(\ld))=(\vep(f),\ld)~~\iff~~ (f,1_A)=\vep(f),~~\ld\neq 0.\nn\\
&&(\eta(\al),a)=(\al,\vep(a))~~\iff~~ (1_{A^\ast},a)=\vep(a),~~\al\neq 0.\nn\\
&&(S(f),a)=(f,S(a)),~~~~\txt{$S$ is self dual}.
\eea

 A left integral $\int \phi~\in H^\ast$ of an element $\phi\in H^\ast$ is a left-invariant linear functional $\int\phi:H\ra F$, ~$\int L^\ast_h\phi=\vep(h)\int\phi~\forall h\in H$, ~where
\bea
&&(L_h a,\phi)=(a,L^\ast_h\phi),\nn\\
&&\txt{ie.}~~L^\ast_h\phi(a)= \phi(L_ha)=\phi(ha)=\phi(\mu(h\otimes a))=\mu_F\circ\Delta(\phi)~(h\otimes a)\nn\\
&&~~~~=\phi_{(1)}(h)~\phi_{(2)}(a).\nn\\
&&\phi\circ\mu_H=\mu_F\circ \Delta(\phi).
\eea
Therefore a left integral $\int$ on $H$ is given by
\bea
&&\int L^\ast_h(\phi)=\vep(h)\int\phi~~~\forall (h\in H,~\phi\in H^\ast)
\eea
and similarly, a left integral in $H$ is any $\I\in H$ such that
\bea
L_h\I\eqv h\I=\vep(h)\I~~~\forall h\in H.
\eea
\chapter{Some math concepts}
\section{Groups, Rings (Algebras), Fields, Vector \\ spaces, Modules}
A group $G$ is a set $S$ with an identity $e$, closed under an associative binary operation $S\times S\ra S,~~(a,b)\mapsto ab$ and in which every element has an inverse. That is, $\forall a,b,c\in S~~~~\exists~e,a^{-1}\in S$ such that
\bea
&&ae=a,~~a(bc)=(ab)c,~~a^{-1}a=e.\nn\\
&&G=(S,S\times S\ra S).
\eea
$G$ is an Abelian group $G_0$ if $ab=ba~~\forall a,b\in G$. The binary operation of the Abelian group is written as $+$ and the identity is written as $0$ and the inverse $a^{-1}$ of $a\in G_0$ is written as $-a$. That is $G_0=(S,+:S\times S\ra S)$.

A ring (or an algebra) $R$ is an Abelian group $G_0$ that is closed under an additional associative binary operation ~$\cdot:G_0\times G_0\ra G_0,~~(a,b)\mapsto ab$~ that is distributive over $+$. That is
\bea
&&a(b+c)=ab+ac,~~(b+c)a=ba+ca,~~a(bc)=(ab)c~~\forall a,b,c\in G_0.\nn\\
&&R=(G_0,\cdot:G_0\times G_0\ra G_0)=(S,+,\cdot:S\times S\ra S)\eqv (G_0,\cdot)\eqv (S,+,\cdot).\nn
\eea
A field $F$ is a ring $(S_0,+,\cdot)$ such that $(S_0\backslash\{0\},\cdot)$ is an Abelian group. That is~ $F=(S_0,+,\cdot)|_{(S_0\backslash\{0\},\cdot)\in \G}$~ where $\G=\{G\}$ is the family of groups.
A field $F$ is ordered iff there is $P\subseteq F$ such that
\bea
&&+,\cdot:P\times P\ra P,~~P\cap -P=\{\},~~P\cup\{0\}\cup -P=F,
\eea
where $A^{-1}=\{a^{-1},~~a\in A\},~~AB=\{ab,~a\in A,~b\in B\},~A^2=AA$.

If $I\subset R_0$ is an ideal (ie. $I+I=I,~R_0I=IR_0=I$) of an Abelian ring $R_0=(S_0,+,\cdot)$  then $F_I=R_0\backslash I=\{a+I,~a\in R_0\}$ is a field.

With notation understood, one defines a vector space $V_F$ over a field $F$ and a module $M_R$ over a ring $R$ as
~$V_F=(V,+,F\times V\ra V)$,\\
~~$M^{\txt{left}}_R=(M,+,R\times M\ra M),~~M^{\txt{right}}_R=(M,+,M\times R\ra M)$~ respectively.

\section{Set commutant algebra}
Let the commutator of two subsets $A,B$ of an algebra ${\A}:=({\A},+,\star)$ be
\bea
&&[A,B]=\{[a,b],~~a\in A,~b\in B\}.
\eea
Let $S\subseteq {\A}$, then the commutant $S'\subseteq\A$ of $S$ in ${\A}$ is defined to be the subset of ${\A}$, with the highest possible number of elements that each commute with every element of $S$.
\bea
S'=\max_{[U,S]=\{0\}} U,~~U\subseteq {\A},~~0+g=g~\forall g\in {\A}.
\eea
If $A,B\subseteq {\A}$ and $A\subseteq B$, then
\bea
\label{subset}B'\subseteq A'
\eea
since some of the elements of $A'$ may fail to commute with $B\backslash A\eqv B\backslash A\cap B$ and hence fail to commute with $B$.

Also, since both $S$ and $S''$ (the commutant of $S'$) commute with $S'$ and $S''$ is supposed to be the maximum of all sets that commute with $S'$, it follows that
\bea
\label{subset2}S\subseteq S''.
\eea
One can then deduce using (\ref{subset}) and (\ref{subset2}) that
\bea
&&S\subseteq S''=S''''=...=S^{(2n)},~~n>2,\nn\\
&&S'=S'''=...=S^{(2n-1)},~~n\geq 1.\nn\\
\eea
As a check~ $S\subseteq S''~~\Ra~~(S'')'\subseteq S'$. Also $S'\subseteq (S')''$ and so $S'=S'''$ ~follows by the identification  $(S'')'=(S')''\eqv S'''$. Therefore
\bea
\A=S' \cup S''.
\eea

Furthermore
\bea
\label{setcompare1}&&(A\cup B)'\subseteq A'\cap B'
\eea
since
$A\subseteq A\cup B,~~B\subseteq A\cup B~~\Ra~~(A\cup B)'\subseteq A',~~(A\cup B)'\subseteq B'$.
Similaly
\bea
&&A\cap B\subseteq A,~~A\cap B\subseteq B~~\Ra~~A'\subseteq (A\cap B)',~~B'\subseteq (A\cap B)'.\nn\\
\label{setcompare2}&&\txt{Hence}~~A'\cup B'\subseteq (A\cap B)'.
\eea
If $A,B$ are Von Neumann algebras $A=A'',B=B''$ then it follows from (\ref{setcompare1}) and (\ref{setcompare2})
that
\bea
 (A\cup B)'=A'\cap B',~~(A\cap B)'=A'\cup B'.
\eea

Also one observes that the center $Z(S):=S\cap S'$ of $S$ is always a commuting set and so $S$ is a commuting set iff $S\subseteq S'$.

\emph{\textbf{Remarks:}}
\begin{itemize}
\item Although $S$ is merely an arbitrary subset, the derived sequence of subsets $S^{(i)},~~i=1,2...,n$ is a sequence of subalgebras (that is, these subsets are closed under $+~\&~\ast$) if $S$ is self adjoint; ie. both $a,a^\ast\in S$.
\item Consequently $S''$ is seen as the closure of $S$ since it is the smallest closed set that contains $S$ in this sense of closure. Thus $S$ is closed (ie. a subalgebra) iff $S''\subseteq S$ and hence iff $S''=S$ since we also know that $S\subseteq S''$. $S$ is open iff its complement $S^c={\A}\backslash S$ is closed (ie. a subalgebra).
\item In particular if $S$ is a single element set with element $a$ then $a'$ is the symmetry algebra of $a$ and $a'\cap a''$ is the largest commutative set that contains $a$. $\sigma(a)\subseteq\sigma(a'\cap a'')$.
\item A representation $R:S\ra B(\H)$, where $B(\H)$ is the set of  bounded linear operators on a Hilbert space $\H$, of a subalgebra $S$ is  irreducible iff $R(S)'=\mathbb{C}~1_{B(\H)}$ meaning that the commutant $R(S)'$ of $R(S)$ is proportional to the identity (ie. trivial) in $B(\H)$.
\end{itemize}

\section{Projector algebra}
Let ~$p\in \A,~p^2=p,~~D_p=pDp\eqv \{pap;~a\in D\},~~D\subseteq \A$. Then in
\bea
&&\mathbb{N}_p(D):=\{D_p{}^n;~n\in \mathbb{N}\},\nn
\eea
one has that
\begin{itemize}
\item $D_p{}^mD_p{}^{n}=D_p{}^{m+n}.$
\item $\bigcup_{n\in \mathbb{N}} D_p{}^n\subseteq p'=\{c\in\A;~pc=cp\}.$
\item $\txt{If}~~p\in D~~\txt{then}~~m\leq n~~\Ra~~D_p{}^m\subseteq D_p{}^n.$
\item $p\in D,~DD=D~~\Ra~~p\in D_p\subseteq D,~D_pD_p=D_p.$
Therefore for $D=\A$ one sees that projectors correspond to ``closed'' subspaces of $\A$.
\end{itemize}
An Abelian group $\mathbb{Z}_p(D)=\{g^n;~n\in \mathbb{Z}\}$ may also be defined with elements
\bea
&&g^n=\{(D_p{}^{m},D_p{}^{m+n});~m\in \mathbb{N}\},\nn\\
&&g^{-n}=\{(D_p{}^{n+m},D_p{}^{m});~m\in \mathbb{N}\},~~(x,y)(z,w):=(xz,yw).\nn
\eea

\section{Matrix-valued functions and BCH formula}
\subsection{Limits}
For complex numbers $\al,\bi...$~ and matrices \\
$A=e^a,~B=e^b,~...$,
\bea
&&\lim_{n\ra\infty}(1+{\al\over n})^n=e^{\lim_{n\ra\infty}n\ln(1+{\al\over n})}=e^{\lim_{n\ra\infty}{\ln(1+{\al\over n})\over {1\over n}}}=e^\al\eqv \lim_{n\ra\infty}(e^{\al\over n})^n. \nn\\
&&\txt{ie}~~\lim_{n\ra\infty}(1+{\al\over n})^n=\lim_{n\ra\infty}(e^{\al\over n})^n= e^\al.\nn\\
&&\lim_{n\ra\infty}(~(1+{\al\over n})^n(1+{\bi\over n})^n~)=\lim_{n\ra\infty}(1+{\al\over n})^n~\lim_{n\ra\infty}(1+{\bi\over n})^n=\lim_{n\ra\infty}(~(e^{\al\over n})^n (e^{\bi\over n})^n~)\nn\\
&&~~~~~=\lim_{n\ra\infty}(e^{\al\over
n})^n~\lim_{n\ra\infty}(e^{\bi\over n})^n=e^\al e^\bi=e^{\al+\beta}.\nn\\
&&\lim_{n\ra\infty}(~(1+{\al\over n})(1+{\bi\over n})~)^n=\lim_{n\ra\infty}(~(1+{\al+\bi\over n}+{\al\bi\over n^2})~)^n=\lim_{n\ra\infty}(~(1+{\al+\bi\over n}~)^n\nn\\
&&=\lim_{n\ra\infty}(e^{\al+\bi\over n})^n=e^{\al+\bi}=\lim_{n\ra\infty}(~e^{\al\over n}e^{\bi\over n}~)^n,~~~~\Ra\nn\\
&&\lim_{n\ra\infty}(~(1+{\al\over n})^n(1+{\bi\over
n})^n~)=\lim_{n\ra\infty}(~(1+{\al\over n})(1+{\bi\over n})~)^n.
\eea
Similarly,
\bea &&\lim_{n\ra\infty}(1+{a\over n})^n=\lim_{n\ra\infty}(e^{a\over n})^n= e^a.\nn\\
&&\lim_{n\ra\infty}(~(1+{a\over n})^n(1+{b\over n})^n~)=\lim_{n\ra\infty}(1+{a\over n})^n~\lim_{n\ra\infty}(1+{b\over n})^n=\lim_{n\ra\infty}(e^{a\over n})^n~\lim_{n\ra\infty}(e^{b\over n})^n\nn\\
&&~~~~~~=\lim_{n\ra\infty}(~(e^{a\over n})^n (e^{b\over n})^n~)=e^ae^b.\nn\\
&&\lim_{n\ra\infty}(~(1+{a\over n})(1+{b\over
n})~)^n=\lim_{n\ra\infty}(~(1+{a+b\over n}+{ab\over
n^2})~)^n=\lim_{n\ra\infty}(1+{a+b\over n})^n=e^{a+b}\nn\\
&&=\lim_{n\ra\infty}(e^{a\over n}e^{b\over n})^n,~~~~\Ra\nn\\
&&\lim_{n\ra\infty}(~(1+{a\over n})^n(1+{b\over
n})^n~)\neq\lim_{n\ra\infty}(~(1+{a\over n})(1+{b\over n})~)^n.
 \eea
\subsection{Matrix functions}
Let $t=$real parameter.
\bea
&& {d\over dt}e^{a+bt}={d\over dt}~\lim_{n\ra\infty}(e^{a\over n}e^{bt\over n})^n=\lim_{n\ra\infty}{d\over dt}(e^{a\over n}e^{bt\over n})^n\nn\\
&&=\lim_{n\ra\infty}\sum^n_{k=1}(e^{a\over n}e^{bt\over n})^{k-1}~{d\over dt}(e^{a\over n}e^{bt\over n})~(e^{a\over n}e^{bt\over n})^{n-k}\nn\\
&&=\lim_{n\ra\infty}\sum^{n-1}_{k=0}(e^{a\over n}e^{bt\over n})^{n-k-1}~{d\over dt}(e^{a\over n}e^{bt\over n})~(e^{a\over n}e^{bt\over n})^{k} \nn\\
&&~~~~=\lim_{n\ra\infty}{1\over n}\sum^n_{k=1}(e^{a\over n}e^{bt\over n})^k~b~(e^{a\over n}e^{bt\over n})^{n-k}=\lim_{n\ra\infty}{1\over n}\sum^{n-1}_{k=0}(e^{a\over n}e^{bt\over n})^{n-k}~b~(e^{a\over n}e^{bt\over n})^{k}\nn\\
\eea
\bea
&&\sum^n_{k=1}\al^k=\al\sum^n_{k=1}\al^{k-1}=\al\sum^{n-1}_{k=0}\al^k=\al(\sum^n_{k=1}\al^k+1-\al^n),~~\Ra\nn\\
&&\sum^n_{k=1}\al^k= {\al(1-\al^n)\over 1-\al}\nn\\
&&
\sum^{n-1}_{k=0}\al^k=\sum^n_{k=1}\al^k+1-\al^n={\al(1-\al^n)\over
1-\al}+1-\al^n={(1-\al^n)\over 1-\al}\eea
Therefore,
\bea &&{d\over dt}e^{a+bt}|_{t=0}=\lim_{n\ra\infty}{1\over
n}\sum^n_{k=1}e^{a{k\over n}}~b~e^{a(1-{k\over
n})}=\lim_{n\ra\infty}{1\over n}\sum^n_{k=1}(e^{[{a\over
n},~]})^k~b~e^a\nn\\
&&~~~~=\lim_{n\ra\infty}{1\over n}{e^{[{a\over
n},~]}(I-e^{[a,~]})\over I-e^{[{a\over n},~]}
}~b~e^a=\lim_{n\ra\infty}{{1\over n}\over I-e^{[{a\over n},~]}
}(I-e^{[a,~]})~b~e^a \nn\\
&&=-{1\over [a,~]}(I-e^{[a,~]})~b~e^a={e^{[a,~]}-I\over
[a,~]}~b~e^a~,\nn\\\nn\\
&&=\lim_{n\ra\infty}{1\over n}\sum^{n-1}_{k=0}e^{a(1-{k\over
n})}~b~e^{a{k\over n}}=e^a\lim_{n\ra\infty}{1\over
n}\sum^{n-1}_{k=0}e^{-a{k\over n}}~b~e^{a{k\over
n}}=e^a\lim_{n\ra\infty}{1\over n}\sum^{n-1}_{k=0}e^{-[{a\over
n},~]}~b\nn\\
&&=e^a\lim_{n\ra\infty}{1\over n}{I-e^{-[a,~]}\over I-e^{-[{a\over
n},~]} }~b=e^a~{I-e^{-[a,~]}\over [a,~] }~b
 \eea
For a general matrix function $f(t)=\sum^\infty_{r=0}{f^{(r)}(a)\over
r!}(t-a)^r, $
\bea && {d\over dt}e^{f(t)}={d\over dt}~\lim_{n\ra\infty}(\prod^\infty_{r=0}e^{{f^{(r)}(a)\over r!}{(t-a)^r\over n}})^n=\lim_{n\ra\infty}{d\over dt}(\prod^\infty_{r=0}e^{{f^{(r)}(a)\over r!}{(t-a)^r\over n}})^n\nn\\
&&=\lim_{n\ra\infty}\sum^{n-1}_{k=0}(\prod^\infty_{r=0}e^{{f^{(r)}(a)\over r!}{(t-a)^r\over n}})^{n-k-1}~{d\over dt}(\prod^\infty_{r=0}e^{{f^{(r)}(a)\over r!}{(t-a)^r\over n}})~(\prod^\infty_{r=0}e^{{f^{(r)}(a)\over r!}{(t-a)^r\over n}})^{k} \nn\\
\eea
Therefore,
\bea &&{d\over
da}e^{f(a)}:={d\over dt}e^{f(t)}|_{t=a}=e^{f(a)}~{I-e^{-[f(a),~]}\over [f(a),~]
}~{df(a)\over da}.\nn\\
&& f(1)=f(0)+\int_0^1dt{ad f(t)\over I-e^{-ad
f(t)}}(e^{-f(t)}{d\over dt}e^{f(t)}) \nn
 \eea
That is,
\bea
 &&de^{f}=e^{f}~{I-e^{-[f,~]}\over [f,~]
}~df={e^{[f,~]}-I\over [f,~] }~df~e^f,~~\Ra\nn\\
&&df={[f,~]\over I-e^{-[f,~]}}~(e^{-f}de^{f})={\ln e^{[f,~]}\over I-e^{-[f,~]}}~(e^{-f}de^{f}),~~~~\txt{in other words}\nn\\
&&e^{-{\ad}{f}} (d)=e^{-f}de^{f}={I-e^{-{\ad}f}\over \ln
e^{{\ad}{f}}}(df)=\int_0^1d\al~e^{-\al{\ad}f}df=\int_0^1d\al~e^{-\al f}df e^{\al f}\nn\\
&&~~~~ =Df=df-d(\int_0^1d\al~e^{-\al{\ad}f})~f=[d-d(\int_0^1d\al~e^{-\al{\ad}f})]~f \nn\\
&&de^f =\int_0^1d\al~e^{(1-\al) f}~df~ e^{\al
f}=\int_0^1d\al~e^fe^{-\al~{\ad}f}~df
=\int_0^1d\al~e^{f-\al~{\ad}f}~df.
\nn\\
 \eea
\newpage
Similarly, the commutator of any operator, $a$, with the exponential, $e^b$, of another operator $b$ can be written as
\bea
&& [a,e^b]=[a,\lim_{n\ra \infty}(e^{b\over n})^n]=\lim_{n\ra \infty}\sum_{k=0}^{n-1}(e^{b\over n})^{n-k-1}~[a,e^{b\over n} ]~(e^{b\over n})^k\nn\\
&&=e^b~\lim_{n\ra \infty}\sum_{k=0}^{n-1}e^{-{k\over n}b}~e^{-{b\over n}}~[a,e^{b\over n} ]~e^{{k\over n}b}\nn\\
&&=e^b~\lim_{n\ra \infty}\sum_{k=0}^{n-1}e^{-{k\over n}\ad b}~e^{-{b\over n}}~[a,e^{b\over n} ]\nn\\
&&=e^b~\lim_{n\ra \infty}\sum_{k=0}^{n-1}e^{-{k\over n}\ad b}~e^{-{b\over n}}~[a,I+{b\over n} ]\nn\\
&&=e^b~\lim_{n\ra \infty}\sum_{k=0}^{n-1}{1\over n}e^{-{k\over n}\ad b}~e^{-{b\over n}}~[a,b]\nn\\
&&[a,e^b]=e^b~{I-e^{-\ad b}\over \ad b}~[a,b]\nn\\
&&ie.~~~~\txt{ad}e^b=e^b~{I-e^{-\ad b}\over \ad b}~\txt{ad}b=e^b(I-e^{-\ad b})
\eea
More generally,
\bea
&&[a,f(b)]=\int_0^1 dt~f'(b-t~\txt{\textbf{ad}} b)~[a,b]=\del_bf(b)~\int_0^1 dt~e^{-\ola{\del}_bt~\txt{\textbf{ad}} b}~[a,b]\nn\\
&& a^{-1}f(b)=f(a^{-1}ba)~a^{-1}=[a^{-1},f(b)]+f(b)~a^{-1}\nn\\
\eea
\bea
&& [f(a),g(b)]=\int_0^1 d\al\int_0^1 d\bi~g'(b-\al~\txt{\textbf{ad}} b)~f'(a-\bi~\txt{\textbf{ad}} a)~[a,b]\nn\\
\eea
\subsection{Symmetric ordered extension}
With the help of the Fourier transform, a general function of a
matrix might be written as
\bea
&&g(f)=\sum^\infty_{n=0}{g^{(n)}(0)\over
n!}f^n=\sum^\infty_{n=0}\al_nf^n=\int
d\mu(k)~\td{g}(k)~e^{-ikf},\nn\\
&&~~f=\txt{matrix},~~\al_n\in
\mathbb{C},~~\td{g}: \mathbb{C}\ral \mathbb{C},
\eea
has differential
\bea\label{chain-rule} &&dg(f)={dg(f)\over
df}~\int^1_0d\al~e^{-\al
{\ola{\del}\over\del f}~{\ad}f}~df=\sum_{n=0}^\infty {d^{n+1}g(f)\over df^{n+1}}~{(-1)^n\over (n+1)!}~({\ad}f)^n(df)  \nn\\
\label{chain-rule}&&  dg(f)~~\sr{?}{=}~~ \int^1_0d\al~{\del g\over
\del f}(f-\al~{\ad}f)~df~~~~,~~~?=(~\txt{if}
~~[f,{\ad}f]=[{\del\over\del f},{\ad}f]=0~). \nn\\
\eea
Note that given any $\varphi$,
\bea &&{\del\over \del f_i}f_j=I\delta^j_i,\nn\\
&&[{\del\over\del f_i},{\ad}f_j]\varphi={\del\over\del
f_i}({\ad}f_j(\varphi))-{\ad}f_j({\del\over\del
f_i}(\varphi))={\del\over\del
f_i}[f_j,\varphi]-[f_j,{\del\over\del
f_i}\varphi]\nn\\
&&~=[f_j,{\del\over\del f_i}\varphi]-[f_j,{\del\over\del
f_i}\varphi]=0,\nn\\
&&~[f_i,{\ad}f_j]\varphi=f_i{\ad}f_j(\varphi)-{\ad}f_j(f_i\varphi)=f_i[f_j,\varphi]-[f_j,f_i\varphi]=[f_i,f_j]\varphi.\nn\\
&&\txt{~~~~That is,}
\nn\\
&&[{\del\over\del f_i},{\ad}f_j]=0,~~~~[f_i,{\ad}f_j]=[f_i,f_j].
\eea
Therefore, $[{\del\over\del f},{\ad}f]=0$ always. However, if we
have only one variable f, then  $[f,f]=0$, but with more than one
f's,~ $[f_i,f_j]\neq 0$. Therefore one needs to write the general
case with care:
\bea
&&  dg(f)= \int^1_0d\al~{\del g\over \del
f}(f-\al~{\ad}f|_{f})~df
\eea
where $~{\ad}f|_{f}~$( a "partial" adjoint, just like the partial
derivative, whose target independent variables are $f$~ and ~ $df$
) is the adjoint action that leaves the $f$, which is in the same function argument as itself, "constant".
In the case where one defines ~~$g(f):=\int
d\mu(k)~\td{g}(k)~e^{-ik_if^i}$,~ then because of the complete
contraction, the chain rule formula,
\bea
&&dg(f)= \int^1_0d\al~{\del g\over \del
f_i}(f-\al~{\ad}f)~df_i ,~~~~f=(f_i)=(f_1,f_2,...,f_n)\nn
\eea
holds without any restriction such as ~$\ad f|_f$~ since \\
$[k_if_i,{\ad}(k_jf_j)]=[k_if_i,k_jf_j]=0$.

\subsection{Baker-Campbell-Hausdorff (BCH) formula}
\vs If one defines $f(t)$ by $e^{f(t)}=e^Ae^{Bt}$~(~
$e^{-f(t)}=(e^{f(t)})^{-1}=(e^Ae^{Bt})^{-1}=e^{-Bt}e^{-A}~)$,~
then
\bea
&&\label{Hausdorff}f(1)=\ln(e^Ae^B)\nn\\
&&=A+\int_0^1dt~({I-(e^{{\ad}A}e^{{\ad}Bt})^{-1}\over \ln
(e^{{\ad}A}e^{{\ad}Bt})})^{-1}(B)=A+\int_0^1dt~{1\over
\int_0^1d\al~e^{-\al
t~ {\ad}B}e^{-\al~{\ad} A}}(B).\nn\\
&&=A+\int_0^1dt~{ \ln
(e^{{\ad}A}e^{{\ad}Bt})\over I-(e^{{\ad}A}e^{{\ad}Bt})^{-1}}(B)\nn\\
&&=A+\int_0^1dt~e^{{\ad}A}e^{{\ad}Bt}~{ \ln
(e^{{\ad}A}e^{{\ad}Bt})\over e^{{\ad}A}e^{{\ad}Bt}-I}(B)\nn\\
&&=A+\int_0^1dt~e^{{\ad}A}e^{{\ad}Bt}~
\sum^\infty_{n=1}{(-1)^{n+1}\over n}(
e^{{\ad}A}e^{{\ad}Bt}-I)^{n-1}(B)\nn\\
&&=A+B+{1\over 2}[A,B]+{1\over 12}[A,[A,B]]-{1\over 12}[B,[A,B]]-{1\over 48}[B,[A,[A,B]]]\nn\\
&&~~~~-{1\over 48}[A,[B,[A,B]]]+...\nn\\
&&=A+B+{1\over 2}[A,B]+{1\over 12}[A,[A,B]]-{1\over 12}[B,[A,B]]-{1\over 24}[B,[A,[A,B]]]+...\nn\\
 \eea
Similarly,
\bea &&\ln(e^Ae^Be^C)=\ln(e^Ae^B)+\int_0^1dt~{ \ln(e^{{\ad}A}e^{{\ad}B}e^{{\ad}Ct})\over I-e^{-{\ad}Ct}e^{-{\ad}B}e^{-{\ad}A}}(C)\nn\\
&&~~~~=A+\int_0^1dt~{\ln(e^{{\ad}A}e^{{\ad}Bt})\over I-e^{-{\ad}Bt}e^{-{\ad}A}}(B)+\int_0^1dt~{\ln(e^{{\ad}A}e^{{\ad}B}e^{{\ad}Ct})\over I-e^{-{\ad}Ct}e^{-{\ad}B}e^{-{\ad}A}}(C),\nn\\
  &&\ln(fe^A)=\ln f+\int_0^1dt~{ \ln(e^{{\ad}(\ln f)}e^{{\ad}At})\over I-e^{-{\ad}At}e^{-{\ad}(\ln f)}}(A),\nn\\
&&\ln(AB)=\ln A+\int_0^1dt~{ \ln(e^{{\ad}(\ln A)}e^{{\ad}(\ln B)t})\over I-e^{-{\ad}(\ln B)t}e^{-{\ad}(\ln A)}}(\ln B)\nn\\
\eea

\section{Complex analytic transforms}
Given a complex function\\
 $f(z,z^\ast)=f_1(x_1,x_2)+if_2(x_1,x_2),~~z=x_1+ix_2,~~z^\ast=x_1-ix_2$ and a closed contour \\
 $C\subset \mathbb{C}\backslash \infty,~~\infty=\lim_{r\ra+\infty}\{re^{i\theta},~~0\leq\theta\leq 2\pi\}$, Stokes's theorem implies
\bea
&&\oint_{\del D}dz ~f(z,z^\ast)=\oint_Cdx_1f_1-dx_2f_2+i(dx_1f_2+dx_2f_1)\nn\\
&&~~~~=\oint_{\del D}(dx_1f_1+dx_2(-f_2)+i(dx_1+dx_2f_1)\nn\\
&&~~~~=\oint_{\del D}(dx_1f_1+dx_2(-f_2)+i(dx_1f_2+dx_2f_1)\nn\\
&&~~~~=\int_{D}d^2x\{(\del_1(-f_2)-\del_2f_1+i(\del_1f_1-\del_2f_2)\}\nn\\
&&~~~~=\int_{D}d^2x\{-(\del_1f_2+\del_2f_1)+i(\del_1f_1-\del_2f_2)\}\nn\\
&&~~~~=2i\int_{D}d^2x~{\del f(z,z^\ast)\over\del z^\ast}\nn\\
&&~~~~=-\int_{D}dz\wedge dz^\ast~{\del f(z,z^\ast)\over\del z^\ast}\nn\\
&&{\del f(z,z^\ast)\over\del z^\ast}={1\over 2}\{\del_1f_1-\del_2f_2+i(\del_1f_2+\del_2f_1)\}.\nn\\
&&\oint_{\del D}dz ~f(z,z^\ast)+\int_{D}dz\wedge dz^\ast~{\del f(z,z^\ast)\over\del z^\ast}=0
\eea
if $f$ has no singularities in $D$.

Therefore if $f$ is nonsingular (has no singularities) inside \\
$C=\del D$ then ${\del f(z,z^\ast)}=0$ iff
\bea
\oint_Cdz ~f(z)=0~~\txt{or}~~\int _{\Gamma=g(C)}dz{dg^{-1}(z)\over dz}f\circ g^{-1}(z)=0
\eea
for any invertible analytic function $g$.

However in $\overline{\mathbb{C}}=\mathbb{C}\cup\{\infty\}\simeq S^2$, the formula must also hold for the ``exterior'' of the closed contour $C$ for any continuation (which can of course be singular) of the function $f$ into the exterior of $C$. Therefore it may be more correct to say: if a closed contour contains either 1) none or 2) all of the singularities of $f$ in $\overline{\mathbb{C}}$ then ${\del f(z,z^\ast)}=0$ iff
\bea
\oint_Cdz ~f(z)=0.
\eea

We also have that
\bea
&&\int_0^{2\pi}e^{in\theta} d\theta=0,~~\forall n\in \mathbb{Z}.
\eea
Therefore if $f=f(z)$ is analytic (ie. can be expanded as a power series in $z$) then
\bea
&&f(z)={1\over 2\pi}\int_0^{2\pi}d\theta~f(re^{i\theta}+z)~~\forall r=const.
\eea
Thus if $\omega$ is a point on a circle of constant radius $C_r$ centered at $z$; ie. $\omega-z=re^{i\theta}$ then
\bea
&&f(z)={1\over 2\pi}\int_0^{2\pi}d\theta~f(re^{i\theta}+z)= {1\over 2\pi i}\int_0^{2\pi }{d(re^{i\theta})\over r e^{i\theta}}~f(re^{i\theta}+z) \nn\\
&&~~~~={1\over 2\pi i}\oint_{C_r}{d(\omega-z)\over \omega-z}~f(\omega-z+z)={1\over 2\pi i}\oint_{C_r}{d\omega\over \omega-z}~f(\omega)\nn\\
&&~~={1\over 2\pi i}\oint_{C_r}d\omega{f(\omega)\over \omega-z}.
\eea
Let $f(\omega)$ be nonsingular inside $C$ and be analytic about $\omega=z$ then $g(\omega)={1\over 2\pi i}{f(\omega)\over \omega-z}$ is nonsingular in the region between $C$ and some circle $C_r$ lying in $C$ and centered at $z$. We will write $C(z)$ to mean that the point $z$ lies inside the closed contour $C$. Therefore
\bea
\oint_{C(z)-C_r(z)}dz ~g(z)=0={1\over 2\pi i}\oint_{C(z)}d\omega{f(\omega)\over \omega-z}-{1\over 2\pi i}\oint_{C_r(z)}d\omega{f(\omega)\over \omega-z}  .
\eea

That is, if a complex function $f=f(z)$ has none of its poles inside any given closed contour $C$ then
\bea
&&f(z)={1\over 2\pi i}\oint_{C(z)} d\omega{f(\omega)\over \omega-z}.
\eea

This easily extends to a nonsingular function in $D$ as
\bea
&&f(z,z^\ast)={1\over 2\pi i}\oint_{\del D(z)}d\omega ~{f(\omega,\omega^\ast)\over \omega-z}+{1\over 2\pi i}\int_{D(z)}d\omega\wedge d\omega^\ast~{1\over \omega-z}{\del f(\omega,\omega^\ast)\over\del \omega^\ast}\nn\\
&&~~~~={1\over 2\pi i}\oint_{\del D(z)}d\omega ~{f(\omega,\omega^\ast)\over \omega-z}+{1\over 2\pi i}\int_{D(z)}d\omega\wedge df(\omega,\omega^\ast)~{1\over \omega-z}
\eea

\subsection{Laurent series}
If f is known to be singular at $a\in \mathbb{C}$ then for any two inner/outer curves $C_1,C_2$ each containing $a$, $f$ in the region between $C_1,C_2$ that excludes $a$ is given by
\bea
&& 2\pi i~f(z)=-\oint_{{C_1(a)}}d\omega_1{f(\omega_1)\over \omega_1-z}+\oint_{{C_2(a,z)}}d\omega_2{f(\omega_2)\over \omega_2-z}\nn\\
&&~~=-\oint_{{C_1(a)}}d\omega_1f(\omega_1){z-a\over \omega_1-a-(z-a)}{1\over z-a}\nn\\
&&~~+\oint_{{C_2(a,z)}}d\omega_2f(\omega_2){\omega_2-a\over \omega_2-a-(z-a)}{1\over \omega_2-a}\nn\\
&&~~=-\oint_{{C_1(a)}}d\omega_1f(\omega_1){1\over {\omega_1-a\over z-a}-1}{1\over z-a}+\oint_{{C_2(a,z)}}d\omega_2f(\omega_2){1\over 1-{z-a\over \omega_2-a}}{1\over \omega_2-a}\nn\\
&&=\oint_{{C_1(a)}}d\omega_1f(\omega_1)\sum_{n=0}^\infty ({\omega_1-a\over z-a})^n{1\over z-a}+\oint_{{C_2(a,z)}}d\omega_2f(\omega_2)\sum_{n=0}^\infty({z-a\over \omega_2-a})^n{1\over \omega_2-a}\nn\\
&&~~~~~~~~~~~~\left\{~|{\omega_1-a\over z-a}|<1~~\forall \omega_1,~~|{z-a\over \omega_2-a}|<1~~\forall \omega_2~\right\}\nn\\
&&~~=\oint_{{C_1(a)}}d\omega_1f(\omega_1)\sum_{n=0}^\infty {(\omega_1-a)^n\over (z-a)^{n+1}}+\oint_{{C_2(a,z)}}d\omega_2f(\omega_2)\sum_{n=0}^\infty{(z-a)^{n}\over (\omega_2-a)^{n+1}},\nn\\
&&~~=\oint_{{C_1(a)}}d\omega_1f(\omega_1)\sum_{n=1}^\infty {(\omega_1-a)^{n-1}\over (z-a)^{n}}+\oint_{{C_2(a,z)}}d\omega_2f(\omega_2)\sum_{n=0}^\infty{(z-a)^{n}\over (\omega_2-a)^{n+1}},\nn\\
&&~~=\oint_{{C_1(a)}}d\omega_1f(\omega_1)\sum_{n=-\infty}^{-1} {(z-a)^{n}\over (\omega_1-a)^{n+1}}+\oint_{{C_2(a,z)}}d\omega_2f(\omega_2)\sum_{n=0}^\infty{(z-a)^{n}\over (\omega_2-a)^{n+1}},\nn\\
&&f(z)=\sum_{n=-\infty}^\infty\al^f_n(a)~(z-a)^n,\\
&&\al^f_n(a)=\left\{
             \begin{array}{ll}
               {1\over 2\pi i}\oint_{{C_1(a)}}d\omega {f(\omega)\over (\omega-a)^{n+1}}, & n\leq -1 \\
                {1\over 2\pi i}\oint_{{C_2(a,z)}}d\omega {f(\omega)\over (\omega-a)^{n+1}}, & n\geq 0
             \end{array}
           \right\}\nn\\
&&~~={1\over 2\pi i}\oint_{{C_1(a)}\theta(-1-n)+{C_2(a,z)}\theta(n)}d\omega {f(\omega)\over (\omega-a)^{n+1}}\nn\\
&&~~\eqv {1\over 2\pi i}\oint_{\Gamma(a)}d\omega {f(\omega)\over (\omega-a)^{n+1}},~~a\in C_1^0\subset \Gamma^0\subset C_2^0,~~n\in \mathbb{Z},\nn\\
&&~~\txt{ie.}~~0<|\omega_1-a|\leq |\omega-a|\leq |\omega_2-a|~~\forall \omega\in \Gamma.\nn\\
&& 0<|\omega_1-a|<|z-a|<|\omega_2-a|~~\forall \omega_1\in C_1,\omega_2\in C_2.
\eea
The condition $|\omega_1-a|<|z-a|<|\omega_2-a|~~\forall \omega_1\in C_1,\omega_2\in C_2$ is satisfied for the case where $C_1,C_2$ are circular so that the domain $D$ of convergence of the series is any strip
\bea
&& D=D(r_1,r_2)=\{z,~~0<r_1<|z-a|< r_2\}
\eea
where $r_1$ is the radius of $C_1$ about $a$ and $r_2$ is the radius of $C_2$ about $a$.

If all $k$ poles of $f$ lie in a region of finite size $L$ and $f$ has no poles at $\infty$ then $C_2$ may be taken to $\infty$ and $C_1$ can be chosen to consist of a chain of ``small'' circles, each of raduis $r_1\ra 0$ and encircling one pole, covering all poles $\{a_i,~i=1,...,k\}$ of $f$.

For the nonholomorphic case
\bea
&&f(z,z^\ast)=\sum_{n=-\infty}^\infty\al^f_n(a,a^\ast)~(z-a)^n.\nn\\
&&\al^f_n(a,a^\ast)={1\over 2\pi i}\oint_{\Gamma(a)}d\omega ~{f(\omega,\omega^\ast)\over (\omega-a)^{n+1}}+{1\over 2\pi i}\int_{\Gamma^0(a)}d\omega\wedge d\omega^\ast~{1\over (\omega-a)^{n+1}}{\del f(\omega,\omega^\ast)\over\del \omega^\ast}.\nn
\eea
\subsection{Fourier series and other derived transforms}
The Laurent series $f(z)=\sum_{n=-\infty}^\infty\al^f_n(a)~(z-a)^n$ may also be rewritten as
\bea
&&f(a+q^z)=\sum_{n=-\infty}^\infty\al^f_n(a)~q^{nz},~~\forall q\in \mathbb{C}
\eea
since it is true in general that $f(g(z))=\sum_{n=-\infty}^\infty\al^f_n(a)~(g(z)-a)^n$ for any function~~$g:\mathbb{C}\ra \A(\mathbb{C})$ and $g(z)=a+q^z$ is an example. On the other hand, letting $f\ra f\circ g^{-1}$, we have
\bea
&&f(z)=\sum_{n=-\infty}^\infty\al^{f\circ g^{-1}}_n(a)~(g(z)-a)^n,\nn\\
&&\al^{f\circ g^{-1}}_n(a)={1\over 2\pi i}\oint_{\Gamma(a)}d\omega {f\circ g^{-1}(\omega)\over (\omega-a)^{n+1}},~~a\in C_1^0\subset \Gamma^0\subset C_2^0,~~n\in \mathbb{Z},\nn\\
&&~~~\txt{ie.}~~0<|\omega_1-a|\leq |\omega-a|\leq |\omega_2-a|~~\forall \omega\in \Gamma,\nn
\eea
for any invertible $g:\mathbb{C}\ra \mathbb{C}$, where we must now restrict the function $f$ to a domain where $g^{-1}$ is single-valued. For example, in the case \\
$g(z)=a+e^{z},~~g^{-1}(z)\in\{g^{-1}_k(z)=2\pi ki+\ln(z-a),~k\in \mathbb{Z}\}$ we must choose only one from the following infinite sequence of regions
\bea
&&D_k=\{z=x+iy,~x\in \mathbb{R},~2\pi k\leq y<2\pi (k+1)\},~~k\in \mathbb{Z}.
\eea

The same trick applied to Cauchy's integral formula implies that
\bea
&&f(z)\eqv f\circ g^{-1}(g(z))={1\over 2\pi i}\oint _{\Gamma(z)}d\omega{f\circ g^{-1}(\omega)\over \omega-g(z)}\nn\\
&&~~~~={1\over 2\pi i}\int _{\Gamma'=g^{-1}(\Gamma(z))}du~{dg(u)\over du}{f(u)\over g(u)-g(z)} ~~\forall g,\nn
\eea
whenever $f\circ g^{-1}$ has no singularities in $D=\Gamma^0\cup\Gamma$ and $g^{-1}$ is single-valued on $\del D=\Gamma$~~(~ie. $g(u_1)=g(u_2)\Ra u_1=u_2~~\forall u_1,u_2\in \Gamma'=g^{-1}(\Gamma)$~).

Thus if $f\circ g^{-1}$ is singular (ie. undetermined) at $0$ (ie. $f$ is singular at $g^{-1}(0)$) [and $g^{-1}$ is single-valued on $\del D=\Gamma$] then $\Gamma^0\cup \Gamma$ must be chosen to avoid this singularity and thus in a strip about $g^{-1}(0)$~~ $f$ will have the Laurent expansion
\bea
&&f(z)=\sum_{n=-\infty}^\infty\al^{f\circ g^{-1}}_n~(g(z))^n\eqv \sum_{n=-\infty}^\infty\td{f}_g(n)~(g(z))^n,\nn\\
&&\al^{f\circ g^{-1}}_n={1\over 2\pi i}\oint_{\Gamma(0)}d\omega {f\circ g^{-1}(\omega)\over \omega^{n+1}}= {1\over 2\pi i}\int_{\Gamma'=g^{-1}(\Gamma(0))}du~{dg(u)\over du} {f(u)\over (g(u))^{n+1}}\eqv\td{f}_g(n),\nn\\
&&~~0\in C_1^0\subset \Gamma^0\subset C_2^0,~~n\in \mathbb{Z},~~\omega=g(u),\nn\\
&&\txt{ie.}~0<|\omega_1|\leq |\omega|\leq |\omega_2|~\forall \omega\in \Gamma.~~0<|\omega_1|< |g(z)|< |\omega_2|~\forall \omega_1\in C_1,~\omega_2\in C_2,\nn\\
\eea
where $\Gamma=\Gamma(0)$ means that $\Gamma$ is a closed curve in a strip $S$ about $0$ [~note that the expansion of $f$ is about $g^{-1}(0)$ and the corresponding image curve is \\
$\Gamma'=g^{-1}(\Gamma)\simeq\Gamma'(g^{-1}(0))$, a curve in or on $g^{-1}(S)$ that may approach but may not reach $g^{-1}(0)$~]. One notes that $\ln 0=\infty$ (ie. in the case of \\
$g(z)=e^z=e^xe^{iy}=e^x\cos(y)+ie^x\sin(y),~~g^{-1}(z)=\ln z=\ln|z|+i\txt{Arg}(z)$). If $\Gamma(0)=\{\omega\}$ is chosen to be any circle of radius $r,~~\vep=|\omega_1|\leq r=|\omega|\leq\rho=|\omega_2|$ centered at 0, then the resulting series is
{\footnotesize\bea
&&f(z)=\sum_{n=-\infty}^\infty \td{f}_n~e^{nz},~~\td{f}_n={1\over 2\pi i}\int_{\ln r-\pi i}^{\ln r+\pi i}du~f(u)~e^{-nu}\eqv{1\over 2\pi i}\int_{\ln r}^{\ln r+2\pi i}du~f(u)~e^{-nu},\nn\\
&&\vep=|\omega_1|\leq |\omega|=|e^{u}|\leq\rho=|\omega_2|,~~~~\vep=|\omega_1|<|e^{z}|=e^{\txt{Re}(z)}=e^{x}\leq \rho=|\omega_2|,\nn\\
&&~~~~\Ra~\ln\vep\leq x=\txt{Re}(z)\leq \ln\rho~~(\txt{convergence requirement}).
\eea}
Therefore if $\vep\ra 0,~~\rho\ra\infty$ then $-\infty <x=\txt{Re}(z)<\infty$ and so
\bea
&&f(z)\eqv f_\gamma(z)=\sum_{n=-\infty}^\infty \td{f}_n~e^{nz},~~\nn\\
&&~~~~\td{f}_n={1\over 2\pi i}\int_{\gamma-\pi i}^{\gamma+\pi i}du~f(u)~e^{-nu}\eqv {1\over 2\pi i}\int_{\gamma}^{\gamma+2\pi i}du~f(u)~e^{-nu},\nn\\
&&-\infty<\gamma<\infty,~~-\infty< x=\txt{Re}(z)< \infty,\nn\\
&&\txt{ie.}~~\forall z\in \mathbb{C}~~\&~~\forall f~~st.~~f ~\txt{is may be undetermined only at $\infty$}.\nn
\eea
The integral ${1\over 2\pi i}\int_{-\pi i}^{\pi i}du ~e^{nu}e^{-mu}={\sin(n-m)\pi\over (n-m)\pi}=\delta_{nm}$ (the analog of\\
 $\oint_{\Gamma(a)}dz~{(z-a)^{n-m}\over z-a}=\delta_{nm}$) is useful for motivating the series from an alternative point of view where $\{e_n(z)=e^{nz},~n\in \mathbb{Z}\}$ may be regarded as a complete set of orthonormal functions in terms of which $f(z)$ can be expanded. One can similarly define a continuous series with the help of the function:
\bea
&&\lim_{a\ra\infty}{1\over 2a}\int_{-a}^a dq ~e^{uq}e^{-vq}=\lim_{a\ra\infty}{\sin(u-v)a\over (u-v)a} =\delta_{uv}.\nn\\
&&\lim_{a\ra\infty}{1\over 2}\int_{-a}^a dq ~e^{uq}e^{-vq}=\lim_{a\ra\infty}{\sin(u-v)a\over (u-v)} =\delta(u-v).
\eea

In this case, if one considers only periodic functions of the form $y\eqv y+2\pi$ then the choice of $\Gamma(0)$ is no longer restricted to the region where $g^{-1}(z)=\ln z$ is single-valued but is only restricted by the singular/non-singular requirement for $f$ as usual.

Notice that in the integral formula with transformed contour $C$
\bea
&& f(z)={1\over 2\pi i}\int _{C=g^{-1}(\Gamma(z))}du~{dg(u)\over du}{f(u)\over g(u)-g(z)},
\eea
setting ${1\over f(z)}={dg(z)\over dz}$ implies that
\bea
{1\over {dg(z)\over dz}}={1\over 2\pi i}\int _{C=g^{-1}(\Gamma(z))}{du\over g(u)-g(z)},
\eea
where $g'\circ g^{-1}$ has no singularities in $D=\Gamma^0\cup\Gamma$ and $g^{-1}$ is single-valued on $\del D=\Gamma$.

In the case $g(z)={az+b\over cz+d},~~g^{-1}(z)=-{zd-b\over zc-a}$ for example one has
\bea
&& f(z)={1\over 2\pi i}\int _{C=g^{-1}(\Gamma(z))}du~{dg(u)\over du}{f(u)\over g(u)-g(z)}\nn\\
&&~~~~={1\over 2\pi i}\int _{C=-{\Gamma(z) d-b\over \Gamma(z) c-a}}du~{cz+d\over cu+d}~{f(u)\over u-z},\nn
\eea
where $f\circ g^{-1}(\omega)=f(-{\omega d-b\over \omega c-a})$ has no singularities in $D(z)=\Gamma^0(z)\cup\Gamma(z)$ and $g^{-1}(\omega)=-{\omega d-b\over \omega c-a}$ is single-valued on $\del D(z)=\Gamma(z)$.

\subsection{Groups of invertible functions and related transforms}
To summerize the properties of the contour integral, let $\Gamma$ be a closed contour with interior $\Gamma^0$ and
\bea
&&\delta_\Gamma(z)=\left\{
                     \begin{array}{ll}
                       1, & z\in D=\Gamma^0\cup\Gamma \\
                       0, & z\not\in D=\Gamma^0\cup\Gamma
                     \end{array}
                   \right.,
\eea
then
\bea
&&\oint_\Gamma{d\omega\over \omega-g(z)}=\delta_\Gamma(g(z)),~~\txt{ie.}~~\oint_{\Gamma(g(z))}{d\omega\over \omega-g(z)}=1~~\forall g.\nn\\
&&\oint_{\Gamma(g(z))}{d\omega\over \omega-g(z)}=\int_{C(z)=g^{-1}(\Gamma(g(z)))}{dg(u)\over du}{du\over g(u)-g(z)}=1.
\eea
If $g^{-1}$ exists and $f\circ g^{-1}$ is non-singular in $D=\Gamma^0\cup\Gamma,~~\Gamma^0\ni g(z);~ \txt{ie.}~\Gamma=\Gamma(g(z))$, then
{\footnotesize\bea
&&f(z)=f\circ g^{-1}(g(z))=\oint_{\Gamma(g(z))}d\omega{f\circ g^{-1}(\omega)\over \omega-g(z)}=\int_{C(z)=g^{-1}(\Gamma(g(z)))}du{dg(u)\over du}{f(u)\over g(u)-g(z)}.\nn\\
\eea}
That is $\forall f,g,C$ such that $g(C)=\Gamma$ is a closed contour, $f\circ g^{-1}$ is non-singular in $D=\Gamma\cup \Gamma^0$ and $g^{-1}$ is single-valued in $D=\Gamma\cup \Gamma^0$ we have
\bea
&&f(z)=f\circ g^{-1}(g(z))=\int_{C(z)=g^{-1}(\Gamma(g(z))}du{dg(u)\over du}{f(u)\over g(u)-g(z)}.\nn\\
\eea
It may also be possible to restrict $f$ and/or $g$ to a class of functions where $C$ would also be a closed contour.

If $G=\{g\in \F(\mathbb{\mathbb{C}}),~~g^{-1}~\exists\}$ is a group of complex invertible functions (maps in general) with function composition as the group product, then any given function $f$ has a $G$-representation $f_G$ for all possible $G$'s and may be decomposed, for each $G$, through the insertion of an identity as follows
\bea
&&f_G(z)={1\over |G|}\sum_{g\in G}f\circ g^{-1}(g(z))\eqv \int_{g\in G}d\mu(g)~f\circ g^{-1}(g(z))\nn\\
&&~~~~=\int_{g\in G}d\mu(g)~\sum_{n=-\infty}^\infty\al_n^{f\circ g^{-1}}(0)~(g(z))^n=\sum\!\!\!\!\!\!\!\!\int_{(n,g)\in \mathbb{Z}\times G}d\mu(g)~\td{f}_n^g~(g(z))^n,\nn\\
&&~~~~\td{f}_n^g=\al_n^{f\circ g^{-1}}(0),~~\int_{g\in G}d\mu(g)~1(g)=1,~~\sum_{g\in G}1(g)=|G|.
\eea
Such decompositions may be used to represent solutions, of differential equations, which typically determine $\td{f}$. Boundary/initial conditions can then be used to determine the actual form or ``shape'' of $G$. Note that $G$ may also be chosen to contain the space of inverses of $g$ if one wishes to extend to domains where $g^{-1}$ is not unique.

If one takes the example $G=\{g:z\mapsto g(z)=e^{\omega z},~~d\mu(g)={d\omega\over\omega}$,\nn\\
$~~\omega\in \Gamma=\Gamma(0)=\del D,~ 0\in D\subset \mathbb{\mathbb{C}}\}$,~ as $\oint_{\Gamma(a)}{d\omega\over \omega-a}=1$,~ then
\bea
&&f_{\Gamma(a)}(z)=\oint_{\Gamma(a)}{d\omega\over \omega-a}~\sum_{n=-\infty}^\infty\al_n^{f\circ g^{-1}}(0)~e^{n\omega z}=\oint_{\Gamma(a)}{d\omega\over \omega-a}~\sum_{n=-\infty}^\infty\al_n^{f\circ g_n^{-1}}(0)~e^{\omega z}\nn\\
&&~~~~=\oint_{\Gamma(a)}d\omega~\td{f}_a(\omega)~e^{\omega z},\nn\\
&&f\circ g^{-1}(z)=f({1\over\omega}\ln z),~~f\circ g_n^{-1}(z)=f({n\over\omega}\ln z).\nn\\
&&\al_n^{f\circ g_n^{-1}}(0)= \oint_{C(0)}dv {f\circ g_n^{-1}(v)\over v^{n+1}}=\oint_{C(0)}dv {f({n\over\omega}\ln(v))\over v^{n+1}}\nn\\
&&~~~~=\int_{C'=g_n^{-1}(C(0))}du{dg_n(u)\over du}{f(u)\over (g_n(u))^{n+1}}=\int_{C'={n\over\omega}\ln(C(0))}du~{\omega\over n}~e^{-{\omega u\over n}}~f(u).\nn\\
&&\td{f}_a(\omega)={1\over\omega-a}\sum_{n=-\infty}^\infty\al_n^{f\circ g_n^{-1}}(0)={1\over\omega-a}\oint_{C(0)}dv \sum_{n=-\infty}^\infty{f({n\over\omega}\ln(v))\over v^{n+1}}\nn\\
&&~~~~={\omega\over\omega-a}\sum_{n=-\infty}^\infty\int_{C'={n\over\omega}\ln(C(0))}du~f(u) {e^{-{\omega u\over n}}\over n}\nn\\
&&~~~~={\omega\over\omega-a}\int_{C'={1\over\omega}\ln(C(0))}du~e^{-\omega u}~\sum_{n=-\infty}^\infty f(nu).\nn\\
&&|v_1|\leq |v|\leq |v_2|,~~0<|v_1|< |e^{\omega z}|<|v_2|.
\eea
This Fourier-like transform verifies the existence of the Fourier transform.

\subsection{Several variables}

 We may also consider $n$ complex variables $Z=(z_1,...,z_n)$ for which case the integral formula applied to each argument, of the holomorphic function, separately becomes
 \bea
&&f(Z)={1\over (2\pi i)^n}\oint_{S(Z)} d^n\Omega{f(\Omega)\over \prod_{i=1}^n(z_i-\omega_i)},~~\Omega=(\omega_1,...,\omega_n),\nn\\
&&\oint_{S(Z)} d^n\Omega~\eqv~\oint_{C_1(z_1)} d\omega_1\oint_{C_2(z_2)} d\omega_2~...\oint_{C_n(z_n)} d\omega_n.
\eea
One may write $Z_1=(z_1,0,...,0),~~Z_2=(0,z_2,0,...,0),..., Z_n=(0,...,0,z_n)$, then for each $i$~ contour $C_i(z_i)$ can be replaced by a $(2n-1)$-dimensional (hollow) cylinder-like hypersurface $\C_i(Z_i)$  in $\mathbb{C}^n$ and thus $S(Z)=\bigcap_{i}\C_i(Z_i)$ is the $(2n-1)$-dimensional hypersurface in $\mathbb{C}^n$ formed by the intersection $\bigcap_{i}\C_i(Z_i)$ of the $(2n-1)$-dimensional (hollow) cylinder-like hypersurfaces.

Similarly one can define a Fourier-like transform
\bea
&&f_{S(A)}(Z)=\oint_{S(A)}d^n\Omega~\td{f}_A(\Omega)~e^{\Omega Z}=\oint_{\bigcap_{i}\C_i(A_i)}d^n\Omega~\td{f}_A(\Omega)~e^{\Omega Z},\nn\\
&&S(A)=\bigcap_i\C_i(A_i),~A_1=(a_1,0,..,0),~A_2=(0,a_2,0,..,0),..,~A_1=(0,..,0,a_n).\nn
\eea

\section{Some inequalities}
\subsection{Young's inequality}
Let~ $\vphi:\mathbb{R}^+\ra \mathbb{R}^+,~\vphi(0)=0,~\lim_{x\ra\infty}\vphi(x)=+\infty$~ be increasing \\
(ie. ${d\vphi(x)\over dx}\eqv\vphi'(x)\geq 0~~\forall x\geq 0$). Then $\vphi^{-1}$ is also increasing as \\ $\vphi^{-1}(\vphi(x))=x~~\Ra~~\vphi^{-1}{}'(\vphi(x))={1\over\vphi'(x)}\geq 0$. We also have
\bea
f(c)=\int_0^c dx~\vphi(x)+\int_0^{\vphi(c)} dx~\vphi^{-1}(x)=c\vphi(c)~~\forall c\geq 0
\eea
since $f'(c)=\vphi(c)+\vphi'(c)~\vphi^{-1}(\vphi(c))=\vphi(c)+c\vphi'(c)=(c\vphi(c))'.$

Therefore the \emph{continuous} function
\bea
&&g:\mathbb{R}^+\ra \mathbb{R}^+,~a\mapsto g(a)={ab\over \int_0^a dx~\vphi(x)+\int_0^{b} dx~\vphi^{-1}(x)}\eqv h(a,b),~~b\in \mathbb{R}^+\nn
\eea
 is stationary at $a=\vphi^{-1}(b)$ ~(~by ~$ g'(a)\eqv \del_ah(a,b)=0$~).

Furthermore one can check that
\bea
&&g(\vphi^{-1}(b))=1,~~\lim_{a\ra 0}g(a)=0=\lim_{a\ra +\infty}g(a),
\eea
hence $g(a)\leq 1~~\forall a$ since $g$ is continuous. That is, we have the inequality

\bea
&&ab\leq \int_0^a dx~\vphi(x)+\int_0^{b} dx~~\forall a,b
\eea
where equality holds when $b=\vphi(a)$.

Setting $\vphi(x)=x^{p-1},~~p\in \mathbb{R}^+$, the conditions $\vphi(0)=0,~\lim_{x\ra\infty}\vphi(x)=\infty$ are satisfied if $p>1$ and one obtains
\bea
&& ab\leq {a^p\over p}+{b^{p\over p-1}\over {p\over p-1}}\eqv {a^p\over p}+{b^{p'}\over p'},~~~~{1\over p}+{1\over p'}=1.\nn\\
\eea
Equality holds iff $a^p=b^{p'}$.

\subsection{Holder's inequality}
With $p>1$ define ~$\|f\|_p=(\int d\mu(x)|f(x)|^p)^{1\over p}\eqv (\int d\mu|f|^p)^{1\over p}$~ and set \\
$a={|f(x)|\over \|f\|_p},~b={|g(x)|\over \|g\|_{p'}}$.
Then
\bea
&&{|f(x)|\over \|f\|_p}{|g(x)|\over \|g\|_{p'}}\leq {1\over p}{|f(x)|^p\over (\|f\|_{p})^p}+{1\over p'}{|g(x)|^{p'}\over (\|g\|_{p'})^{p'}}\nn\\
&&{|f(x)g(x)|\over \|f\|_p\|g\|_{p'}}\leq {1\over p}{|f(x)|^p\over (\|f\|_{p})^p}+{1\over p'}{|g(x)|^{p'}\over (\|g\|_{p'})^{p'}}\nn\\
&&{1\over \|f\|_p\|g\|_{p'}}\int d\mu |fg|\leq {1\over p}{\int d\mu|f|^p\over (\|f\|_{p})^p}+{1\over p'}{\int d\mu|g|^{p'}\over (\|g\|_{p'})^{p'}}={1\over p}+{1\over p'}=1,\nn\\
&&\int d\mu |fg|\leq \|f\|_p\|g\|_{p'}.
\eea
Equality holds iff ~$|f(x)|^p=\al~|g(x)|^{p'},~~\al\in \mathbb{R}^+,~\al\neq 0$.

For ~$0<p<1,~~q={1\over p}>1$,~ writing ~$f=u^{-p}=u^{-{1\over q}},$~~\\
$g= u^{1\over q}v^{1\over q}=(uv)^{1\over q},~~u(x)\geq 0,~v(x)\geq 0~~\forall x$~ one obtains
\bea
\int d\mu~uv\geq (\int d\mu~ v^p)^{1\over p}~(\int d\mu~ u^{p'})^{1\over p'}.
\eea

\subsection{Minkowski's inequality}
For $p>1$
{\small\bea
&&(\|f+g\|_p)^p=\int d\mu |f+g|^p=\int d\mu |f+g|^{p-1}|f+g|\leq \int d\mu |f+g|^{p-1}(|f|+|g|)\nn\\
&&~~~~=\int d\mu |f+g|^{p-1}|f|+\int d\mu |f+g|^{p-1}|g|\nn\\
&&~~~~\leq \|f\|_p~\int (d\mu |f+g|^{(p-1)p'})^{1\over p'}+\|g\|_p~\int d\mu |f+g|^{(p-1)p'})^{1\over p'}\nn\\
&&~~~~=(\|f\|_p+\|g\|_p)~\int (d\mu |f+g|^{p})^{1\over p'}=(\|f\|_p+\|g\|_p)~(\|f+g\|_p)^{p\over p'},\nn\\
&&(\|f+g\|_p)^{p-{p\over p'}}=\|f+g\|_p \leq \|f\|_p+\|g\|_p.
\eea}
For $0<p<1$ the same argument and Holder's inequality for $0<p<1$ gives
\bea
\|f+g\|_p \geq \|f\|_p+\|g\|_p.
\eea

\section{Map continuity}
A map $f:\A\ra \B$ between two linear metric spaces ~$\A=(\A,||),~\B=(\B,||)$~ is continuous iff any of the following is true

\bea
&&(1)~\txt{(uniformly)}~x\ra y~~\Ra~~f(x)\ra f(y).\nn\\
&&(2)~\txt{(uniformly)}~|x-y|\ra 0~~\Ra~~|f(x)-f(y)|\ra 0.\nn\\
&&(3)~\txt{(uniformly)}~|x-y|<\vep\ra 0^+\nn\\
&&~~~~~~~~\Ra~~\exists \delta=\delta(\vep)\sr{\vep\ra 0^+}{\ral} 0^+~~st~~|f(x)-f(y)|<\delta(\vep).\nn\\
&&(4)~~\forall B_\vep(x),~\vep\ra 0^+,~~\exists~~B_{\delta(\vep)}(f(x))~~\nn\\
&&~~~~~~~~st~~ f(B_\vep(x))\subseteq B_{\delta(\vep)}(f(x)),~\delta(\vep)\sr{\vep\ra 0^+}{\ral} 0^+\nn\\
&&\txt{where}~~~~B_\vep(x)=\{y,~~|x-y|<\vep\}.\nn\\
&&(5)~[\txt{if $f^{-1}~\exists$}]~~\forall B_\vep(x),~\vep\ra 0^+,~~\exists~~B_{\delta(\vep)}(f(x))~~\nn\\
&&~~~~~~~~st~~ B_\vep(x)\subseteq f^{-1}(B_{\delta(\vep)}(f(x))),~\delta(\vep)\sr{\vep\ra 0^+}{\ral} 0^+.\nn\\
\eea
It follows that a composition $f\circ g$ of two continuous maps $f,g$ is continuous since
\bea
&&|x-y|<\vep~~\Ra~~|g(x)-g(y)|<\delta(\vep)\eqv\vep_g~~\nn\\
&&~~~~\Ra~~|f\circ g(x)-f\circ g(y)|<\delta(\vep_g).\nn
\eea
The same is true for sums and products of continuous maps by the triangle inequality:
\bea
&&|x-y|<\vep~~\Ra\nn\\
&&|f(x)+g(x)-(f(y)+g(y))|\leq |f(x)-f(y)|+|g(x)-g(y)|\nn\\
&&~~~~<\delta_f(\vep)+\delta_g(\vep)\eqv \delta(\vep),\nn\\
&&|f(x)g(x)-f(y)g(y)|=|f(x)g(x)-f(y)g(x)+f(y)g(x)-f(y)g(y))|\nn\\
&&~~~~ \leq |f(x)-f(y)||g(x)|+|f(y)||g(x)-g(y)|\nn\\
&&~~~~~~~~<\delta_f(\vep)|g(x)|+\delta_g(\vep)|f(y)|\eqv \delta(\vep).\nn\\
\eea

A set $S$ is open iff for any $x\in S$ one can find $B_\vep(x)\subseteq S$.  Notice that in the definition of map continuity if $\B$ is open then $B_{\delta(\vep)}(f(x))\subseteq \B$ for sufficiently small $\vep$. But if $f^{-1}~\exists$ then $B_{\delta(\vep)}(f(x))\subseteq \B~~\Ra~~f^{-1}(B_{\delta(\vep)}(f(x)))\subseteq f^{-1}(\B)=\A$ and hence $B_\vep(x)\subseteq f^{-1}(B_{\delta(\vep)}(f(x)))$ guarantees that $\A$ must also be open since one has $B_\vep(x)\subseteq f^{-1}(B_{\delta(\vep)}(f(x)))\subseteq \A$ and this is true for any $x\in\A$. Therefore the map continuity condition implies that the inverse image of any open set is open. For the converse, if the inverse image of every open set is open under $f$ then for any $x\in \A$~  $f^{-1}(B_{\delta}(f(x)))$~ is open for all $\delta>0$ since $B_{\delta}(f(x))$ is open. Now since $x\in f^{-1}(B_{\delta}(f(x)))$ one can find $\vep>0$ such that $B_\vep(x)\subseteq f^{-1}(B_{\delta}(f(x)))$ and in particular, since $\delta>0$ was arbitrary, one can choose $\delta=\delta(\vep)\sr{\vep\ra 0^+}{\ral}0^+$, which is the condition for continuity. Hence a map $f$ is continuous iff the inverse image of every open set is open. One observes here that map continuity can also be stated as: for any nbd $B_{\delta}(f(x))$ one can find $\vep=\vep(\delta)$ such that $f(B_{\vep(\delta)}(x))\subseteq B_{\delta}(f(x))$~~~OR~~~for any nbd $B_{\delta}(u)$ of $u\in\B$ one can find $\vep=\vep(\delta)$ such that $f(B_{\vep(\delta)}(f^{-1}(u)))\subseteq B_{\delta}(u)$

A map $f:D\subseteq\A\ra \B$ is said to be (uniformly) bounded if
\bea
&&\forall x,y\in D,~~|f(x)-f(y)|\leq M,~~0\leq M<\infty.
\eea

A map $f:D\subseteq\A\ra \B$ is (uniformly) differentially bounded iff
\bea
&&\forall x,y\in D,~~|f(x)-f(y)|\leq M|x-y|,~~0\leq M<\infty.
\eea
It is clear that a differentially bounded map is continues as one may simply set $\delta(\vep)=M\vep$. The composition or sum of two differentially bounded maps is differentially bounded.

In a general metric space $({\S},d)$ rather than a linear metric space $(\H,|~|)$ one needs to replace $|a-b|$ by $d(a,b)$.
\subsection{Uniform continuity in terms of sets}
Uniform continuity means
\bea
B_\vep(x)\cap B_\vep(y)\neq \{\}~~\Ra~~B_{\delta(\vep)}(f(x))\cap B_{\delta(\vep)}(f(y))\neq \{\}.
\eea
On the other hand continuity requires
\bea
f(B_\vep(x))\subseteq B_{\delta(\vep)}(f(x)),~~f(B_\vep(y))\subseteq B_{\delta(\vep)}(f(y))
\eea
which implies
\bea
f(B_\vep(x))\cap f(B_\vep(y))\subseteq B_{\delta(\vep)}(f(x))\cap B_{\delta(\vep)}(f(y)).
\eea
Since $f(B_\vep(x)\cap B_\vep(y))\subseteq f(B_\vep(x))\cap f(B_\vep(y))$ we identify the condition for uniform continuity as
\bea
f(B_\vep(x)\cap B_\vep(y))\subseteq B_{\delta(\vep)}(f(x))\cap B_{\delta(\vep)}(f(y))
\eea
which reflects the fact that a uniformly continuous maps is continuous but the converse may not be true.
Let a set $A$ be \emph{\textbf{uniformly open}} iff $\forall x,y\in A$ one can find $\vep >0$ such that $B_\vep(x)\cap B_\vep(y)\subseteq A$. Then a uniformly open set is open but the converse may not be true. Uniform continuity/openness and continuity/openness are equivalent in a separable space ( one in which every pair $(x,y;~x\neq y)$ of distinct points  have disjoint neighborhoods ~$B_{\vep_1}(x),~B_{\vep_2}(y),~~B_{\vep_1}(x)\cap B_{\vep_2}(y)=\{\}$ ~) since an open set would be automatically uniformly open if one decides that $\{\}\subseteq A$ for any set $A$, but not necessarily in a nonseparable space.

One can check as in the case of continuity that a map $f$ is uniformly continuous iff the inverse image $f^{-1}(\breve{O})$ of every uniformly open set $\breve{O}$ is uniformly open.

The fact that the intersection $A\cap B$ of two open sets is open follows because for $a\in A\cap B$,~~$\exists~\vep_1,\vep_2 >0$ such that
\bea
B_{\vep_1}(a)\subseteq A,~~B_{\vep_2}(a)\subseteq B
\eea
which implies that
\bea
B_{\vep_1}(a)\cap B_{\vep_2}(a)\subseteq A\cap B.
\eea
But this means that $B_{\min(\vep_1,\vep_2)}(a)\subseteq B_{\vep_1}(a)\cap B_{\vep_2}(a)\subseteq A\cap B$ and hence
$A\cap B$ is open. One can similarly check that $B_{\max(\vep_1,\vep_2)}(a)\subseteq B_{\vep_1}(a)\cup B_{\vep_2}(a)\subseteq A\cup B$ and hence $A\cup B$ is open. With the same steps one can show that the unions and intersections of uniformly open sets are uniformly open.

\section{Sequences and series}
A sequence $s$ in a set ${\S}$ is an ordered selection of objects in ${\S}$; ie. a map from the natural numbers $\mathbb{N}$ to a set of objects ${\S}$.
\bea
s:\mathbb{N}\ra {\S},~~n\mapsto s_n.
\eea
The sequence $s$ is bounded iff one can find $M\in \mathbb{R}$ such that
\bea
d(s_n,s_m)\leq M~~\forall n,m\in \mathbb{N}.
\eea
Define the $\vep$ neighborhood $\N_\vep(A)$ of a set $A\subseteq {\S}$ by
\bea
\N_\vep(A)=\{y\in {\S},~~d(a,y)<\vep,~~\forall a\in \A\}\eqv \bigcup_{a\in A}\N_\vep(a).
\eea

A sequence in a metric space $\S$ is convergent (ie. converges to a point $L\in \S$) iff
\bea
&&1)~~\forall \vep >0,~~\exists N=N(\vep)<\infty ~~s.t.~~d(s_n,L)<\vep~~\forall n>N(\vep).\nn\\
&&2)~~\forall \N_\vep(L)~~\exists N(\vep)<\infty~~s.t.~~s_n\in \N_\vep(L)~~\forall n>N(\vep).\nn\\
&&(3)~~\forall N<\infty,~~\exists~0<\vep=\vep(N)\sr{N\ra\infty}{\ral}0~~s.t.~~\forall n>N,~~d(s_n,L)<\vep(N).\nn\\
&&(3)~~\forall N<\infty,~~\exists~0<\vep=\vep(N)\sr{N\ra\infty}{\ral}0~~s.t.~~n>N~~\Ra~~d(s_n,L)<\vep(N).\nn\\
&&(4)~~\forall N<\infty,~~\exists 0<\vep=\vep(N)\sr{N\ra\infty}{\ral}0~~s.t.~~n>N~~\Ra~~s_n\in \N_{\vep(N)}(L).\nn
\eea

A sequence in a metric space is (uniformly) converging or Cauchy iff
\bea
&&1)~~\forall \vep >0,~~\exists N=N(\vep)<\infty ~~s.t.~~d(s_m,s_n)<\vep~~\forall m,n>N(\vep).\nn\\
&&2)~~\forall \vep >0~~\exists N=N(\vep)~~s.t.~~\N_\vep(s_m)\cap  \N_\vep(s_n)\neq \{\}~~\forall~m,n>N(\vep).\nn\\
&&(3)~~\forall N<\infty,~~\exists~0<\vep=\vep(N)\sr{N\ra\infty}{\ral}0~~s.t.~~\forall n,m>N,~~d(s_n,s_m)<\vep(N).\nn\\
&&(3)~~\forall N<\infty,~~\exists~0<\vep=\vep(N)\sr{N\ra\infty}{\ral}0~~s.t.~~ n,m>N~~\Ra~~d(s_n,s_m)<\vep(N).\nn\\
&&(4)~~\forall N<\infty,~~\exists 0<\vep=\vep(N)\sr{N\ra\infty}{\ral}0~~s.t.\nn\\
&&~~~~~~~~~~~~~~m,n>N~~\Ra~~\N_{\vep(N)}(s_m)\cap  \N_{\vep(N)}(s_n)\neq \{\}.\nn
\eea
Every convergent sequence, $\lim_{n\ra\infty}s_n=L$, is (uniformly) converging since
\bea
d(s_m,s_n)\leq d(s_m,L)+d(L,s_n)<\vep +\vep=2\vep~~\forall n,m >N(\vep).
\eea

Every Cauchy sequence is bounded; one simply needs to set $M=\max_{N\in \mathbb{N}}\vep(N)$. One can also check that sums and products of Cauchy sequences are Cauchy sequences.

A metric space ${\S}$ is complete if every Cauchy sequence in ${\S}$ converges to a point in ${\S}$.
The Cauchy completion of a space ${\S}$ is the union of the space ${\S}$ and the set consisting of the limit points of all Cauchy sequences in ${\S}$. That is, a space $\S$ is complete iff any (uniformly) converging sequence in $\S$ converges to a point in $\S$.

A series $S=S(s)$ is the sum of the terms of a sequence $s$,
\bea
S(s)=\sum_{k=1}^\infty s_k.
\eea
A series $S(s)$ is convergent iff the sequence of partial sums \\
~$S_n=S_n(s)=\sum_{k=1}^n s_k$~ is convergent.

 One can check that \emph{\textbf{a set $S$ is open iff only a finite number of points of any sequence that converges to a point $L\in S$ can lie outside of $S$}}.

 A Cauchy sequence in a closed set $C$ must converge to a point in $C$ for if it converges to a point in the complement $\widetilde{C}$ which is open (ie. $CP$ is closed) then that sequence lies in $\widetilde{C}$ instead as it would then have only a finite number of points in $C$. Thus \emph{\textbf{a closed set is complete}}. Also the complement $\widetilde{CP}$ of a complete set $CP$ is open for if $\widetilde{CP}$ were not open then one can find a point $b\in \widetilde{CP}$ such that $\N_\vep(b)\cap CP\neq\{\}~~\forall \vep >0$ meaning that one can construct a Cauchy sequence in $CP$ that converges to $b\not\in CP$ in contradiction to the completeness of $CP$. Thus \emph{\textbf{a complete set is closed}} and hence \emph{\textbf{a set is closed iff it is complete}}.

 The \emph{\textbf{closure}} $\overline{A}$ of a set $A$ is its Cauchy \emph{\textbf{completion}}.

\section{Connectedness and convexity}
A space $S$ is \emph{connected} iff for any two points $x,y\in\S$ one can find a continuous path $\Gamma:[0,1]\ra \S,~t\mapsto\Gamma(t)$ such that
$\Gamma(0)=x,~\Gamma(t)=y$. The points $x,y$ are said to be connected by the path $\Gamma$. The space $\S$ is topologically trivial iff its power set ~$\P(\S)=\{A;~A\subseteq \S\}$~ is connected.

 A metric space $(\S,d)$ is \emph{convex} iff any two points $x,y\in \S$ can be connected by a \emph{unique} continuous path \bea
 \Gamma_0\in [x,y]=\{\gamma:[0,1]\ra\S,~t\mapsto\gamma(t),~\gamma(0)=x,~\gamma(1)=y\}
  \eea
  such that
  \bea
  \min_{\gamma\in [x,y]}l[\gamma]=l[\Gamma_0]=d(x,y),~~l[\gamma]=\int_{0}^1d(\gamma(t),\gamma(t+dt)).
  \eea

\section{Some topology}
A \emph{\textbf{set}} is a collection of objects where each object individually satisfies a certain basic condition.

The inverse or complement $\wt{A}$ of a set $A$ is given by
\bea
B=A\cap B~\cup~ \wt{A}\cap B~~~~\forall B.
\eea
 A set $O$ is (uniformly) \emph{\textbf{open}} (or [uniformly] \emph{\textbf{continuous}}) iff its complement $\wt{O}$ is (uniformly) \emph{\textbf{complete}}.\textbf{ \emph{The union or intersection of an arbitrary number of open sets is also an open set.}}

A set $A$ is said to be \emph{\textbf{closed}}, $A\in \C=\{C\}$, iff its is \emph{\textbf{complete}} (or iff its inverse is open, $\wt{A}\in \O$). \textbf{\emph{The union or intersection of any finite number of closed sets is also a closed set.}}

A \emph{\textbf{neighborhood}} (nbd) $\N(A)$ of a set $A$ is any open superset of $A$. That is
\bea
A\subseteq \N(A)\in \O.
\eea
Equality is possible only when $A$ is open.
The \emph{\textbf{closure}} $\overline{A}$ of a set $A$  is the intersection of all closed supersets of $A$ and is thus the smallest closed superset of $A$,
\bea
\overline{A}=\bigcap_{A\subseteq C\in \C}C=\min_{A\subseteq C\in \C}C
\eea
and the \emph{\textbf{interior}} $A^0$ of $A$ is the union of all open subsets of $A$ and is thus the largest open subset of $A$,
\bea
A^0=\bigcup_{A\supseteq O\in \O}O=\max_{A\supseteq O\in \O}O.
\eea
The \emph{\textbf{boundary}} $\del A$ of $A$ is given by
\bea
&& \del A=\wt{A^0}\cap \overline{A}.
\eea
A point $x\in A$ is a \emph{\textbf{limit point}} of $A$ iff ~$\N(x)\cap A\neq \{\}~~\forall \N(x)$. A set $A$ is closed iff ~$\overline{A}=A$~ iff $A$ contains all its limit points. A set $A$ is open iff $A=A^0$.

A collection $\Sigma=\{\sigma\}$ of (open) sets such that $$A\subseteq \bigcup_{\sigma\in\Sigma}\sigma $$ is called a \emph{\textbf{cover}} $\Sigma(A)$ of $A$.

A set $A$ is \emph{\textbf{compact}}, $A\in \K=\{K\}$, if every cover $\Sigma(A)$ contains a finite subcover $\O_n(A)$,
\bea
\Sigma(A)\supseteq\O_n(A)=\{O_k\in \O,~k=1,...,n,~~A\subseteq \bigcup_{k=1}^nO_k\}.
\eea
Since any cover $\Sigma(A\cup B)$ for $A\cup B$ is also a cover of $A$ and of $B$, it follows that \emph{\textbf{the union of a finite number of compact sets is also a compact set}}.

A \emph{\textbf{space}} is a structured collection of one or more sets whose elements are known as points; the elements or points of the space are obtained through well defined interactions between the elements of the defining sets.
A topology $\T(\S)$ for a space $\S$ is any subfamily (ie. is closed under union and intersection) of the family of open subsets of $\S$ that covers $\S$ and which contains both $\S$ and $\{\}$. ie. $\T(\S)\subseteq \O(\S)=\{O\subseteq \S;~O\in \O\},~\cup,\cap:\T(\S)\times \T(\S)\ra \T(\S),$\\
$~~S\subseteq \bigcup \T(\S)=\bigcup_{O\in \T(\S)}O,~\S,\{\}\in \T(\S)$.
A \emph{\textbf{topological space}} is any given pair $X=(\S,\T(\S))$. ~ $\{\},\S$ are both open and closed as they are members of $\T(\S)$~ and  ~$\S=\widetilde{\{\}},~\{\}=\widetilde{\S}$.

A topological space $X=(\S,\T(\S))$ is \emph{\textbf{separable}} (or \emph{\textbf{Hausdorff}}) iff for any $A,B\subseteq X$ such that $A\cap B=\{\}$ one can find nbds $\N_1(A),~\N_2(B)$ such that
\bea
\N_1(A)\cap\N_2(B)=\{\}.
\eea
A space $\S$ is \emph{\textbf{uniformly open}} iff for any $A,B\subseteq \S$ one can find nbds $\N_1(A),\N_2(B)$ whose intersection lies in $\S$,
\bea
&&\N_1(A)\cap\N_2(B)\subseteq \S.
\eea
A space $\S$ is \emph{\textbf{locally compact}} iff every point $x\in\S$ has a nbd $\N(x)$ whose closure $\overline{\N(x)}$ is compact.

A subset $D\subseteq \S$ is \emph{\textbf{dense}} in $\S$ iff $\overline{D}=\S$. A set $\S$ is \emph{\textbf{countable}} iff its is isomorphic to $\mathbb{N}$;~ ie.~ $\exists~~i:\S\ra \mathbb{N}$.

A \emph{\textbf{map}} $m$ between two topological spaces $m:S\ra T$ is \emph{\textbf{continuous}} iff the inverse image $m^{-1}(O)$ of every open set $O\subseteq T$ is open, ie. $m^{-1}(O)$ belongs to $\O(S)$.

A \emph{\textbf{sequence}} $s:\mathbb{N}\ra S,~n\mapsto s_n$ in a topological space $S$ \emph{\textbf{converges}} to a point $L$ iff
\bea
&&\forall \N(L)~~\exists N=N(\N(L))<\infty~~s.t.~~s_n\in \N(L)~~\forall n>N.
\eea
That is, every nbd of $L$ contains an infinite number of points of the sequence since there is an infinite number of terms between $\infty$ and any $N<\infty$.

A \emph{\textbf{sequence}} on a topological space $S$ is \emph{\textbf{Cauchy}} iff
\bea
&&\forall~O\in\O(S)~~\exists N=N(O)<\infty~~s.t.~~\N_O(s_m)\cap  \N_O(s_n)\neq \{\}~~\forall~m,n>N(O)\nn
\eea
where $\N:\O(S)\ra \O(S),~O\mapsto \N_O$ is a map that assigns $O$ as a nbd $\N_O$ of a point or set. That is, each point $s_n$ of the sequence becomes increasingly nonseparable from its neighbors as $n$ increases. If the set of all nbds  of $A\subseteq S$ is $\N[A]$ then
\bea
\N[A]~\ni~\N_O(A)=\left\{
          \begin{array}{ll}
            \{\}, & A\not\subseteq O \\
            O, & A\subseteq O.
          \end{array}
        \right.
\eea

Every convergent sequence, $\lim_{n\ra\infty}s_n=L$, is a Cauchy sequence since
\bea
\N_O(s_m)\cap  \N_O(s_n)\supseteq \N_O(s_m)\cap  \N_O(L)~\cup~\N_O(L)\cap  \N_O(s_n)\neq \{\}.
\eea
A topological space $S$ is \emph{\textbf{complete}} iff every Cauchy sequence in $S$ converges to a point in $S$.
The \emph{\textbf{Cauchy completion}} of a space $S$ is the union of the space $S$ and the set consisting of the limit points of all Cauchy sequences in $S$.

A map $m$ is a \emph{\textbf{$P$-map}} iff the image $m(A)$ has the property $P$ whenever $A$ has the property $P$; ie. $m$ preserves the property $P$. For example one has singular/nonsingular maps, open/closed maps, measurable/nonmeasurable maps, bounded/unbounded maps, compact/noncompact maps, connected/nonconnected, convex/nonconvex, etc.

Since the identity map (or linear map in general) is both invertible and open it follows that continuity of a space and its open topology are equivalent concepts. That is, a continuous or topological space is one that has an open topology and continuity of a map $m$ measures how much of the continuity or topology of a space is preserved by the inverse map $m^{-1}$. Thus reassigning continuity to sets means that a map $m$ is said to be continuous iff $m^{-1}$ is a continuous map.

\section{More on compactness and separability}
We work in a Hausdorff space where any two disjoint sets have disjoint nbds. For simplicity we will denote $A\cap B$ as $AB$ and $A\cup B$ as $A+B$.
\begin{itemize}
\item Let $K$ be compact and $C$ be closed. Then $\widetilde{C}$ is open. If $\Sigma(K C)$ is any cover for $K C$; ie. $$K C\subseteq \bigcup_{\sigma\in\Sigma(K C)}\sigma$$ then
    $$K=K C+K \widetilde{C}\subseteq \bigcup_{\sigma\in\Sigma(K C)}\sigma+K\widetilde{C}\subseteq \bigcup_{\sigma\in\Sigma(K C)}\sigma+\widetilde{C}$$ and so $\{\Sigma(K C),\widetilde{C}\}$ is a cover for $K$ and therefore has a finite subcover $\O_n(K)=\{O_1,...,O_n,\widetilde{C}\}$ as $K$ is compact. That is $K\subseteq \bigcup_{k=1}^nO_k+ \widetilde{C}$, which implies that $K C\subseteq \bigcup_{k=1}^nO_k$ which means that $\{O_1,...,O_n\}$ is a finite subcover for $K C$ and hence $KC$ is also compact. That is, if $K$ is compact and $C$ is closed then $KC$ is compact. It follows that \emph{\textbf{every closed subset of a compact set is also compact}}.

    \item Let $K$ be compact and $O,O'$ be open and $K\subseteq O+O'$. Then \\
    $K\subseteq O+O'~\Ra~\widetilde{K}\supseteq \widetilde{O}\widetilde{O'}~\Ra~ K\widetilde{K}\supseteq K\widetilde{O}~K\widetilde{O'}~\Ra~\{\}\supseteq K\widetilde{O}~K\widetilde{O'}$\\
        $~~\Ra~~ K\widetilde{O}~K\widetilde{O'}=\{\}$. Therefore $K\widetilde{O}$ and $K\widetilde{O'}$ are disjoint compact sets, since $\widetilde{O},\widetilde{O'}$ are closed and $K$ is compact, and since we are in a Hausdorff space we can find disjoint open sets $O_1,O_2$ such that $K\widetilde{O}\subseteq O_1$ and $K\widetilde{O'}\subseteq O_2$.
        \bea
        &&K\widetilde{O}\subseteq O_1~~\Ra~~\widetilde{O_1}\subseteq \widetilde{K}+O~~\Ra~~K_1=K\widetilde{O_1}\subseteq O,\nn\\
        &&K\widetilde{O'}\subseteq O_2~~\Ra~~\widetilde{O_2}\subseteq \widetilde{K}+O'~~\Ra~~K_2=K\widetilde{O_2}\subseteq O'.\nn\\
        \eea
        Therefore we have found compact sets $K_1,K_2$ such that $ K_1\subseteq O,~K_2\subseteq O'$ and $K_1+K_2=K\widetilde{O_1}+K\widetilde{O_2}=K(\widetilde{O_1}+\widetilde{O_2})=K~\widetilde{O_1O_2}=K~\widetilde{\{\}}=K$.

    \item Let $A\subseteq B$ in a Hausdorff space. Since $A\subseteq B~~\Ra~~A\widetilde{B}=\{\}$, one can find disjoint open sets $O_1,O_2$ such that $A\subseteq O_1,~~\widetilde{B}\subseteq O_2$ where \\
        $\widetilde{B}\subseteq O_2~~\Ra~~\widetilde{O_2}\subseteq B$.

        But $O_1O_2=\{\}~~\Ra~~O_1\subseteq \widetilde{O_2}$ and therefore one has the sandwich relations
        \bea
        A\subseteq O_1\subseteq \widetilde{O_2}\subseteq B
        \eea
        which can also be iterated to obtain sequences of inclusions. Thus given any two sets $A,B$ in a Hausdorff space one can connect them with sequences through $A\cap B$ and/or $A\cup B$ since
        \bea
        &&A\cap B\subseteq A\subseteq A\cup B,~~A\cap B\subseteq B\subseteq A\cup B.
        \eea
\end{itemize}

\section{On the realization of compact spaces}
Here ''cover'' will mean ''open cover''.

Let a \emph{\textbf{minimal or essential cover}} for a set $\S$ be one that contains no proper subcovers.
That is $\Sigma_0(\S)$ is minimal iff
\bea
\Sigma(\S)\subseteq \Sigma_0(\S)~~\Ra~~\Sigma(\S)=\Sigma_0(\S).
\eea
It follows that any cover of $\S$ can be generated from one or more minimal covers. That is the set of all minimal covers of $\S$ is basic and generates the rest of the nonminimal covers.

Then compactness of $\S$ implies that every minimal cover of $\S$ is a finite cover since every cover contains a finite subcover and this also implies that any subcover of the finite cover must in turn contain a finite subcover. Conversely, suppose that every minimal cover of a space $\S$ is finite.
Then $\S$ must be compact since (by the generating property of the set of minimal covers) every cover of $\S$ contains at least one minimal cover, which is finite (a finite subcover) by the supposition. This means that \emph{\textbf{a set $\S$ is compact iff every minimal cover of $\S$ is finite}}. In other words a compact set is one that is essentially finite in the topological sense.

A nonempty open set $O\neq\{\}$ in a (separable) metric space cannot be compact since \\$\lim_{\vep\ra 0}\{\N_\vep(a),~a\in O\}$ is a minimal cover of $O$ that is not finite. Partially open sets cannot be compact either since an open set, which is not compact, can be obtained through the union of a finite number of partially open sets. Also unbounded sets are isomorphs of open and partially open sets and so cannot be compact. Hence \emph{\textbf{a compact set in a metric space must be closed and bounded}}.

One also notes that the image $m(\S)$ of a compact space $\S$ under an isomorphic map $m:\S\ra m(\S)$ is also compact. It follows therefore that a (free) compact space is actually an equivalence class of all such spaces under all possible isomorphisms. In particular \emph{\textbf{a space is compact iff it is compact as a subspace}}. To see this, one notes that a compact space $\S$ is a compact subspace of itself. Conversely if $\S$ is a compact subspace of some space then the equivalence class \\
$[\S]=\{i(\S),~i:\S\ra i(\S),~i\in I \}$ of its images under all possible isomorphisms $I=\{i\}$ generates the (free) compact space.

[~~To verify that the image of a compact set $\S$ under an isomorphism $i$ is compact, one notes that in general \\
$f(A\cup B)\subseteq f(A)\cup f(B),~~f(A\cap B)\subseteq f(A)\cap f(B)$ since either of $A\cup B$ and $A\cap B$ has less points to transform than has $A$ and $B$ separately.
But under an isomorphism ($i(a)=i(b)$~~iff~~$a=b$) one has
\bea
i(A\cup B)= i(A)\cup i(B),~~i(A\cap B)= i(A)\cap i(B)~~\forall A,B
\eea
and so all the structures and/or statements that characterize compactness are preserved implying that if $\S$ is compact then so is $i(\S)$. It may also be worth recalling that $A\subseteq B$~~iff~~$A\cap B=B$~~iff~~$A\cup B=B$. Also \emph{\textbf{the direct product of a finite number of compact spaces is also a compact space}}.~~]

Thus whenever possible one can check noncompactness of a space $\S$ by embedding it into a (separable) metric space $(\H,d),$ ~$i:\S\ra S\subseteq (\H,d)$ and using the fact that a compact subspace $S$ of a (separable) metric space $(\H,d)$ must be closed ($\td{S}$ is open in $(\H,d:\H\times\H\ra \mathbb{R}^+)$) and bounded ($\max_{x,y\in\S} d(x,y)<\infty$).

The arguments concerning compactness may be adapted to other properties such as openness (or continuity), closedness (or completeness), connectedness, convexity, measurability and so on.

\section{Metric topology of $\mathbb{R}$}
In $\mathbb{R}$ the finite interval $I_0(a,b)=\{c,~~a<c<b\}$ is the basic open subset and every open set can be written as a union and/or intersection of finite open intervals. The finite interval  $I(a,b)=\{c,~~a\leq c\leq b\}$ is the basic closed subset which is also the closure $\overline{I_0(a,b)}$. The finite closed interval is compact since the only possible noncompact sets are open and half open intervals and their isomorphs. $\mathbb{R}$ is open in that every point has a  finite open interval as a neighborhood, and since the closure of any finite open interval is the compact interval it means that $\mathbb{R}$ is a locally compact space. The direct product space $\mathbb{R}^n=\{x=(x^1,x^2,...,x^n),~~x^1,x^2,...,x^n\in \mathbb{R}\},~n\in \mathbb{N}^+$ inherits the topological properties of $\mathbb{R}$ alongside additional ones. One has as possible metrics
\bea
&&d_1(x,y)=\max_{i\in \mathbb{N}_n}|x^i-y^i|,~~d_2(x,y)=\sqrt{\sum_{i=1}^n(x^i-y^i)^2}.
\eea
 \emph{\textbf{A subset of $\mathbb{R}^n$ is compact iff it is closed (complete) and bounded}}.

Consider the real maps $\F(\mathbb{R},\mathbb{R}^n)=\{f:\mathbb{R}^n\ra \mathbb{R}\}$. Then a subspace $S$ of $\mathbb{R}^n$ may be specified through implicit relations imposed pointwise (ie. simultaneously) on a sequence of functions
\bea
S=\{x\in \mathbb{R}^n,~~f_i(x)\sim 0,~~f_i\in \F(\mathbb{R},\mathbb{R}^n),~~i\in \mathbb{N}\}.
\eea

where $\sim$ includes relations such as ~$=,~<,~\leq,~>,~\geq,$ ~etc.

\section{On Measures I}
A \emph{content} $\ld$ is a finite, positive, subadditive, additive, and monotone function on the set of compact sets $\K=\{K\}$.
\bea
&&\ld:\K\ra R^+\backslash\{\infty\},\nn\\
&&K_1\subseteq K_2~~\Ra~~\ld(K_1)\leq \ld(K_2).\nn\\
&&\ld(K_1\cup K_2)\leq \ld(K_1)+\ld(K_2)~~(\txt{subadditivity}).\nn\\
&&K_1\cap K_2=\{\}~~\Ra~~\ld(K_1\cup K_2)= \ld(K_1)+\ld(K_2)~~(\txt{additivity}).\nn
\eea
Additivity implies that $\ld(\{\})=0$.

An \emph{inner content} $\ld_\ast$ induced by $\ld$;
\bea
\ld_\ast(A)=\sup_{K\subset A}\ld(K),~~~~\ld_\ast(\{\})=\ld(\{\})=0,
\eea
is the \emph{content of the biggest compact subset} of $A$.

If $\O=\{O\}$ is the set of open sets, the \emph{outer measure} $\mu_o$;
\bea
&&\mu_o(A)=\inf_{A\subset O}\ld_\ast(O),~~~~\mu_o(\{\})=\ld_\ast(\{\})=0,
\eea
is the \emph{inner content of the smallest open superset} of $A$.

\textbf{Remarks}
\begin{itemize}

\item The content (measure) of a set is unique if the inner and outer contents (measures) coincide.

\item Let $A\subseteq B$ then
\bea
&&\ld_\ast(A)=\sup_{K\subseteq A}\ld(K),~~\ld_\ast(B)=\sup_{K\subseteq B}\ld(K)\geq \sup_{K\subseteq A}\ld(K)=\ld_\ast(A)\nn\\
&&~~\Ra~~ \ld_\ast(A)\leq \ld_\ast(B).
\eea
Similarly,
\bea
&&\mu_o(A)=\inf_{A\subseteq O}\ld_\ast(O),\nn\\
&&\mu_o(B)=\inf_{B\subseteq O}\ld_\ast(O)\geq \inf_{A\subseteq O}\ld_\ast(O)=\mu_o(A)\nn\\
&&~~\Ra~~ \mu_o(A)\leq \mu_o(B).
\eea

\item From these inequalities one sees that
\bea
\label{crossineq1}\ld_\ast(K)=\sup_{K'\subseteq K}\ld(K')\leq \ld(K)~~~~\forall K\in \K
\eea
and
\bea
\label{crossineq2}&&\mu_o(A)=\inf_{A\subseteq O}\ld_\ast(O)\geq \ld_\ast(A)~~~~\forall A.
\eea
In particular
\bea
&&\ld(K)\leq \ld_\ast(K)\leq \mu_o(K)~~~~\forall K\in \K.
\eea

\item Also
\bea
&&\mu_o(O)=\inf_{O\subseteq O'}\ld_\ast(O')\leq \ld_\ast(O)~~~~\forall O\in \O
\eea
since $O\subseteq O$. But from  (\ref{crossineq2}) $\mu_o(O)\geq \ld_\ast(O)$. Therefore
\bea
\mu_o(O)= \ld_\ast(O)~~~~\forall O\in \O.
\eea
Similarly
\bea
\ld_\ast(K)=\sup_{K'\subseteq K}\ld(K')\geq \ld(K)~~~~\forall K\in \K
\eea
since $K\subseteq K$. But from  (\ref{crossineq1}) $\ld_\ast(K)\leq \ld(K)$. Therefore
\bea
\ld_\ast(K)= \ld(K)~~~~\forall K\in \K.
\eea

\item One also deduces that
\bea
&&\mu_o(\txt{int} K)=\ld_\ast(\txt{int} K)\leq \ld_\ast(K)=\ld(K)\leq \mu_o(K).
\eea

\item For open sets $O_1,O_2$,~~ $\ld_\ast(O_1+O_2)\leq \ld_\ast(O_1)+\ld_\ast(O_2)$. This follows because for any compact $K\subseteq O_1+O_2$ one can find compacts \\
    ~$K_1\subseteq O_1,~~K_2\subseteq O_2$~ such that $K\subseteq K_1+K_2$. Therefore
    \bea
    &&K\subseteq K_1+K_2~~\Ra~~\nn\\
    &&\ld(K)\leq \ld(K_1+K_2)\leq \ld(K_1)+\ld(K_2)\nn\\
    &&~~\Ra~~\sup_{K\subseteq O_1+O_2}\ld(K)\leq \sup_{K_1\subseteq O_1}\ld(K_1)+\sup_{K_2\subseteq O_2}\ld(K_2)\nn\\
    &&~~~~\Ra~~~~\ld_\ast(O_1+O_2)\leq \ld_\ast(O_1)+\ld_\ast(O_2).
    \eea
Furthermore if $O_1O_2=\{\}$ then since by construction $K_1=K\widetilde{O}_1,~K_2=K\widetilde{O}_2$ one sees that
\bea
&&K_1K_2=K\widetilde{O}_1K\widetilde{O}_2=K~\widetilde{O}_1\widetilde{O}_2=K\widetilde{(O_1+O_2)}=\{\},\nn\\
&&K_1+K_2=K\widetilde{O}_1+K\widetilde{O}_2=K\widetilde{(O_1O_2)}=K\{\}^c=K.
\eea
Therefore
\bea
    &&K= K_1+K_2~~\Ra~~\nn\\
    &&\ld(K)=\ld(K_1+K_2) = \ld(K_1)+\ld(K_2)\nn\\
    &&~~\Ra~~\sup_{K\subseteq O_1+O_2}\ld(K)= \sup_{K_1\subseteq O_1}\ld(K_1)+\sup_{K_2\subseteq O_2}\ld(K_2),\nn\\
    &&~~~~\Ra~~~~\ld_\ast(O_1+O_2)= \ld_\ast(O_1)+\ld_\ast(O_2).
    \eea
That is, $O_1O_2=\{\}~~\Ra~~\ld_\ast(O_1+O_2)= \ld_\ast(O_1)+\ld_\ast(O_2)$.

These results are automatically valid for $\mu_o$ since $\mu_o(O)=\ld_\ast(O)$.

Given any $A,B$ then for any open supersets $O_1\supseteq A,~O_2\supseteq B$ one can find $\vep>0$ such that
\bea
&&\mu_o(O_1)\leq \mu_o(A)+{\vep_1\over 2},~~\mu_o(O_2)\leq \mu_o(B)+{\vep_2\over 2}\nn\\
&&~~\Ra~~\mu_o(A+B)\leq\mu_o(O_1+O_2)\leq \mu_o(O_1)+\mu_o(O_2)\nn\\
&&~~~~~~~~~~~~~~~~\leq \mu_o(A)+\mu_o(B) +{\vep_1+\vep_2\over 2}.\nn
\eea
Due to separability one can continue to generate smaller and smaller intermediate subsets~ $O_1\supseteq \widetilde{O}_{1i}\supseteq O_{1i'}\supseteq A,~~O_2\supseteq \widetilde{O}_{2i}\supseteq O_{2i'}\supseteq B$~ until $\vep_1,\vep_2\ra 0$. Thus ~$\mu_o(A+B)\leq \mu_o(A)+\mu_o(B)~~~~\forall A,B$.

These results can be iterated and verified through induction.

\end{itemize}

\subsection{Measurability}
The set $A$ is $\mu_o$-measurable iff
\bea
&&\mu_o(B)=\mu_o(A B)+\mu_o(\widetilde{A} B)
\eea
for any set $B$. ie. measurability is defined by requiring that the additivity property holds for $\mu_o$ as $A^c$ is defined by
\bea
&&B=B A+ B \widetilde{A}~~\forall B,~~A \widetilde{A}=\{\}.
\eea

\section{On Measures II}
A \emph{\textbf{measure}} $\mu$ on sets is defined as
\bea
&&\mu(A)\geq 0,~~~~A\subseteq B~~\Ra~~\mu(A)\leq \mu(B),\nn\\
&&\mu(A+ B)\leq \mu(A)+\mu(B),\nn\\
&& A B=\{\}~~\Ra~~\mu(A+ B)= \mu(A)+\mu(B).
\eea
Upon making the replacements ~$A\ra A B,~~B\ra \wt{A} B$~ in \\
~$\mu(A+ B)\leq \mu(A)+\mu(B)$~ one obtains
\bea
\label{meas.rel1}\mu(B)\leq\mu(A B)+\mu(\wt{A} B)
\eea
 using the assumptions that
 \bea
 &&B=B \widetilde{\{\}}=B (\widetilde{A\wt{A}})= B(\wt{A}+ A)=A B~+~\wt{A} B.\nn
 \eea
A set $A$ is $\mu$-\emph{\textbf{measurable}} iff equality holds in (\ref{meas.rel1}) for all sets $B$. That is, $A$ is $\mu$-measurable iff
\bea
\label{meas.rel2}\mu(B)=\mu(A B)+\mu(\wt{A} B)~~~~\forall B.
\eea

Any collection ~$\Sigma_0=\{\sigma\} $~ of nonintersecting sets ~$\sigma_1\sigma_2=\{\}~~\forall \sigma_1,\sigma_2\in\Sigma_0$~ is a partition and the partition $\Sigma_0$ is a $\mu$-\emph{\textbf{measuring scale}} (or simply $\mu$-\emph{\textbf{measurable}}) iff
\bea
\mu(B)=\sum_{\sigma\in\Sigma_0}\mu(\sigma B)~~~~\forall B.
\eea
Thus a set $A$ is $\mu$-measurable iff the partition $\{A,\widetilde{A}\}$ is $\mu$-measurable.

Measurability can also be expressed entirely in terms of open sets: a set $M$ is $\mu_o$-measurable iff
\bea
\mu_o(O)\geq \mu_o(OM)+\mu_o(O\wt{M})~~~~\forall O\in\O.
\eea
This is because one has that
\bea
&&\mu_o(O)\geq \mu_o(OM)+\mu_o(O\wt{M})~~\Ra~~\nn\\
&&\mu_o(A)=\inf_{A\subseteq O}\ld_\ast(O)=\inf_{A\subseteq O}\mu_o(O)\geq \inf_{A\subseteq O}\{\mu_o(OM)+\mu_o(O\wt{M})\}\nn\\
&&~~~~\geq \mu_o(AM)+\mu_o(A\wt{M})\},
\eea
and $\mu_o(A)\leq \mu_o(AM)+\mu_o(A\wt{M})$ by subadditivity and so\\
 $\mu_o(A)= \mu_o(AM)+\mu_o(A\wt{M})~~\forall A$.
Conversely, if $M$ is $\mu_o$-measurable, ie. \\
$\mu_o(A)= \mu_o(AM)+\mu_o(A\wt{M})~~\forall A$ then for any open set $O$ in particular we have ~$\mu_o(O)= \mu_o(OM)+\mu_o(O\wt{M})$~ which satisfies ~$\mu_o(O)\geq \mu_o(OM)+\mu_o(O\wt{M})$.

The product ~$\Sigma_A\Sigma_B=\{A_iB_j;~A_i\in\Sigma_A,~B_j\in \Sigma_B\}$~ of two $\mu$-measurable partitions $\sigma_A=\{A_i\},~~\Sigma_B=\{B_i\}$ is $\mu$-measurable:
\bea
\mu(M)=\sum_{i}\mu(A_i M)=\sum_{i}\sum_j\mu(B_jA_i M)= \sum_{ij}\mu(B_jA_i M)~~~~\forall B.\nn\\
\eea

If $\sigma_A$ is a measurable partition and $\Sigma_A\leq \Sigma_B$ (meaning that each $A_i\in \Sigma_A$ is a subset of some $B_j\in \Sigma_B$) then $\Sigma_B$ is also measurable:
\bea
&&\sum_i\mu(B_iM)=\sum_i\sum_j\mu(A_jB_iM)=\sum_i\sum_j\mu(A_jB_iM)=\sum_j~\sum_{i,~A_j\subseteq B_i}\mu(A_jM)\nn\\
&&~~~~=\sum_j\mu(A_jM)=\mu(M).
\eea

A partition $\Sigma=\{A_i;~\forall i\}$ is measurable iff each of $A_i$ is measurable:  $A_i$ is measurable $\forall i$ iff $\Sigma_i=\{A_i,\wt{A}_i\}$ is measurable $\forall i$ and so is their product;
\bea
&&\prod_i\{A_i,\wt{A}_i\}=(A_1,A_2,...,A_n,\prod_{i=1}^n \widetilde{A}_i)~~\txt{measurable}~~\Ra~~\nn\\
&&\mu(M)=\sum_{i=1}^n\mu(A_iM)+\mu(\prod_{i=1}^n \widetilde{A}_i~M)\geq \sum_{i=1}^n\mu(A_iM).
\eea
But we also have  ~$\mu(M)\leq \sum_{i=1}^n\mu(A_iM)$~ and so ~$\mu(M)= \sum_{i=1}^n\mu(A_iM)$ and thus $\Sigma$ is measurable.
Conversely if $\Sigma$ is measurable then $\Sigma_i=\{A_i,\wt{A}_i\}$ is also measurable for each $i$ since for any given $i$, $ \Sigma\leq \Sigma_i$.

If $A,B$ are measurable then so are $\wt{A},AB,A+B$ since
\bea
&&A=\widetilde{\widetilde{A}},~~~~AB\in \{A,\wt{A}\}\{B,\wt{B}\}=\{AB,A\widetilde{B},\widetilde{A}B,\widetilde{A}\widetilde{B}\},\nn\\
&&A+B=\widetilde{\widetilde{A}\widetilde{B}}.
\eea

If the sets $A_1,A_2,...,A_n,~~\forall n\in \mathbb{N}$ are measurable, then so is \\
$A=\sum_{i=1}^nA_i~~\forall n\in N$ by induction.
One notes that one can write $A$ in terms of disjoint sets:
\bea
&&A=\sum_{i=1}^nA_i=A_1+\widetilde{A}_1A_2+\widetilde{A}_1\widetilde{A}_2A_3+...+\widetilde{A}_1\widetilde{A}_2...\widetilde{A}_{n-1}A_n\nn\\
&&~~~~=\sum_{i=1}^nA_i\prod_{ 1\leq j\leq i-1}\wt{A}_j.
\eea

For any two sets $A,B$ where one of them, say $A$, is measurable \\
(ie. $\mu(M)=\mu(MA)+\mu(M\widetilde{A})~~\forall M$), the we have
\bea
&& \mu(B)=\mu(BA)+\mu(B\widetilde{A}),\nn\\
&&\mu(A+B)=\mu((A+B)A)+\mu((A+B)\widetilde{A})=\mu(A+BA)+\mu(B\widetilde{A})\nn\\
&&~~~~=\mu(A)+\mu(B\widetilde{A})=\mu(A)+\mu(A)-\mu(BA).\nn\\
&&\mu(A+B)=\mu(A)+\mu(A)-\mu(BA).
\eea
One notes that the second line could have simply been expressed as
\bea
\mu(A+B)=\mu(A+\widetilde{A}B)=\mu(A)+\mu(\widetilde{A}B)
\eea
without using measurability of $A$.

Every open set is $\mu_o$-measurable: Given $O_1,O_2\in\O$  consider $K_1,K_2\in\K$ such that ~$K_1\subseteq O_1O_2~(ie.~\widetilde{O}_1+\widetilde{O}_2\subseteq \widetilde{K}_1~~or~~\widetilde{O}_1O_2\subseteq \widetilde{K}_1O_2),$\\
$~~K_2\subseteq \widetilde{K}_1O_2~(ie.~K_2\subseteq \widetilde{K}_1O_2)$~ then $K_1K_2=\{\}$ (as $K_2\subseteq \widetilde{K}_1$) and
\bea
&&K_1,K_2\subseteq O_2~~\Ra~~K_1+K_2\subseteq O_2,\nn\\
&&\mu_o(O_2)=\ld_\ast(O_2)=\sup_{K\subseteq O_2}\ld(K)\geq \sup_{K=K_1+K_2\subseteq O_2}\ld(K_1+K_2)\nn\\
&&~~~~=\sup_{K_1\subseteq O_2}\ld(K_1)+\sup_{K_2\subseteq O_2}\ld(K_2)\geq \ld_\ast(O_1O_2)+\ld_\ast(\widetilde{K}_1O_2)\nn\\
&&~~~~=\mu_o(O_1O_2)+\mu_o(\widetilde{K}_1O_2)\geq \mu_o(O_1O_2)+\mu_o(\widetilde{O}_1O_2).\nn\\
&&~~\Ra~~\mu_o(O_2)\geq \mu_o(O_1O_2)+\mu_o(\widetilde{O}_1O_2)~~\forall O_1,O_2\in \O.
\eea

\subsection{Haar measure: existence and uniqueness}\label{haar-measure}
Let $\S$ be a measurable space and $\ld_1,\ld_2$ be two contents defined on compact subsets $\K(\S)$ of $\S$.
If $h:\S\ra\S$ is a self homeomorphism of $\S$ such that $\ld_2=\ld_1\circ h$ then the induced measures $\mu_1,\mu_2$ of $\ld_1,\ld_2$ are also related as $\mu_2=\mu_1\circ h$, where $\circ$ denotes map composition.

Let $G$ be a locally compact topological group (Topological in that~ \\
$()\cdot():G\times G\ra G$,~ or equivalently ~$L_u:G\ra G,~g\mapsto ug,$\\
$~~R_u:G\ra G,~g\mapsto gu~~\forall u\in G$,~ and ~$()^{-1}:G\ra G,~g\mapsto g^{-1}$~ are continuous maps). Thus in $G$ the existence of a left-invariant content $\ld$ will imply the existence of a left-invariant measure $\mu$ since left translation $L_u:G\ra G,~g\mapsto L_u(g)=ug$~ by $u\in G$ is a homeomorphism. One simply needs to set \\
$\ld_1=\ld,~h=L_u,~~\ld_2\eqv\ld_u=\ld\circ L_u$; then $\ld_2=\ld_1\circ L_u~~\Ra~~\mu_2=\mu_1\circ L_u$. Thus $\ld_1=\ld_2~~\Ra~~\mu_1=\mu_2$ or, equivalently, that $\ld=\ld\circ L_u~~\Ra~~\mu=\mu\circ L_u$.

Let $K\in\K(G)$ be a compact subset of $G$ and $o\in \O(G),~o\neq\{\}$ be a \emph{small} nonempty open subset of $G$. It is possible to form a cover \\
$\Sigma^o_n(K)=\{g_io;~i=1,2,...,n\}$ for $K$ (this is possible for any $A\subseteq G$) made up of a number of translated copies $g_io$  of $o$ by some elements $\{g_i\in G;~i=1,2,...,n\}$. That is
\bea
\label{haar-incl}&&K\subseteq \bigcup_{i=1}^ng_io\eqv \bigcup \Sigma^o_n(K).
\eea
Consider the map
\bea
&&n_o:\K(G)\ra \mathbb{R}_+,~~A\mapsto n_o(K)=\min_{K\subseteq \bigcup \Sigma^o_n(K)}n.
\eea
That is \emph{\textbf{$n_o(K)$ is the smallest number of copies of $o$ needed to just cover $K$}}.
Since only the number $n$ of copies of $o$, and not the elements $\{g_i\}$, is important we may simply write the inclusion (\ref{haar-incl}) as
\bea
K\subseteq n_o(K)~o
\eea
and use heuristics to deduce the following properties. Let $A\subseteq G,~A^0\neq \{\}$. Then
\bea
&&K\subseteq n_A(K)~A,~~A\subseteq n_o(A)~o~~\Ra~~ K\subseteq n_A(K)~A\subseteq n_A(K)n_o(A)~o\nn\\
\label{content-obsv}&&~~~~\Ra~~n_o(K)\leq n_A(K)n_o(A)~~\Ra~~{n_o(K)\over n_o(A)}\leq n_A(K).\\
&&K\subset K_1~~\Ra~~n_o(K)\leq n_o(K_1).\nn\\
&&K_1+K_2\subseteq n_o(K_1+K_2)~o,~K_1\subseteq n_o(K_1)~o,~K_2\subseteq n_o(K_2)~o\nn\\
&&~~\Ra~~K_1+K_2\subseteq (n_o(K_1)+n_o(K_2))~o~~\Ra~~n_o(K_1+K_2)\leq n_o(K_1)+n_o(K_2).\nn\\
&&(n_o\circ L_u)(K)=n_o(u\cdot K)=\min_{u\cdot K\subseteq \bigcup \Sigma^o_n(u\cdot K)}n=\min_{K\subseteq \bigcup \Sigma^o_n(K)}n\nn\\
&&~~~~~~~~~~~~=n_o(K)~~~~\forall u\in G.
\eea

If $K_1K_2=\{\}$ then it is possible to find nbds ~$\N(K_1),\N(K_2)$~ such that $\N(K_1)\N(K_2)=\{\}$. Consequently, $o$ can be chosen arbitrarily small so that
\bea
&&K_1K_2=\{\}~~\Ra~~n_o(K_1+K_2)=n_o(K_1)+n_o(K_2).
\eea
Thus we have additivity. However the number $n_o(K)$ can clearly diverge and so we now define a reqularized version (see~ \ref{content-obsv})
\bea
\ld_o(K)={n_o(K)\over n_o(A)}\leq n_A(K).
\eea
Then $\ld_o$ clearly inherits all the essential properties of $n_o$ and is bounded by $ n_A(K)~~\forall o$. One can define the desired content $\ld$ as
\bea
&&\ld(K)=\min_{ \{\}\neq o\in\O(G)}\ld_o(K).
\eea

To check uniqueness, let $\mu,\nu$ be two left invariant measures on $G$ and consider two continuous functions $\al,\beta:K\in \K(G)\ra \mathbb{C}$. Then
\bea
&&\int_{K} d\mu(x)~\al(x)~\int_{K} d\nu(y)~\beta(y)=\int_{K} d\mu(x)d\nu(y)~\al(x)\beta(y)\nn\\
&&~~~~=\int_{K} d\mu(x)d\nu(y)~\al(y^{-1}x)\beta(y)\nn\\
&&~~~~=\int_{K} d\mu(x)d\nu(y)~\al((x^{-1}y)^{-1}))\beta(x(x^{-1}y))=\int_{K} d\mu(x)d\nu(y)~\al(y^{-1})\beta(xy)\nn\\
&&~~~~=\int_K d\nu(y)~\al(y^{-1})~\int_{K} d\mu(x)\beta(xy).\nn
\eea

One can left translate $\beta$ to obtain
\bea
&&\int_{K} d\mu(x)~\al(x)~\int_{K} d\nu(y)~\beta(yg)=\int_K d\nu(y)~\al(y^{-1})~\int_{K} d\mu(x)\beta(xyg).\nn
\eea
Now integrate over $g$ to obtain
\bea
&&\int_{K} d\mu(x)~\al(x)~\int_{K} d\nu(y)~\int_K d\rho(g)\beta(yg)\nn\\
&&~~~~~~~~~~~~=\int_K d\nu(y)~\al(y^{-1})~\int_{K} d\mu(x)~\int_K d\rho(g)\beta(xyg)\nn
\eea
where $\rho$ can be either $\mu$ or $\nu$. Thus
\bea
&&\int_{K} d\mu(x)~\al(x)~\int_{K} d\nu(y)~\int_K d\rho(g)\beta(g)\nn\\
&&~~~~~~~~~~~~=\int_K d\nu(y)~\al(y^{-1})~\int_{K} d\mu(x)~\int_K d\rho(g)\beta(g)\nn\\
&&\Ra~~|K|_\nu\int_{K} d\mu(x)~\al(x)=|K|_\mu\int_K d\nu(y)~\al(y^{-1}),~~|K|_\mu=\int_K d\mu(x).\nn
\eea
In particular for any function $\al$ such that $\al(y^{-1})=\al(y)$, ~eg \\
~$\al(y)=\gamma(y)+\gamma(y^{-1})$~ or ~$\al(y)=\gamma(y)\gamma(y^{-1})$,~ one has that
\bea
&&\int_{K} d\mu(g)~\al(g)={|K|_\mu\over |K|_\nu}\int_K d\nu(g)~\al(g)~~\forall \al:K\ra \mathbb{C},~\al(g)=\al(g^{-1}).\nn\\
&&~~~~\mu=c~\nu,~~~~c={|K|_\mu\over |K|_\nu}.
\eea

 Also since
 \bea
&& \int_{K} d\mu(x)~\al(xa)=\int_Kd\mu(xaa^{-1})~\al(xa)=\int_Kd\mu(xa^{-1})~\al(x),\nn\\
&&\int_{K} d\mu(x)~\al(uxa)=\int_{K} d\mu(x)~\al(xa)=\int_Kd\mu(xaa^{-1})~\al(xa)=\int_Kd\mu(xa^{-1})~\al(x),\nn
 \eea
one sees that if $\mu$ is a left invariant measure, ie $\mu\circ L_G=\mu$ then \\
~$\mu_a:=\mu\circ R_a~~\forall a\in G$,~ where $R_a$ denotes right translation, is another left invariant measure. This means that $\mu$ and $\mu_a$, by uniqueness, differ only by a constant.
\bea
&&\mu_a=\mu\circ R_a =c(a)~\mu~~~~\forall a\in G.\nn\\
&&\mu_{ab}=c(b)\mu_a=c(b)~c(a)~\mu=c(ab)~\mu~~\Ra~~c(ab)=c(a)c(b).\nn
\eea

\subsection{Invariant linear maps}
Let $\A,\B$ be two algebras and \\
$L(\B\slash\A)\subset \A\slash\B=\{\pi:\A\ra \B,~~\pi(a_1+a_2)=\pi(a_1)+\pi(a_2)\}$ be the set of all linear maps from $\A$ to $\B$.

Also let $J=J(L(\B\slash\A)):\A\ra \B~$ be the addition/composition  $(+,\circ)$ algebra of these linear maps, which is a $\B$-module as
$\B J,JB\subset J$.

One notes that each $\pi:\A\ra\B$ is equivalent to a bilinear pairing \\
$P_\pi:\A\times \B\ra \A\ot\B$.

 The induced relative central set $\A^\ast_\B$ of $\A$ is the ``kernel'' of $J$ given by
\bea
\A^\ast_\B=\{s\in J,~s:\A\ra Z(\B)= \B\cap \B'\}
\eea
where $\B'$ is the commutant of $\B$. That is, $\pi\in \A^\ast_\B$ if $\pi(a)\in Z(\B)=\B\cap \B'~~\forall a\in \A$.

Let another algebra $\C=\{c\}$ act on $\A$ through a linear representation $\rho$, ie.
\bea
&&\rho:\C\ra O(\A),~c\ra \rho_c:\A\ra\A,\nn\\
&&\rho_c\rho_{c_1}=\rho_{cc_1},~~\rho_c(a+a_1)=\rho_c(a)+\rho_c(a_1).
\eea
Then a linear map $\int$,
 \bea
\int~:\A^\ast_\B\ra \A^\ast_\B,~~s\ra \int s,~~\int (s+s_1)=\int s+\int s'
\eea
is a $\rho$-integral if there is some $s_0\in \A^\ast_\B$ such that
\bea
\int s\circ \rho_c=s_0(c)\int s~~~~\forall (c\in \C,~s\in \A^\ast_\B).
\eea
That is,
\bea
\int s\circ \rho_c(a)=\int s(\rho_c(a)) =s_0(c)\int s(a)~~~~\forall (a\in A,~c\in \C,~s\in \A^\ast_\B).\nn
\eea

More generally, $\int:L(\B\slash\A)\ra L(\B\slash\A)$ is a $\rho$-integral if there is some $\pi_0\in \A^\ast_\B$ such that
\bea
\int \pi\circ \rho_c=\pi_0(c)\int \pi~~~~\forall (c\in \C,~\pi\in L(\B\slash\A)).
\eea

On the other hand, $a\in \A$ is a $\rho$-integral element under the map $\pi$ if there is some $s_0\in \A^\ast_\B$ such that
\bea
&&\pi\circ\rho_c(a)=\pi(\rho_c(a))=s_0(c)~\pi(a)~~\forall c\in C.
\eea

Even more generally, if there is an equivalence relation $\sim$  among the elements of $L(\B\slash\A)$ which separate into equivalence classes $\{[\pi]\}$ then   \\
$\int:L(\B\slash\A)\ra L(\B\slash\A)$ is a $\rho$-integral  if there is some $\pi_0\in \A^\ast_\B$ such that
\bea
[\int \pi\circ \rho_c]=[\int \pi]~~~~\forall (c\in \C,~\pi\in L(\B\slash\A)).
\eea

\chapter{$C^\ast$-algebras}

\section{Cauchy-Schwarz inequality}
   Let us define a ${\ast}$-algebra $\mathcal{A}$ to be an associative algebra over the field of complex numbers $\mathbb{C}$ that is closed under an operation $^\ast$ (that is, $a^\ast\in \mathcal{A}~~\forall a\in \mathcal{A}$) with the following properties.
\bea
&&a^\ast{}^\ast=a,~~(ab)^\ast=b^\ast a^\ast,~~(a+b)^\ast=a^\ast+b^\ast,\nn\\
&&\al^\ast=\bar{\al}~~~~\forall~a,b\in \mathcal{A},~~\al\in \mathbb{C},
\eea
where $\bar{\al}$ denotes the complex conjugate of $\al$.

Let $\A$ be a $\ast$-algebra.
Consider any $A\in \mathcal{A}$,~ a collection \\
$\{B_i\in \mathcal{A},~~i=1,2,...,p\}$ and $\phi\in \mathcal{A}^\ast_+$, the set of positive linear functionals on $\A$.
\bea
&\phi(a+b)=\phi(a)+\phi(b),~~~~\phi(a^\ast)=\overline{\phi(a)},\nn\\
&\phi(\ld a)=\ld\phi(a),~~\phi(a^\ast a)\geq 0~~\forall a,b\in \A,~\ld\in C.
\eea

Also define the function ~$f:\mathbb{C}^p\ra \mathbb{R}^+,~\ld\mapsto f(\ld,\overline{\ld})$,

\bea
&&f(\ld,\overline{\ld})=\phi((A+\ld_i B_i)^\ast(A+\ld_i B_i))\nn\\
&&~~~~=\phi(A^\ast A)+\ld_i\phi(A^\ast B_i)+\overline{\ld}_i\phi(B_i^\ast A )+\overline{\ld}_i\ld_j\phi(B_i^\ast B_j)\nn\\
&&~~~~\eqv \phi(A^\ast A)+\ld_i \overline{N}_i+\overline{\ld}_iN_i+\overline{\ld}_i\ld_j M_{ij}\geq 0,\\
&&N_i=\phi(B_i^\ast A ),~~M_{ij}=\phi(B_i^\ast B_j),~~\overline{M}_{ij}=M_{ji}.
\eea
The value of $f$ at its extreme point gives the Cauchy-Schwarz inequality. That is,
\bea
\label{lin-indep}&&{\del\over\del \overline{\ld}_i}f(\ld',\overline{\ld}')=0~~\iff~~ \ld'_i=-M^{-1}_{ij}N_j,~~\det \overline{M}_{ij}> 0,\\
\label{cauchy-inequal}&&~~\Ra~~f(\ld',\overline{\ld}')=\phi(A^\ast A)-\overline{N}_i\overline{M}^{-1}_{ij}N_j\geq 0.
\eea

If we write~ $B_I=(A,B_i)\eqv (B_0,B_i),~~M_{IJ}= \phi(B_I^\ast B_J),~~I,J=0,1,...,p$~ then the inequality (\ref{cauchy-inequal}) becomes
\bea
\label{cauchy-inequal-2}&&f(\ld',\overline{\ld}')={\det \overline{M}_{IJ}\over \det \overline{M}_{ij}}\geq 0~~\Ra~~ \det \overline{M}_{IJ}\geq 0.
\eea
In the case $p=1$ one has
\bea
\phi(A^\ast B)\phi(B^\ast A)=\phi(A^\ast B)\overline{\phi(A^\ast B)}\leq \phi(A^\ast A)\phi(B^\ast B).
\eea

\section{Hilbert space and operator norm}
Define the operator norm $\|a\| $  for a bounded operator $a\in B(\H)=\{b:\H\ra \H\}$ on a Hilbert space $\H$ (~inner product vector space $(V,\lang|~\rang:V\times V\ra \mathbb{C})$ which is complete with respect to the inner product norm $\|~\|:\xi\mapsto \|\xi\|=\sqrt{\lang\xi|\xi\rang}$~) as
\bea
\label{operator-norm}&&\tc{red}{\|a\|=\sup_{\xi\in\H}{\|a\xi\|\over \|\xi\|},~~~~\|\xi\|=\sqrt{\langle \xi|\xi\rangle}}.
\eea

It follows from the definition (\ref{operator-norm}) that
\bea
&& \tc{red}{\|a\xi\|\leq \|a\|\|\xi\|~~\forall \xi\in \H}
\eea
and since $b\H=\H~~\forall b\in \H$ we then have that
\bea
&& \|ab\xi\|\leq \|a\|\|b\xi\|\leq \|a\|\|b\|\|\xi\|
\eea
and therefore
\bea
&&\|ab\|=\sup_{\xi\in\H}{\|ab\xi\|\over \|\xi\|}\leq \sup_{\xi\in\H}{\|a\|\|b\|\|\xi\|\over \|\xi\|}=\sup_{\xi\in\H}\|a\|\|b\|=\|a\|\|b\|.\nn\\
\label{norm-ineq}&&\txt{ie.}~~\tc{red}{\|ab\|\leq \|a\|\|b\|}.
\eea
This in turn implies that one could also define the norm
\bea
&&\tc{red}{\|a\|_{B(\H)}=\sup_{b\in B(\H)}{\|ab\|\over\|b\|}\leq \|a\|}.
\eea

If one defines $a^\ast$ by ~$\langle a^\ast\eta|\xi\rangle=\langle\eta|a\xi\rangle$~ then
\bea
&&(ab)^\ast=b^\ast a^\ast,~~a^{\ast\ast}=a,~~(\ld a)^\ast=\overline{\ld} a^\ast~~\forall a,b\in B(\H)~\&~\ld\in \mathbb{C},\nn\\
&&\|a^\ast\|=\sup_{\xi\in\H}{\|a^\ast\xi\|\over\|\xi\|}= \sup_{\xi\in\H}{\sqrt{|\lang a^\ast\xi|a^\ast\xi\rang|}\over\|\xi\|}= \sup_{\xi\in\H}{\sqrt{|\lang\xi|aa^\ast\xi\rang|}\over\|\xi\|},\nn\\
&&\|a\|=\sup_{\xi\in\H}{\|a\xi\|\over\|\xi\|}= \sup_{\xi\in\H}{\sqrt{|\lang a\xi|a\xi\rang|}\over\|\xi\|}= \sup_{\xi\in\H}{\sqrt{|\lang\xi|a^\ast a\xi\rang|}\over\|\xi\|}.
\eea
Thus the mean-center inequality (\ref{mean-center1},~\ref{mean-center2}) and the norm inequality $(\ref{norm-ineq})$ give
\bea
 \|a\|^2\leq \|a^\ast a\|\leq \|a^\ast\|\|a\|,~~\Ra~~\|a\|\leq \|a^\ast\|.
\eea
And $a^{\ast\ast}=a$ therefore gives
\bea
\|a\|\leq \|a^\ast\|\leq \|a^{\ast\ast}\|=\|a\|~~\Ra~~~~\|a^\ast\|= \|a\|,~~\|a^\ast a\|=\|a\|^2.
\eea

\textbf{Remarks}
\begin{itemize}
\item Each vector $\xi\in \H$ corresponds to an operator $\xi_o={1\over \sqrt{\lang\xi|\xi\rang}}|\xi\rang\lang\xi|$ whose norm with respect to a positive linear functional $\phi_\xi$, defined below in terms of the operator $\rho_\xi\eqv {1\over \lang\xi|\xi\rang}|\xi\rang\lang\xi|\in B(\H)$, gives the norm of $\xi$. That is
\bea
&&\phi_\xi( a)=\Tr(\rho_\xi a )={1\over \lang\xi|\xi\rang}\Tr(|\xi\rang\lang\xi| a )={\lang\xi|a|\xi\rang\over \lang\xi|\xi\rang},\nn\\
&&\|a\|_\xi=\sqrt{\phi_\xi(a^\ast a)},~~~~\|a\|= \sup_{\xi\in \H}\|a\|_\xi,\nn\\
&&\|\xi_o\|_\xi=\sqrt{\phi_\xi(\xi^\ast_o\xi_o)}=\sqrt{\phi_\xi(|\xi\rang\lang\xi|)} =\sqrt{\Tr \rho_\xi|\xi\rang\lang\xi|}\nn\\
&&~~~~=\sqrt{\Tr|\xi\rang\lang\xi|} =\sqrt{\lang\xi|\xi\rang}=\Tr\xi_o=\|\xi\|.\nn
\eea

One has the Cauchy-Schwarz inequality
\bea
\label{cre}&&\phi_\xi(a^\ast b)\overline{\phi_\xi(a^\ast b)}=|\phi_\xi(a^\ast b)|^2\leq \phi_\xi(a^\ast a) \phi_\xi(b^\ast b).
\eea

Setting $b=1$ in (\ref{cre}) gives
\bea
&&\phi_\xi(a^\ast)\phi_\xi(a)(2-\phi(1))\leq \phi_\xi(a^\ast a),\nn\\
\label{mean-center1}&& \txt{or}~~\phi_\xi((a-\phi_\xi(a))^\ast(a-\phi_\xi(a)))\geq 0,
\eea
which in turn gives
\bea
\label{mean-center2}&&\tc{red}{\phi_\xi(a^\ast b)\overline{\phi_\xi(a^\ast b)}(2-\phi(1))\leq \phi_\xi((a^\ast b)^\ast a^\ast b)=\phi_\xi( b^\ast aa^\ast b)},
\eea
as well as
\bea
\label{cre-full}&&\tc{red}{\phi_\xi(a^\ast b)\overline{\phi_\xi(a^\ast b)}\leq   \phi_\xi(a^\ast a) \phi_\xi(b^\ast b)\leq {1\over (2-\phi(1))} \phi_\xi(a^\ast ab^\ast b )}.
\eea

\item
The triangular inequality also follows thus:
\bea
&&\phi_\xi((a+b)^\ast(a+b))=\phi_\xi(a^\ast a)+\phi_\xi(a^\ast b)+\phi_\xi(b^\ast a)+\phi_\xi(b^\ast b)\nn\\
&&~~~~=\phi_\xi(a^\ast a)+2\Re\phi_\xi(a^\ast b)+\phi_\xi(b^\ast a^\ast)\leq \phi_\xi(a^\ast a)+2\sqrt{\phi_\xi(a^\ast b)\overline{\phi_\xi(a^\ast b)}}+\phi_\xi(b^\ast b)\nn\\
&&~~~~\leq  \phi_\xi(a^\ast a)+2\sqrt{\phi_\xi(a^\ast a)\phi_\xi(b^\ast b)}+\phi_\xi(b^\ast b)=(\phi_\xi(a^\ast a)+\phi_\xi(b^\ast b))^2,\nn\\
&&\|a+b\|_\xi\leq \|a\|_\xi+\|b\|_\xi~~\forall \xi\in\H,\nn\\
&&~~\Ra~~\tc{red}{\|a+b\|\leq \|a\|+\|b\|}.
\eea

Using  $\|a\|_\xi\leq \sup_{\xi\in\H}\|a\|_\xi=\|a\|$ one can also check that
\bea
&&\|a\|_\xi\|b\|_\xi\leq \sup_{\xi\in\H}(\|a\|_\xi\|b\|_\xi)\leq \sup_{\xi\in\H}(\|a\|_\xi \|b\|)=\|a\|\|b\|.
\eea

\item Since $a\H=\H$, $\|a\|$ is the same for all elements in the conjugacy class \\
$[a]=\{b,~ \exists c\in \H,~~b=c^\ast ac\}$.
\item For the finite dimensional case, $a^\ast a$ may be diagonalized: ie. $a^\ast a=P\Ld P^{-1}$ and if one chooses a basis $\{|i\rang\}$ for $\H$ then
\bea
&&\xi= |i\rang \xi_i,~~a^\ast=(a^\ast a)_{ij}|i\rang \lang j|,~\Ld=\ld_i~ |i\rang \lang i|,~~\lang i|j\rang=\delta_{ij},  \nn\\
&&\|a\|=\sup_{\xi\in\H}~{\sqrt{\sum_i\ld_i|\xi_i|^2}\over \sqrt{\sum_j|\xi_j|^2}}=\sup_{\xi\in\H}\|a\|_\xi .\nn\\
\label{extrema1}&&{\del\over\del |\xi_i|}\|a\|_\xi=|\xi_i|{1\over \|a\|_\xi}\{\ld_i- \|a\|_\xi^2 \}=0~~\forall i.
\eea
(\ref{extrema1}) may have several different solutions but one should take the one on which $\|a\|$ is biggest.
In particular one can choose $\xi$ to be in the direction $i_{\max}$ where $\ld_{i_{\max}}=\max_i\ld_i$. That is
\bea
&&\tc{red}{\xi_i=|\xi|~\delta_{i i_{\max}}~~\Ra~~\|a\|=\sqrt{\max_i\ld_i}}.
\eea

The eigenvalue character may be defined more generally as
\bea
&&\ld_\xi(a)=\txt{Extr}_{\eta\in\H}{\lang\eta|a\xi\rang\over \lang\eta|\xi\rang},
\eea
where $\txt{Extr}_{\eta\in\H}$ refers to extremization in $\H$.

\item A $C^\ast$-algebra given abstractly as
\bea
\label{c-star-algebra}&&\tc{red}{\A=(B=\{a,b,c,...\},~B\sr{\ast,\mathbb{C}}{\ral}B,~B\sr{\|\|}{\ral}\mathbb{R}^+ )}
 \eea
has the following defining properties
\bea
\label{convol}&&\tc{red}{(ab)^\ast=b^\ast a^\ast,~(a+b)^\ast=a^\ast+b^\ast,~(\ld a)^\ast=\overline{\ld}a^\ast ,~a^{\ast\ast}=a},\\
&&~~~~(\txt{involutive algebra}),\nn\\
\label{norm}&&\tc{red}{\|a\|\geq 0,~~\|ab\|\leq \|a\|\|b\|},\\
&&~~\txt{also with}~~\tc{red}{\|a^\ast\|=\|a\|}~~(\txt{normed involutive algebra}),\nn\\
&&~~\txt{also}~~\tc{red}{\|\|-\txt{complete}}~~(\txt{normed involutive Banach algebra}),\nn \\
\label{c-star}&&\tc{red}{\|a^\ast a\|= \|a\|^2}~~(\txt{$C^\ast$-algebra})
\eea
(\ref{norm}) and (\ref{c-star}) give
\bea
\label{pre-c-star}&&\|a^\ast a\|= \|a\|^2\leq \|a^\ast\|\|a\|~~\Ra~~\|a\|\leq \|a^\ast\|.
\eea
 (\ref{convol}) and (\ref{pre-c-star}) then give
\bea
\|a^\ast\|\leq \|a^{\ast\ast}\|=\|a\|.~~\txt{ie}~~\|a^\ast\|\leq \|a\|
\eea
and therefore $\|a^\ast\|=\|a\|$. Therefore the algebra of bounded operators $B(\H)$ is a $C^\ast$-algebra with the operator norm.
In certain cases it may also be possible that one can obtain the norm inequality (\ref{norm}) when given only the $C^\ast$-condition (\ref{c-star}) and the Cauchy-Schwarz inequality.

\item Examples of $C^\ast$-algebras are given by pointwise product (denoted $\star$ ) algebras $B(\H)=\F_\mu(X)=\{\mu_f:\F(X)\ra \F(X),~g\ra (f\star g)(x),~~f\in \F(X)\}$ of complex functions $\H=\F(X)=\{\xi:X\ra \mathbb{C}\}$ over a topological space $X$ under a suitably defined operator norm.

\bea
&& \|\xi\|=\sqrt{\sum_{x\in X}|\xi(x)|^2},~~~~\|\mu_f\|=\sup_{\xi\in \H}{\|\mu_f\xi\|\over \|\xi\|},\nn\\
&&~~\Ra~~\|\mu_f\xi\|\leq \|\mu_f\|\|\xi\|~~\forall \xi\in \H.
\eea

For case of a separable (ie. local) pointwise product any $f\in \H=\F(X)$ corresponds to an operator $\mu_f\in \F_\mu(X)$ that acts on $\H=\F(X)$ linearly as
\bea
&&\mu_f\xi(x)=(f\star\xi)(x)=f(x)\xi(x),\nn\\
&&\|\mu_f\|=\sqrt{\max_{x\in X}|f(x)|^2}=\max_{x\in X}|f(x)|
\eea
where in the norm we have used the fact that for each $x\in X$,  $f(x)$ is regarded as an eigenvalue of $\mu_f$. Also the last step is due to the fact that ${\del |f(x)|^2\over\del x}=2|f(x)|{\del |f(x)|\over\del x}$.

One also has the pointwise convolution product algebra $\F_\nu(X)$
\bea
&&\nu_f\xi(x)=\sum_{y\in X}f(x-y)\xi(y)\eqv \sum_{k\in X}e^{ikx}\widetilde{f}(k)\widetilde{\xi}(k),\nn\\
&& \|\xi\|=\sqrt{\sum_{x\in X}|\xi(x)|^2},\nn\\
&&\|a\|=\sup_{\widetilde{\xi}\in\H}~{\sqrt{\sum_{k\in X}|\widetilde{f}(k)|^2|\widetilde{\xi}_{k}|^2}\over \sqrt{\sum_{k'\in X}|\widetilde{\xi}_{k'}|^2}}=\max_{k\in X}|\widetilde{f}(k)|,
\eea
where $\widetilde{f}$ is the Fourier transform of $f$.

One notes that if the product is noncommutative, then there are two possible and independent representations $\mu^L,~\mu^R$ of the product corresponding to left and right multiplication respectively
\bea
&&\mu^L_f\xi(x)=(f\star\xi)(x),~~~~\mu^R_f\xi(x)=(\xi\star f)(x),\nn\\
&&\mu^L_f\mu^L_g=\mu^L_{f\star g},~~\mu^R_f\mu^R_g=\mu^L_{g\star f},~~\mu^L_f\mu^R_g=\mu^R_g\mu^L_f.
\eea
Therefore the derived multiplication $\mu^\ld=\al\mu^L+\beta \mu^R$, where $\ld=(\al,\beta)$ are commuting or central numbers, has the commutation relation
\bea
[\mu^\ld_f,\mu^\ld_g]=\al^2~\mu^L_{f\star g-g\star f}+\beta^2~\mu^R_{-f\star g+g\star f}.
\eea
For a subset of elements $A=\{a\}\subseteq \F(X)$ where $a\star b-b\star a$ is central for all $a,b\in A$ one has  $\mu^L_{a\star b-b\star a}=-\mu^R_{-f\star g+g\star f}$ and so $\mu^\ld$ will give a commutative representation $\mu_{a_c}=\mu^\ld_a~~\forall a\in \A$ of $A$ on $ \H=\F(X)$ with $\al^2=\beta^2$.

\item For a self adjoint operator $a^\ast=a$, the $C^\ast$ condition $\|a^\ast a\|=\|a\|^2 $ becomes $\|a^2\|=\|a\|^2$. Thus if one defines $\sqrt{a}$ then
\bea
&&\|a\|=\|\sqrt{a}^2\|=\|\sqrt{a}\|^2=(\sup_{\xi\in\H}{\sqrt{|\lang\xi|\sqrt{a}^\ast \sqrt{a}\xi\rang|}\over\|\xi\|})^2=(\sup_{\xi\in\H}{\sqrt{|\lang\xi|a\xi\rang|}\over\|\xi\|})^2\nn\\
&&~~~~=\sup_{\xi\in\H}{\lang\xi|a\xi\rang\over \lang\xi|\xi\rang}.
\eea

\item The following names are used:
\bea
&& a^\ast a=1~~~~(\txt{$a$ is an isometry}),\nn\\
&& (a^\ast a)^2=a^\ast a~~~~(\txt{$a$ is a partial isometry}),\nn\\
&& a^\ast a=aa^\ast~~~~(\txt{$a$ is a normal element}),\nn\\
&& a^\ast a=aa^\ast=1~~~~(\txt{$a$ is a normal isometry or a unitary element}).\nn
\eea

\end{itemize}

\section{Convex subspaces}
In a Hilbert space one has the basic expansion identity
\bea
&&\tc{red}{\|\xi+\eta\|^2=\|\xi\|^2+\lang\xi|\eta\rang +\lang\eta|\xi\rang+ \|\eta\|^2}.
\eea
$C\subset \H$ is convex if $\forall~c,c'\in C,~~\al c+(1-\al)c'\in C,~\forall~ 0<\al<1$.
Alternatively $C\subset \H$ is convex if
\bea
\label{convex-distance}&&\tc{red}{d(\xi,C)=\min_{c\in C}\|\xi-c\|=\|\xi-\xi_C\|}
\eea
 is unique for any $\xi\in \H$. Consider the collection $\del C=\{\xi_C,~~\xi\in \H\}$ of extreme points of $C$. Then one deduces from the primary definition that any point $\eta\in C$ can be expanded as
  \bea
  \eta=\sum_{b\in \del C}\eta_bb,~~\sum_{b\in \del C}\eta_b=1,~~0\leq \eta_b\leq 1,
  \eea
  The points of $\del C=\{b\}$ are pure in that any element $b$ defines a unique equivalence class of elements of $\H$ given by $[b]=\{\xi\in \H,~d(\xi,C)=d(\xi,b)> 0\}$. The impure elements of $C$ are those that do not belong to $\del C$.

From the definition (\ref{convex-distance}) it follows that
\bea
&&\tc{red}{\|\xi-\xi_C\|\leq\|\xi-c\|~~\forall c\in C}.
\eea
In particular~ $\xi_C+{\lang \xi-\xi_C|c\rang\over \|c\|^2}c\eqv c_p\in C$~  because of the fact that both \\
~$\xi_C~~\&~~{\lang \xi-\xi_C|c\rang\over \|c\|^2}c~\in C$~~ assuming that $C$ is a closed linear subspace (Here ~$ {\lang \xi-\xi_C|c\rang\over \|c\|^2}c$~ is the projection of $\xi-\xi_C$ in the direction of an arbitrary $c\in C,~c\neq 0$).

Therefore
\bea
&&\|\xi-\xi_C\|^2\leq \|\xi-c_p\|^2\nn\\
&&~~~~=\|\xi-(\xi_C+{\lang \xi-\xi_C|c\rang\over \|c\|^2}c)\|^2\nn\\
&&~~~~=\|\xi-\xi_C-{\lang \xi-\xi_C|c\rang\over \|c\|^2}c\|^2\nn\\
&&~~~~=\|\xi-\xi_C\|^2-{|\lang \xi-\xi_C|c\rang|^2\over \|c\|^2}\nn\\
&&~~\Ra~~\lang \xi-\xi_C|c\rang=0~~\forall c\in C,~c\neq 0.
\eea
That is~ $d(\xi,C)=\|\xi-\xi_C\|$~ implies that $\xi-\xi_C$ is orthogonal or normal to $C$ and therefore if $P_C\in B(\H)$ is the orthogonal projection unto $C$ then
\bea
&&\tc{red}{d(\xi,C)=\|\xi-\xi_C\|=\|\xi-P_C\xi\|=d(\xi,P_C\xi)},\\
&&P_C\H=C,~~P_C^\ast=P_C^2=P_C,~~\|P_C\|=1.\nn
\eea

In particular
\bea
&&\lang \xi-\xi_C|\xi_C\rang=0,~~\Ra~~\lang \xi|\xi_C\rang=\lang\xi_C|\xi_C\rang=\|\xi_C\|^2
\eea
and therefore
\bea
&& \|\xi\|^2=\|\xi-\xi_C\|^2+\lang\xi|\xi_C\rang+\lang\xi_C|\xi\rang-\|\xi_C\|^2.\nn\\
&&\tc{red}{\|\xi\|^2=\|\xi_C\|^2+\|\xi-\xi_C\|^2}.
\eea

\subsection{States of a  $\ast$-algebra}
A linear functional $\phi:\A\ra \mathbb{C},~~\phi\in \A^\ast_+$ on an algebra $\A=\{a\}$ is said to be positive it maps positive elements $P(\A)=\{p=a^\ast a,~~a\in \A\}$ to positive numbers. Consider the possibility of introducing a basis $\{\vphi_i\}$ for the set of positive linear functionals (plfs). Then a general plf may be expanded as
\bea
\phi=\sum_i\ld_i\vphi_i.
\eea
Then ~$\phi(p)=\sum_i\ld_i\vphi_i(p)\geq 0~~\forall p\in P(\A)$~ implies that ~$\ld_i\geq 0~~\forall i$.
Therefore for the set ~$S(\A)=\{\rho\in \A^\ast_+,~~\rho(1)=1\}$~ of normalized plfs one has\\
~$\rho=\sum_i\ld_i\vphi_i,~~\ld_i\geq 0$~ and  ~$\rho(1)=1$~ gives ~$\sum_i\ld_i=1,~~0\leq \ld_i\leq 1$~ if~ $\vphi_i(1)=1~~\forall i$. Therefore $S(\A)$ is a convex set generated by the basis elements $\{\vphi_i\}$  which are known as pure states due to their role as the extreme points in the convexity of $S(\A)$.

\section{Spectral theory}\label{spectral-theory}
Spectrum $\sigma(a)$/spectral radius $\rho(a)$
\bea
&&\sigma_\A(a)=\{\ld\in C,~~(\ld-a)^{-1}=\sum_{n=0}^\infty\ld^{-(n+1)} a^n~\nexists~~\txt{in $\A$}\}\nn\\
\eea
The spectral radius or radius of convergence
\footnote
{
This utilizes L'Hospital's rule: if $f,g$ are differentiable functions and \\
~$\lim_{x\ra a}f(x)=f(a)=0=\lim_{x\ra a}g(x)=g(a)$~ then
\bea
&&\lim_{x\ra a}{f(x)\over g(x)}=\lim_{x\ra a}\lim_{b\ra x}{{f(b)-f(a)\over b-a}\over {g(b)-g(a)\over b-a}}=\lim_{x\ra a}{\lim_{b\ra x}{f(b)-f(a)\over b-a}\over \lim_{c\ra x}{g(c)-g(a)\over c-a}}=\lim_{x\ra a}{f'(x)\over g'(x)},
\eea
and similarly for ~$\lim_{x\ra a}f(x)=\infty=\lim_{x\ra a}g(x)$.
}
of $\sum_{n=0}^\infty\ld^{-(n+1)} a^n$ is
\bea
&&\rho(a)=\lim_{n\ra\infty}\|a^n\|^{1\over n}=e^{\lim_{n\ra\infty}{\ln \|a^n\|\over n}}\leq e^{\lim_{n\ra\infty}{\ln \|a\|^n\over n}}\nn\\
&&~~~~=e^{\lim_{n\ra\infty}\ln \|a\|}=\|a\|.
\eea
The series is convergent (~ie. $\ld-a$~ is invertible or $\ld\not\in \sigma_\A(a)$~) if
\bea
\lim_{n\ra\infty}\|\ld^{-(n+1)} a^n\|^{1\over n}=|\ld|^{-1}\lim_{n\ra\infty} \|a^n\|^{1\over n}=|\ld|^{-1}~\rho(a)< 1,
\eea
and similarly the series is not convergent (ie. $\ld\in \sigma_\A(a)$~) if
\bea
&&\rho(a)< |\ld|.
\eea
Therefore as $\rho_\A(a)$ is finite, $\sigma_\A(a)$ cannot be empty in $\mathbb{C}$. \\
That is $\forall a\in\A,~~\sigma_\A(a)\neq \{\}$.

Corresponding to any single variable function $f$, one  can define an $\A$-valued function $f:\A\ra \A,~a\mapsto f(a)$. In particular one can make use of holomorphic functions
\bea
&&f(a)={1\over 2\pi i}\oint_{\Gamma(\sigma_\A(a))} dz~{f(z)\over z-a },
\eea
where $\Gamma(\sigma_\A(a))$ is any closed curve in $C$ that encloses $\sigma_\A(a)$.

\begin{itemize}
\item $aa^\ast=a^\ast a~~\Ra~~\rho(a^\ast a)=\|a\|^2$ since
\bea
&& \rho(a^\ast a)=\lim_{n\ra\infty}\|(a^\ast a)^n\|^{1\over n}=\lim_{n\ra\infty}\|(a^n)^\ast a^n\|^{1\over n}\nn\\
&&~~~~=\lim_{n\ra\infty}\| a^n\|^{2\over n}=\|a\|^2.
\eea

\item $a=a^\ast~~\Ra~~\rho(a)=\|a\|$ ~since by the $C^\ast$ condition (\ref{c-star})
\bea
&&\rho(a)=\lim_{n\ra\infty}\|a^n\|^{1\over n}=\lim_{n\ra\infty}\|a^{2^n}\|^{1\over 2^n}=\|a\|.
\eea

\item One may also verify that ~$\sigma(ab)=\sigma(ba)~~\forall a,b\in\A$~ due to the following identity:
\bea
&&(1-ab)^{-1}=1+ab+(ab)^2+(ab)^3+...\nn\\
&&~~~~=1+a(1+ba+(ba)^2+(ba)^3+...)b\nn\\
&&~~~~=1+a(1-ba)^{-1}b.
\eea

\end{itemize}

\subsection{Gelfand-Mazur theorem}
If $\A$ has unit then $\ld-a\in\A~~\forall \ld\in \mathbb{C},~\forall a\in \A$. Therefore if every element $a\in\A$ is invertible except when $a=0$ then so does $(a-\ld)^{-1}~\exists$ except when $a-\ld=0$.

But $\sigma(a)=\{\ld,~(a-\ld)^{-1}~\nexists\}\neq \{\}$ and therefore for each $a\in \A$, $\exists~\ld\in \mathbb{C}$ ~such that ~$a-\ld=0$. That is if $\A$ has a unit and if every element $a\in\A$ is invertible except when $a=0$ then $\A\simeq\mathbb{C}$.

It follows that if $\A$ has unit and $I\subset \A$ is a maximal (having no proper subs) two-sided ideal, $I\A=\A I=I,~I+I=I$, then the quotient $\A\slash I\simeq \mathbb{C}$ where $\A\slash I=\{a+I;~a\in\A\}$.

\subsection{Gelfand-Naimark theorem}
A character of an abelian algebra $\A_0$ is defined by
\bea
&&\chi:\A_0\ra \mathbb{C}\backslash \{0\},~~\chi(ab)=\chi(a)\chi(b),~\chi(a+b)=\chi(a)+\chi(b).
\eea
If $\A_0$ is unitary with identity $1$ the $\chi(1)=\chi(1)^2~\Ra~\chi(1)=1$. \\
Thus $\chi(\al a)=\al\chi(a)~\forall \al\in \mathbb{C}$.
This coincides with the definition of the eigenvalue and generalizes the fact that any two commuting operators can be simultaneously diagonalized (ie. have a common set of eigenvectors).

Recall that the spectrum $\sigma(a)$ is given by $\sigma(a)=\{\ld(a)\in \mathbb{C},~(\ld(a)-a)^{-1}\nexists\}$ and satisfies $\ld(a^\ast)=\overline{\ld(a)},~~\ld(f(a))=f(\ld(a)),~~|\ld(a)|\leq\rho(a)\leq \|a\|,$~~etc. Thus $\ld:\A_0\ra \mathbb{C}$ is a character on $\A_0 \backslash \{0\}$ and by uniqueness of $\ld$,~ $\ld~\&~\chi$ coincide on $f(a)~\forall f$ and hence $\chi(a)\in \sigma(a)$ which means that
\bea
&&|\chi(a)|\leq\|a\|~~\forall a\in \A_0,~\forall \chi\in \sigma(\A_0)=\{\X:\A_0\ra \mathbb{C}\backslash \{0\}\}.\nn\\
&&\rho(a)=\sup_{\chi\in \sigma(\A_0)}|\chi(a)|.
\eea
Define the spectrum~ $\sigma(\A_0)=\{\chi:\A_0\ra \mathbb{C}\backslash \{0\}\}$~ of $\A_0$. Then the map $a\mapsto\td{a}:\sigma(\A_0)\ra \mathbb{C},~~\td{a}(\chi)=\chi(a)$\\
isomorphically an isometrically identifies abelian $C^\ast$-algebra $\A_0$ with the commutative product algebra $\F_\mu(\sigma(\A_0))$ of complex functions $\F(\sigma(\A_0))$ since
\bea
&& \chi(ab)=\chi(a)\chi(b)=\td{a}(\chi)\td{b}(\chi)=(\td{a}\td{b})(\chi),\nn\\
&&\chi(a+b)=\chi(a)+\chi(b)=\td{a}(\chi)+\td{b}(\chi)=(\td{a}+\td{b})(\chi).
\eea
That is,
\bea
&&\A_0\simeq \td{\A}_0\simeq \F(\sigma(\A_0))\simeq \F_\mu(\sigma(\A_0)),\nn\\
&&\td{\A}_0=\{\td{a},~a\in \A_0\}
\eea
and also the spectrum $\sigma(\mu_{\td{a}})=\sigma(a)$ since if $(\ld-a)^{-1}~\nexists$ then\\
$\chi((\ld-a)^{-1}b)~\nexists~~\forall \chi\in \sigma(\A_0),~0\neq b\in \A$ where
\bea
&&\chi((\ld-a)^{-1}b)=(\chi(\ld)-\chi(a))^{-1}\chi(b)=(\ld-\td{a}(\chi))^{-1}\td{b}(\chi)\nn\\
&&~~~~=((\ld-\mu_{\td{a}})^{-1}\td{b})(\chi)~\nexists~~\forall \chi,\td{b}\nn
\eea
and vice versa.

Thus $\rho(\mu_{\td{a}})=\rho(a)$.
\bea
&&\chi(a^\ast)=\overline{\chi(a)}=\overline{\td{a}(\chi)}=\td{a}^\ast(\chi).
\eea

For the function multiplication algebra $\F_\mu(X)$ the spectrum of the multiplication operator is
\bea
&&\sigma(\mu_f)=\{f(x),~x\in X\}=f(X),\nn\\
&&\|\mu_f\|=\sup_{x\in X}|f(x)|,~~\|\mu_f\mu_g\|\leq \|\mu_f\|\|\mu_g\|,\nn\\
&&\|\mu^\ast_f\mu_f\|=\|\mu_f\|^2.
\eea
Similarly,
\bea
&&\sigma(\mu_{\td{a}})=\{\td{a}(\chi),~\chi\in \sigma(\A_0)\}=\{\chi(a),~\chi\in \sigma(\A_0)\}=\td{a}(\sigma(\A_0))\simeq \sigma(\A_0) ,\nn\\
&&~~\Ra~~\sigma(a)=\sigma(\mu_{\td{a}})\simeq\sigma(\A_0),\nn\\
&&\|\mu_{\td{a}}\|=\sup_{\chi\in \sigma(\A_0)}|{\td{a}}(\chi)|=\sup_{\chi\in \sigma(\A_0)}|\chi(a)| ,~~\|\mu_{\td{a}}\mu_{\td{b}}\|\leq \|\mu_{\td{a}}\|\|\mu_{\td{b}}\|,\nn\\
&&\|\mu^\ast_{\td{a}}\mu_{\td{a}}\|=\|\mu_{\td{a}}\|^2,~~\mu_{\td{a}}^\ast=\mu_{\td{a}^\ast}.
\eea
To check that the map $a\ra \mu_{\td{a}}$ is an isometry
\bea
&&\|\mu_{\td{a}}\|^2=\|\mu^\ast_{\td{a}}\mu_{\td{a}}\|=\rho(\mu^\ast_{\td{a}}\mu_{\td{a}})\nn\\
&&~~~~=\rho(\mu_{\td{a}^\ast\td{a}})=\rho(\mu_{\widetilde{a^\ast a}})=\rho(a^\ast a)=\|a\|^2,\nn\\
&&~~\Ra~~\|\mu_{\td{a}}\|=\|a\|^2.
\eea

\section{Ideals and Identities}
Given an algebra $\A$, the concept of its ideals (or its invariant subalgebras in general) is a generalization of the zero element meanwhile the concept of its identities (or its symmetry groups in general) is a generalization of the unit element. Let $\P(\A)=\{S\subseteq \A\}~(~P(\A)=\{S\subset \A\}=\P(\A)\backslash \A$~) be the set of all subsets (proper subsets) of $\A$.

\tc{green}{Definition:} Consider $A,B\in \P(\A)$ and define $ AB=\{ab,~a\in A,~b\in B\}$.

It \tc{red}{follows} that $Ab,aB\subseteq AB~~\forall a\in A,~b\in B$. Also $ A+B=\{a+b,~a\in A,~b\in B\}$ from which it \tc{red}{follows} that $A+b,a+B\subseteq A+B~~\forall a\in A,~b\in B$.

\tc{green}{Definition:} $A$ is a left (right) ideal, denote it $I_l~(I_r)\in P(\A)$, if
\bea
\label{dfn1}&& I_l\A=I_l,~I_l+I_l=I_l~~( \A I_r=I_r,~I_r+I_r=I_r).
\eea
It \tc{red}{follows} that
\bea
\label{ob2}&& I_lS\subseteq I_l~~\forall S\in \P(\A),~~(~SI_r \subseteq I_r,~~\forall S\in \P(\A)~).
\eea

That is $I_l(I_r)$ is a left(right) $\A$-invariant abelian (ie. additive) proper subgroup (a proper subset that is closed under addition).

\tc{green}{Definition:} An ideal is two-sided if it is both a left and a right ideal.

\tc{green}{Definition:} A subset of $\A$ is said to be nonideal if it is not a subset of any ideal.

\tc{green}{Definition:} Similarly $A$ is a left (right) identity, denote it $E_l~(E_r)\in P(\A)$, if
\bea
\label{dfn2}&&E_l\A=\A,~E_lE_l=E_l~~( \A E_r=\A,~E_rE_r=E_r).
\eea
That is $E_l(E_r)$ is a multiplicative proper subgroup (a proper subset that is closed under multiplication) under which $\A$ is left(right)-invariant.

\tc{green}{Definition:} An identity is two-sided if it is both a left and a right identity.

\tc{green}{Definition:} A subset of $\A$ is said to be nonidentity if it is not a subset of any identity.

\tc{red}{Observe} that by definitions (\ref{dfn1}) and (\ref{dfn2})
\bea
\label{ob3}&&I_l(\A)\cap E_l(\A)=\{\}=I_r(\A)\cap E_r(\A)
\eea
where $I(\A)$ is the set of all ideals in $P(\A)$ and $E(\A)$ is the set of all identities in $P(\A)$.

\tc{green}{Definition:} A multiplicative left(right) inverse $\td{S}^{l}\in P(\A)~(\td{S}^{r}\in P(\A))$ of a subset $S\in \P(\A)$ is any subset such that $\td{S}^{l}S~(S\td{S}^{r})$ is a left (right) identity; ie. $\td{S}^{l}S\in E_l(\A)~(S\td{S}^{r}\in E_r(\A)$, where $E_l(\A)~(E_r(\A))$ is the set of left(right) identities of $\A$.

\tc{red}{Observe}(1)~~ that $\td{z}^l_l~\txt{may}~\nexists~~\forall z_l\in \P(I_l),~~\forall I_r\in I_r(\A)~(~\td{z}^r_r~\txt{may}~\nexists~~\forall z_r\in \P(I_r),~~\forall I_r\in I_r(\A)~)$ by (\ref{ob2}) and (\ref{ob3}).
This is because $z_lS\subseteq I_l~~\forall S\subseteq \A$ by definition and if it has a left inverse $\td{z}^l_l$ then one can find a left identity $E^z_l$ such that $E^z_l=\td{z}^l_l z_l$. That is, one has the two conditions $\td{z}^l_l z_l\A=\A $ and $z_lS\subseteq I_l~~\forall S\subseteq \A$ but $z_l\A\subseteq I_l~~\Ra~~\td{z}^l_lz_l\A\subseteq \td{z}^l_lI_l$ and so for $\td{z}^l_l$ ($\td{z}^r_r$) to exist we must have \\$\td{z}^l_lI_l=\A$~~ ($I_r\td{z}^r_r=\A$). In particular $\td{z}^l_l$ ($\td{z}^r_r$) cannot exist if $I_l$~($I_r$) is also a left ideal [ie. if $I_l$ ($I_r$) is a two-sided ideal].

\tc{red}{Observe}(2)~~ that $\td{z}^r_l~\txt{may}~\nexists~~\forall z_l\in \P(I_l),~~\forall I_l\in I_l(\A)~(~\td{z}^l_r~\txt{may}~\nexists~~\forall z_r\in \P(I_r),~~\forall I_r\in I_r(\A)~)$ by (\ref{ob2}) and (\ref{ob3}).
This is because $z_lS\subseteq I_l~~\forall S\subseteq \A$ by definition and if it has a right inverse $\td{z}^r_l$ then one can find a right identity $E^z_r$ such that $E^z_r= z_l\td{z}^r_l$. That is, one has the two conditions $\A z_l\td{z}^r_l=\A $ and $z_lS\subseteq I_l~~\forall S\subseteq \A$ which together imply that $\A I_l=\A$ ($I_r\A=\A$). Thus for $\td{z}^r_l$ ($\td{z}^l_r$) to exist $I_l$ ($I_r$) must also be a right ideal and thus a two-sided ideal. Thus $\td{z}^r_l$ ($\td{z}^l_r$) cannot exist unless $I_l$ ($I_r$) is a two-sided ideal. It follows that if $\td{z}^r_l$ ($\td{z}^l_r$)  can exist then $\td{z}^l_l$ ($\td{z}^r_r$) cannot exist. Putting results together and removing labels one finds that \emph{\textbf{a subset $z$ of a two-sided ideal $I$ cannot have an inverse}}.

\tc{green}{Definition:} An ideal $I$ is maximal if it is not a subset of any other ideal; ie. if $I\not\in \P(I')~~\forall I'\in I(\A)$.

\tc{green}{Definition:} Also an ideal is simple if it contains no proper subideal(s).

The spectrum of an element is defined as
\bea
&&\sigma(a)=\{\ld\in \mathbb{C},~~(a-\ld 1)^{-1}~\nexists\}.
\eea
Thus obviously, if $0\in \sigma(a)$ then $(a-0)^{-1}\nexists= a^{-1}~\nexists$.
That is, inversion of an element $a$ (even if $a\neq 0$, which is all that is required for the elements of $\mathbb{C}$) is not possible whenever the spectrum $\sigma(a)$ contains $0$. For the abelian, ie. $\A_0$, case where the spectrum
of an element is
\bea
\sigma(a)\simeq\sigma(\mu_{\td{a}})=\td{a}(\sigma(\A_0))\simeq \sigma(\A_0),
\eea
one can quotient $\A_0$ for one chosen character $\chi$, by (ie. remove) those elements $I_\chi(\A_0)=Ker_{\A_0}(\chi)=\{a\in\A_0,~\td{a}(\chi)=0\}$ that can take on zero values at $\chi$. One can check that $I_\chi$ is a maximal ideal in $\A_0$ for any $\chi\in\sigma(\A_0)$. The quotienting is consistent only if $I_\chi$ is an ideal and this is the case for abelian algebras.
The space
\bea
\A_0\slash I_\chi=\{c=a+I_\chi,~a\in\A_0\},
\eea
in which every nonideal (ie. non-$I_\chi$) element $I_\chi\neq c\in \A_0\slash I_\chi$ is now invertible, is by the Gelfand-Mazur theorem equivalent to $\mathbb{C}$. ie. $\A_0\slash I_\chi\simeq \mathbb{C}~~\forall \chi\in \sigma(\A_0)$.

One needs to check that every element $I_\chi\neq a+I_\chi\in\A_0\slash I_\chi$ is invertible. That is
\bea
0\not\in\sigma(a+I_\chi)=\td{a}(\sigma(\A_0))+I_\chi(\sigma(\A_0)).
\eea
Assume on the contrary that $\exists y\in \sigma(\A_0),~y\neq\chi$ such that \\
$\td{a}(y)+I_\chi(y)=0,~\td{a}(\chi)\neq 0$. Since $I_\chi$ is a ``large'' set and this must hold for all its elements the only possibility is $\td{a}(y)=0,~~I_\chi(y)=0$ which in turn means that $\{a,I_\chi\}\subseteq I_y$ in contradiction to the fact that $I_\chi$ is a maximal ideal. Therefore each character $\chi$ is uniquely specified in $\sigma(\A_0)$ by the maximal ideal $I_\chi$.

Therefore given $\A_0$, all one needs is knowledge of (a means to construct) the space $\I(\A_0)=\{I\}$ of its (maximal) ideals, from which characters can then be defined as projections
\bea
&&\sigma(\A_0)=\{\chi_I:\A_0\ra {\A_0\over I}\backslash \{0\},~I\in \I(\A_0)\}.
\eea

Thus the maximal ideals of an arbitrary $C^\ast$-algebra $\A$ may be used to define its (noncommutative) point-like topology/geometry. The space of maximal ideals $\I(\A)$ may be written as
\bea
\I(\A)=\{I_u\subset \A;~\A I_u=I_u=I_uI_u=I_u+I_u,~\bigcup_uI_u=\A,~I_u\not\subset I_v,~\forall u,v\in S\},\nn
\eea
where $S$ is a parameter space ($S\simeq \sigma(\A_0)$ in the commutative algebra $\A_0$ case).

\section{GNS construction}
A state on $\A$ is a (normalized) positive linear functional
\bea
&&\phi(a^\ast a)\geq 0,~~\phi(1)=1.
\eea
It follows that
\bea
&&\label{cre2} |\phi(a^\ast b)|^2\leq \phi(a^\ast a)\phi(b^\ast b).
\eea
Any null element $n\in \A,~~\phi(n^\ast n)=0$ is completely orthogonal to $\A$ with respect to $\A$ since (\ref{cre2}) implies that
\bea
|\phi(n^\ast b)|^2\leq \phi(n^\ast n)\phi(b^\ast b)=0~~\forall b\in\A.
\eea
That is, $\phi(n^\ast a)=0=\phi(a^\ast n)~~\forall a\in \A$~~ or simply $\phi(\A n)=0$ or
\bea
&&\phi(\A N_\phi)=0,~~N_\phi=N_\phi(\A)=\{n\in \A,~\phi(\A n)=0\}.
\eea
 Thus $N_\phi$ is a left ideal ($\A N_\phi=N_\phi$) in $\A$ and \\
 $\H_1=\A/N_\phi(\A)=\{\xi=a+N_\phi(\A),~a\in\A\}$ is a prehilbert space (to be completed to a hilbert space $\H_\phi$) with inner product
\bea
&&\phi(\xi^\ast\eta)\eqv\lang \xi|\eta \rang=\lang a+N_\phi(\A)|b+N_\phi(\A) \rang=\phi(a^\ast b).
\eea
This induces the norm
\bea
&&\|\xi\|=\sqrt{\lang\xi|\xi\rang}=\sqrt{\phi(b^\ast b)},~~\xi=b+N_\phi.\nn\\
&&|\lang\eta|\xi\rang|^2\leq \|\eta\|\|\xi\|
\eea
which gives the operator norm
\bea
&&\|\pi_\phi(a)\|=\sup_{\xi\in \H_\phi}{\|\pi_\phi(a)\xi\|\over\|\xi\|},\nn\\
&&~~\Ra~~\tc{red}{\|\pi_\phi(a)\xi\|\leq \|\pi_\phi(a)\|\|\xi\|~~\forall \xi\in \H}
\eea
can be witten.

Thus one can define a representation
\bea
&&\pi_\phi:\A\ra B(\H_\phi),~~\pi_\phi(\A)\xi_\phi\simeq \H_\phi,~~\xi_\phi=1_\A+N_\phi(\A)
\eea
such as that provided by the left action \\$\pi_\phi(a)=L_a:~b+N_\phi(\A)\mapsto L_a(b+N_\phi(\A))=a(b+N_\phi(\A)) =ab+N_\phi(\A), $
\bea
&&\lang\xi_\phi|\pi_\phi(a)\xi_\phi\rang=\phi(a),\nn\\
\eea
where the boundedness of $L_a$ needs to be checked. From the definition of the operator norm
\bea
&&\|L_a\eta\|^2=\leq\|a\|^2\|\eta\|^2~~\Ra~~\|L_a\|=\sup_{\eta\in\H_\phi}{\|L_a\eta\|^2\over \|\eta\|^2}\leq \|a\|.
\eea

 The system $(\pi_\phi,\H_\phi,\xi_\phi)$,~~up to unitary isomorphisms, is unique due to cyclicity of the vector $\xi_\phi$. The unitary isomorphism with any new system $(\pi,\H,\xi)$ may be written as
\bea
&& U:\H_\phi\ra \H,~~\pi(a)=U\pi(a)U^\ast,~~\xi=U(\xi_\phi),\nn\\
&&\pi:\A\ra B(\H).
\eea

\section{Algebra Homomorphisms (Representations)}
Let $\pi:\A\ra \A',~~\pi(ab)=\pi(a)\pi(b),~~\pi^\ast(a)=\pi(a^\ast)$.
The expansion ~$\sigma(\pi(a))\subseteq\sigma(a)$ since $(\ld 1'-\pi(a))^{-1}~\nexists=(\ld \pi(1)-\pi(a))^{-1}~\nexists=\pi((\ld 1-a)^{-1})~\nexists  $ ~shows that if $\ld\in \sigma(\pi(a))$ then $\ld\in \sigma(a)$ also.

Therefore
\bea
&&\|\pi(a)\|^2=\|\pi^\ast(a)\pi(a)\|=\|\pi(a^\ast)\pi(a)\|=\|\pi(a^\ast a)\|=\rho(\pi(a^\ast a))\nn\\
&&~~~~\leq \rho(a^\ast a)\leq \|a^\ast a\|= \|a\|^2.\nn\\
&&~~\txt{ie.}~~\|\pi(a)\|\leq \|a\|.
\eea

\section{Geometry/algebra dictionary}
{\tiny
\begin{tabular}{|l|l|}
   GEOMETRY& ALGEBRA \\
  \hline
   points $X=\{x\}$ of a topological space & characters ${\X}=\{\ld:\F(X)\ra \mathbb{\mathbb{C}}\backslash\{0\}\}$ \\
   \hline
   group $X=(G,\circ),~\circ:G\times G\ra G=\{x\}$ & characters $({\X},\circ),~{\X}=\{\ld:\F(G)\ra \mathbb{\mathbb{C}}\backslash\{0\}\}$\\
   \hline
   complex functions~ $\F(G)=\{f:G\ra \mathbb{\mathbb{C}}\}$, &  $(\F(G),\Delta,\txt{pt-wise-conv}),~\Delta:\F(G)\ra \F(G)\ot\F(G)$  \\
   $f(x\circ x')=\lang f|x\circ x'\rang=\lang f|\circ(x, x')\rang=\lang \Delta(f)|(x,x')\rang$ &~~~~$\mu_\B:\B\ot\B\ra\B~~\forall\B$,\\
   ~~~~$=\sum_\al f_\al(x)f^\al(x')\eqv\mu_\mathbb{C}\sum_\al (f_\al\ot f^\al)(x\ot x') $& $\lang f g|x\circ x'\rang=\lang \Delta(f g)|(x,x')\rang=\lang \Delta(f)\Delta( g)|(x,x')\rang$ \\
   \hline
   complex functions $\F(X)=\{f:X\ra \mathbb{C}\}$ & function $\ast$-algebra $(\F(X),\txt{pt-wise})$ \\
   \hline
   map $m:X\ra Y$& $\ast$-homomorphism $h:\F(X)\ra \F(Y)$  \\
   \hline
   symmetry of $S:X\ra X$ & $\ast$-automorphism $U:\F(X)\ra \F(X) $  \\
   \hline
   direct product $X\times Y$& tensor product $ \A\ot\B,~\A=(\F(X),\txt{pt-wise})$,\\
     &~~~~~~~~$\B=(\F(Y),\txt{pt-wise})$ \\
     \hline
   probability measures & normalized positive linear functionals  \\
   \hline
   sections $\Gamma(E)=\{s:X\ra E\simeq X\times V\}$  & projective module $M(\A)$ over $\A=(\F(X),\txt{pt-wise})$,  \\
   ~~~~~~~~of a vector bundle $E$  over $X$. &~~~~~~~~$\A\times M(\A)\ra M(\A)$ \\
   \hline
   directed (Lie) differential $L_\xi:\Gamma(E)\ra \Gamma(E)$  & Liebnitz differential $D:M(\A)\ra M(\A),$  \\
   ~~~~~along a smooth vector field $\xi: X\ra V$  & $~D(mm')=D(m)m'+mD(m')$   \\
    $L_\xi=|_{\delta x\ra\xi}\circ\delta\eqv \delta|_{\delta x\ra\xi},~~\delta s(x)=s(x+\delta x)-s(x)$\\
   \hline
   differential forms $\Omega^n(X)=\{\omega_n:L(V\slash X)^n\ra \F(X)$.& $\Omega^n(\A)=\{\omega_n:\txt{Der}(\A)^n\ra \A ,~ \A=(\F(X),\txt{pt-wise})\}$, \\
   ~~~~$L(V\slash X)=\{L_\xi,~\xi:X\ra V\}$     & $\txt{Der}(A)=\{D:M(\A)\ra M(\A)\}$ \\
   \hline
   exterior differentials $d,d^\ast:\Omega^n(X)\ra \Omega^{n\pm 1}(X)$ & graded differentials. $d_g,d_g^\ast:\Omega^n(\A)\ra \Omega^{n\pm 1}(\A)$ \\
     & $d_g(mm')=d_g(m)~m'+\pi_g(m)~d_g(m')$,\\
     &$\pi:G\times M\ra M,~(g,m)\mapsto \pi(g,m)\eqv \pi_g(m)$,\\
     & $G=S_n,~~\Delta(d)=d\ot id+\pi\ot d$ \\
  \hline
\end{tabular}
}

\chapter{Sets and Physical Logic}
\section{Exclusive sets}
A (an exclusive) set $S$ is a selection or conditional collection of objects
 \bea
 S=S(X)=\{x\in X;~C_S(x)\}
 \eea
where ~$C_S:X\ra \{\txt{True},\txt{False},\txt{Unsure}\}\subset X,~x\mapsto C_S(x)$~   is a condition that $x\in X$ needs to satisfy in order to be a member of the set $S$. That is, ~$x\in S$~ iff ~$C_S(x)=\txt{True}$~ and ~$x\not\in S$~ iff ~$C_S(x)=\txt{False}$.
The collection $X$ can be arbitrary or not. We will simply write ``$F$'' for ``False'', ``$T$'' for ``True'' and ``$U$'' for ``Unsure''. The result ``Unsure'' is obtained whenever $C_S(x)$ neither evaluates to $T$ nor to $F$ due to whatever reasons all of which we will refer to as Uncertainty.

The \emph{\textbf{complement}} or \emph{\textbf{negation}} ${S^\sim}$ of the set $S$ is given by
\bea
{S^\sim}={S^\sim}(X)=\{x\in X;~{C^\sim}_S(x)\}.
\eea
where ${C^\sim}_S$ is the statement that evaluates to $F$ whenever $C_S$ evaluates to $T$ and vice versa. That is we write $F^\sim=T,~T^\sim=F,~U^\sim=U$.

In anticipation of situations where it can be much more difficult to determine when two sets are equal than to determine when one includes the other, one could introduce inclusion ~$\subseteq$~ where ~$A\subseteq B$~ iff ~$x\in A~~\Ra~~x\in B$. In terms of intersection and sum/union
\bea
&&A B=\{x\in A;~C_B(x)\}=\{x\in B;~C_A(x)\}\nn\\
&&~~~~= \{x\in X;~C_B(x)C_A(x)\},\nn\\
&&A+B= \{x\in X;~C_B(x)+C_A(x)\},\nn\\
&&A\cap B=AB,\nn\\
&&A\cup B=A+B+AB=\{x\in X;~C_B(x)+C_A(x)+C_B(x)C_A(x)\},\nn
\eea
where we have introduced point-wise multiplication/addition of conditions; ie. $(C_1C_2)(x)=C_1(x)C_2(x),~~(C_1+C_2)(x)=C_1(x)+C_2(x)$.

We have the following conditions
\bea
&& A\subseteq B~~\txt{iff}~~AB=A~~\txt{iff}~~A+B=B\nn\\
&&~~\txt{iff}~~C_A+C_B=C_B~~\txt{iff}~~C_AC_B=C_A.\nn\\
&&A=B~~\txt{iff}~~(A\subseteq A)(B\subseteq A).
\eea
When the condition of a set evaluates to either $F$ or $U$ for all $x\in X$ for example in
$S^\sim S=\{x\in X;~C^\sim_S(x)C_S(x)\}$ we leave the set blank and call it the empty set denoted $\{\}$.
\bea
&&S^\sim S=\{x\in X;~C^\sim_S(x)C_S(x)\}=\{x\in X;~C(x)=U\}\eqv\{\},\nn\\
&&S^\sim+ S=\{x\in X;~C^\sim_S(x)+C_S(x)\}=\{x\in X;~C(x)\neq U\},\nn\\
&&S^\sim\cap S=S^\sim S,~~~~S^\sim\cup S=S^\sim+ S.
\eea

\subsection{Conditional algebra}
All \emph{\textbf{statements}} are (composite) conditions involving \emph{\textbf{\textbf{implication}}} $C_1\Ra C_2$ and \emph{\textbf{equality}} $C_1\iff C_2$, \emph{\textbf{negations}} and so on. The operations such as implication, equality, negation and so on, may be written in terms of an algebra system on the set of conditions $\C$.

In order to compare, compose, decompose, ..., sets one needs to have a means to do similar manipulations on the set of conditions.
Let $$\C=\{C\in X;~C:X\ra \{T,F,U\}\}$$ be the set of conditions.
Then the value set $\{T,F,U\}$ behaves as follows:

 Boolean system:
\bea
&&TT=T,~~FT=F,~~FF=F,\nn\\
&&T+T=T,~~F+T=T,~~F+F=F.\nn\\
\eea
Now if $C_1(x)$ is True but $C_2(x)$ is Unsure then intersection is Unsure meanwhile sum or union is True; ie.
\bea
TU=U,~~T+U=T.
\eea
If $C_1(x)$ is False but $C_2(x)$ is Unsure then intersection is False meanwhile sum or union is Unsure; ie.
\bea
FU=F,~~F+U=U.
\eea
Finally if both $C_1(x),C_2(x)$ are Unsure then both intersection and sum/union are Unsure; ie.
\bea
&&UU=U,~~U+U=U.
\eea

Thus the summary of the operations is as follows:
\bea
&&TT=T,~~FT=F,~~FF=F,\nn\\
&&T+T=T,~~F+T=T,~~F+F=F.\nn\\
&&TU=U,~~T+U=T.\nn\\
&&FU=F,~~F+U=U.\nn\\
&&UU=U,~~U+U=U.\nn\\
&&F^\sim=T,~T^\sim=F,~U^\sim=U.
\eea
One may write the algebra system as $\A=(\{T,F,U\},+,\cdot,\sim)$ where multiplication $\cdot$ may be thought of as the $\sim$-conjugate $+^\sim$ of addition $+$ since \\
$(a+b)^\sim=a^\sim b^\sim,~~(ab)^\sim=a^\sim+b^\sim ~~\forall a,b\in \A$. This special property  will be lost when an arbitrary $\sim$-algebra system is considered. One also has the ``exclusive or'' operation $\op$
\bea
a\op b=ab^\sim+a^\sim b,~~(a\op b)^\sim=ab+a^\sim b^\sim,~~~~a,b\in \A.
\eea

Thus \emph{\textbf{\textbf{implication}}} $C_1\Ra C_2$ is equivalent to $C_1C_2=C_2$~ or~ \\
$(C_1C_2)(x)=C_2(x)~~\forall x$ and \emph{\textbf{equality}}   $C_1\iff C_2$ ~is equivalent to~ $C_1=C_2$ ~or~ $C_1(x)=C_2(x)~~\forall x$.

\emph{\textbf{The algebra of sets has now been reduced to the algebra of the corresponding set generation conditions.}}

\subsection{Maps and bundling}
Given two sets $A,B$ one can form another set $C=A\times B$ by pairing elements thus
\bea
C=\{c;~c=(a,b),~a\in A,~b\in B\}\eqv A\times B=\{(a,b);~a\in A,~b\in B\}.\nn
\eea
This operations can be iterated to form ~$A_1\times A_2\times A_3\times...$~ given $A_1,~A_2,~A_3,~...$
In general one can form
\bea
\label{excl-prod}A\star B=\{x\in X;~C_\star(x,C_A,C_B)\},
\eea
where $C_\star(x,C_A,C_B)$ can for example consist of the sequence of conditions\\
$x=(a,b),~a\in A,~b\in B$ ~~or ~~ $x=(a,b),~C_A(a)C_B(b)$  corresponding to the direct product $A\times B$.
That is, we have the conditions~\\
 $C_\star(x,C_A,C_B)\mapsto  x=(a,b),~C_A(a)C_B(b)$ ~iff~ $A\star B\mapsto AB$.
We can also have the conditions $C_\star(x,C_A,C_B)\mapsto C_A(x)C_B(x)$ ~iff~ $A\star B\mapsto AB$
and similarly for the sum we have  $C_\star(x,C_A,C_B)\mapsto C_A(x)+C_B(x)$ ~iff~ $A\star B\mapsto A+B$. This general product can be iterated as well.

If $M(\A)=\{m\in X;~m:\A\ra X\}$ is the space of maps on $\A$ and \\
$P(\A)=\{A\in X;~A\subseteq \A\}$ is the set of subsets of $\A$ then one can define a bundle twisting map
\bea
&&[~]:\A\times M(\A)\ra P(\A),~(a,m)\mapsto [a]_m=\{b\in \A;~m(a)=m(b)\}\nn
\eea
which makes a twisted bundle
\bea
&&[~](\A\times M(\A))=[\A]_{M(\A)}\simeq\A\times [~]_{M(\A)}\eqv [\A]\times M(\A),\nn\\
&&[~]_{M(\A)}:\A\ra (P(\A))^{|M(\A)|},~a\mapsto [a]_{M(\A)},\nn\\
&&[~]_{m}:\A\ra P(\A),~a\mapsto [a]_{m}~~\forall m\in M(\A),\nn\\
&&[\A]:M(\A)\ra (P(\A))^{|\A|},~~m\mapsto [\A]_m,\nn\\
&&[a]:M(\A)\ra (P(\A)),~~m\mapsto [a]_m~~\forall a\in \A
\eea
from the trivial bundle $\A\times M(\A)$.

In general,
\bea
&&m:A\times B\ra C,~(a,b)\mapsto m(a,b)=c,\nn\\
&&A\times B\sr{m}{\ral} C,~(a,b)\sr{m}{\longmapsto} m(a,b)=c,\nn\\
&&m:A\times B|_{(a,b)}\mapsto C|_{c=m(a,b)},\nn\\
&&A\times B|_{(a,b)}\sr{m}{\ral} C|_{c=m(a,b)}\nn
\eea
may be written as
\bea
&&m:A\ra M(B),~a\mapsto m(a,~):B\ra C,~b\mapsto m(a,~)(b)= m(a,b)=c,\nn\\
&&m:B\ra M(A),~b\mapsto m(~,b):A\ra C,~a\mapsto m(~,b)(a)= m(a,b)=c,\nn\\
&&A|_{a}\sr{m}{\ral}M(B)|_{m(a,~)}\sr{B|_{b}}{\ral} C|_{c=m(a,b)},\nn\\
&&B|_{b}\sr{m}{\ral}M(A)|_{m(~,b)}\sr{A|_{a}}{\ral} C|_{c=m(a,b)},\nn\\
&&M(A\times B)|_{m=m(~,~)}\sr{A|_a}{\ral}M(B)|_{m(a,~)}\sr{B|_b}{\ral} C|_{c=m(a,b)}.\nn
\eea

In bundle form
\bea
&&m(A\times B)\simeq A\times m^B\eqv m_A\times B,\nn\\
&&m^B\subseteq M(B\ra C)\eqv C\slash B\eqv M(B,C)\subset M(B),\nn\\
&&m_A\subseteq M(A\ra C)\eqv C\slash A\eqv M(A,C)\subset M(A),\nn\\
&&m^B:A\ra C^{|B|},~~~~m_A:B\ra C^{|A|},\nn\\
&&m^b:A\ra C,~a\mapsto m(a,b)~~~~\forall b\in B,\nn\\
&&m_a:B\ra C,~b\mapsto m(a,b)~~~~\forall a\in A,\nn
\eea
where $|A|,|B|$ are the number of elements in $A,B$ respectively.

\subsection{Counting isomorphisms ?}
If $|A|$ denotes the number of elements in the set $A$ then the number $|I(A,B)|$ of isomorphic maps $I(A,B)\subseteq \F(A,B)=\{m;~m:A\ra {\D}\subseteq B,~|{\D}|=|A|\}$~~  is
\bea
&&|I(A,B)|={\Gamma(|B|)\over \Gamma(|A|)\Gamma(|B|-|A|)}|I(A,{\D})|={\Gamma(|B|)\over \Gamma(|B|-|A|)},\nn\\
&&~~~~~~\Gamma(n)=n\Gamma(n-1),~~\Gamma(1)=1,\nn\\
&&|I(A,{\D})|=|I(A,A)|=|({\D},{\D})|=\Gamma(|A|).
\eea

\section{Nonexclusive sets: Generalizations}
The sets we have defined so far have absolute or rigid rules for choosing their members and thus we can only have members and nonmembers. However in practice there can be intermediate situations with different levels or steps of membership. Therefore we will consider sets for which the set generation conditions (sgc's) can take values in an arbitrary $\ast$-algebra system\footnote{Other examples of algebra systems include natural numbers $\mathbb{N}$ (which arose due to the need to count things), fractional numbers $\mathbb{Q}$ (which arose due to the need to compare countings) and real or continuous numbers $\mathbb{R}$ (which arose due to the need to compare uncountable characteristics such as lengths). Various products of these number systems (or number sets) also arose due to the need to compare (geometric) shapes and sizes of things.} $\A$.

The operations of multiplication and addition will simply parallel those of the $\ast$-algebra system $\A$. Note that the $\ast$-algebra system $\A$ may \emph{\textbf{neither}} be \emph{\textbf{commutative nor associative in general}} and the sets will directly inherit these properties as well. However we will assume associativity, but not commutativity, for simplicity.

A set $S$ is a selection or conditional collection of objects
 \bea
 S=S(X)=\{x\in X;~C_S(x)\}
 \eea
where ~$C_S:X\ra \A\subset X,~x\mapsto C_S(x)$~ is a condition that determines the degree (or probability amplitude) of membership in $S$ of each and every $x\in X$.

\subsection{G1}
In one means of generalization we suppose that ~$x\in S$~ with degree or amplitude of membership (aom) $a\in \A$ iff ~$C_S(x)=a$~ and ~$x\not\in S$~ with degree or amplitude of nonmembership (aon) $a^\sim\in \A$ iff ~$C_S(x)=a^\sim$. The collection $X$ can be arbitrary.

Each $a\in \A$ corresponds to a selection of elements $[a]^S=\{x\in X;~C_S(x)=a\}$ so that $S=\bigcup_{a\in\A}[a]^S$. Whether $\A$ is represented as an algebra of operators on a Hilbert space, ie. $\A\ra O(\H)$, or not one may use the characters $\X(\A)=\{\ld\in \A^\ast;~\ld:\A\ra \mathbb{C}\backslash \{0\}\}$ to measure the degrees or amplitude of membership (aom) or nonmembership (aon) carried by each $a\in \A$. The uncertain elements which are those with the property $a^\sim= a$ have a degree of uncertainty or unsureness of membership and their values may be conveniently measured with the help of real linear functionals $\A^\ast_R=\{\phi\in \A^\ast;~a^\sim=a~~\Ra~~\phi(a)\in \mathbb{R}\}$. [Note that in the Boolean algebra system $\A=\{T,F\}$ the elements $a=T,F$ obey $a^2=a$ and so the characters are given by~ $\ld(a^2)=\ld(a)^2=\ld(a)~~\Ra~~\ld(a)=0,1$~ and one usually chooses $\ld_1(T)=1,~\ld_1(F)=0$ although the only other alternative choice $\ld_2(T)=0,~\ld_2(F)=1$ is equally valid.]

Observe that for a real algebra system where $a^\sim=a~~\forall a\in \A$ membership of a set is completely determined by the degree or amplitude of unsureness (aou) of membership.

The \emph{\textbf{complement}} or \emph{\textbf{negation}} ${S^\sim}$ of the set $S$ is given by
\bea
{S^\sim}={S^\sim}(X)=\{x\in X;~{C^\sim}_S(x)\}.
\eea
where ${C^\sim}_S$ is the statement that evaluates to $a^\sim$ whenever $C_S$ evaluates to $a$ and vice versa. That is we have $a^{\sim\sim}=a$. One should not confuse the logic operation $\sim$ with the $\ast$ operation with properties
\bea
&&a^{\ast\ast}=a,~~(ab)^\ast=b^\ast a^\ast,~~(a+b)^\ast=a^\ast+b^\ast.
\eea
[{\footnotesize A $\sim$-algebra system $\A$ with the properties $$(a+b)^\sim=a^\sim b^\sim,~~(ab)^\sim=a^\sim+b^\sim ~~\forall a,b\in \A$$
such as the \emph{\textbf{commutant operation}} in \textbf{set commutant algebra}, and similar types of analysis, is closer to that of exclusive set theory. A set $\S=\{A\}$ consisting of Von Neumann algebras for example has such properties:
\bea
&& A''=A,~~(A\cap B)'=A'\cup B',~~(A\cup B)'=A'\cap B'.~~]
\eea
}

The set operations are as before given by
\bea
&&A B=\{x\in A;~C_B(x)\}=\{x\in B;~C_A(x)\}\nn\\
&&~~~~= \{x\in X;~C_B(x)C_A(x)\},\nn\\
&&A+B= \{x\in X;~C_B(x)+C_A(x)\},\nn\\
&&A\cap B=AB,\nn\\
&&A\cup B=A+B+AB=\{x\in X;~C_B(x)+C_A(x)+C_B(x)C_A(x)\},\nn
\eea
where the point-wise multiplication/addition of conditions, \\
~$(C_1C_2)(x)=C_1(x)C_2(x),~~(C_1+C_2)(x)=C_1(x)+C_2(x)$,~
is used but this time ~$(C_1C_2)(x)\neq (C_2C_1)(x)$~ in general.

\subsection{G2}
Here we maintain that in ~$S=\{x\in X;~C_S(x)\}$~ each and every $x\in X$ is a member of $S$ with degree or amplitude of membership (dom or aom) $C_S(x)$ and degree or amplitude of nonmembership (don or aon)  $C^\sim_S(x)$. That is, \emph{\textbf{there is no ``sharp'' distinction between ``members'' and ``nonmembers''}}. There will be uncertainty, with degree or amplitude of uncertainty (dou or aou) $C_S(x)$, in the membership of $x$ if $C^\sim_S(x)=C_S(x)$ even when $C^\sim\neq C$ on all of $X$.

To illustrate, let $C_S\in \A$ where $\A$ is an arbitrary $\ast$-algebra and \\
$X=\A^\ast_{1+}\subseteq A^\ast_+\subseteq A^\ast$ be the set of normalized positive linear functionals (nplf's) of $\A$. Then every element $a\in\A$ represents a set generation condition (sgc) for an associated set ~$S_a=\{\phi\in \A^\ast_{1+};~\phi(a)\}$~ and each $\phi\in \A^\ast_{1+}$ is a member of $S_a$ with aom $\phi(a)$ and aon $\phi(a^\sim)$. To any such $\phi$  satisfying $\phi(a^\sim)=\phi(a)$ even when $a^\sim\neq a$ we rather associate an aou $\phi(a)$.

It is important to mention that the set $S_a$ actually corresponds to an equivalence class $[a]=\{b\in\A;~\phi(b)=\phi(a)~~\forall\phi\in \A^\ast_{1+}\}$ of sgc's since every member of $[a]$ generates exactly the same set. That is $S_a\eqv S_{[a]}$.

Set addition and multiplication are straightforward and given by
\bea
&&S_aS_b=\{\phi\in \A^\ast_{1+};~\phi(ab)\}=S_{ab}\eqv S_a\cap S_b,\nn\\
&&S_a+S_b=\{\phi\in \A^\ast_{1+};~\phi(a+b)\}=S_{a+b},\nn\\
&&S_a\cup S_b=\{\phi\in \A^\ast_{1+};~\phi(a+b+ab)\}=S_{a+b+ab}
\eea
and the $\sim$-complement or conjugate $S^\sim$ of $S$ is given by
\bea
S^\sim_a=S_{a^\sim}=\{\phi\in \A^\ast_{1+};~\phi(a^\sim)\}.
\eea

Regarding set inclusion, we would like that a set includes itself in which case we must have $S_aS_a=S_{a^2}=S_a$. Thus a further condition for $a\in \A$ to be a \emph{\textbf{``pure set''}} generation condition (psgc) is for it to be a projector~ $a^2=a$. Consequently one has pure and impure sets. If one has a collection of projectors $\P=\{p\in \A;~p^2=p\}$ such that ~$p_1p_2\in \P~~\forall p_1,p_2\in \P$,~that is $(p_1p_2)^2=p_1p_2$ so that the product of any two sets gives another set, then one has a closed
    \footnote{One notes that for any given projector $p\in\A$,~ $vpu$ is another projector $\forall u,v\in \A$ such that $uv=1$. One also has partial projectors: if $p$ is a projector then $\forall u\in \A$,~ $q$ in the relation $pu=uq$,~ $q=q_{(u,v)}$ is a partial projector. Thus corresponding to any projector $p$ is the class of projectors ~$\P_p(\A)=\{vpu;~u,v\in \A,~uv=1\}$~ noting that given $\{u_1,u_1,...,u_n\}\subset\A$ and $\{\td{u}_1,\td{u}_2,...,\td{u}_n\}\subset\A$ such that ~$u_i\td{u}_i=1~~\forall i$~ one has ~$U_nV_n=1$~ where ~$U_n=\prod_{j=1}^n u_j,~~V_n=\prod_{i=n}^1\td{u}_i$.

Also, given a projector $p$, $\td{p}p$ is a projector for any $\td{p}\in p'=\{a\in\A;~[a,p]=0\},$ the commutant of $p$.

For any given $a\in \A$, $p^L_a=aa_L^{-1},~~p^R_a=a_R^{-1}a$~~ are projectors, where $a_L^{-1}a=1_\A=aa_R^{-1}$. Also $a(1_\A-p^R_a)=0= (1_\A-p^L_a)a$.

Given $\phi\in \A^\ast_+$ a system of projectors ~$P=\{p_i\in\A;~p_i^2=p_i~\forall i\}$~ is \emph{\textbf{right $\phi$-measurable}} iff $\phi(a)=\sum_i\phi(ap_i)~~\forall a\in \A$. One may refer to an orthogonal system of projectors \\ $P_1=\{p_i;~p_ip_j=\delta_{ij}p_j~\forall i,j\}$ as a \emph{\textbf{partition}}. In a complete system of projector\\ ~$P=\{p_i\in\A;~p_i^2=p_i\}$~ any given $a\in\A$ may be expanded as
\bea
&&a=\al+\al^ip_i+\al^{ij}p_ip_j+...+\al^{i_1...i_k}p_{i_1}...p_{i_k}+...=\sum_{k}\al^{i_1...i_k}p_{i_1}...p_{i_k},\nn\\
&&~~\al^{i_1...i_k}\in \mathbb{C}.
\eea
One may orthogonalize a given system of projections $\Pi=\{{\pi}_i\in\A;~{\pi}_i^2={\pi}_i,~{\pi}^\ast_i={\pi}_i~\forall i\}$ when $\A$ is represented on a Hilbert space $\H,~~\A\ra O(\H)\simeq \H\ot\H$~ where one can write~ \\ ${\pi}_i={|\xi_i\rang\lang\xi_i|\over\lang\xi_i|\xi_i\rang}$~ and the set $\{|\xi_i\rang\}$ can then be orthogonalized. Corresponding to $\{|\xi_i\rang\}$ is the dual set $\{|\xi^\ast_i\rang=|\xi_j\rang \lang\xi_j|\xi_i\rang^{-1[j,i]}\}$,~ with ~$\lang\xi_i|\xi^\ast_j\rang=\delta_{ij}$, ~from which one obtains the orthogonal set ~$\{|\hat{\xi}_i\rang=|\xi_j\rang \lang\xi_j|\xi_i\rang^{-{1\over 2}[j,i]}\}$~ with $\lang \hat{\xi}_i|\hat{\xi}_j\rang=\delta_{ij}$. Hence $\hat{{\pi}}_i=|\hat{\xi}_i\rang\lang\hat{\xi}_i|$ will satisfy $\hat{{\pi}}_i\hat{{\pi}}_j=\delta_{ij}\hat{{\pi}}_j$.
Similarly for projectors written as $p_i={|\xi_i\rang\lang\eta_i|\over\lang\eta_i|\xi_i\rang}$ corresponds the orthogonal system\\ $\hat{p}_i=|\xi_k\rang \lang\eta_k|\xi_i\rang^{-{1\over 2}[k,i]} \lang\eta_i|\xi_l\rang^{-{1\over 2}[i,l]} \lang\eta_l|,~~\hat{p}_i\hat{p}_j=\delta_{ij}\hat{p}$. Summation convention is used and the inverse (and square root) is partial in that they are of the matrix in the index types ~$i,j,k,...$~ only.

The representation of the projectors $p_i$ in terms of subsets of the Hilbert space $\H$ will take the general form
\bea
&&p_i=\sum_{(\eta,\xi)\in\H_1\times \H_2}|\xi_i\rang \lang\eta_i|\xi_i\rang^{-1[\eta,\xi]}\lang\eta_i|,~~\H_1,\H_2\subseteq \H,\nn\\
&&\hat{p}_i=\sum_{(\eta,\xi)\in\H_1\times \H_2}|\xi_k\rang \lang\eta_k|\xi_i\rang^{-{1\over 2}[k,i][\eta,\xi]} \lang\eta_i|\xi_l\rang^{-{1\over 2}[i,l][\eta,\xi]} \lang\eta_l|,\nn\\
&&~~~~\eqv \sum_{(\eta,\xi)\in\H_1\times \H_2}|\xi_k\rang \lang\eta_k|\xi_i\rang^{-{1\over 2}} \lang\eta_i|\xi_l\rang^{-{1\over 2}} \lang\eta_l|,~~~~\hat{p}_i\hat{p}_j=\delta_{ij}\hat{p}
\eea
where $[\eta,\xi]$ simply indicates a partial inverse which is that of a $|\H_1|\times|\H_2|$ matrix, $|\H_1|$ being the size of  $\H_1$ and $[i,j]$ has a similar meaning meanwhile $[i,j][\eta,\xi]$ indicates a full inverse where both index types are involved. The case $\H_1=\H_2$ corresponds to projections or equivalently $\H_1=\H_2^\ast,~~\lang\xi_i|\eta_i\rang=\delta(\eta-\xi)~\lang\xi_i|\xi_i\rang$. Thus \emph{\textbf{projectors correspond to subspaces of $\H^2=\H\times\H$ meanwhile projections (real projectors) correspond to subspaces of the Hilbert space $\H$}}. \emph{\textbf{The sum of any number of orthogonal projectors is also a projector}}. An orthogonal system of projectors $\{\hat{p}_i\}$ spans a commutative algebra with elements \\ $c=\sum_i\td{c}~\hat{p}_i,~~\Tr(\hat{p}_i)=1~~\Ra~~\td{c}_i=\Tr(c\hat{p}_i)$.

Corresponding to each projector $p$ with additional property $\Tr(pp^\ast)=1$ one can define a state $\phi_p$ given by
\bea
&&\Tr(a)=\sum_{\xi\in\H}{\lang\xi|a|\xi\rang\over\lang\xi|\xi\rang},~~~~\phi_p(a)=\Tr(pap^\ast),\nn\\
&&\Tr(|u\rang\lang v|)=\sum_{\xi\in\H}{\lang\xi|u\rang\lang v|\xi\rang\over\lang\xi|\xi\rang}=\sum_{\xi\in\H}{\lang v|\xi\rang \lang\xi|u\rang\over\lang\xi|\xi\rang}=\lang v|\sum_{\xi\in\H}{|\xi\rang \lang\xi|\over\lang\xi|\xi\rang}|u\rang=\lang v|u\rangle,\nn\\
&&\Tr_p([a,b]_p)=0,~~where~~ [a,b]_p:=apb-bpa,~~\Tr_p(a)=\Tr(pa).
\eea
A projector $p$ can be written as a sum ~$p=\pi_1+\pi_2,~~\pi_1^\ast=\pi_1,~\pi_2^\ast=-\pi_2$~ of a hermitian and an antihermitian operator with the following properties
\bea
&&\pi_1^2=\pi_1-\pi_2^2,~~\pi_1\pi_2+\pi_2\pi_1=\pi_2\nn\\
&&~~~~\Ra~~\pi_2\pi_1\pi_2=\pi_1(1-\pi_1)^2.
\eea
 A tensor product $p_1\ot p_2\ot ...$ of projectors $p_1,p_2,...$ is also a projector. In general one may form a $\Delta$-deformed tensor product (compare with \ref{excl-prod}) of two sets as
 \bea
 &&S_a\ot_\Delta S_b=\{\phi\in \A^\ast_{1+};~\pi_2\Delta(\phi)(a\ot b)\}\eqv S_{a\ot_\Delta b},~~\Delta:(\A^\ast)^n\ra \bigoplus_{k=0}^N(\A^\ast)^{\ot k},~~n,N\in \mathbb{N},\nn\\
 &&(id\ot\pi_2\Delta)\circ\pi_2\Delta=(\pi_2\Delta\ot id)\circ\pi_2\Delta~~\txt{(may not necessarily hold in general)},\nn\\
 &&\pi_2\Delta(\phi)(a\ot b)=\lang \phi_\al\ot\phi^\al|a\ot b \rang=\phi_\al(a)\phi^\al(b).\nn\\
 &&\pi_n\Delta=(~(id\ot)^k~\pi_2\Delta~(\ot id)^{n-k-1}~)\circ \pi_{n-1}\Delta^{n-1}~~\forall 1\leq k\leq n-1,\nn\\
 &&\pi_3\Delta(\phi)=(id\ot\pi_2\Delta)\circ\pi_2\Delta(\phi)=(id\ot\pi_2\Delta)(\phi_\al\ot\phi^\al)=\phi_\al\ot\pi_2\Delta(\phi^\al)\nn\\
 &&~~~~=\phi_\al\ot(\phi^\al)_\beta\ot (\phi^\al)^\beta.\nn\\
 &&\pi_1\Delta(\phi)(a)=\phi(a),~~~~\pi_3\Delta(\phi)(a\ot b\ot c)=\phi_\al(a)~(\phi^\al)_\beta(b)~(\phi^\al)^\beta(c),~~...
 \eea
 The tensor product that corresponds (ie. is dual) to the product in $\A$ is a particular case $\Delta_1$ of $\Delta$ defined by
 ~$\phi(ab)=\lang\phi|ab\rang=\lang\pi_2\Delta_1(\phi)|a\ot b\rang\eqv \pi_2\Delta_1(\phi)(a\ot b).$~
 On the other end the $\Delta$ that corresponds to the usual (undeformed) tensor product $\ot$ is given by \\
 ~$\pi_2\Delta_0(\phi)(a\ot b) =\lang\pi_2\Delta_0(\phi)|a\ot b\rang=\lang\phi\ot\phi|a\ot b\rang=\phi(a)\phi(b).$ Thus possible $\Delta$'s interpolate between $\Delta_0$ and $\Delta_1\eqv \Delta_0^\ast$. The actual $\Delta$'s to be considered may be determined by the way one or more physical systems behave (or evolve) relative to (ie. interact or correlate with) one another. $S_a$ may be interpreted as the amplitude distribution or ``painting'' in $\A^\ast$ of the system represented by $a\in\A$.
 }
 system of sets.

 In a commutative algebra where $\phi(ab)=\phi(a)\phi(b)$, one has that for a projector $p$, $\phi(p^2)=\phi(p)\phi(p)=\phi(p)~~\Ra~~\phi(p)=0,1~~\forall \phi$. That is, \emph{\textbf{membership is of the exclusive type for a projector (projective sgc) in a commutative algebra}} $\A_0$. Thus projectors indeed generalize sgc's from exclusive (ie. commutative) to nonexclusive (ie. noncommutative) logic. The generalization of the logic operation $\sim$ is ~$p^\sim=1_\A-p$.

\emph{\textbf{Addition/union of sets is possible but not essential}} since every set contains the same elements without any exclusions and two sets can only differ in the aom, aon or aou of individual elements. That is, in this sense, all sets are already united.

We may now say that $A$ is a left (right) or two-sided subset of $B$ ~iff~ $AB$ ($BA$) is more related to $A$ than it is to $B$ or any other set. ie.
\bea
AB\simeq A~~(BA\simeq A)~~\txt{or}~~AB\simeq BA\simeq A.
\eea
In particular for each given projector $p\in\A$ which is a sgc for $S_p$ any other projector $p'\in\A$ such that $p'p=p'$ or $pp'=p'$ or $p'p=pp'=p'$ generates a subset $S_{p'}$ of $S_p$ with  ~$S_{p'}S_{p}=S_{p'}$~ or ~$S_{p}S_{p'}=S_{p'}$~ or ~$S_{p'}S_{p}=S_{p}S_{p'} =S_{p'}$ respectively.

Since every set now has the same members one may introduce a measure on the sets and compare their sizes as for example
\bea
\mu_1(S_a)=\sqrt{\sum_{\phi\in \A^\ast_{1+}}|\phi(a)|^2},~~\mu_2(S_a)=\max_{\phi\in \A^\ast_{1+}}|\phi(a)|,~...
\eea

We will define a \emph{\textbf{family of open sets}} to be one in which the intersection (ie. product) of any number of open sets is another open set. Since summation/union is not necessary so is the concept of a cover for a space $\S$ unnecessary.
The existence of one or more ``closed'' or ``complete'' systems of projectors (ie. the existence of one or more families of open sets) in $\A$ as described above is sufficient to account for results that could require completeness/compactness in terms of covers. Any closed collection of projectors \\
~$\P=\{a\in \A;~a^2=a\},~\P\P=\{ab;~a,b\in \P\}=\P$~ generates a family of open sets which may be considered to define a topology on ~$\A^\ast$~(one can choose to work with the whole set of linear functionals). Thus the number of such $\P$ collections will give the number of possible topologies available for one to work with. Due to the duality between $\A$ and $\A^\ast$ any given topology on $\A^\ast$ automatically induces an equivalent topology on $\A$.

\section{Physics: The logic of quantum theory}
At any given time, the \emph{conditional presence or state} $S_a$ of a physical system in $\A^\ast$ is determined (or generated) by the \emph{creation or existence condition} $a\in \A$ of the physical system. That is \emph{\textbf{a physical system, with conditional presence or state $S_a$ in $\A^\ast$, is defined (by a community of physical observers) by specifying a creation or existence condition $a\in\A$ for the physical system}} .

We will consider the set generating projectors $p\in\A$ to represent creation or existence conditions of actual physical systems living or operating in the space $\A^\ast$ and each closed collection of projectors $\P$ will represent a collection of basic or elementary physical systems [eps's] (where the systems are basic or elementary in that the product of any two of them gives another).  The set $S_p$, or equivalently $\phi(p),~~\forall \phi\in \A^\ast$, determines the amplitude distribution, at a given time, of the elementary physical system (eps) represented by $p\in\A$. That is $S_a$ is interpreted as the (probability) amplitude distribution or ``painting'' in $\A^\ast$ of the system represented by $a\in\A$.

As time progresses the eps can change $p=p(t)$ and thus its amplitude distribution $S_{p(t)}$ changes and maps out a ``path'' (time parametrized set of amplitude distributions) in the space $\A^\ast$. For $p(t)$ to remain a projector (ie. for system to remain an eps) during the time evolution the time evolution needs to be in the form ~$p(t)=U(t,t_0)p(t_0)U^{-1}(t,t_0),~~U(t,t)=1_\A~~\forall t$~(More generally $p(t)=U(t,t_0)p(t_0)V(t,t_0),~V(t,t_0)U(t,t_0)\in Z(\A)=\A\cap\A'$).~ Moreover, for the product ~$p_1(t)p_2(t)$~ of any ~$p_1(t),p_2(t)\in \P$~ to also remain in $\P$ ~(ie. $ (p_1(t)p_2(t))^2=p_1(t)p_2(t)$) the time evolution $U(t,t_0)$ must be common to all elements of $\P$ (ie. for all eps's).
 An infinitesimal time evolution, for such a \emph{\textbf{pure or elementarity preserving time evolution}}, may be effected using a directional derivation along a hermitian variable $h(t)\in \A,~~h(t)^\ast=h(t)$~ which generates unitary time evolution; ie. with~ $U^\ast(t,t_0)=U^{-1}(t,t_0)$.
\bea
&&[{d\over dt},p(t)]a(t)=-i[h(t),p(t)]a(t)~~\forall~ a:\mathbb{R}\ra \A,\nn\\
&&{dp(t)\over dt}=-i[h(t),p(t)]=-iD_h~ p(t),\nn\\
&&{d\phi(p(t))\over dt}=-i\phi([h(t),p(t)]).
\eea
Even though these equations were derived by considering projective classes, nonprojective solutions may be possible and all possible solutions can be physically significant as \emph{\textbf{any given solution either describes pure time evolution or describes impure time evolution}}.

Although the actual eps is described by $p(t)$, different observers experimenting on the eps may use different methods and/or parameters (or coordinates) to construct or represent $p(t)$ and $h(t)$. In addition measurements are carried out during experiments and the measurement parameters are the functionals $\phi\in \A^\ast$ and consequently \emph{\textbf{different observers may also use different functionals}}.

For a particular observer, if we imagine the projector $p(t)$ and $h(t)$ to be constructed from \emph{\textbf{auxiliary variables}} $q(t),~~q:\mathbb{R}\ra \A^N=\A\times \A^{N-1}$, which we will refer to as coordinates,
$p(t)=P(t,q),~~h(t)=H(t,q)$ then we have
\bea
&&{\del P(t,q)\over \del t}=-iD_H~ P(t,q)=-i[H(t,q),P(t,q)]~~\forall P~~\Ra\nn\\
&&{dq^i(t)\over dt}=-iD_H~ q^i(t)=-i[H(t,q),q^i(t)],~~i=1,2,...,N.
\eea
It is important to realize that there can be more than one choice of the variables $q^i(t)$, say $q^i_1(t)$ and $q^i_2(t)$ as well as the choice of functionals, say $\phi_1$ and $\phi_2$, that give the same projector $P(t,q)$ and same $H(t,q)$. The transformation \\ $q^i_1(t)\ra q^i_2(t),~\phi_1\ra \phi_2$~ is a symmetry of $P(t,q)$ and $H(t,q)$, or simply a symmetry of the eps that $p$ represents. The center $\Z(G_\ast)=G_\ast\cap G'_\ast$ of the algebra $G_\ast$ of the symmetry group $G\subset \A$ of the transformations commutes with all of $G_\ast\subset \A$. However, when two transformations commute \emph{\textbf{they share the same spectrum}} and are therefore equivalent in a sense. For this reason, the spectrum of the center $Z(G_\ast)$ \emph{\textbf{represents properties that are shared by all}} of $G_\ast$ and hence by all observers and in particular $Z(G_\ast)$ may therefore be considered to be intimately related to the most important (ie. basic or elementary) physical (ie. observer independent) characteristics of the eps. The spectrum of $Z(G_\ast)$ (ie. its spectral orbit in $\A^\ast$) can be used to predict, including yet unobserved, basic or elementary characteristics which the eps will eventually display under suitable conditions and which each and every observer will be able to detect even with their different coordinate or parameter systems.

From the point of view of the community of observers, \emph{\textbf{specifying an eps is equivalent to specifying its elementary physical properties}} (epp's). Hence elementary time evolution (ete) of the eps must also preserve any symmetry group (or equivalently any symmetry group of the eps should preserve the ete of the eps) in order that the epp's be maintained.

Possible conditions that can be imposed by physical observations on the variables ~$q^i(t)$~ include
\bea
&&(1)~[q^i(t),q^j(t)]=i\Omega^{ij}\in Z(\A)=\A\cap\A',\nn\\
&&(2)~[q^i(t),q^j(t)]=C^{ij}{}_kq^k(t),~~C^{ij}{}_k\in Z(\A)\nn\\
&&~~~~etc.
\eea
One notes that it is also possible to have \emph{\textbf{more general impure time evolution during which a physical system can tunnel from one pure sgc class to a different pure sgc class in a dynamically projective manner}}. That is, intermediate stages of time evolution involve impure sgc's (ie. nonprojective sgc's). Thus \emph{\textbf{different pure sgc classes may be associated with inequivalent physical vacua}}. The form of the infinitesimal time evolution equation in this case can be more general (\emph{\textbf{nonlinear}}) than the simple (\emph{\textbf{linear}}) form  considered so far.
To see how, consider the \emph{\textbf{linear ansatz}}
\bea
&&\dot{p}=hp+ph_1,~~p^2=p~~\Ra~~p\dot{p}+\dot{p}p=\dot{p}~~(\Ra~p\dot{p}p=0)\nn\\
&&~~\Ra~~php+ph_1p=0~~\Ra~~h_1=-h.
\eea
[~Note: These results show that ~$p+\dot{p}p$~ and ~$p+p\dot{p}$~ are also projectors for any given projector $p$. One notes also that ~$p^2=p~~\Ra~~p\dot{p}+\dot{p}p=\dot{p}$~ but ~$p\dot{p}+\dot{p}p=\dot{p}~~\not\Ra~~p^2=p$~ and so we will simply consider the operators obeying ~$p\dot{p}+\dot{p}p=\dot{p}$~ as a dynamical generalization of those obeying $p^2=p$ and refer to them as \emph{\textbf{dynamical projectors}}.]

The \emph{\textbf{nonlinear ansatz}}  ~$\dot{p}=h_1p-h_2p+ph_3p$~ implies~ $p(h_1-h_2+h_3)p=0$ and so
\bea
\label{projector-evol}&&{d p(t)\over dt}=h_1(t)p(t)-p(t)h_2(t)+p(t)(h_2(t)-h_1(t))p(t)\\
&&~~~~=[h_1,p]+p(-\delta h_1+\delta h_1~p),~~\delta h_1=h_2-h_1,\nn\\
&&{d p_1\over dt}=-p_1V_1+p_1V_1p_1,~V_1=U_1^{-1}\delta h_1U_1,~p_1=U_1^{-1}pU_1,~U_1=T(e^{\int^th_1}),\nn\\
&&{d p_2\over dt}=V_2p_2-p_2V_2p_2,~~~~{d \T_{12}\over dt}=\T_{12}V_{21}\T_{12},\\
&&\T_{12}=U_1^{-1}pU_2,~~V_{21}=U_2^{-1}(h_2-h_1)U_1,~~U_1=T(e^{\int^th_1}),~~U_2=T(e^{\int^th_2})\nn
\eea
\emph{\textbf{will describe dynamically projective impure time evolution}} of $p$. [One notes once more that \emph{\textbf{dynamically nonprojective solutions}} (which are of course impure) \emph{\textbf{are possible}}.]

Thus
\bea
&&{d [p]_\al\over dt}=V_{0\al}~[p]_\al-[p]_\al~V_{0\al}~[p]_\al,\nn\\
&&~~V_{0\al}=U^{-1}_\al (h_0-h_\al)U_\al,~~U_\al=T(e^{\int^th_\al})
\eea
describes tunneling between any given pure sgc class $[p]_\al$ and a reference pure sgc class $[p]_0$ with $V_{0\al}$ being the ``tunneling potential''.
$V_{0\al}=0$~ corresponds to zero tunneling or pure time evolution in the class $[p]_\al$.

Since ~$dQ^{-1}=-Q^{-1}dQ~Q^{-1}$~ the solution to ~${d \T_{12}\over dt}=\T_{12}V_{21}\T_{12}$~ is

\bea
&&\T_{12}=-(\int^tV_{21})^{-1}~\eqv~U_1^{-1}pU_2~~~~\Ra~~\nn\\
&&p=U_1\T_{12}U_2^{-1}=-U_1~(\int^tV_{21})^{-1}~U_2^{-1}=-U_1~{1\over \int^tU_2^{-1}(h_2-h_1)U_1 }~U_2^{-1},\nn\\
&&U_1=T(e^{\int^th_1}),~~~~U_2=T(e^{\int^th_2}).
\eea
In the limit $h_2\ra h_1\eqv h$ one obtains the linear solution
\bea
&&p=Up_0U^{-1},~~p_0=\lim_{h_2\ra h_1\eqv h}{-1\over \int^tU_2^{-1}(h_2-h_1)U_1 }=p_0(h),
\eea
where one may check that~ ${dp_0\over dt}=0$.

Writing $p=U_1p_{12}U^{-1}_1=U_2p_{21}U^{-1}_2$ one identifies the \emph{\textbf{``directed'' tunneling operators}}
\bea
&&p_{12}=-{1\over \int^tU_2^{-1}(h_2-h_1)U_1 }~U_2^{-1}U_1,\nn\\
&&p_{21}=-U^{-1}_2U_1~{1\over \int^tU_2^{-1}(h_2-h_1)U_1 }.
\eea

Represented on a Hilbert space, a particular projective solution of \\
~$\dot{p}=h_1p-ph_2+p(h_2-h_1)p$~ takes the form
\bea
&&p={|\eta\rang\lang\xi|\over \lang\xi|\eta\rang},~~{d|\eta\rang\over dt}=h_1|\eta\rang,~~{d\lang\xi|\over dt}=-\lang\xi|h_2.\nn\\
&&\Tr~p^\ast p={\lang\eta|\eta\rang\lang\xi|\xi\rang\over |\lang\xi|\eta\rang|^2}\eqv {1\over \cos\theta_{\eta\xi}}.
\eea

 When one wishes that $p^\ast$ be described by the same equation as $p$ (which is not necessary if $p^\ast$ describes an independent [anti-]system) we must have \\
 $h_1^\ast=-h_2,~~h_2^\ast=-h_1$.

The possible kinds of dynamics may be classified as follows:
\vspace{.5cm}

\begin{tabular}{|l|l|}
  \hline
  dynamics (time evolution) &  \\
  \hline
  linear~~ $\dot{p}=[h,p]$ & projective (pure) $p^2=p$ \\
   & nonprojective (impure) $p^2\neq p$ \\
   \hline
   nonlinear  & dynamically projective (pure/impure) \\
    ~~~~$\dot{p}=h_1p-ph_2+p(h_2-h_1)p$ & $\dot{p}p+p\dot{p}=\dot{p}$ \\
    & dynamically nonprojective (impure) \\
    & $\dot{p}p+p\dot{p}\neq\dot{p}$ \\
  \hline
\end{tabular}
\vspace{.2cm}

where one notes that the projective linear dynamics is always dynamically projective, and also that the linear nonprojective dynamics can either be dynamically projective or dynamically nonprojective.

One may also regard \emph{\textbf{interactions}} within/without a given physical system as \emph{\textbf{some kind of tunneling}} where $\delta h=h-h_0=h_{I}$ is the interaction Hamiltonian. However one should emphasize that this is only a particular case which can exhaust neither the applicability of the nonlinear tunneling equation \\
~$\dot{p}=h_1p-ph_2+p(h_2-h_1)p$~ nor that of any possible generalizations of the equation.

If one defines a \emph{\textbf{finite evolution process}} as the (time) ordered (tensor) product \\
~$Q_{if}[p]=\prod_{t=t_i}^{t_f} \ot p(t)$~~where one can also multiply/add processes to obtain new ones,~ then the amplitude $A^\phi$ of involvement or participation of a particular functional $\phi\in \A^\ast$ in the process is
\bea
&&A^\phi_{if}=\Delta(\phi)(Q_{if}[p])=\Delta(\phi)(\prod_{t=t_i}^{t_f}\ot p(t)),
\eea
where $p(t)$ may be interpreted as an \emph{\textbf{instantaneous evolution process}} an \emph{\textbf{infinite evolution process}} will involve an infinite time interval. The overall process amplitude is
\bea
A_{if}=\sum_{\phi}A^\phi_{if}=\sum_{\phi}\Delta(\phi)(\prod_{t=t_i}^{t_f} \ot p(t)).
\eea
An \emph{\textbf{example}} of a physical process amplitude (\emph{\textbf{in coordinate representation}}) is the \emph{\textbf{path integral}} in quantum theory.

One notes that an evolution process may involve the switching on and off of interactions in specific time intervals $[t_r,t_s]$: eg. in the case of linear time evolution one may have
\bea
&&h(t)=h_0(t)+\sum_{rs}\theta(t-t_r)\theta(t_s-t)~V_{rs}(t)\eqv h_0(t)+h_I(t).
\eea

Process classes can be named according to the class of dynamics that determines $p(t)~~\forall t$.

One may  also relate $h_1$ and $h_2$ by imposing either the conservation of $h_1$
\bea
&&\dot{h}_1=h_1h_1-h_1h_2+h_1(h_2-h_1)h_1=0\nn\\
&&~~\Ra~~h_1=h_2~~\txt{or}~~h_1=1
\eea
 or the conservation of $h_2$
 \bea
&&\dot{h}_2=h_1h_2-h_2h_2+h_2(h_2-h_1)h_2=0\nn\\
&&~~\Ra~~h_1=h_2~~\txt{or}~~h_2=1.
\eea

One can similarly derive an evolution equation for a system of sgc's satisfying $p_ip_j=f_{ij}{}^kp_k$.
\bea
&&p_ip_j=f_{ij}{}^kp_k~~\Ra~~\dot{p}_ip_j+p_i\dot{p}_j=f_{ij}{}^k\dot{p}_k,
\eea
Then linear time evolution
\bea
&&\dot{p}_i=hp_i-p_ih
\eea
needs no modification. However, nonlinear evolution will be in the form
\bea
\dot{p}_i=h_1p_i-p_ih_2+c_i{}^{jk}~p_j(h_2-h_1)p_k
\eea
where the $c$'s obey some contraction identities with the $f$'s.

One can have more general nonlinear time evolutions (which would describe tunneling from a given vacuum into more than one different vacua simultaneously) as for example:
\bea
&&\dot{p}=h_1p+ph_2+ph_3p+h_4ph_5p+ph_6ph_7,\\
&&\dot{p}p+p\dot{p}=\dot{p}~~~~\Ra\nn\\
&&h_1+h_2+h_3=0,~h_4=h_6,~h_5+h_7=0\nn\\
&&\txt{or}~~~~h_1+h_2+h_3=0,~h_4+h_6=0,~h_5=h_7\nn
\eea
and
\bea
&&\dot{p}=h_1p+ph_2+ph_3p+h_4ph_5p+ph_6ph_7+ph_8ph_9p,\\
&&\dot{p}p+p\dot{p}=\dot{p}~~~~\Ra\nn\\
&&h_1+h_2+h_3=0,~h_4=h_6=h_8,~h_5+h_7+h_9=0\nn\\
&&\txt{or}~~~~h_1+h_2+h_3=0,~h_4+h_6+h_8=0,~h_5=h_7=h_9\nn
\eea
and so on.

\subsection{Coordinate types}
We will consider only linear time evolution.

The ``mechanical'' choice of coordinates, \\~$q:\mathbb{R}\ra \A^N,~t\mapsto q(t)=\{q^i(t);~i=1,2,...,N\}\eqv (q^1(t),q^2(t),...,q^{N}(t))$~~\\
or ~~$q:\mathbb{N}_N\times \mathbb{R}\ra \A,~(i,t)\mapsto q^i(t),~\mathbb{N}_N=\{1,2,...,N\}$,~~ which we made earlier is of course only for illustration.
In principle both the number and choices of coordinates (ie. observers), and hence of the corresponding symmetries, is arbitrarily diverse. The following are a few other examples of coordinate choices:

\begin{itemize}
\item Scalar fields
\bea
&&q:\mathbb{R}^{d}\times \mathbb{R}\eqv \mathbb{R}^{d+1}\ra \A,~x=(t,\vec{x})\mapsto q^{\vec{x}}(t)\eqv q(t,\vec{x}).\nn\\
&&{\del q(t,\vec{x})\over \del t}=-i[H(t,q),q(t,\vec{x})]=-iD_Hq(t,\vec{x}).
\eea
\item Vector fields
\bea
&&q:\mathbb{N}_{d+1}\times\mathbb{R}^{d+1}\ra \A,~(\mu,x)\mapsto q^\mu(t,\vec{x}).\nn\\
&&{\del q^\mu(t,\vec{x})\over \del t}=-i[H(t,q),q^\mu(t,\vec{x})]=-iD_Hq^\mu(t,\vec{x}).
\eea
\item $p$-Tensor fields
\bea
&&q:(\mathbb{N}_{d+1})^p\times\mathbb{R}^{d+1}\ra \A,~(\al,x)\mapsto q^\al(t,\vec{x}).\nn\\
&&{\del q^\al(t,\vec{x})\over \del t}=-i[H(t,q),q^\al(t,\vec{x})]=-iD_Hq^\al(t,\vec{x}).
\eea
\item Spinor fields
\bea
&&q:\mathbb{N}_{2^{d\over 2}}\times\mathbb{R}^{d+1}\ra \A,~(\sigma,x)\mapsto q^\sigma(t,\vec{x}).\nn\\
&&{\del q^\sigma(t,\vec{x})\over \del t}=-i[H(t,q),q^\sigma(t,\vec{x})]=-i(D_H)^\sigma{}_{\sigma'}~ q^{\sigma'}(t,\vec{x}),\nn\\
&&~~(D_H)^\sigma{}_{\sigma'}=(\gamma^\mu)^\sigma{}_{\sigma'}~D_{H^\mu}=D_{H^\mu}(\gamma^\mu)^\sigma{}_{\sigma'} ,~~\gamma^\mu\gamma^\nu+\gamma^\nu\gamma^\mu=2g^{\mu\nu},\nn\\
&&\gamma^\mu=\gamma^\mu(t,\vec{x}),~~g^{\mu\nu}=g^{\mu\nu}(t,\vec{x}).
\eea

\item $r$-Gauge fields
\bea
&&q:(\mathbb{N}_{d+1})^m\times(\mathbb{N}_{2^{d\over 2}})^n\times (\mathbb{N}_N)^r\times\mathbb{R}^{d+1}\ra \A,~(u,x)\mapsto q^u(t,\vec{x}).\nn\\
&&{\del q^u(t,\vec{x})\over \del t}=-i[H(t,q),q^u(t,\vec{x})],~~u=(\mu_1,...,\mu_m,\sigma_1,...,\sigma_n,a_1,...,a_r).\nn
\eea

\item Composite coordinates:~ In general $q=(q_1,q_2,...)$ can be made up of one or more of the coordinate systems above meaning $H$ depends on the whole composite as well; \\
    $H=H(t,q)=H(t,q_1,q_2,...)$.
\end{itemize}

The (more basic) coordinate types $q$ above are thought to correspond to irreducible representations of a symmetry group whose action may be expressed in a coordinate dependent way as
\bea
&&q'{}^{u}(x')=U^{-1}(\Ld,b)q^u(x)U(\Ld,b)=S^u{}_v(\Ld,b)~q^{v}(\Ld x+b),\nn\\
&&~~b\in \mathbb{R}^{d+1},~\Ld\in \mathbb{R}^{d+1}\ot \mathbb{R}^{d+1},\nn
\eea
or in a coordinate independent way as
\bea
U(\Ld,b)U(\Ld',b')=U(\Ld\Ld',\Ld b'+b).
\eea
This is also the isometry group of $\mathbb{R}^{d+1}$ as a metric space
\bea
&&\H(\mathbb{R}^{d+1})=(\txt{Der}(\mathbb{R}^{d+1}),\lang,\rang=\mu_\mathbb{C}\circ\eta),~~\eta\in\txt{Der}^\ast(\mathbb{R}^{d+1})\ot \txt{Der}^\ast(\mathbb{R}^{d+1}),\nn\\
&& \lang U(\Ld,b)\xi|U(\Ld,b)\xi_1\rang=\lang\xi|\xi_1\rang~~\forall \xi,\xi_1\in \H(\mathbb{R}^{d+1}),\nn\\
&&\txt{Der}(\mathbb{R}^{d+1})=\{D:\F(\mathbb{C},\mathbb{R}^{d+1})\ra \F(\mathbb{C},\mathbb{R}^{d+1}),~D(f+h)=D(f)+D(h),\nn\\
&&~~~~~D(fh)=D(f)~h+f~D(h)~~\forall f,h\in \F(\mathbb{C},\mathbb{R}^{d+1})\}\nn\\
&&~~~~\eqv \{D_v;~v\in \F(\mathbb{C},\mathbb{R}^{d+1})^{d+1}\},~~(D_v(f))(x)=v^i(x)\del_if(x),\nn\\
&&(\lang D_u|D_v\rang)(x)=\eta_{ij}~u^i(x)v^j(x),\nn\\
&&(\lang U(\Ld,b)D_u|U(\Ld,b)D_v\rang)(x)=\eta_{ij}~u^i(\Ld x+b)v^j(\Ld x+b)=\eta_{ij}~u^i(x)v^j(x)\nn\\
&&~~~~\Ra~~u^i(x)=v^i(x)=dx^i~(~\Ra~~\txt{Der}(\mathbb{R}^{d+1})\simeq \mathbb{R}^{d+1}~) ,~~\eta_{\al\beta}\Ld^\al{}_i\Ld^\beta{}_j=\eta_{ij},\nn
\eea
where $\txt{Der}(\mathbb{R}^{d+1})$ is the space of all directional derivatives in $\mathbb{R}^{d+1}$.

Dynamically (ie. $p(t)=P(t,q)$),~ $p$ is determined by $H=H(t,q)$ and therefore the possible types of dynamics (including interactions) of various physical systems are described, by observers, by specifying various functional forms of $H(t,q)$.

\subsection{On Gravity}
One may want to ``enlarge'' the isometry group of $\mathbb{R}^{d+1}$ to that of an  $\mathbb{R}^{d+1}$-manifold $\M(\mathbb{R}^{d+1})$ in order to treat gravity which is believed to be related to the metric/curvature of some $\mathbb{R}^{d+1}$-manifold.
That is, gravity is related to the isometry group \\
(~$U(\vphi)U(\vphi')=U(\vphi\circ\vphi'),~\vphi,\vphi':\D\subseteq\M\ra \C\subseteq\M$~) of $\M=\M(\mathbb{R}^{d+1})$ with its tangent fiber as a metric space
\bea
&&\H({\M})=(\txt{Der}({\M}),\lang,\rang=\mu_\mathbb{C}\circ{g}),~~{g}\in\txt{Der}^\ast({\M})\ot \txt{Der}^\ast({\M}),\nn\\
&& \lang U(\vphi)\xi|U(\vphi)\xi_1\rang=\lang\xi|\xi_1\rang~~\forall \xi,\xi_1\in \H({\M}),\nn\\
&&\txt{Der}({\M})=\{D:\F(\mathbb{C},{\M})\ra \F(\mathbb{C},{\M}),~D(f+h)=D(f)+D(h),\nn\\
&&~~~~~D(fh)=D(f)~h+f~D(h)~~\forall f,h\in \F(\mathbb{C},{\M})\}\nn\\
&&~~~~\eqv \{D_v;~v\in \F(\mathbb{C},{\M})^{d+1}\},~~(D_v(f))(x)=v^i(x)\del_if(x),\nn\\
&&(\lang D_u|D_v\rang)(x)={g}_{ij}(x)~u^i(x)v^j(x)~~\forall u,v,\nn\\
&&(\lang U(\vphi)D_u|U(\vphi)D_v\rang)(x)={g}_{ij}(\vphi^{-1}(x))~u^i(\vphi(x))v^j(\vphi(x))={g}_{ij}(x)~u^i(x)v^j(x)\nn\\
&&~~~~=\lang D_u|D_v\rang)(x),
\eea
where $\txt{Der}({\M})$ is the space of all directional derivatives on ${\M}$. One can apply the functional operator ~$Q_{\al\beta}=\int d\mu(x)d\mu(z)~{\delta\over\delta u^\al(y)}{\delta\over\delta u^\beta(z)},~d\mu(x)=\sqrt{\det g(x)}d^{d+1}x$~  on the equation ~${g}_{ij}(\vphi^{-1}(x))~u^i(\vphi(x))v^j(\vphi(x))={g}_{ij}(x)~u^i(x)v^j(x)~~\forall u,v$~ to remove the $u,v$ dependence (Check the infinitesimal form ~$\vphi^i(x)=x^i+\delta x^i(x)\eqv x^i+\xi^i(x)$). There is the constraint
\bea
g_{ij}(\vphi^{-1}(x))~\det g(\vphi(x))=g_{ij}(x)~\det g(x)~~\Ra~~ \det g(\vphi^{-1}(x))= \det g(x).\nn
\eea
Once the metric $g$ has been determined (eg. by postulating the matter energy momentum tensor as the source of the curvature $R$ generated by $g$, or by some other means) then the transformations $\vphi$, and hence the irreducible representations of \\ ~$U(\vphi)U(\vphi')=U(\vphi\circ\vphi')=U(\vphi\circ \vphi'\circ\vphi^{-1}\circ\vphi'{}^{-1})~U(\vphi')U(\vphi)$~  can then be determined as well.

However this ``enlargement'' effect can also be realized in different ways: by dimensional increase/reduction, coordinate spectrum increase/decrease (eg. by making $x^i$ noncommutative), etc. Then gravity can arise as a physical effect induced by dimensional reduction, coordinate spectrum increase, etc. And since the usual (general relativistic) gravity theory is acceptable as an effective theory, any other fundamental theory of gravity needs to be compatible with it.

The \emph{\textbf{time evolution of a gravitational system may involve tunneling between inequivalent physical vacua}} and hence the nonlinear impure time evolution equation (\ref{projector-evol}) may be more suitable for describing a physical gravitational system.

\subsection{Projectors on Self Hilbert Spaces}

\bea
&&\H^\phi=\H^\phi(\A)=(\A,\lang,\rang_\phi=\phi\circ\mu_\A\circ(\ast\ot id))\eqv \{|\xi\rang_\phi;~\xi\in \A\},~~\phi\in \A^\ast,\nn\\
&&~~~~\mu_{\A}\circ(\ast\ot id):\A\ot\A\ra\A,~a\ot b\mapsto a^\ast b.\nn\\
&&\H^{\A^\ast}=\H^{\A^\ast}(\A)=(\A,\lang,\rang_{\A^\ast}={\A^\ast}\circ\mu_\A\circ(\ast\ot id))\simeq \A\times\A^\ast.\nn
\eea
One can have
multiplication operator representations:
\bea
&&{m}^L:\A\ra O(\H^\phi(\A)),~a\ra {m}^L_a:\H^\phi(\A)\ra \H^\phi(\A),~|\xi\rang\mapsto |a\xi\rang.\nn\\
&&{m}^R:\A\ra O(\H^\phi(\A)),~a\ra {m}^R_a:\H^\phi(\A)\ra \H^\phi(\A),~|\xi\rang\mapsto |\xi a\rang.\nn
\eea
and/or matrix representations:
\bea
&&\pi^\phi:\A\ra O(\H^\phi(\A)),~a\mapsto \pi^\phi(a)=\sum_{(\xi,\eta)\in\H_1^\phi\times\H_2^\phi}|\xi_\al\rang a^{\al\beta}\lang\eta_\beta|\nn\\
&&~~~~~~~~~~~~\eqv \sum_{(\xi,\eta)\in\H_1^\phi\times\H_2^\phi}a^{\al\beta}~\xi_\al\ot\eta^\ast_\beta,\nn\\
&&p^\phi=\pi^\phi(p)=\sum_{(\xi,\eta)\in\H_1^\phi\times\H_2^\phi}|\xi_\al\rang\lang\eta_\al|\xi_\beta\rang^{-1}_\phi\lang\eta_\beta|.\nn\\
&&p^{\phi_i}(t_i)=U^{-1}(t_i,t)p^\phi(t)U(t_i,t)=U^{-1}(t_i,t_f)p^{\phi_f}(t_f)U(t_i,t_f).\nn
\eea

For example:
\bea
&&\A=\A_\theta(\mathbb{R}^{D})=\{a_f=W(f)=\sum_{x\in \mathbb{R}^{D}}f(x)\hat{\delta}_x;~f:\mathbb{R}^{D}\ra \mathbb{C}\},\nn\\
&&\A^\ast=\A^\ast_\theta(\mathbb{R}^{D})=\{\phi_x=\Tr\circ {m}_{\hat{\delta}_x};~x\in \mathbb{R}^{D} \},\nn\\
&&\hat{\delta}_x=\sum_{k\in \mathbb{R}^{D}}e^{ik(\hat{x}-x)},~~[\hat{x}^\mu,\hat{x}^\nu]=i\theta^{\mu\nu}.\nn\\
&&\A_\delta=\{W(\delta_y)=\sum_{x\in \mathbb{R}^{D}}\delta_y(x)\hat{\delta}_x=\sum_{x\in \mathbb{R}^{D}}\delta(y-x)\hat{\delta}_x=\hat{\delta}_y;~y\in\mathbb{R}^{D}\}\subset \A,\nn\\
&&\A_e=\{W(e_k)=\sum_{x\in \mathbb{R}^{D}}e_k(x)\hat{\delta}_x=\sum_{x\in \mathbb{R}^{D}}e^{ikx}\hat{\delta}_x=e^{ik\hat{x}};~k\in\mathbb{R}^{D}\}\subset \A.\nn\\
&&\H^{\phi_u}(\A)=(\A,\lang,\rang_{\phi_u}),~~~~\H^{\phi_u}_{e,\delta}(\A)=(\A_{e,\delta}~,\lang,\rang_{\phi_u})\subset \H^{\phi_u}(\A).\nn\\
&&\lang,\rang_{\phi_u}=\phi_u\circ\mu_\A\circ(\ast\ot id)=\Tr\circ{m}_{\hat{\delta}_u}\circ\mu_\A\circ(\ast\ot id).
\eea

\section{Primitivity: The logic of human society}
The logic can be exclusive, nonexclusive or both.

The analysis in the previous section (Physics: The logic of quantum theory)  is a reflection of the primitivity or science of human society. The algebra $\A$ is the collection of all possible human emotions (the language of \emph{Eternity} or \emph{Greed}, known otherwise as \emph{God}). The number field, such as the field of complex numbers $\mathbb{C}$, in which the linear functionals $\phi\in \A^\ast$ take values is the set of all possible \emph{Gold} (or money) amplitudes or potentials. Here $\phi\in \A^\ast$ represents an individual being and $\phi(a)$ is the gold amplitude of $\phi$ to the primitive system represented by the emotion $a\in \A$. A high gold amplitude is supposedly a blessing from Eternity meanwhile a low gold amplitude would mean Eternity's disapproval.

At any given time, the \emph{emotional presence or state} $S_a$ of a primitive system in $\A^\ast$ is determined (or generated) by the \emph{creation or existence emotion} $a\in \A$ of the primitive system. That is \emph{\textbf{a primitive system, with emotional presence or state $S_a$ in $\A^\ast$, is defined (by a community of primitive observers [explicitly or implicitly prophets/messengers of Eternity]) by specifying a creation or existence emotion $a\in\A$ for the primitive system}}.

We will consider the set generating projectors $p\in\A$ to represent creation or existence emotions of actual primitive systems living or operating in the space $\A^\ast$ and each closed collection of projectors $\P$ will represent a collection of basic or elementary primitive systems [eps's] (where the systems are basic or elementary in that the product of any two of them gives another).  The set $S_p$, or equivalently $\phi(p),~~\forall \phi\in \A^\ast$, determines the amplitude distribution (or configuration), at a given time, of the elementary primitive system (eps) represented by $p\in\A$. That is $S_a$ is interpreted as the (probability) amplitude distribution or configuration in $\A^\ast$ of the system represented by $a\in\A$.

The dynamics of a primitive system may be described in parallel to the previous section with the following replacements:\\
physics$\ra$primitivity,~ physical$\ra$primitive,~ condition$\ra$emotion,\\
observer$\ra$prophet/messenger of Eternity,~ and so on.

In the dynamics of primitive systems there can be interactions (a case of tunneling), involving one or more primitive systems. During an interaction process the prophets or messengers of Eternity (the observers) make various readjustments or redefinitions (known as ``offerings or sacrifices'' willed by Eternity) of the primitive systems. Thus an interaction may result in the conversion (as decided by Eternity through the observers) of some of the initial primitive systems involved into other primitive systems which were not involved initially.

\addcontentsline{toc}{part}{Bibliography}

\begin{thebibliography}{999999}

\bibitem{szabo2} R. J. Szabo. {\it Quantum Gravity, Field theory and Signature of Non-commutativity}. (arXiv:0906.2913v2 [hep-th], Jul 09 2009)

\bibitem{wigner} E. Wigner, {\it On Unitary Representations of  the Inhomogeneous Lorentz Group}, Ann. Math. 40 149 (Jan 1939).

\bibitem{nair} V. P. Nair, {\it Quantum Field Theory A Modern Perspective}, Springer [ISBN 0-387-21386-4]
\bibitem{masson} T. Masson, {\it An informal introduction to the ideas and concepts of noncommutative geometry}. [arXiv:math-ph/0612012v3 15 Dec 2006]
\bibitem{myers} Robert C. Myers. {\it Dielectric Branes}. arXiv:hep-th/9910053 (1999).
\bibitem{BBS} Katrin Becker, Melanie Becker, John H. Schwarz: String Theory and M-Theory, A Modern Introduction. (sect. 6.5). Cambridge University Press.

\bibitem{madore} J. Madore, {\it An introduction to noncommutative differential geometry and its physical applications} 2nd ed, Cambridge, Cambridge Univ. Press, (1999); J. Madore, {\it Noncommutative geometry for pedestrians}, [arXiv:gr-qc/9906059].

\bibitem{fuzzybook} A. P. Balachandran, S. Vaidya and S. Kurkcuoglu, {\it Lectures on fuzzy and fuzzy SUSY physics}, World Scientific (2007).


\bibitem{doplicher} S. Doplicher, K. Fredenhagen and J. E. Roberts, {\it Spacetime quantization induced by classical gravity}, Phys. Lett. {\bf B 331}, 33-44 (1994).

\bibitem{thooft} G. `t Hooft, {\it Quantization of point particles in (2+1)-dimensional gravity and spacetime discreteness}, Class. and Quant. Grav. {\bf 13} (1996) 1023.

\bibitem{qhe} A. P. Balachandran, Kumar S. Gupta and Seckin Kurkcuoglu, {\it Interacting quantum topologies and the quantum Hall effect}, [arXiv:0708.0069].

\bibitem{Snyder} H. S. Snyder, {\it Quantized space-time} Phys. Rev. {\bf 71} (1947) 38. {\it The Electromagnetic field in Quantized Space-Time}, Phys. Rev. {\bf 72} (1947) 68.

\bibitem{yang} C. N. Yang, {\it On quantized space-time}, Phys. Rev. {\bf 72} (1947) 874.

\bibitem{connesA} A. Connes, {\it Non-commutative differential geometry}, Publ. Math. l'IHES 62, pp. 41--144 (1985).

\bibitem{woronowicz} S. L. Woronowicz, {\it Twisted SU(2) group, an example of a non-commutative differential calculus}, Publ. RIMS, Kyoto Univ. {\bf 23} (1987) 399.

\bibitem{ConnesC} A. Connes, {\it Noncommutative geometry}, Academic Press (1994).

\bibitem{Madore} J. Madore, {\it An introduction to noncommutative geometry and its physical applications}, Cambridge University Press (1999).

\bibitem{Landi} G. Landi, {\it An introduction to noncommutative spaces and their geometries}, Springer Verlag (1997).

\bibitem{Bondia} J. M. Gracia-Bond\'ia, J. C. V\'arilly and H. Figueora, {\it Elements of noncommutative geometry}, Birkh\"{a}user (2001).

\bibitem{RGCai} R. G. Cai and N. Ohta, {\it Lorentz transformation and light-like noncommutative SYM}, JHEP 10:036 (2000), [arXiv: hep-th/0008119].

\bibitem{Ydri} B. Ydri, {\it Fuzzy physics} (2001), [arXiv: hep-th/0110006].

\bibitem{queiroz} A. P. Balachandran, A. R. Queiroz, A. M. Marques and P. Teotonio-Sobrinho, {\it Deformed Kac-Moody and Virasoro algebras}, J. Phys. A: Math. Theor. 40 (2007) 7789-7801, [arXiv:hep-th/0608081].

\bibitem{BPQ} A. P. Balachandran, A. Pinzul, A. R. Queiroz, Twisted Poincare Invariance, Noncommutative Gauge Theories and UV-IR Mixing , Phys.Lett.B668:241-245,2008 [arXiv:0804.3588].

\bibitem{lipkin} H. J. Lipkin, {\it Lie groups for pedestrians}, Dover Publications, 2002.

\bibitem{weyl} H. Weyl, {\it Gruppentheorie und quantenmechanik: The theory of groups and quantum mechanics}, New York, Dover Publications, 1950; H. Weyl, {\it Quantum mechanics and group theory}, Z. Phys. 46, 1 (1927).

\bibitem{grosse-pres} H. Grosse and P. Presnajder, {\it The Construction on non-commutative manifolds using coherent states}, Lett. Math. Phys. 28, 239 (1993).

\bibitem{Haag} R. Haag, {\it Local quantum physics : fields, particles, algebras}, Berlin, Springer-Verlag (1996).

\bibitem{Kontsevich} M. Kontsevich, {\it Deformation quantization of Poisson manifolds, I}, Lett. Math. Phys. 66, 157 (2003), [arXiv:q-alg/9709040].

\bibitem{chaichian} M. Chaichian, P. P. Kulish, K. Nishijima and A. Tureanu, {\it On a Lorentz-invariant interpretation of noncommutative space-time and its implications on noncommutative QFT}, Phys. Lett. {\bf B 604}, 98 (2004), [arXiv:hep-th/0408069]; M. Chaichian, P. Presnajder and A. Tureanu, {\it New concept of relativistic invariance in NC space-time: Twisted Poincar\'e symmetry and its implications}, Phys. Rev. Lett. 94, 151602 (2005), [arXiv:hep-th/0409096].

\bibitem{mack1} G. Mack and V. Schomerus, {\it Quasi Hopf quantum symmetry in quantum theory}, Nucl. Phys. {\bf B 370}, 185 (1992).

\bibitem{mack2} G. Mack, V. Schomerus, {\it Quantum symmetry for pedestrians}, preprint DESY - 92 - 053, March 1992; G. Mack and V. Schomerus, J. Geom. Phys. {\bf 11}, 361 (1993).

\bibitem{drinfeld} V. G. Drinfeld, {\it Quasi-Hopf algebras}, Leningrad Math. J.
{\bf 1} (1990), 1419-1457.

\bibitem{majid} S. Majid, {\it Foundations of quantum group theory}, Cambridge
University Press, 1995.

\bibitem{fiore2} G.~Fiore and P.~Schupp, {\it Identical particles and quantum symmetries}, Nucl.\ Phys.\ B {\bf 470}, 211 (1996), [arXiv:hep-th/9508047].

\bibitem{fiore1} G. Fiore and P. Schupp, {\it Statistics and quantum group symmetries}, [arXiv:hep-th/9605133]. Published in {\em Quantum groups and quantum spaces, Banach Center Publications vol. 40, Inst. of Mathematics, Polish Academy of Sciences, Warszawa} (1997), P. Budzyski, W. Pusz, S. Zakrweski Editors, 369-377.

\bibitem{fioresolo1} G.~Fiore, {\it Deforming maps and Lie group covariant creation and annihilation operators}, J.\ Math.\ Phys.\  {\bf 39}, 3437 (1998), [arXiv:q-alg/9610005].

\bibitem{fioresolo2} G.~Fiore, {\it On Bose-Fermi statistics, quantum group symmetry, and second quantization}, [arXiv:hep-th/9611144].

\bibitem{watts1} P.~Watts, {\it Noncommutative string theory, the R-Matrix, and Hopf algebras}, Phys. Lett. {\bf B 474}, 295--302 (2000), [arXiv:hep-th/9911026].

\bibitem{Oeckl:2000eg} R. Oeckl, {\it Twisting noncommutative ${\mathbb R}^{d}$ and the equivalence of quantum field theories}, Nucl.\ Phys.\ B {\bf 581}, 559 (2000), [arXiv:hep-th/0003018].

\bibitem{watts2} P.~Watts, {\it Derivatives and the role of the Drinfel'd twist in noncommutative string theory}, [arXiv:hep-th/0003234].

\bibitem{gms} H.~Grosse, J.~Madore and H.~Steinacker, {\it Field theory on the q-deformed fuzzy sphere II: Quantization}, J.\ Geom.\ Phys.\  {\bf 43}, 205 (2002), [arXiv:hep-th/0103164].

\bibitem{Dimitrijevic:2004rf} M. Dimitrijevic and J. Wess, {\it Deformed bialgebra of diffeomorphisms}, [arXiv:hep-th/0411224].

\bibitem{matlock} P.~Matlock, {\it Non-commutative geometry and twisted conformal symmetry}, Phys.\ Rev.\ D {\bf 71}, 126007 (2005), [arXiv:hep-th/0504084].


\bibitem{bal-unitary} A. P. Balachandran, T. R. Govindarajan, C. Molina and P.
Teotonio-Sobrinho, {\it Unitary quantum physics with time-space noncommutativity},
JHEP 0410 (2004) 072, [arXiv:hep-th/0406125].

\bibitem{bal} A. P. Balachandran, A. Pinzul and S. Vaidya, {\it  Spin and statistics
on the Groenewold-Moyal plane: Pauli-forbidden levels and transitions}, Int. J. Mod.
Phys. A {\bf 21} 3111 (2006), [arXiv:hep-th/0508002].

\bibitem{uv-ir} A. P. Balachandran, A. Pinzul and B. A. Qureshi, {\it UV-IR Mixing
in noncommutative plane}, Phys. Lett. {\bf B 634} 434 (2006), [arXiv:hep-th/0508151].

\bibitem{bal-sasha-babar} A. P. Balachandran, B. A. Quereshi, A. Pinzul and S. Vaidya, {\it Poincar\'e invariant gauge and gravity theories on Groenewold-Moyal plane}, [arXiv:hep-th/0608138]; A. P. Balachandran, A. Pinzul, B. A. Quereshi and S. Vaidya, {\it Twisted gauge and gravity theories on the Groenewold-Moyal plane}, [arXiv:0708.0069 [hep-th]]; A. P. Balachandran, A. Pinzul, B. A. Quereshi and S. Vaidya, {\it S-matrix on the Moyal plane: Locality versus Lorentz invariance}, [arXiv:0708.1379 [hep-th]]; A. P. Balachandran, A. Pinzul and B. A. Quereshi, {\it Twisted Poincar\'e Invariant Quantum Field Theories}, [arXiv:0708.1779 [hep-th]].

\bibitem{bal-stat} A. P. Balachandran, T. R. Govindarajan, G. Mangano, A. Pinzul, B.
A. Qureshi and S. Vaidya, {\it Statistics and UV-IR mixing with twisted Poincar\'e
invariance}, Phys. Rev. D 75 (2007) 045009, [arXiv:hep-th/0608179].

\bibitem{cpt-paper} E. Akofor, A. P. Balachandran, S. G. Jo and A. Joseph, {\it Quantum fields on the Groenewold-Moyal plane: C, P, T and CPT}, JHEP 08 (2007) 045, [arXiv:0706.1259 [hep-th]].

\bibitem{twistd} A. P. Balachandran, A. Pinzul and B. A. Qureshi, {\it Twisted Poincar\'e invariant quantum field theories}, [arXiv:0708.1779 [hep-th]].

\bibitem{Grosse} H. Grosse, {\it On the construction of M\"oller operators for the nonlinear Schr\"odinger equation}, Phys. Lett. {\bf B 86}, 267 (1979).

\bibitem{Faddeev-Zamolodchikov} A. B. Zamolodchikov and Al. B. Zamolodchikov, Ann. Phys. 120, 253 (1979); L. D. Faddeev, Sov. Sci. Rev. 0 1 (1980) 107.

\bibitem{queiroz1} A. P. Balachandran, A. R. Queiroz, A. M. Marques and P. Teotonio-Sobrinho, {\it Quantum fields with noncommutative target spaces}, [arXiv:0706.0021 [hep-th]]. L. Barosi, F. A. Brito and A. R. Queiroz, {\it Noncommutative field gas driven inflation}, JCAP 04 (2008) 005, [arXiv:0801.0810 [hep-th]].

\bibitem{cmb} E. Akofor, A. P. Balachandran, S. G. Jo, A. Joseph and B. A. Qureshi, {\it Direction-dependent CMB power spectrum and statistical anisotropy from noncommutative geometry}, [arXiv:0710.5897 [astro-ph]].

\bibitem{Carmona} J. M. Carmona, J. L. Cortes, J. Gamboa and F. Mendez, {\it Noncommutativity in field space and Lorentz invariance violation}, Phys. Lett. B565 (2003) 222-228, [arXiv:hep-th/0207158].

\bibitem{Carmona1} J. M. Carmona, J. L. Cortes, J. Gamboa and F. Mendez, {\it Quantum theory of noncommutative fields}, JHEP 0303 (2003) 058, [arXiv:hep-th/0301248].

\bibitem{Bal-Queiroz} A. P. Balachandran and A. R. Queiroz, In preparation.

\bibitem{sk} Y. Suzuki {\it et al.}  [SuperKamiokande collaboration], Phys. Lett. {\bf B 311}, 357 (1993).

\bibitem{borexino} H. O. Back {\it et al.}  [Borexino collaboration], {\it New experimental limits on violations of the Pauli exclusion principle obtained with the Borexino counting test facility},
Eur.\ Phys.\ J.\ C {\bf 37}, 421 (2004), [arXiv:hep-ph/0406252].

\bibitem{Gianpiero} A. P. Balachandran, G. Mangano and B. A. Quereshi, In preparation.

\bibitem{pathria} R. K. Pathria, {\it Statistical mechanics}, Butterworth-Heinemann Publishing Ltd (1996), 2nd edition.

\bibitem{Uhlenbeck} G. E. Uhlenbeck and L. Gropper, {\it The equation of state of a non-ideal Einstein-Bose or Fermi-Dirac gas}, Phys Rev {\bf 41}, 79 (1932).

\bibitem{correlation} B. Chakraborty, S. Gangopadhyay, A. G. Hazra and F. G. Scholtz, {\it Twisted Galilean symmetry and the Pauli principle at low energies}, J. Phys. A 39 (2006) 9557-9572, [arXiv:hep-th/0601121].

\bibitem{Aschieri} P. Aschieri, C. Blohmann, M. Dimitrijevic, F. Meyer, P. Schupp and J. Wess, {\it A gravity theory on noncommutative spaces}, Class. Quant. Grav. 22, 3511 (2005), [arXiv:hep-th/0504183]; P. Aschieri, M. Dimitrijevic, F. Meyer, S. Schraml and J. Wess, {\it Twisted gauge theories}, Lett. Math. Phys. {\bf 78} 61 (2006), [arXiv:hep-th/0603024].


\bibitem{Bogoliubov} N. N. Bogoliubov and D. V. Shirkov, {\it Introduction to the theory of quantized fields}, Interscience Publishers, New York (1959).

\bibitem{Weinberg1} S. Weinberg, {\it Feynman rules for any spin}, Phys. Rev. {\bf 133}, B1318 (1964).

\bibitem{Weinberg2} S. Weinberg, {\it The quantum theory of fields, Vol 1: Foundations}, Cambridge University Press, Cambridge (1995).

\bibitem{pct} {\it PCT, spin and statistics, and all that}, R. F. Streater and A. S.
Wightman, Benjamin/Cummings, 1964; O. W. Greenberg, {\it Why is CPT fundamental?},
[arXiv:hep-ph/0309309].

\bibitem{cpt} G. Luders, Dansk. Mat. Fys. Medd. 28 (1954) 5; W. Pauli, {\it
Niels Bohr and the development of physics}, W. Pauli (ed.), Pergamon Press, New York, 1955.

\bibitem{dimitrijevic} M. Dimitrijevic and J. Wess, {\it Deformed bialgebra of diffeomorphisms}, [arXiv:hep-th/0411224].

\bibitem{bal-sasha-queiroz} A. P. Balachandran, A. Pinzul and A. R. Queiroz, {\it Twisted Poincare invariance, noncommutative gauge theories and UV-IR mixing}, [arXiv:0804.3588 [hep-th]].

\bibitem{guth} A. H. Guth, {\it Inflationary universe: A possible solution to the horizon and flatness problems}, Phys. Rev. D {\bf 23}, 347 (1981).

\bibitem{Linde} A. D. Linde, {\it A new inflationary universe scenario: a possible solution of the horizon, flatness, homogeneity, isotropy and primordial monopole problems}, Phys. Lett. B {\bf 108}, 389 (1982).

\bibitem{Albrecht} A. Albrecht and P. J. Steinhardt, {\it Cosmology for grand unified theories with radiatively induced symmetry breaking}, Phys. Rev. Lett. {\bf 48}, 1220 (1982).

\bibitem{Greene} C. S. Chu, B. R. Greene and G. Shiu, {\it Remarks on inflation and noncommutative geometry}, Mod. Phys. Lett. A16 2231-2240 (2001), [arXiv: hep-th/0011241].

\bibitem{Lizzi} F. Lizzi, G. Mangano, G. Miele, M. Peloso, {\it Cosmological perturbations and short distance physics from noncommutative geometry}, JHEP 0206 (2002) 049, [arXiv: hep-th/0203099].

\bibitem{Brandenberger1} R. Brandenberger and P. M. Ho, {\it Noncommutative spacetime, stringy spacetime uncertainty principle, and density fluctuations}, Phys.Rev. D66 023517 (2002) [arXiv:hep-th/0203119].

\bibitem{Huang}  Q. G. Huang, M. Li, {\it CMB power spectrum from noncommutative spacetime}, JHEP 0306 (2003) 014, [arXiv:hep-th/0304203]; Q. G. Huang, M. Li, {\it Noncommutative inflation and the CMB Multipoles}, JCAP 0311 (2003) 001, [arXiv:astro-ph/0308458].

\bibitem{Brandenberger2} S. Tsujikawa, R. Maartens and R. Brandenberger, {\it Non-commutative inflation and the CMB}, Phys. Lett. B574 141-148 (2003) [arXiv:astro-ph/0308169].


\bibitem{Fatollahi2} A. H. Fatollahi, M. Hajirahimi, {\it Noncommutative black-body radiation: implications on cosmic microwave background}, Europhys. Lett. 75 (2006) 542-547, [arXiv:astro-ph/0607257].

\bibitem{Fatollahi1} A. H. Fatollahi, M. Hajirahimi, {\it Black-body radiation of noncommutative gauge fields}, Phys. Lett. {\bf B} 641 (2006) 381-385, [arXiv:hep-th/0611225].








\bibitem{chaichian2} M. Chaichian and A. Demichev, {\it Introduction to quantum groups}, World Scientific, Singapore (1996).

\bibitem{chari} V. Chari and A. Pressley, {\it A guide to quantum groups}, Cambridge University Press, Cambridge (1994).


\bibitem{bal} A. P. Balachandran, A. Pinzul and S. Vaidya, {\it  Spin and statistics
on the Groenewold-Moyal plane: Pauli-forbidden levels and transitions}, Int. J. Mod.
Phys. A {\bf 21} 3111 (2006) [arXiv:hep-th/0508002].

\bibitem{uv-ir} A. P. Balachandran, A. Pinzul and B. A. Qureshi, {\it UV-IR mixing
in noncommutative plane}, Phys. Lett. {\bf B 634} 434 (2006) [arXiv:hep-th/0508151].

\bibitem{bal-sasha-babar} A. P. Balachandran, B. A. Quereshi, A. Pinzul and S. Vaidya, {\it Poincar\'e invariant gauge and gravity theories on Groenewold-Moyal plane}, [arXiv:hep-th/0608138]; A. P. Balachandran, A. Pinzul, B. A. Quereshi and S. Vaidya, {\it Twisted gauge and gravity theories on the Groenewold-Moyal plane}, [arXiv:0708.0069 [hep-th]]; A. P. Balachandran, A. Pinzul, B. A. Quereshi and S. Vaidya, {\it S-matrix on the Moyal plane: Locality versus lorentz invariance}, [arXiv:0708.1379 [hep-th]]; A. P. Balachandran, A. Pinzul and B. A. Quereshi, {\it Twisted Poincar\'e invariant quantum field theories}, [arXiv:0708.1779 [hep-th]].

\bibitem{bal-statuv-ir} A. P. Balachandran, T. R. Govindarajan, G. Mangano, A. Pinzul, B. A. Qureshi, S. Vaidya, {\it Statistics and UV-IR mixing with twisted Poincare invariance}, Phys. Rev.  D {\bf 75} 045009 (2007), [arXiv:hep-th/0608179].



\bibitem{gauge-gravity} A. P. Balachandran, B. A. Qureshi, A. Pinzul, S. Vaidya, {\it Poincare invariant gauge and gravity theories on the Groenewold-Moyal plane}, [arXiv:hep-th/0608138].



\bibitem{dodelson} S. Dodelson, {\it Modern cosmology}, Academic Press, San Diego (2003).

\bibitem{Mukhanov} V. Mukhanov, {\it Physical foundations of cosmology}, Cambridge University Press, (2005).

\bibitem{mukhanov} V. F. Mukhanov, H. A. Feldman and R. H. Brandenberger, {\it Theory of cosmological perturbations}, Phys. Rept. {\bf 215}, 203 (1992).

\bibitem{Ackerman} L. Ackerman, S. M. Carroll, M. B. Wise, {\it Imprints of a primordial preferred direction on the microwave background}, Phys. Rev. D 75, 083502 (2007), [arXiv:astro-ph/0701357].

\bibitem{Kamionkowski} A. R. Pullen, M. Kamionkowski, {\it Cosmic microwave background statistics for a direction-dependent primordial power spectrum}, [arXiv:0709.1144 [astro-ph]].

\bibitem{Christian} C. Armendariz-Picon, {\it Footprints of Statistical Anisotropies}, JCAP 0603 (2006) 002, [arXiv:astro-ph/0509893].

\bibitem{Mota1} C. G. Boehmer and D. F. Mota, {\it CMB anisotropies and inflation from non-standard spinors}, [arXiv:0710.2003 [astro-ph]].

\bibitem{Mota2} T. Koivisto and D. F. Mota, {\it Accelerating cosmologies with an anisotropic equation of state}, [arXiv:0707.0279 [astro-ph]].

\bibitem{Bal-locality} A. P. Balachandran, A. Pinzul, B. A. Qureshi, S. Vaidya, {\it S-Matrix on the Moyal plane: locality versus Lorentz invariance}, [arXiv:0708.1379 [hep-th]].

\bibitem{numerical} Earnest Akofor, A.P. Balachandran, Anosh Joseph, Larne Pekowsky, Babar A. Qureshi, Constraints from CMB on Spacetime Noncommutativity and Causality Violation,  Phys. Rev.D79: 063004, 2009.

\bibitem{cmbpaper} E. Akofor, A. P. Balachandran, S. G. Jo, A. Joseph and B. A. Qureshi, 
JHEP 05 092  (2008), arXiv:0710.5897 [astro-ph].

\bibitem{Starobinsky79} A. A. Starobinsky, JETP Lett. 30:682-685 (1979), Pisma Zh. Eksp. Teor. Fiz. 30:719-723 (1979).

\bibitem{Starobinsky82} A. A. Starobinsky, Phys. Lett. B117:175-178 (1982).




\bibitem{WMAP1} E. Komatsu, {\it et.al}, ApJS (2008) 
arXiv:0803.0547 [astro-ph].

\bibitem{WMAP2} M. R. Nolta {\it et al.}, ApJS (2008), arXiv:0803.0593 [astro-ph].

\bibitem{WMAP3} J. Dunkley {\it et al.}, ApJS (2008), arXiv:0803.0586 [astro-ph].

\bibitem{ACBAR1} C. L. Reichardt, {\it et al.}, arXiv:0801.1491 [astro-ph].

\bibitem{ACBAR2} C. L. Kuo {\it et al.}, 
Astrophys. J. 664:687-701 (2007), arXiv:astro-ph/0611198.

\bibitem{ACBAR3} C. L. Kuo {\it et al.}, 
Astrophys. J. 600:32-51 (2004), arXiv:astro-ph/0212289.

\bibitem{CBI1} B.S. Mason {\it et al.}, 
Astrophys. J. 591:540-555 (2007), arXiv:astro-ph/0205384.

\bibitem{CBI2} J. L. Sievers {\it et al.}, 
Astrophys. J. 660:976-987 (2007), arXiv:astro-ph/0509203.

\bibitem{CBI3} J. L. Sievers {\it et al.}, 
Astrophys. J. 591: 599-622 (2003), arXiv:astro-ph/0205387.

\bibitem{CBI4} T. J. Pearson, {\it et al.}, 
Astrophys. J. 591:556-574 (2003), arXiv:astro-ph/0205388.

\bibitem{CBI5} A. C. S. Readhead {\it et al.}, 
Astrophys. J. 609:498-512 (2004), arXiv:astro-ph/0402359.

\bibitem{Sachin} A. P. Balachandran, A. Pinzul, B. A. Qureshi and S. Vaidya, Phys. Rev. D77:025020 (2008), arXiv:0708.1379 [hep-th]; A. P. Balachandran, B. A. Qureshi, A. Pinzul and S. Vaidya, arXiv:hep-th/0608138.



\bibitem{Doran} M. Doran, 
JCAP 0510:011 (2005), arXiv:astro-ph/0302138.

\bibitem{Seljak} U. Seljak and M. Zaldarriaga, 
Astrophys. J. 469:437-444 (1996), arXiv:astro-ph/9603033.

\bibitem{brandenberger} R. H. Brandenberger, Lect. Notes Phys. 646:127-167 (2004), arXiv:hep-th/0306071.


\bibitem{cuba} T. Hahn, 
Comput. Phys. Commun. 168:78-95 (2005), arXiv:hep-ph/0404043.

\bibitem{sorkin-sinha} Supurna Sinha, Rafael D. Sorkin. Brownian motion at absolute zero; Physical Review B, {\bf 45}, 8123 (1992); (arXiv:cond-mat/0506196v1).
\bibitem{qft-us} Earnest Akofor, A. P. Balachandran, Anosh Joseph. Quantum Fields on the Groenewold-Moyal Plane (	 arXiv:0803.4351v2 [hep-th]), Int.J.Mod.Phys. A23:1637-1677 (2008).
\bibitem{fradkin}  Eduardo Fradkin. http://webusers.physics.uiuc.edu/~efradkin/phys582/LRT.pdf
\bibitem{trodden-vachaspati} Tanmay Vachaspati and Mark Trodden. Causality and Cosmic Inflation (arXiv:gr-qc/9811037), What is homogeneity of our universe telling us? (arXiv:gr-qc/9905091).
\bibitem{fdt..}  E. Akofor, A.P. Balachandran. Finite Temperature Field Theory on the Moyal Plane. Phys. Rev.D80: 036008, 2009.
\end{thebibliography}

\end{document}